\documentclass[twoside]{cybernets-thesis}
\usepackage[T1]{fontenc}
\usepackage[utf8]{inputenc}
\usepackage{dsfont}
\usepackage{mathtools}
\usepackage{amsthm, amssymb}
\usepackage[binary-units=true]{siunitx}
\DeclareSIUnit{\bits}{bits}
\usepackage{subcaption} 
\usepackage{tikz}
\usepackage[draft]{hyperref}
\usepackage{pgfplots}
\usetikzlibrary{pgfplots.groupplots}
\pgfplotsset{compat=1.17}
\usepgfplotslibrary{fillbetween}
\usepgfplotslibrary{groupplots,dateplot}
\usepackage[toc=False]{glossaries-extra}
\setabbreviationstyle[acronym]{long-short}

\makeglossaries
\usepackage[ruled,vlined]{algorithm2e}
\usepackage{longtable}
\usepackage{hhline}
\usepackage[export]{adjustbox}
\usepackage{indentfirst}
\usepackage{pdfpages}
\usepackage[style=ieee, dashed=false]{biblatex}
\renewcommand{\vec}[1]{\mathbf{#1}}
\newcommand{\vecs}[1]{\boldsymbol{#1}}

\newcommand{\av}{\vec{a}}
\newcommand{\bv}{\vec{b}}
\newcommand{\cv}{\vec{c}}
\newcommand{\dv}{\vec{d}}
\newcommand{\ev}{\vec{e}}

\newcommand{\gv}{\vec{g}}
\newcommand{\hv}{\vec{h}}
\newcommand{\iv}{\vec{i}}

\newcommand{\mv}{\vec{m}}
\newcommand{\nv}{\vec{n}}

\newcommand{\pv}{\vec{p}}
\newcommand{\qv}{\vec{q}}
\newcommand{\rv}{\vec{r}}
\newcommand{\sv}{\vec{s}}
\newcommand{\tv}{\vec{t}}
\newcommand{\uv}{\vec{u}}
\newcommand{\vv}{\vec{v}}
\newcommand{\wv}{\vec{w}}
\newcommand{\xv}{\vec{x}}
\newcommand{\yv}{\vec{y}}
\newcommand{\zv}{\vec{z}}
\newcommand{\zerov}{\vec{0}}

\newcommand{\thetav}{\vecs{\theta}}
\newcommand{\psiv}{\vecs{\psi}}
\newcommand{\kappav}{\vecs{\kappa}}

\newcommand{\Am}{\vec{A}}
\newcommand{\Bm}{\vec{B}}
\newcommand{\Cm}{\vec{C}}
\newcommand{\Dm}{\vec{D}}
\newcommand{\Em}{\vec{E}}
\newcommand{\Fm}{\vec{F}}
\newcommand{\Gm}{\vec{G}}
\newcommand{\Hm}{\vec{H}}
\newcommand{\Id}{\vec{I}}
\newcommand{\Jm}{\vec{J}}
\newcommand{\Km}{\vec{K}}
\newcommand{\Lm}{\vec{L}}
\newcommand{\Mm}{\vec{M}}
\newcommand{\Nm}{\vec{N}}

\newcommand{\Qm}{\vec{Q}}
\newcommand{\Rm}{\vec{R}}
\newcommand{\Sm}{\vec{S}}
\newcommand{\Tm}{\vec{T}}
\newcommand{\Um}{\vec{U}}
\newcommand{\Vm}{\vec{V}}
\newcommand{\Wm}{\vec{W}}
\newcommand{\Xm}{\vec{X}}
\newcommand{\Ym}{\vec{Y}}

\newcommand{\Thetam}{\vecs{\Theta}}
\newcommand{\Sigmam}{\vecs{\Sigma}}
\newcommand{\Omegam}{\vecs{\Omega}}

\newcommand{\Psim}{\vecs{\Psi}}

\newcommand{\Cc}{{\cal C}}

\newcommand{\Hc}{{\cal H}}

\newcommand{\Lc}{{\cal L}}

\newcommand{\Nc}{{\cal N}}

\newcommand{\Pc}{{\cal P}}

\newcommand{\CC}{\mathbb{C}}

\newcommand{\RR}{\mathbb{R}}

\newcommand{\htp}{^{\mathsf{H}}}
\newcommand{\tp}{^{\mathsf{T}}}

\newcommand{\LB}{\left(}
\newcommand{\RB}{\right)}
\newcommand{\LP}{\left\{}
\newcommand{\RP}{\right\}}
\newcommand{\LSB}{\left[}
\newcommand{\RSB}{\right]}

\renewcommand{\ln}[1]{\mathop{\mathrm{ln}}\LB #1\RB}
\renewcommand{\log}[1]{\mathop{\mathrm{log}}\LB #1\RB}
\renewcommand{\exp}[1]{\mathop{\mathrm{exp}}\LB #1\RB}

\newcommand{\EE}{{\mathbb{E}}}

\makeatletter
\newcommand{\doublehat}[1]{%
    \begingroup%
      \let\macc@kerna\z@%
      \let\macc@kernb\z@%
      \let\macc@nucleus\@empty%
      \hat{\raisebox{.35ex}{\vphantom{\ensuremath{#1}}}\smash{\hat{#1}}}%
    \endgroup%
    }
\makeatother

\newcommand{\gpeak}{\gamma_{\text{peak}}}
\newcommand{\bleak}{\beta_{\text{leak}}}
\newacronym{3GPP}{3GPP}{3rd Generation Partnership Project}
\newacronym{ACM}{ACM}{adaptive coding and modulation}
\newacronym{ACLR}{ACLR}{adjacent channel leakage ratio}
\newacronym{ADC}{ADC}{analog-to-digital conversion}
\newacronym{AI}{AI}{artificial intelligence}
\newacronym{AMP}{AMP}{approximate message passing}
\newacronym{AGC}{AGC}{automatic gain control}
\newacronym{AWGN}{AWGN}{additive white Gaussian noise}
\newacronym{BER}{BER}{bit error rate}
\newacronym{BS}{BS}{base station}
\newacronym{BLER}{BLER}{block error rate}
\newacronym{BCE}{BCE}{binary cross-entropy}
\newacronym{BICM}{BICM}{bit-interleaved coded modulation}
\newacronym{BMD}{BMD}{bit-metric decoding}
\newacronym{BP}{BP}{backpropagation}
\newacronym{CE}{CE}{cross-entropy}
\newacronym{CFO}{CFO}{carrier frequency offset}
\newacronym[longplural={convolutional neural networks}]{CNN}{CNN}{convolutional neural network}
\newacronym[longplural={cyclic prefixes}]{CP}{CP}{cyclic prefix}
\newacronym{CCDF}{CCDF}{complementary cumulative distribution function}
\newacronym{CSI}{CSI}{channel state information}
\newacronym{DAC}{DAC}{digital-to-analog conversion}
\newacronym{DPD}{DPD}{digital pre-distortion}
\newacronym{DFT}{DFT}{discrete Fourier transform}
\newacronym{DL}{DL}{deep learning}
\newacronym{ELU}{ELU}{exponential linear unit}
\newacronym{FFT}{FFT}{fast Fourier transform}
\newacronym{FBS}{FBS}{frequency baseband symbol}
\newacronym{GAN}{GAN}{generative adversarial network}
\newacronym{GRU}{GRU}{gated recurrent unit}
\newacronym{iid}{i.i.d.\@}{independent and identically distributed}
\newacronym{IFFT}{IFFT}{inverse fast Fourier transform}
\newacronym{IDFT}{IDFT}{inverse discrete Fourier transform}
\newacronym{ISI}{ISI}{intersymbol interference}
\newacronym{KL}{KL}{Kullback-Leibler}
\newacronym{LLR}{LLR}{log-likelihood ratio}
\newacronym{LSTM}{LSTM}{long short-term memory}
\newacronym{LDPC}{LDPC}{low-density parity-check}
\newacronym{LMMSE}{LMMSE}{linear minimum mean squared error}
\newacronym{MAP}{MAP}{maximum a posteriori}
\newacronym{MAC}{MAC}{medium access control}
\newacronym{MDP}{MDP}{Markov decision process}
\newacronym{ML}{ML}{machine learning}
\newacronym{MLP}{MLP}{multilayer perceptron}
\newacronym{MIMO}{MIMO}{multiple-input multiple-output}
\newacronym{MU-MIMO}{MU-MIMO}{multi-user multiple-input multiple-output}
\newacronym{MU}{MU}{multi-user}
\newacronym{MSE}{MSE}{mean squared error}
\newacronym[longplural={neural networks}]{NN}{NN}{neural network}
\newacronym{NIRE}{NIRE}{nearest interpolated resource element}
\newacronym{NR}{NR}{new radio}
\newacronym{NLOS}{NLOS}{non-line of sight}
\newacronym{OFDM}{OFDM}{orthogonal frequency-division multiplexing}
\newacronym{pdf}{pdf}{probability density function}
\newacronym{pmf}{pmf}{probability mass function}
\newacronym{PA}{PA}{power amplifier}
\newacronym{PAPR}{PAPR}{peak-to-average power ratio}
\newacronym[longplural={power spectral densities}]{PSD}{PSD}{power spectral density}
\newacronym{PRT}{PRT}{peak reduction tone}
\newacronym{QPSK}{QPSK}{quadrature phase-shift keying}
\newacronym[longplural={quadrature amplitude modulations}]{QAM}{QAM}{quadrature amplitude modulation}
\newacronym{PSNR}{PSNR}{Peak Signal to Noise Ratio}
\newacronym{RBF}{RBF}{Rayleigh block-fading}
\newacronym{RB}{RB}{resource block}
\newacronym{RE}{RE}{resource element}
\newacronym{RG}{RG}{resource grid}
\newacronym{ReLU}{ReLU}{rectified linear unit}
\newacronym{RTN}{RTN}{radio transformer network}
\newacronym{RL}{RL}{reinforcement learning}
\newacronym[longplural={recurrent neural networks}]{RNN}{RNN}{recurrent neural network}
\newacronym{SFO}{SFO}{sampling frequency offset}
\newacronym{SER}{SER}{symbol error rate}
\newacronym{SNR}{SNR}{signal-to-noise ratio}
\newacronym{SINR}{SINR}{signal-to-interference-plus-noise ratio}
\newacronym{SGD}{SGD}{stochastic gradient descent}
\newacronym{SISO}{SISO}{single-input single-output}
\newacronym{SIMO}{SIMO}{single-input multiple-output}
\newacronym{SVD}{SVD}{singular value decomposition}
\newacronym{SU}{SU}{single-user}
\newacronym{TDD}{TDD}{time-division duplexing}
\newacronym{TR}{TR}{tone reservation}
\newacronym{UE}{UE}{user equipment}
\newacronym{ULA}{ULA}{uniform linear array}
\newacronym{UMi}{UMi}{urban microcell}
\newacronym{wrt}{w.r.t.\@}{with respect to}

\DeclareFontFamily{OT1}{pzc}{}
\DeclareFontShape{OT1}{pzc}{m}{it}{<-> s * [1.100] pzcmi7t}{}
\DeclareMathAlphabet{\mathscr}{OT1}{pzc}{m}{it}

\newcommand{\shortminus}{\scalebox{0.75}[1.0]{\( - \)}}

\DeclareMathAlphabet{\mathscr}{OT1}{pzc}{m}{it}

\newcommand{\nocontentsline}[3]{}
\newcommand{\tocless}[2]{\bgroup\let\addcontentsline=\nocontentsline#1{#2}\egroup}

\setcapindent{0pt}

\begin{document}
\spacing{1.15}
\ordernumber{----}
\DSnumber{160}
\university{University of Lyon}
\institution{INRIA - MARACAS Team}
\company{Nokia Bell Labs}
\DSname{Electronics, Electrotechnics and Automation\\(ED EEA)}
\specialization{Signal and Image Processing}
\title{Ph.D. Thesis Title}
\author{Mathieu Goutay}
\titre{Machine Learning for the Physical Layer of Communication Systems}

\titre{Applications of Deep Learning to the Design of Enhanced Wireless Communication Systems}
\titrefr{Applications de l'Apprentissage Profond à la Conception de Systèmes de Communication Sans Fil Améliorés}

\universite{Université de Lyon}
\institutionfr{INRIA - \MakeUppercase{é}quipe MARACAS}
\companyfr{Nokia Bell Labs}
\EDnom{\MakeUppercase{é}lectronique, \MakeUppercase{é}lectrotechnique et Automatique\\(ED EEA)}
\specialite{Traitement du Signal et de l'Image}


\addreviewer{Le Ruyet, Didier}{Professeur}{Professor}{CNAM, France}{CNAM, France}
\addreviewer{Langlais, Charlotte}{Chargée de Recherche, HDR}{Permanent Research Staff, HDR}{IMT Atlantique, France}{IMT Atlantique, France}
\addexaminer{Fijalkow, Inbar}{Professeure}{Professor}{ENSEA, France}{ENSEA, France}
\addexaminer{ten Brink, Stephan}{Professeur}{Professor}{Université de Stuttgart, Allemagne}{University of Stuttgart, Germany}

\advisor{Gorce, Jean-Marie}{Professeur}{Professor}{Université de Lyon, France}{University of Lyon, France}
\addcoadvisor{Hoydis, Jakob}{Chargé de Recherche}{Principal Research Scientist}{Nvidia, France}{Nvidia, France}
\addguest{Nom}{Titre}{Grade}{Institution}

\frontmatter



\begin{frontpage}
\begin{center}
\usefont{T1}{pnc}{l}{n}
\hfill
\hfill
\hfill
\\
\vspace{.7cm}
{\Large \MakeUppercase{\@company} -- \Large \MakeUppercase{\@university}\\
Doctoral School of \@DSname\\}
\vspace{1.cm}
{\usefont{T1}{pbk}{l}{n} \Huge THESIS\\}
\vspace{1.25cm}
{\large presented on January 28, 2022  by \\}
\vspace{.25cm}
{\usefont{T1}{pbk}{l}{n}\fontsize{16}{16}\selectfont \@author\\}
\vspace{.25cm}
{\large for the degree of\\}
\vspace{.25cm}
{\normalfont {\Large \textit{Doctor of Philosophy}}\\}
\vspace{.25cm}
\vspace{.5cm}
{Specialization: \@specialization\\}
\vfill
\definecolor{rule}{rgb}{0.5,0.5,0.5}
{\color{rule} \rule{\textwidth}{0.7pt}}\\
\vspace{.5cm}
{\usefont{T1}{pbk}{m}{n}\fontsize{25}{25}\selectfont \@titre\\} 
\vspace{.5cm}
{\color{rule} \rule{\textwidth}{0.7pt}}\\
\vfill
\usefont{T1}{ppl}{l}{n}
{\small Members of the Jury:}\\
\vspace{.8cm}

\small

\begin{tabular}{hllll}
&\textbf{Thesis supervisor:}& \\
&\getadvisor{name} & \getadvisor{gradeen} & \getadvisor{labelen}\\
&\textbf{Thesis co-supervisors:}& \\
\getcoadvisors
&Ait Aoudia, Fayçal & Senior Research Scientist & Nvidia, France\\
&\textbf{Reviewers:}& \\
\getreviewers
&\textbf{Examiners:} & \\
\getexaminers
\end{tabular}
\end{center}
\end{frontpage}
\begin{otherlanguage}{french}
\begin{frontpage}
\begin{center}
\usefont{T1}{pnc}{l}{n}
\hfill
\hfill
\hfill
\\
\vspace{.7cm}
{\Large \MakeUppercase{\@company} -- 
\Large \MakeUppercase{\@universite}\\
\MakeUppercase{é}cole Doctorale d'\@EDnom\\}
\vspace{1.cm}
{\usefont{T1}{pbk}{l}{n} \Huge TH\MakeUppercase{è}SE\\}
\vspace{1.25cm}
{\large présentée le 28 janvier 2022  par \\}
\vspace{.25cm}
{\usefont{T1}{pbk}{l}{n}\fontsize{16}{16}\selectfont \@author\\}
\vspace{.25cm}
{\large en vue de l'obtention du\\}
\vspace{.25cm}
{\normalfont {\Large \textit{Doctorat de l'Université de Lyon}}\\}
\vspace{.5cm}
{Spécialité: \@specialite\\}
\vfill
\definecolor{rule}{rgb}{0.5,0.5,0.5}
{\color{rule} \rule{\textwidth}{0.7pt}}\\
\vspace{.5cm}
{\usefont{T1}{pbk}{m}{n}\fontsize{23}{23}\selectfont \@titrefr\\}
\vspace{.5cm}
{\color{rule} \rule{\textwidth}{0.7pt}}\\
\vfill
\usefont{T1}{ppl}{l}{n}
{\small Devant le jury composé de:}\\
\vspace{.8cm}
\small

\begin{tabular}{hllll}
&\textbf{Directeur de th\`{e}se:}& \\
&\getadvisor{name}  & \getadvisor{gradefr} & \getadvisor{labelfr}\\
&\textbf{Co-encadrants de thèse:}& \\
\getcoadvisorsfr
&Ait Aoudia, Fayçal & Chargé de Recherche & Nvidia, France\\
&\textbf{Rapporteurs:}& \\
\getreviewersfr
&\textbf{Examinateurs:}&\\
\getexaminersfr
\end{tabular}
\end{center}
\end{frontpage}
\end{otherlanguage}

\begin{nopageskip}
\chapter*{\centering \fontsize{30}{30}\selectfont{Acknowledgements}} 

First, I would like to express my deepest appreciation to Dr. Jean-Marie Gorce, who has guided me through the last six years of my studies, to Dr. Jakob Hoydis, who gave my this incredible Ph.D. opportunity three years ago and has always inspired me since, and to Fayçal Ait Aoudia, for his invaluable help, motivation, and support.
It has been a great privilege and honor to work under your shared supervision.
In addition, I would like to thank all my colleagues who have contributed to creating a fantastic work environment.

Secondly, I am extremely grateful to my parents Bernard and Marie-Hélène, and to my stepfather Fabrice, for their constant love and affection.
The education I received is one of my greatest assets to navigate this world, and so I would like to dedicate this manuscript to you.
Special thanks to my brother Nicolas, for showing me the way and proving me that everything was possible after high school. 
Thanks also to my grandparents, Jeanne-Marie, Joseph, Monique, and Robert, for always being there for me.

Finally, I would like to express my gratitude to my girlfriend, Camille, who has always believed in me.
Your love, humor and support have been a steady light in the lonely moments of the pandemic.
I would also like to give warm thanks all my friends, starting with \emph{les anciens}, who have been with me for so many years, and all the other friends from INSA and elsewhere who helped me grow and become who I am.

\newpage


\chapter*{\centering \fontsize{30}{30}\selectfont{Abstract}} 

Innovation in the physical layer of communication systems has traditionally been achieved by breaking down the transceivers into sets of processing blocks, each optimized independently based on mathematical models.
This approach is now challenged by the ever-growing demand for wireless connectivity and the increasingly diverse set of devices and use-cases.
Conversely, \gls{DL}-based systems are able to handle increasingly complex tasks for which no tractable models are available.
By learning from the data, these systems could be trained to embrace the undesired effects of practical hardware and channels instead of trying to cancel them.
This thesis aims at comparing different approaches to unlock the full potential of \gls{DL} in the physical layer.

First, we describe a \gls{NN}-based block strategy, where an \gls{NN} is optimized to replace one or multiple block(s) in a communication system.  
We apply this strategy to introduce a \gls{MU-MIMO} detector that builds on top of an existing \gls{DL}-based architecture.
The key motivation is to replace the need for retraining on each new channel realization by a hypernetwork that generates optimized sets of parameters for the underlying \gls{DL} detector.
Second, we detail an end-to-end strategy, in which the transmitter and receiver are modeled as \glspl{NN} that are jointly trained to maximize an achievable information rate.
This approach allows for deeper optimizations, as illustrated with the design of waveforms that achieve high throughputs while satisfying \gls{PAPR} and \gls{ACLR} constraints.
Lastly, we propose a hybrid strategy, where multiple DL components are inserted into a traditional architecture but trained to optimize the end-to-end performance.
To demonstrate its benefits, we propose a \gls{DL}-enhanced MU-MIMO receiver that both enable lower \glspl{BER} compared to a conventional receiver and remains scalable to any number of users.

Each approach has its own strengths and shortcomings.
While the first one is the easiest to implement, its individual block optimization does not ensure the overall system optimality.
On the other hand, systems designed with the second approach are computationally complex and do not comply with current standards, but allow the emergence of new opportunities such as high-dimensional constellations and pilotless transmissions.
Finally, even if the block-based architecture of the third approach prevents deeper optimizations, the combined flexibility and end-to-end performance gains motivate its use for short-term practical implementations.

\newpage

\begin{otherlanguage}{french}
\chapter*{\centering \fontsize{30}{30}\selectfont{Résumé}}

L'innovation dans la couche physique des systèmes de communications a traditionnellement été réalisée en modélisant les émetteurs-récepteurs comme une suite de blocs, chacun étant optimisé indépendamment sur la base de modèles mathématiques. 
Cette approche est aujourd'hui remise en question par la demande croissante de connectivité et la diversité des cas d'utilisation.
À l'inverse, les systèmes basés sur l'apprentissage profond (deep learning, DL) sont capables de traiter des tâches de plus en plus complexes en apprenant à partir de données. 
Cette thèse vise donc à comparer différentes approches pour exploiter le plein potentiel du DL dans la couche physique. 

Tout d'abord, nous décrivons une stratégie basée sur un réseau neuronal (neural network, NN) qui est optimisé pour remplacer un ou plusieurs blocs consécutifs dans un système de communication. 
Nous appliquons cette stratégie pour introduire un détecteur multi-utilisateurs à entrées et sorties multiples (multi-user multiple-intput multiple-output, MU-MIMO) qui s'appuie sur un détecteur existant basé sur du DL. 
L'idée est d’utiliser un hyper-réseau de neurones pour générer des paramètres optimisés pour le détecteur DL sous-jacent. 
Deuxièmement, nous détaillons la stratégie de bout en bout, dans laquelle les émetteurs-récepteurs sont modélisés comme des NNs qui sont entraînés conjointement pour maximiser un taux d'information réalisable. 
Cette approche permet des optimisations plus profondes, comme l'illustre la conception de formes d'onde qui atteignent des débits élevés tout en satisfaisant des contraintes sur le signal et son spectre.
Enfin, nous proposons une stratégie hybride, où plusieurs composants DL sont insérés dans une architecture traditionnelle mais entraînés pour optimiser les performances de bout en bout. 
Pour démontrer ses avantages, nous proposons un récepteur MU-MIMO amélioré par DL qui permet à la fois de réduire les taux d'erreur binaire (bit error rate, BER) par rapport à un récepteur classique et de rester adaptable à un nombre variable d'utilisateurs. 

Chaque approche a ses propres forces et faiblesses. 
Si la première est la plus facile à implémenter, l'optimisation individuelle de chaque bloc ne garantit pas l'optimalité du système entier. 
En revanche, les systèmes conçus selon la seconde approche sont souvent trop complexes et ne sont pas conformes aux standards actuels, mais ils permettent l'émergence de nouvelles possibilités telles que des constellations de grande dimension et des transmissions sans pilote. 
Enfin, même si l'architecture par blocs de la troisième approche empêche des optimisations plus poussées, la combinaison de sa flexibilité et de son optimisation de bout en bout motive son utilisation pour des implémentations à court terme.

\bigskip

\emph{Un résumé de la thèse en français est disponible dans l'Appendix B de ce manuscript.}

\end{otherlanguage}
\end{nopageskip}
\newpage

\begin{nopageskip}
\chapter*{\centering \fontsize{30}{30}\selectfont{Publications}} 

\section*{Journal papers published during my Ph.D. studies}
\begin{refsection} 
\nocite{goutay2020machine, goutay2021learning_} 
\printbibliography[heading=none] 
\end{refsection}

\section*{Conference papers published during my Ph.D. studies}
\begin{refsection} 
\nocite{goutay2021endtoend_, goutay2021machine_, goutay2020deep} 
\printbibliography[heading=none] 
\end{refsection}

\section*{Conference papers published during my master studies}
\begin{refsection} 
\nocite{9144089, 8464894} 
\printbibliography[heading=none] 
\end{refsection}

\newpage

\chapter*{\centering \fontsize{30}{30}\selectfont{Patents and Tutorials}} 

\section*{Patents granted at the time of publication}
\begin{refsection} 
\nocite{patent1, patent2, patent3} 
\printbibliography[heading=none] 
\end{refsection}

\section*{Tutorials}
\begin{refsection} 
\nocite{tutorial1, tutorial2} 
\printbibliography[heading=none] 
\end{refsection}

\newpage

\end{nopageskip}

\begin{nopageskip}
\tableofcontents
\pagebreak
\end{nopageskip}

\vfill
\pagebreak
\thispagestyle{empty}
\cleardoublepage


\pagestyle{headings}
\mainmatter

\glsresetall

\begin{refsection}

\chapter{Introduction}
\label{ch:Intro}

\section{When Machine Learning Meets Signal Processing}
\label{sec:intro_1}
The first \gls{NN} model was introduced in 1943~\cite{mcculloch1943logical}, but sixty years of research and of processing power increase were required to enable a major adoption of \gls{ML}  by the industry~\cite{Bennett07thenetflix}.
In particular, the 2010s have seen significant improvements in parallel computing, leading to  the advent of \gls{DL} and to breakthroughs in computer vision~\cite{krizhevsky2017imagenet}, speech recognition~\cite{6854661}, and many other domains~\cite{thies2016face2face, devlin2018bert, silver2016mastering}.
\Gls{DL} is especially useful when the task at hand is difficult to formalize mathematically or when the mathematical models are untractable.
By shifting from model-driven to data-driven algorithms, \gls{DL} techniques are able to circumvent this problem as long as a sufficient dataset is available.
Typically, massive progresses have been possible in the field of image recognition thanks to the publication of the ImageNet dataset in 2009~\cite{5206848}, containing more than 3 millions labelled images.

\begin{figure}[h]
    \centering
    \includegraphics[width=0.70\textwidth]{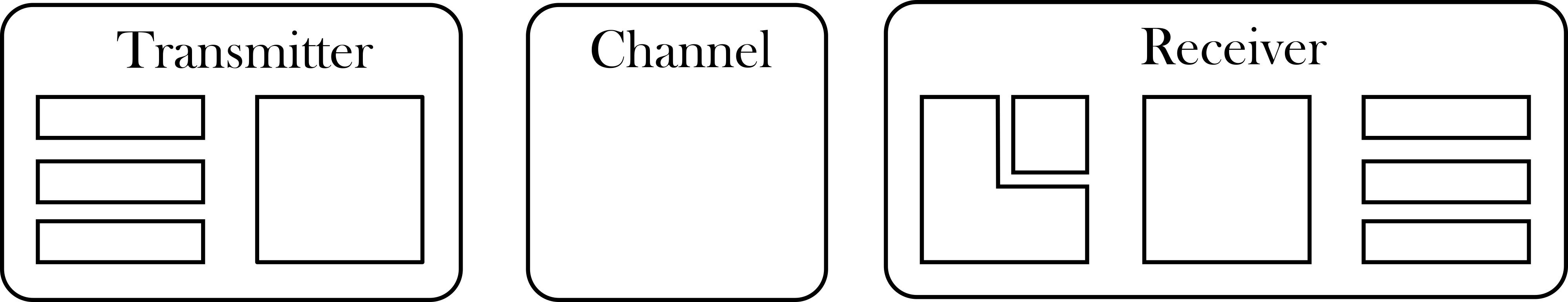}
    \caption{A traditional block-based communication system.}
    \label{fig:bkg_regular_system}
\end{figure}
In the meantime, new generations of cellular communication systems have emerged every ten years, starting from 1979~\cite{rappaport1996wireless}. 
Each generation brings multiple connectivity improvements, such as faster and more reliable communications, in part thanks to a better modeling of the wireless channel.
These mathematical models allowed the design of algorithms that can take advantage of the available knowledge in information theory and signal processing.
As the transmitters and receivers became more and more complex, tractability was achieved by splitting the transmit and receive processing chains into small components, usually referred to as \emph{processing blocks} and illustrated in Fig.~\ref{fig:bkg_regular_system}.
Such bloc-based communication systems suffer from multiple drawbacks.
On the one hand, simplistic channel models fails at capturing all the specificities of the underlying hardware and
propagation phenomenons.
On the other hand, the joint optimization of the transmitter and receiver quickly becomes intractable when more realistic channels models are derived, and therefore the optimization of each block is typically performed independently.
This does not ensure the optimality of the resulting system, as it can be shown for the channel coding and modulation blocks~\cite{141453}.
Finally, signalling is often required between the transmitter and the receiver, which introduces an overhead that reduces the system throughput. 

\begin{figure}[h]
    \centering
       \begin{subfigure}{1\textwidth}
        \centering
       \includegraphics[width=0.70\textwidth]{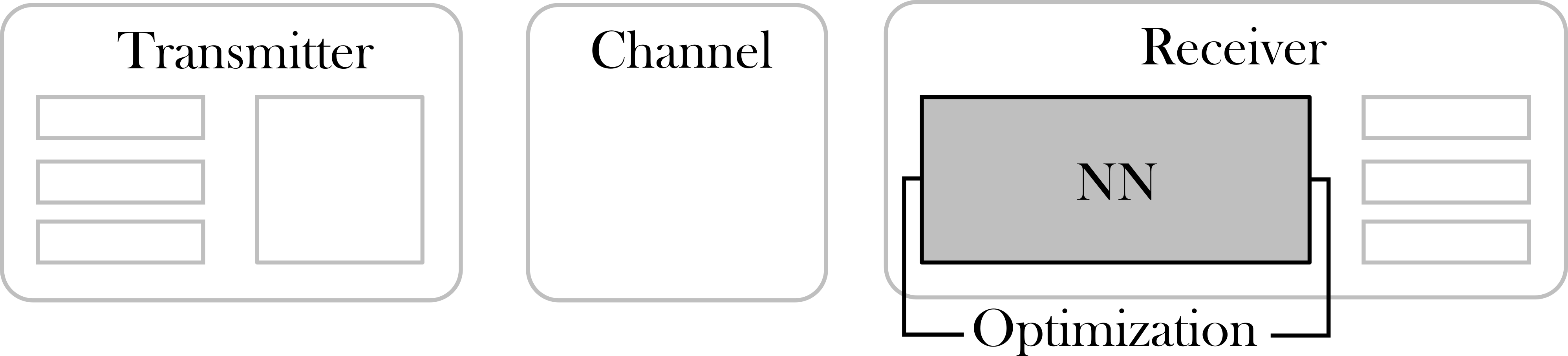}
       \caption{NN-based block optimization: an \gls{NN} is optimized to replace one or multiple block(s) in a communication system.}
       \label{fig:intro_ml_syst_1} 
       \vspace{10pt}
    \end{subfigure}
    \begin{subfigure}{1\textwidth}
        \centering
       \includegraphics[width=0.7\textwidth]{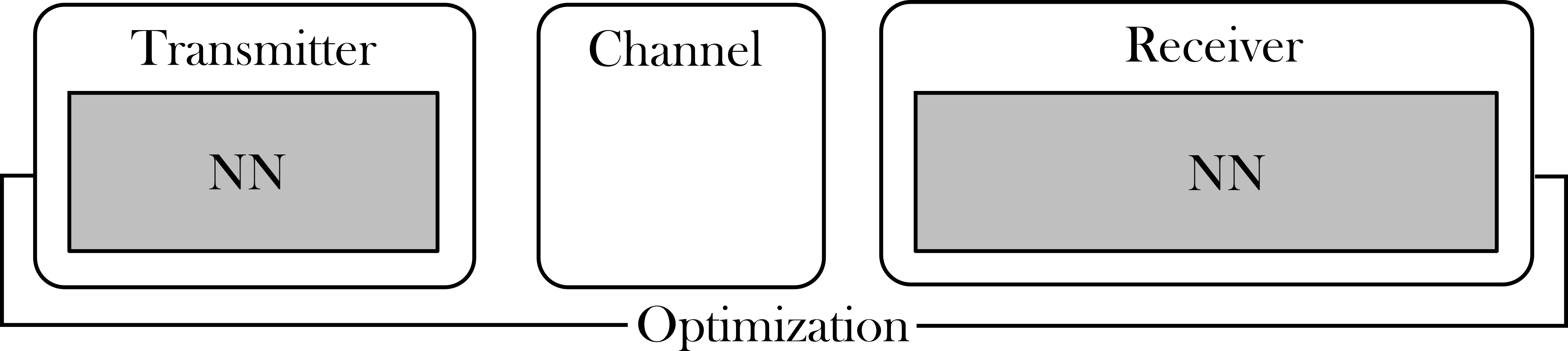}
       \caption{End-to-end optimization: NN-based transceivers are optimized to maximize the end-to-end performance of a system.}
       \label{fig:intro_ml_syst_3}
    \end{subfigure}
    \caption{Different level of \gls{NN} integrations into communication systems.}
    \label{fig:intro_ml_syst}
    \end{figure}

\gls{DL} for the physical layer was already studied in the nineties~\cite{563540}, but a renewed interest started in 2016-1017 thanks to the publication of multiple promising papers.
One of them was published in 2016 by Be'ery et al., who represented the channel decoding algorithm as an \gls{NN} to improve the \glspl{BER} of systems using various codes~\cite{7852251}.
This approach corresponds to an \emph{NN-based block optimization strategy} as shown in Fig.~\ref{fig:intro_ml_syst_1}, in which one or multiple consecutive processing blocks are replaced by an \gls{NN}.
To handle such \gls{NN}-based blocks, the next generation of communication system needs to be \emph{designed for \gls{DL}}, in a way that allows for practical training and testing of these components.
Although this idea is interesting, the true \gls{DL} revolution started when O'Shea and Hoydis introduced end-to-end learning for communication systems in their seminal paper from 2017~\cite{8054694}.
Such approach is often referred to the \emph{end-to-end optimization strategy}, as depicted in Fig.~\ref{fig:intro_ml_syst_3}, and experimental gains were quickly shown by Hoydis, Dörner, Cammerer et al. with over-the-air transmissions~\cite{8335670}.
This strategy allows the systems to be entirely optimized from real-world data, therefore enabling efficient handling of hardware impairments and other channel distortions without requiring any mathematical model~\cite{8792076}.
Moreover, they can contribute to reducing the signaling overhead, either by removing the pilots required for channel estimation at the receiver~\cite{pilotless20} or by learning optimal \gls{MAC} protocols~\cite{valcarce2021joint}.
For these reasons, end-to-end systems are often seen as the next big step in the evolution of the physical layer, as the transmit and receive processing would be \emph{designed by \gls{DL}}~\cite{Hoydis2021}.

The study of the two strategies presented in Fig.~\ref{fig:intro_ml_syst} have lead to the discovery of deep connections between the fields of DL, information theory, and of signal processing in communication systems~\cite{8839651}.
For example, the transmitter-receiver pair can be modeled as an autoencoder, in which the estimation of the transmitted bits becomes a binary classification problem~\cite{8490882}.
Moreover, an achievable rate of a communication system can be expressed in terms of cross entropy~\cite{9118963}, which is a well-known metric in information theory and \gls{DL}.
These connections, along with performance improvements demonstrated on multiple systems and environments, indicate that \gls{DL} will play an important role in future generations of communication systems~\cite{9078454, 9061001, ali20206g}.
However, each strategy has its shortcomings: the block-based \glspl{NN} (Fig.~\ref{fig:intro_ml_syst_1}) are not trained to maximize the overall performance of the system, and the fully learned transceivers (Fig.~\ref{fig:intro_ml_syst_3}) lack interpretability and scalability.
This thesis therefore aims at providing some answers to the question of the optimal integration of \gls{DL} components into wireless communication systems.

\section{Current Challenges and Contributions of this Work}
\label{sec:intro_challenges}

The next generation of cellular networks will need to support a growing number of different services and devices~\cite{9040431}. 
To that aim, the available resources need to be more efficiently shared among users.
One key technique is the use of \glsdesc{MU}\glsunset{MU} \glsdesc{MIMO}\glsunset{MIMO} (MU-MIMO) systems, where spatial multiplexing is exploited to increase both the channel capacity and the number of users that can be served simultaneously~\cite{massivemimobook}.
One of the main challenges related to the deployment of such systems is the complexity of the symbol detection algorithm, which grows with the number of antennas and users. 
For example, \gls{MAP} detection is optimal but known to be NP-hard, and sphere decoders have exponential worst-case complexity~\cite{7244171}.
The conventional solution to tackle this problem is to use linear detectors that are computationally tractable, but suffer from performance degradation on ill-conditioned channels.
In the past years, several approaches tried to address those challenges by implementing the detector as an \gls{NN}, which corresponds to the NN-based block optimization strategy.
However, they either still achieve unsatisfying performance on spatially correlated channels, or are computationally demanding since they require retraining for each channel realization.
In this work, we address both issues by training an additional \gls{NN}, referred to as the hypernetwork, which takes as input the channel matrix and generates the weights of the NN-based detector. 
Results show that the proposed approach achieves near state-of-the-art performance without the need for re-training.

Another key research direction is the improvement of the \gls{OFDM} waveform, used in most modern communication systems such as 4G, 5G, and Wi-Fi.
Indeed, conventional \gls{OFDM} suffers from multiple drawbacks, such as a high \gls{PAPR} and \gls{ACLR}.
To tackle these problems, we leverage the end-to-end optimization strategy and model the transmitter and receiver as \glspl{NN} that respectively implement a high-dimensional modulation scheme and estimate the transmitted bits.  
We then propose a learning-based approach to design OFDM waveforms that satisfy selected constraints while maximizing an achievable information rate, with the additional advantage that no pilots are needed during transmissions.
Evaluated with \gls{ACLR} and \gls{PAPR} targets, the trainable system is able to satisfy the constraints while enabling significant throughput gains compared to a \gls{TR} baseline.

\begin{figure}[h]
    \centering
    \includegraphics[width=0.65\textwidth]{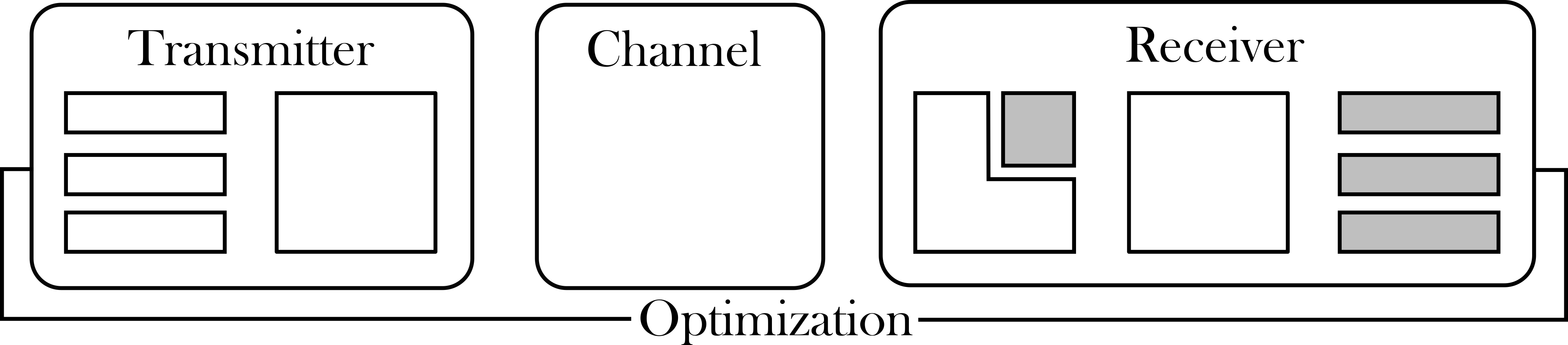}
    \caption{A hybrid training strategy.}
    \label{fig:into_ml_syst_4}
\end{figure}
As mentioned in Section~\ref{sec:intro_1}, such end-to-end systems lack interpretability, as the black-box design prevents the capture of intermediate data such as channel estimates.
They also lack scalability, which is especially important in \gls{MU}-\gls{MIMO} transmissions since the receive algorithm must allow for easy adaptation to a varying number of users.
It is therefore still unclear if this strategy is competitive with respect to conventional MU-MIMO receivers in realistic scenarios and under practical constraints.
For this reason, we propose a DL-enhanced MU-MIMO receiver that builds on top of a conventional architecture to preserve its interpretability and scalability, but is trained to maximize an achievable transmission rate.
This approach can be seen as a \emph{hybrid strategy}, in which multiple \gls{DL}-based components are inserted in a traditional block-based architecture but are optimized to maximize the end-to-end system performance (Fig.~\ref{fig:into_ml_syst_4}).
The resulting system can be used in the up- and downlink and does not require hard-to-get perfect \gls{CSI} during training, which contrasts with existing works.
Simulation results demonstrate consistent performance improvements over a \gls{LMMSE} baseline which are especially pronounced in high mobility scenarios.

\section{Thesis Outline}
The remainder of this thesis is organized as follows.
Chapter 2 provides a background on \gls{DL}, on the physical layer, and on the interconnection between the two fields.
\gls{OFDM} is presented first, with a derivation of the channel models corresponding to both \gls{SISO} and \gls{MU}-\gls{MIMO} transmissions.
We then detail the concept of backpropagation, \glspl{NN}, and of \gls{SGD}, and describe the optimization of \gls{DL}-enhanced systems.
Chapter 3 introduces the hypernetwork-based MIMO detector, which is an example of the block-based optimization strategy depicted in Fig.~\ref{fig:intro_ml_syst_1}.
The traditional iterative detection framework and the concept of hypernetworks are presented, followed by a description of the HyperMIMO system and of evaluations on spatially correlated channels.
The fully NN-based transceiver strategy of Fig.~\ref{fig:intro_ml_syst_3} is discussed in Chapter 4, where we design OFDM waveforms that both maximize an achievable rate and satisfy \gls{PAPR} and \gls{ACLR} constraints.
The system model and the baseline are described, and both the \gls{NN} architectures and the learning-based approach used for waveform design are detailed.
Finally, simulation results and insights are provided.
Chapter 5 is dedicated to the presentation of the hybrid strategy (Fig.~\ref{fig:into_ml_syst_4}) for DL-enhanced MU-MIMO receivers.
First, we develop the traditional architectures corresponding to uplink and downlink transmissions.
Second, we highlight two limitations of these architectures and detail how we address them using \glspl{CNN}.
Third, we provide simulation results to compare the DL-enhanced receiver to the baseline.
Finally, Chapter 6 concludes this manuscript.

\section{Notations}
$\RR$ ($\CC$) denotes the set of real (complex) numbers.
Tensors and matrices are denoted by bold upper-case letters and vectors are denoted by bold lower-case letters.
We respectively denote by $\mv_a$ and $m_{a,b}$ the vector and scalar formed by slicing the matrix $\Mm$ along its first and second dimensions.
Note that the notation $[\Mm]_a$ and $[\Mm]_{a,b}$ is also used in Section~\ref{subsec:layers_act_func} for clarity.
Similarly, we denote by $\Tm_{a, b} \in \CC^{N_c \times N_d}$ ($\tv_{a, b, c} \in \CC^{N_d}$, $t_{a, b, c, d} \in \CC$) the matrix (vector, scalar) formed by slicing the tensor $\Tm \in \CC^{N_a \times N_b \times N_c \times N_d}$ along the first two (three, four) dimensions.
The notation $ \Tm^{(k)}$ indicates that the quantity at hand is only considered for the $k^{\text{th}}$ user, and $\vv_{\shortminus a}$ corresponds to the vector $\vv$ from which the $a^{\text{th}}$ element was removed.
$||\Mm||_\text{F}$ denotes the Frobenius norm of $\Mm$.
$\text{Card}(\mathcal{S})$ denotes the number of elements in a set $\mathcal{S}$, $\text{vec}\LB \cdot \RB$ the vectorization operator, and $\odot$ and $\oslash$ the element-wise product and division, respectively.
$(\cdot)\tp$, $(\cdot)\htp$, and $(\cdot)^\star$ respectively denote the transpose, conjugate transpose, and element-wise conjugate operator.
$I(\xv; \yv)$ and $P(\xv, \yv)$ represent the mutual information and joint conditional probability of $\xv$ and $\yv$, respectively.
$\Id_N$ is the $N \times N$ identity matrix and $\mathds{1}_{N \times M}$ is the $N\times M$ matrix where all elements are set to $1$.
Finally, the imaginary unit is $j$, such that $j^2 = -1$.

\chapter{Background on the Physical Layer and Deep Learning}
\label{ch:background}

\section{OFDM Systems}
\label{sec:OFDM_systems}

\begin{figure}[h]
    \centering
    \includegraphics[width=0.8\textwidth]{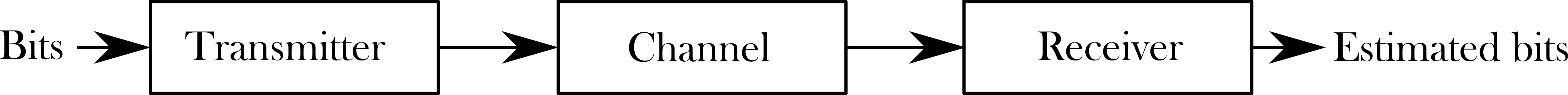}
    \caption{A digital communication system.}
    \label{fig:bkg_comm_system}
\end{figure}

A digital communication system aims at transmitting bits from a transmitter to a receiver by modulating an electromagnetic wave that is transmitted through a channel (Fig.~\ref{fig:bkg_comm_system}).
Multiple waveforms can be used to carry the information, and the waveform choice is usually dictated by the channel distortions that need to be dealt with.
In the following, we present \glsreset{OFDM} \gls{OFDM}, a transmission technique used in most modern communication systems thanks to its ability to handle difficult channel conditions such as selective fading.

\subsection{Transmit Processing}

\begin{figure}
    \centering
    \includegraphics[width=0.35\textwidth]{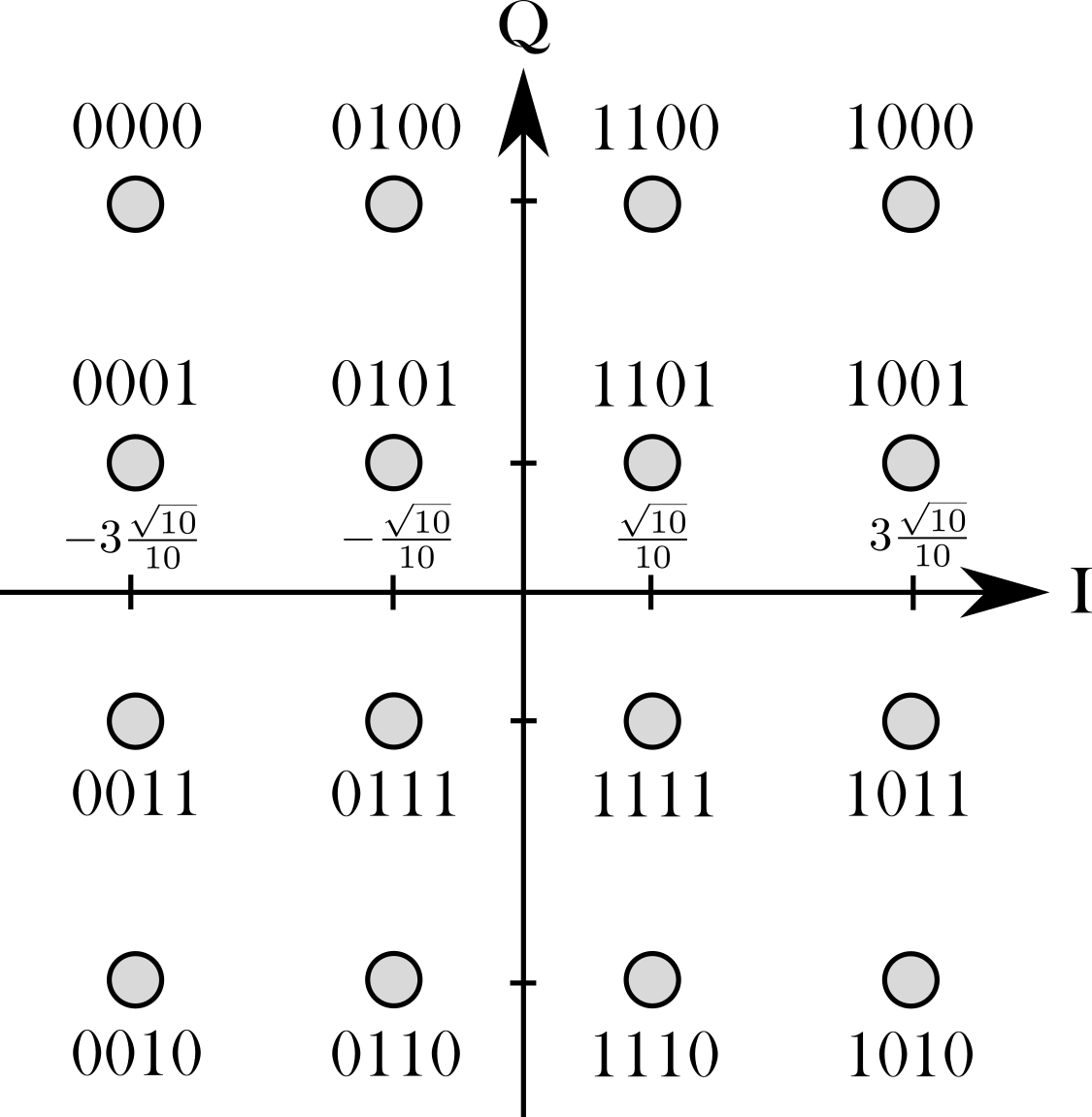}
    \caption{Constellation diagram corresponding to a 16-\gls{QAM} constellation.}
    \label{fig:bkg_16_qam}
\end{figure}

The first operation that is performed by the transmitter is the \emph{bit mapping}, in which vectors  of $Q$ bits  $\bv \in \{0, 1\}^Q$ are mapped to $2^Q$ different \emph{symbols} $x \in \Cc$, where $\Cc$ is referred to as the constellation. 
\Glspl{QAM} are among the most used constellations, and are identified by the number of different symbols that can be transmitted.
As an example, Fig.~\ref{fig:bkg_16_qam} shows the symbols and associated bits for a $2^4$-QAM where the $4$ bits are arranged according to a Gray labeling.
After modulation, the symbols are converted into an electromagnetic wave that is transmitted through the channel.

\begin{figure}[t]
	\centering
	\input{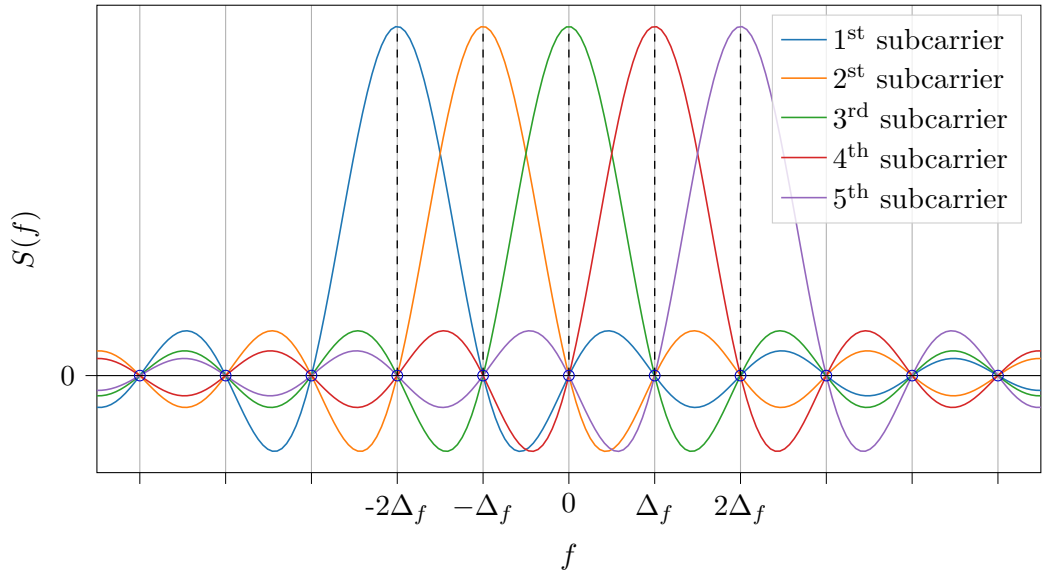}
	\caption{Representation of the amplitude of an OFDM spectrum $S(f)$ with $N=5$ subcarriers centered around 0. Each subcarrier is null at the frequencies corresponding to other subcarriers, ensuring orthogonality.}
	\label{fig:bkg_ofdm}
\end{figure}

\begin{figure}
    \centering
    \includegraphics[width=0.25\textwidth]{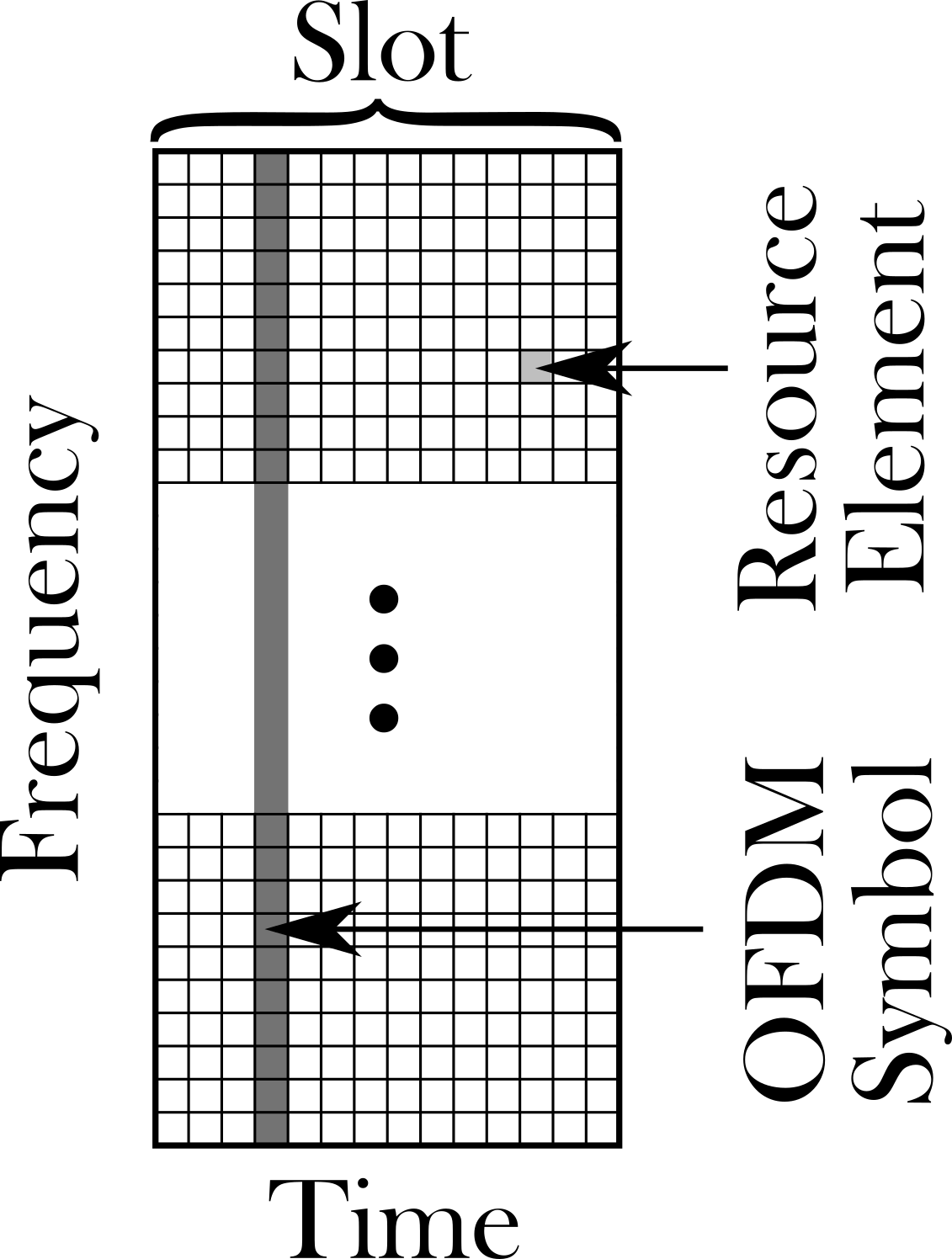}
    \caption{An OFDM resource grid.}
    \label{fig:bkg_rg}
\end{figure}

In \gls{OFDM}, the available bandwidth is divided into a set of $N$ sub-band referred to as \emph{subcarriers}. 
Orthogonality is achieved in the frequency domain by selecting a subcarrier spacing of $\Delta_f$ and applying a matching sinc-shaped pulse.
The resulting spectrum is represented in Fig.~\ref{fig:bkg_ofdm}, where it can be seen that each subcarrier is null at the frequencies corresponding to other subcarriers.
In the time-domain, the duration of the corresponding signal is denoted by $T=\frac{1}{\Delta_f}$, and is referred to as the duration of an \emph{OFDM symbol}.
The entire time-frequency grid, formed by $N$ subcarriers and $M$ OFDM symbols, is referred to as the \emph{\gls{RG}}, while a single element in that grid is referred to as a \emph{\gls{RE}} (Fig.~\ref{fig:bkg_rg}).

%
Transmission over the \gls{RG} is achieved by grouping the flow of symbols $x$ to be transmitted into vectors of symbols $\xv_m \in \Cc^{N}, m\in\{1, \cdots, M\}$ that are transmitted in parallel over all $N$ subcarriers, effectively mapping each $x_{m,n} \in \Cc$ to the \gls{RE} $(m,n)$.
To avoid any confusion between an OFDM symbol designating a column in the \gls{RG} and a symbol that indicates a point in a constellation, the latter will also be referred to as a \gls{FBS}, as the symbol mapping is carried out at the baseband over the available subcarriers.
%

\subsection{The Wireless Channel}
\label{sec:the_wireless_channel}

The channel model is the relation between the transmitted \gls{FBS} $x$ and the received \gls{FBS} $y$, and can be expressed as
\begin{align}
    \label{eq:bkg_ch}
    y = \text{ch}(x),
\end{align}
where $\text{ch}(\cdot)$ represents the distortions caused by the channel or the transceivers imperfections.
In the following, we propose a derivation of the OFDM channel model inspired by \cite{edfors1996introduction}.

\begin{figure}[h]
    \centering
    \includegraphics[width=0.43\textwidth]{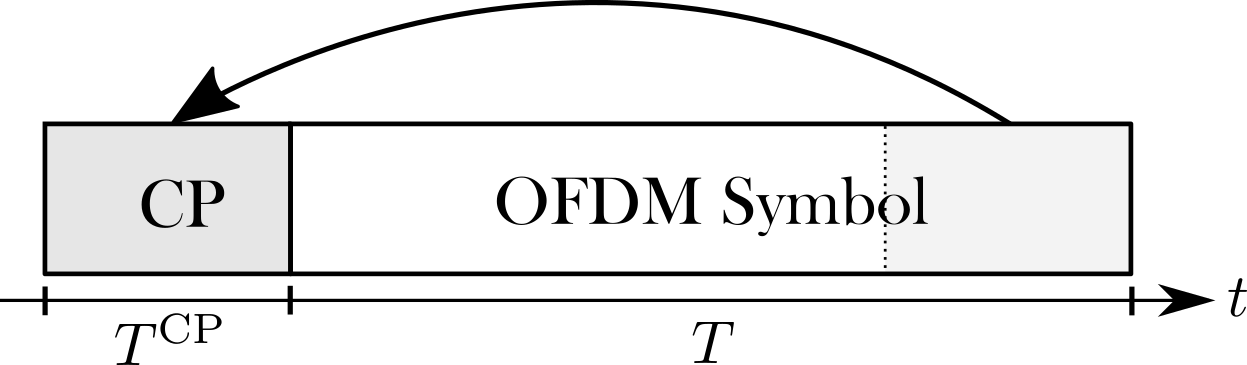}
    \caption{An OFDM symbol with its cyclic prefix appended.}
    \label{fig:bkg_cp}
\end{figure}

\subsubsection*{Transmit filtering}

The time-domain signal $s_m(t)$ corresponding to an OFDM symbol $\xv_m$ is obtained by modulating each $x_{m,n}$ with a different transmit filer $\phi_n$.
During this modulation, a \emph{\gls{CP}} is prepended to the symbol, and contains a copy of the last part of that symbol (Fig.~\ref{fig:bkg_cp}).
The length of the \gls{CP} is denoted by $T^{\text{CP}}$, and total length is $T^{\text{tot}} = T^{\text{CP}} + T$.

If we denote by $\mathcal{N}$ the set of available subcarriers, the transmitted signal corresponding to the $m^{\text{th}}$ \gls{OFDM} symbol is
\begin{align}
    s_m(t) = \sum_{n\in \mathcal{N}} x_{m,n} \phi_n(t - mT^{\text{tot}})
\end{align}
where the filters $\phi_n$ are chosen such that $\phi_k (t) = \phi_k(t+T)$ when $t$ is within the duration of the \gls{CP}, i.e., when $t \in [0, T^{\text{CP}}]$:
\begin{align}
    \phi_{n}(t) &= \begin{cases}\frac{1}{\sqrt{T^{\text{tot}}}} e^{j 2 \pi n \frac{ t-T^{\text{CP}}}{T}} & \text { if } t \in[0, T^{\text{tot}}] \\ 0 & \text { otherwise }\end{cases}
    \\
    &= \frac{1}{\sqrt{T^{\text{tot}}}} \text{rect}\LB \frac{t}{T^{\text{tot}}} -\frac{1}{2} \RB e^{j 2 \pi n \frac{ t-T^{\text{CP}}}{T}}.
\end{align}
Note that we choose the transmit filters so that the average energy per OFDM symbol and per subcarrier is one.
The transmitted signal corresponding to an OFDM slot is
\begin{align}
    \label{eq:intro_time_signal}
    s(t)=\sum_{m=0}^{M-1} s_{m}(t)=\sum_{m=0}^{M-1} \sum_{n \in \mathcal{N}} x_{m,n} \phi_{n}(t-mT^{\text{tot}}).
\end{align}
Without \glspl{CP} ($T^{\text{CP}}=0$), the spectrum of each filter $\phi_n(t)$ corresponds to a subcarrier as shown in Fig.~\ref{fig:bkg_ofdm}.
The removal of the \glspl{CP} at the receiver-side therefore ensures the preservation of the orthogonality between subcarriers.

\subsubsection*{Channel}
Let us denote by $g(\tau, t)$ the response of the channel at time $t$ when excited with a impulse transmitted at time $t- \tau$.
For a multipath channel, we have
\begin{align}
    g(\tau, t)=\sum_{p=0}^{P-1} a_{p}(t) \delta\left(\tau-\tau_{p}(t) \right)
\end{align}
where $P$ is the number of different paths, $a_{p}(t)$ and $\tau_{p}(t)$ respectively denote the complex amplitude
and time delay associated with the $p^{\text{th}}$ path at time $t$, and $\delta (\cdot)$ is the Dirac function.
The length of the \gls{CP} is chosen to be at least equal to the longest delay, such as the impulse response of the channel is restricted to the interval $[0, T^{\text{CP}}]$.
The received signal can be expressed as 
\begin{align}
    r(t) &= \int_{-\infty}^{\infty} g(\tau, t) s(t-\tau) d \tau+ \widetilde{n}(t) \\
        &= \int_{0}^{T^{\text{CP}}} g(\tau, t) s(t-\tau) d \tau+ \widetilde{n}(t) \label{eq:bkg_001}
\end{align}
where  $\widetilde{n}(t)$ is a complex \gls{AWGN} process with \gls{PSD} $N_0$ satisfying 
\begin{align}
    \EE \LSB n(t) n^{\star}(t+\tau) \RSB = N_0 \delta(t-\tau).
\end{align}

As $a_p(t)$ and $\tau_p(t)$ typically vary slowly, it is common to assume that the channel is pseudo-stationnary, i.e., that it is constant over the duration of an \gls{OFDM} symbol.
On the $m^{\text{th}}$ OFDM symbol, we therefore have $g(\tau, t) = g(\tau, t_m)$ with $t_m = m T^{\text{tot}}$, and \eqref{eq:bkg_001} can be written as
\begin{align}
    r(t) = \int_{0}^{T^{\text{CP}}} g(\tau, t_m) s(t-\tau) d \tau+ \widetilde{n}(t) 
\end{align}

\subsubsection*{Receive filtering}

At the receiver, the signal corresponding to the $k^{\text{th}}$ subcarrier is obtained by filtering the received signal $r(t)$ with
\begin{align}
    \psi_k (t) &= \left\{\begin{array}{cl}
        \phi_{k}^{\star}(T^{\text{tot}}-t) & \text { if } t \in\left[0, T \right] \\
        0 & \text { otherwise }
        \end{array}
        \right. \\
           &= \text{rect} \LB \frac{t}{T} - {\frac{1}{2}} \RB \phi_k^{\star}(T^{\text{tot}}-t).
\end{align}
This receive filter, of duration $T$, is matched to the part of the transmit filter carrying the OFDM symbol, thus removing the \gls{CP}.
Since the channel impulse response was shorter than the \gls{CP}, the filtered signal contains no \gls{ISI}, and therefore we can focus on a single OFDM symbol.
The output of the receive filter on the $0^{\text{th}}$ OFDM symbol and $k^{\text{th}}$ subcarrier is 
\begin{align}
    y_{0,k} &=\int_{-\infty}^{\infty} r(t) \psi_{k}(T^{\text{tot}}-t) d t = \int_{T^{\text{CP}}}^{T^{\text{tot}}}r(t) \phi_{k}^{\star}(t) d t \\
        &= \int_{T^{\text{CP}}}^{T^{\text{tot}}}\left(\int_{0}^{T^{\text{CP}}} g(\tau, t_0)\left[\sum_{n \in \mathcal{N}} x_{0, n} \phi_{n}(t- \tau)\right] d \tau\right) \phi_{k}^{\star}(t) d t +\int_{T^{\text{CP}}}^{T^{\text{tot}}} \widetilde{n}(t) \phi_{k}^{\star}(t) dt \\
        &=\sum_{n \in \mathcal{N}} x_{0, n} \int_{T^{\text{CP}}}^{T^{\text{tot}}}\left(\int_{0}^{T^{\text{CP}}} g(\tau, t_0) \phi_{n}(t-\tau) d \tau\right) \phi_{k}^{\star}(t) d t+\int_{T^{\text{CP}}}^{T^{\text{tot}}} \widetilde{n}(t) \phi_{k}^{\star}(t) dt. \label{eq:bkg_002}
\end{align}
The inner integral can be expressed as
\begin{align}
        \int_{0}^{T^{\text{CP}}} g(\tau, t_0) \phi_{n}(t-\tau) d \tau &=\int_{0}^{T^{\text{CP}}} g(\tau, t_0) \frac{1}{\sqrt{T^{\text{tot}}}} e^{j 2 \pi  \frac{n}{T} t-\tau-T^{\text{CP}}} d \tau \\
        &=\frac{e^{j 2 \pi \frac{n}{T} \left(t-T^{\text{CP}}\right)}}{\sqrt{T^{\text{tot}}}} \underbrace{\int_{0}^{T^{\text{CP}}} g(\tau, t_0) e^{-j 2 \pi \frac{n}{T} \tau } d \tau}_{h_{0,n}}, \quad T^{\text{CP}}<t<T^{\text{tot}}
\end{align}
where $h_n$ is the frequency response of the channel at the $n^{\text{th}}$ subcarrier.
The filtered signal \eqref{eq:bkg_002} can now be written as
\begin{align}
    y_{0,k} &=\sum_{n\in\mathcal{N}} x_{0, n} \int_{T^{\text{CP}}}^{T^{\text{tot}}} \frac{e^{j 2 \pi \frac{n}{T} \left(t-T^{\text{CP}}\right) }}{\sqrt{T^{\text{tot}}}} h_{n} \phi_{k}^{\star}(t) d t+ \underbrace{\int_{T^{\text{CP}}}^{T^{\text{tot}}} \tilde{n}(t) \phi_{k}^{\star}(t) d t }_{n_{0,n}}\\
    &=\sum_{n\in\mathcal{N}} x_{0, n} h_{0, n} \int_{T^{\text{CP}}}^{T^{\text{tot}}} \phi_{n}(t) \phi_{k}^{\star}(t) d t+n_{0,n}.
\end{align}
The transmit filters are chosen to be orthogonal, i.e.,
\begin{align}
    \int_{T^{\text{CP}}}^{T^{\text{tot}}} \phi_{n}(t) \phi_{k}^{\star}(t) &= \left\{\begin{array}{cl}
        1 & \text { if } k=n \\
        0 & \text { otherwise }
        \end{array}
        \right. 
\end{align}
and therefore the \gls{FBS} corresponding to the $n^{\text{th}}$ subcarrier is 
\begin{align}
    y_{0, n} = h_{0, n} x_{0, n} + n_{0, n}
\end{align}
where $n_{0, n}$ is an \gls{AWGN}.
The same derivation can be carried out for all subcarriers and OFDM symbols $m\in \{0, \cdots, M-1 \}$, resulting in 
\begin{align}
    \label{eq:intro_mimo_vectors}
    \yv_{m} = \hv_{m} \odot \xv_{m} + \nv_{m}.
\end{align}
where the vectors $\yv_m, \hv_m$, $\xv_m$, and $\nv_m$ respectively contain the values of $y_{m,n}, h_{m,n}, x_{m,n}$, and $n_{m,n}$ for all subcarriers $n \in \mathcal{N}$.

A discrete-time model can be derived by replacing the transmit and receive filters by the \gls{IDFT} and \gls{DFT} operators:
\begin{align}
    \mathbf{y}_{m} &=\text{DFT}\left(\text{IDFT}\left(\mathbf{x}_{m}\right) \circledast \mathbf{g}_{m}+\widetilde{\mathbf{n}}_{m}\right) \\
    &=\text{DFT}\left(\text{IDFT}\left(\mathbf{x}_{m}\right) \circledast \mathbf{g}_{m}\right)+\mathbf{n}_{m},
    \end{align}
where the  $\circledast$ operator denotes a cyclic convolution, $\nv_m = \text{DFT}\LB \widetilde{\nv}_m \RB$ is a vector of uncorrelated Gaussian noise, and $\gv_m$ corresponds to the channel impulse response, i.e., $\gv_m = \text{IDFT}(\hv_m)$.

\subsection{Receive Processing}

The main benefit of OFDM systems is the simplified receiving process. 
If $h_{m,n}$ is known at the receiver, the transmitted symbols can be estimated from the received symbols:
\begin{align}
    \widehat{x}_{m,n} &= \frac{y_{m,n}}{h_{m,n}}\\
                    &= x_{m,n} + \underbrace{\frac{n_{m,n}}{h_{m,n}}}_{n'_{m,n}}.
\end{align}
This process is referred to as the \emph{equalization}, and $n'_{m,n}$ is the post-equalization received noise with variance $\rho_{m,n}^2$.
The next step is the \emph{demapping}, in which the transmitted bits are estimated from $\hat{x}_{m,n}$.
Hard demapping finds the symbol $\doublehat{x}_{m,n} \in \Cc$ that is the closest to $\hat{x}_{m,n}$, i.e.,
\begin{align}
    \doublehat{x}_{m,n} = \underset{c \in \Cc}{\mathrm{argmin}} |c - \hat{x}_{m,n}|^2
\end{align}
and recovers the corresponding bits from the used constellation (as shown in Fig.~\ref{fig:bkg_16_qam} for 16-QAM).
Soft demapping aims at providing probabilities over the transmitted bits in the form of \glspl{LLR}. The \gls{LLR} corresponding to the $q^{\text{th}}$ bit on the \gls{RE} $(m,n)$ is given by
\begin{align}
    \text{LLR}_{m,n}(q) &= \ln{ \frac{P(b_{m,n,q}=1)}{P(b_{m,n,q}=0)} }\\
                        &= \ln{\frac{
                            \sum_{c\in\mathcal{C}_{q,1}} \exp{  - \frac{1}{\rho_{m, n}^2} \abs{ \hat{x}_{m, n} - c}^2 }
                            }{
                            \sum_{c\in\mathcal{C}_{q,0}} \exp{  - \frac{1}{\rho_{m, n}^2} \abs{\hat{x}_{m, n} - c}^2 }
                            } }
\end{align}
where $\mathcal{C}_{q,0}$~($\mathcal{C}_{b,1}$) is the subset of $\mathcal{C}$ which contains all symbols with the $q^{\text{th}}$ bit set to 0~(1).

\bigskip
\emph{Note on channel estimation:} Although is it sometimes assumed that the channel coefficients $h_{m,n}$ are known to the reciever, in practice only channel estimates $\hat{h}_{m,n}$ are available.
Such estimations are usually obtained by transmitting pre-determined \emph{pilot signals} $p_{m,n}$ on a set of fixed \glspl{RE}.
The receiver can then estimate the channel on the \glspl{RE} $(m,n)$ carrying pilots with \mbox{$\hat{h}_{m,n} = \frac{y_{m,n}}{p_{m,n}}$}, and extrapolate these channel estimates to the all remaining \glspl{RE}.
To alleviate the overhead associated with pilot transmissions, end-to-end systems are able to perform pilotless communication by learning constellation that are not circularly symmetrical.
More details on the channel estimation process will be presented in Chapter~\ref{ch:2} and~\ref{ch:3}.

\subsection{Uplink Multiple-Input Multiple-Output Systems}
\label{sec:mimo_systems}

In \gls{MU-MIMO} systems, $K$ single antenna users communicate with a \gls{BS} equipped with $L$ antennas.
In uplink transmissions, the vectors of transmitted and received symbols on each \gls{RE} $(m,n)$ are respectively denoted by $\xv_{m,n} \in \Cc^{K}$ and $\yv_{m,n} \in \CC^{L}$.
The channel model between all users and the \gls{BS} antennas is 
\begin{align}
    \label{eq:_ofdm_mimo_model}
    \yv_{m,n} = \Hm_{m,n} \xv_{m,n} + \nv_{m,n}
\end{align}
where the noise vector is denoted by $\nv_{m,n} \sim \Cc \Nc \LB \mathbf{0}, \sigma^2 \Id_{L} \RB$ and $\Hm_{m,n} \in \CC^{L \times K}$ is the matrix of channel coefficients.
Although the transmit process is unchanged for each user, the detection of $K$ users per \gls{RE} by the \gls{BS} requires a new equalization algorithm.
Assuming that the channel matrix $\Hm_{m,n}$ is known, the optimal hard detection algorithm to minimize the probability of symbol error is the maximum likelihood detector:
\begin{align}
    \label{eq:intro_ml_detector}
    \doublehat{\xv}_{m,n} = \underset{\xv_{m,n} \in \Cc^{K}}{\mathrm{argmin}} || \yv_{m,n} - \Hm_{m,n} \xv_{m,n} ||^2.
\end{align}
However, its complexity being exponential in $K$ often prevents any practical implementations.
A simple soft-detection algorithm is zero forcing, in which the constellation constraint on $\xv_{m,n}$ is removed:
\begin{align}
    \hat{\xv}_{m,n} &= \underset{\xv_{m,n} \in \CC^{K}}{\mathrm{argmin}} || \yv_{m,n} - \Hm_{m,n} \xv_{m,n} ||^2 \label{eq_intro_zero_forcing_obj}\\
                &= \LB\Hm_{m,n}\htp \Hm_{m,n} \RB^{-1} \Hm_{m,n} \htp \yv_{m,n} \label{eq:intro_zero_forcing} \\
        &= \xv_{m,n} + \LB\Hm_{m,n}\htp \Hm_{m,n} \RB^{-1} \Hm_{m,n} \htp \nv_{m,n}
\end{align}
resulting in estimated symbols $\hat{\xv}_{m,n}$ with zero intersymbol interferences.
Note that the notations $\doublehat{\cdot}$ and $\hat{\cdot}$ respectively denote the outputs obtained through hard and soft detection.
The main drawback of this detector is that the noise can be significantly amplified on ill-conditionned channels, resulting in poor performances. 
To understand the cause of this effect, let us factorize the channel matrix $\Hm_{m,n}$ into its \gls{SVD} decomposition:  $\Hm_{m,n} = \Um \boldsymbol{\Lambda} \Vm\htp$, where $\Um \in \CC^{L\times L}$ and $\Vm \in \CC^{K\times K}$ are unitary matrices, and $\boldsymbol{\Lambda} \in \RR^{L\times K}$ is a rectangular diagonal matrix with element $[ \lambda_1, \cdots, \lambda_K ]$ on the diagonal.
The symbol estimate can be re-written as
\begin{align}
    \hat{\xv}_{m,n} = \xv_{m,n} + \Vm \boldsymbol{\Lambda}^{-1} \Um\htp \nv_{m,n}
\end{align}
where $\boldsymbol{\Lambda}^{-1} $ has elements $[ \frac{1}{\lambda_1}, \cdots, \frac{1}{\lambda_K} ]$ on the diagonal.
On ill-conditionned channels, some singular values $\lambda_l, l\in \{1, \cdots, L \}$ are very small, resulting in a matrix $\boldsymbol{\Lambda}^{-1}$ containing large entries that amplify the noise accordingly.  
To reduce the sensitivity of the detector to the sigular values of the channel, a regularization term $\sigma^2 ||\xv_{m,n}||^2$ can be added to the objective function~\eqref{eq_intro_zero_forcing_obj}:
\begin{align}
    \hat{\xv}_{m,n} &= \underset{\xv_{m,n} \in \CC^{K}}{\mathrm{argmin}} || \yv_{m,n} - \Hm_{m,n} \xv_{m,n} ||^2 + \sigma^2 ||\xv_{m,n}||^2.
\end{align}
The linear solution is equivalent to a Wiener filter and is referred to as the \gls{LMMSE} detector. It will be discussed in more details in Chapters 3 and 5.

\begin{figure}[b]
    \centering
    \includegraphics[height=0.85\textheight]{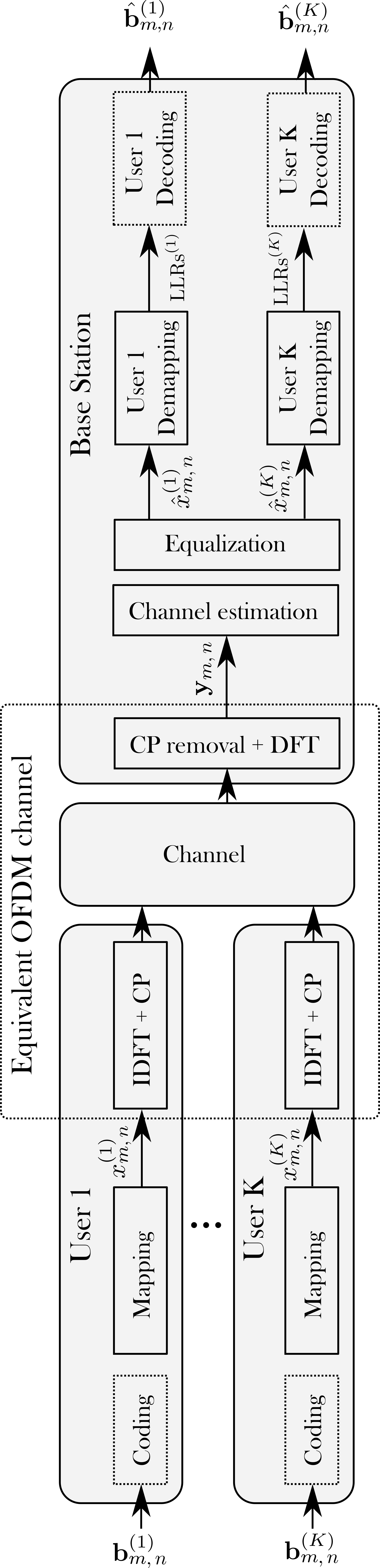}
    \caption{An uplink \gls{MU-MIMO} system.}
    \label{fig:bkg_mimo}
\end{figure}

After equalization, the LLR on the transmitted bit $q$ can be estimated for an OFDM symbol $m$, subcarrier $n$, and user $k$ independently:
\begin{align}
    \text{LLR}_{m,n,k}(q) = \ln{\frac{
                            \sum_{c\in\mathcal{C}_{q,1}} \exp{  - \frac{1}{\rho_{m, n, k}^2} \abs{ \hat{x}_{m, n, k} - c}^2 }
                            }{
                            \sum_{c\in\mathcal{C}_{q,0}} \exp{  - \frac{1}{\rho_{m, n, k}^2} \abs{\hat{x}_{m, n, k} - c}^2 }
                            } }
\end{align}
where $\rho_{m,n,k}^2$ denotes the post-equalization noise variance corresponding to the user $k$ on the \gls{RE} $(m,n)$.
The transmission by $K$ users of the bits corresponding an \gls{FBS} $(m,n)$ is depicted in Fig.~\ref{fig:bkg_mimo}, where the superscript $(k)$ indicates that the quantity at hand is only considered for user $k$.
Note that channel coding (decoding) blocs can be used at the transmitter (receiver) to detect and correct errors on the estimated bits. 

\clearpage
\section{Deep Learning for the Physical Layer}


\subsection{A General Introduction to Deep Learning}

\begin{figure}
    \centering
    \includegraphics[width=0.6\textwidth]{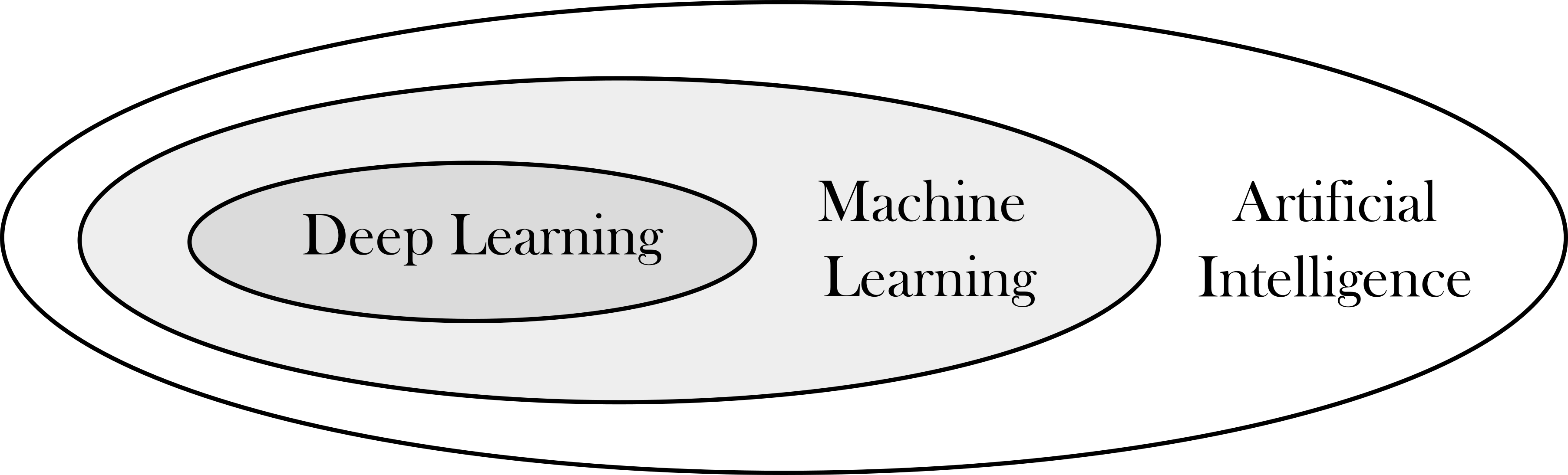}
    \caption{Deep Learning is a subset of Machine Learning, itself a subset of Artificial Intelligence.}
    \label{fig:bkg_dl}
\end{figure}

The concept of \emph{\gls{AI}} can be traced back to 1842, when Ada Lovelace and other mathematicians started wondering whether machine could become intelligent.
The notion of a computer "intelligence" is typically related to its ability to perform tasks commonly associated with intelligent beings.
For example, one of the first successes of \gls{AI} is the victory of IBM's Deep Blue computer against the chess world champion Garry Kasparov in 1997.
The term of \glsreset{ML}\emph{\gls{ML}} refers to \gls{AI} systems capable of extracting patterns from raw data, and can be therefore considered as a subset of \gls{AI}~\cite{Goodfellow-et-al-2016}.
The performance of an \gls{ML}-based algorithm typically depends on the quality of the information given, also known as the \emph{features}.
For example, whereas the height, age, or weight of a medical patient are numbers that represent relevant information, the pixel values of a scan are more difficult to interpret.
A subset of \gls{ML} algorithms therefore focuses on learning useful representations from raw input data.
Among other solutions, \glsreset{DL}\emph{\gls{DL}} systems use a suite of simple mathematical functions to learn as many intermediate representations of the input data.
These simple functions are usually referred to as \emph{layers}, and contain parameters that needs to be optimized.
The relation between \gls{DL}, \gls{ML}, and \gls{AI} is illustrated in Fig.~\ref{fig:bkg_dl}.

\subsubsection*{The concept of gradient descent}

\begin{figure}
    \centering
    \includegraphics[width=0.65\textwidth]{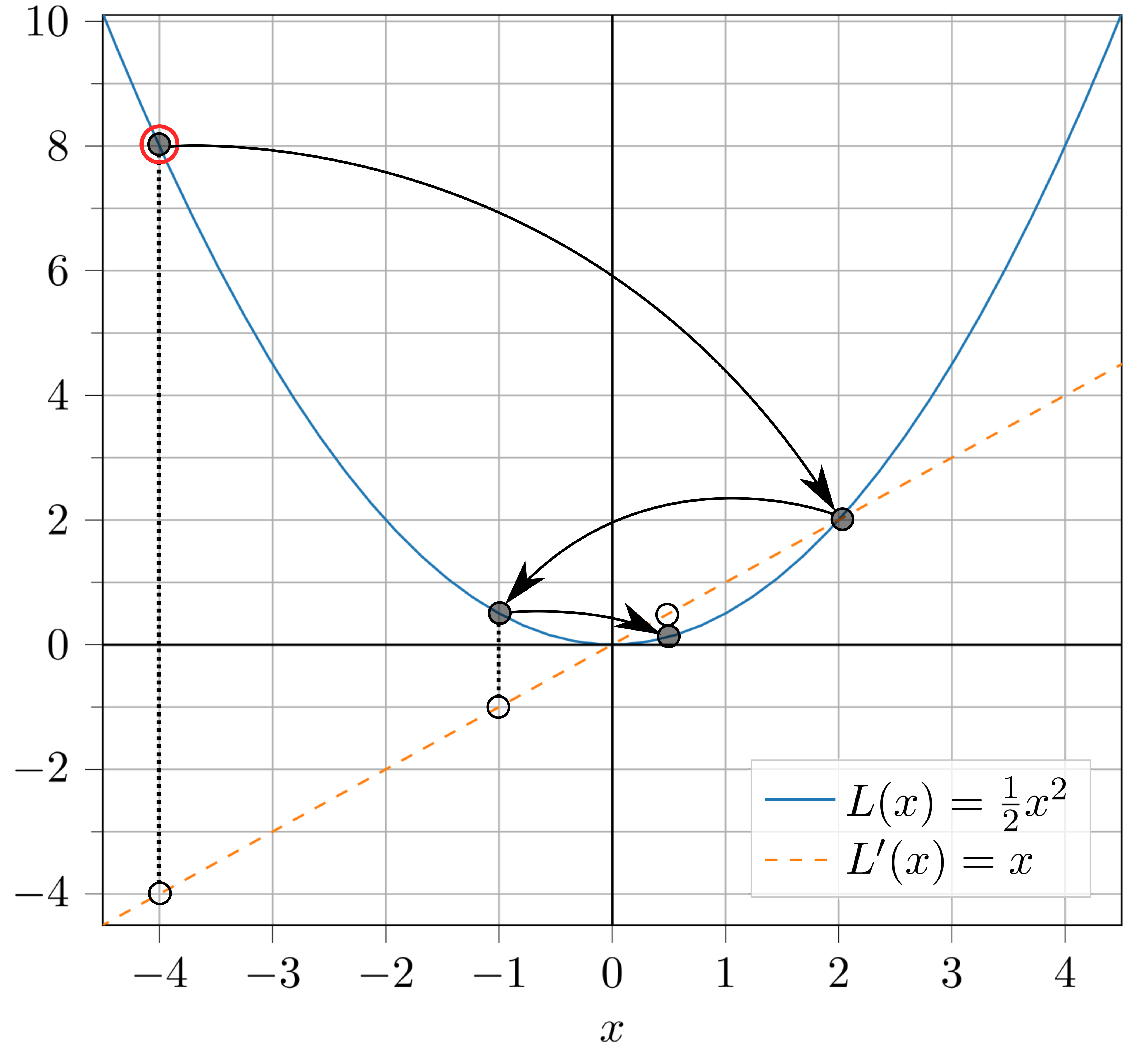}
    \caption{A visual representation of a gradient descent.}
    \label{fig:bkg_sgd}
\end{figure}

The vast majority of \gls{DL}-based systems use gradient-based algorithms to optimize the parameters of every layer.
Such optimization problems involve a \emph{loss function} that needs to be minimized, and that is related to the performance of the system.
Let us define a simple loss function 
\begin{align}
    l = L(x) = \frac{1}{2}x^2, \quad x \in \RR
\end{align}
that needs to be minimized, i.e. we want to find the optimal $x^*$ such that $x^* = \arg \min L(x)$.
To that aim, we often use the derivative function, which gives the slope of the function $L(x)$ at any point $x$.
The derivative of $L(x)=\frac{1}{2}x^2$ is simply given by
\begin{align}
    L'(x) = x.
\end{align}
$L(x)$ and $L'(x)$ are both represented in Fig.~\ref{fig:bkg_sgd}, where one can see that $L(x)$ is decreasing for negative values of $x$, and therefore the associated derivatives are negative.
By opposition, $L(x)$ is increasing for positive values of $x$, and therefore the derivatives are positive.

It is clear that one should update $x$ in the direction opposite to the derivative in order to minimize $L(x)$.
From this observation, the following optimization step can be derived
\begin{align}
    \label{eq:bkg_sgd_simple}
    x^{(i+1)} = x^{(i)} - \eta L'\LB x^{(i)} \RB
\end{align}
where the superscript $(i)$ denote the $i^{\text{th}}$ iteration of the algorithm and $\eta$ is a hyperparameter that defines the size of the optimization step, also known as the \emph{learning rate}.
For the first iteration, $x^{(0)}$ is usually chosen randomly.
To better grasp the intuition behind the algorithm, let us perform some iteration steps, starting with $x^{(0)}=-4$ and using $\eta=\frac{3}{2}$. 
These steps are illustrated in Fig.~\ref{fig:bkg_sgd}, where $x^{(0)}$ is represented by a red circle.
\begin{enumerate}
    \setcounter{enumi}{-1}
    \item We begin with $x^{(0)}=-4$. 
    \item At the first iteration, we start by computing $L'\LB x^{(0)} \RB= L'(-4) = -4 $. Then, one optimization step can be performed: $x^{(1)} = x^{(0)} - \frac{3}{2} L'\LB x^{(0)} \RB = -4 + \frac{3}{2} \times 4 = 2$.
    \item At the second iteration, we have $x^{(1)}=2, L'(2) = 2$, and we can compute $x^{(2)} = x^{(1)} - \frac{3}{2} L'\LB x^{(1)} \RB = 2 - 3 = -1$.
    \item At the third iteration, $x^{(2)}=-1, L'(-1) = -1$, and $x^{(3)} = x^{(2)} - \frac{3}{2} L'\LB x^{(2)} \RB  = 0.5$.
\end{enumerate}
Through these three iterations, $L(x)$ evolved from $L \LB x^{(0)} \RB = 8$ to $L \LB x^{(3)} \RB = 0.125$, thus becoming closer to the minimum value $L \LB x^* \RB = 0$ attainable with  $x^* = 0$.
Please note that the value $\eta=\frac{3}{2}$ is only used here to visualize the different optimization steps, as practical values are usually in the range $\LSB 10^{-5}, 10^{-2} \RSB$.
Finally, the derivative $L'(x)$, also denoted by $\frac{dl}{dx}$, can be extended to a \emph{gradient} if multiple parameters are optimized.
Such gradient is denoted by 
\begin{align}
    \nabla_{\xv} l  = \LSB \frac{\partial l}{\partial x_1}, \cdots, \frac{\partial l}{\partial x_K} \RSB\tp
\end{align}
and is a vector containing the partial derivative of $l$ \gls{wrt} the $K$ parameters to be optimized $\LSB x_1, \cdots, x_K \RSB \tp = \xv$. 
The algorithm performing~\eqref{eq:bkg_sgd_simple}, where $x$ is updated in the direction opposite to the gradient, is therefore known as a \emph{gradient descent} algorithm.

\subsubsection*{Gradient backpropagation}

\begin{figure}
    \centering
    \includegraphics[width=0.7\textwidth]{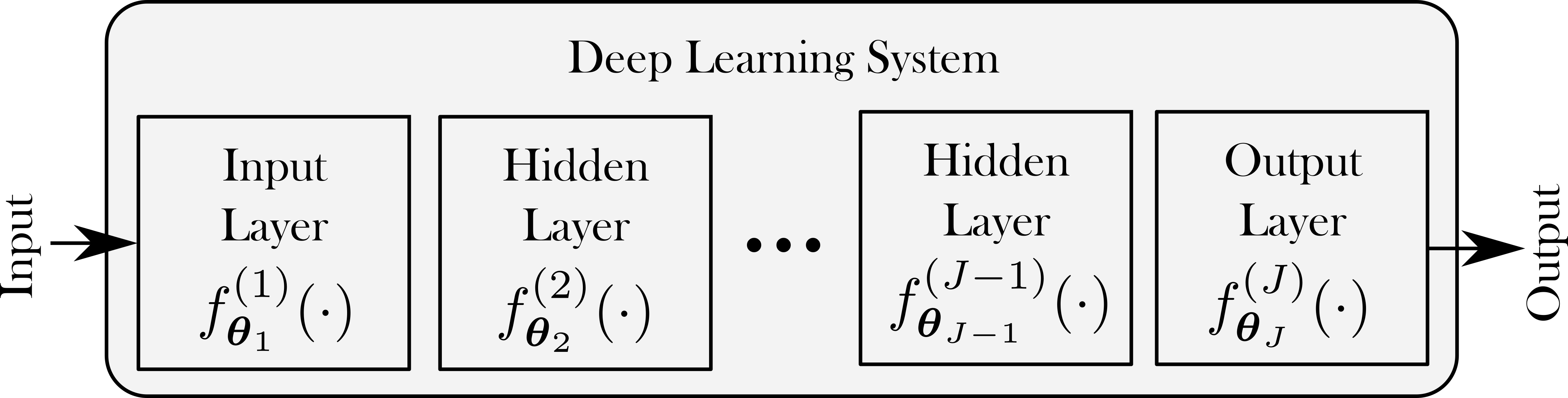}
    \caption{A \gls{DL} system with $J$ layers.}
    \label{fig:bkg_nn_simple}
\end{figure}

As defined previously, \gls{DL} systems use a suite of simple mathematical functions, referred to as layers, to learn different representations of the input data.
The first layer, to which the input data is fed, is the \emph{input layer}, and the last layer, which outputs the estimated quantities, is the \emph{output layer}. 
In between them, "deep" systems typically use multiple \emph{hidden layers}.  
All these layers have trainable parameters that needs to be optimized so that the \gls{DL} system can perform the desired task.
Fig.~\ref{fig:bkg_nn_simple} gives a visual representation of a \gls{DL} system with $J$ layers, each layer implementing a function $f^{(j)}_{\thetav_j} (\cdot)$ with trainable parameters $\thetav_j$.

For example, let us consider a \gls{DL} system that learns to predict the time of flight $t$ of a projectile thrown with an initial velocity $v$, and angle of launch $\alpha$, and being thrown at a height $h$ from the ground.
For simplicity, the system is only composed of one input layer $ y = f^{(1)}_{\thetav_1} (\xv)$ and one output layer $z = f^{(2)}_{\theta_2} (y)$, where $\thetav_1 = \LSB \theta_{1, 1}, \theta_{1, 2}, \theta_{1, 3} \RSB \tp \in \RR^3 $ and $\theta_2 \in \RR$ are the parameters that needs to be optimized.
The system input is denoted by $\xv = [v, \alpha, h]\tp \in \RR^3$, and $ y \in \RR, z \in \RR$ are the output of the first and second layers, respectively.
The loss function calculates an error between the estimated time $z$ and the true time of flight $t$ and is defined by $L(z, t)$.
Please note that we now aim at minimizing $L(z, t)$ \gls{wrt} the parameters $\thetav_1$ and $\theta_2$, in comparison with the minimization illustrated in~Fig.~\ref{fig:bkg_sgd} that was carried out \gls{wrt} $x$.
The time of flight estimated by the \gls{NN} is given by
\begin{align}
    \label{eq:bkg_double_func}
    z = f^{(2)}_{\theta_2} \LB y \RB  = f^{(2)}_{\theta_2} \LB f^{(1)}_{\thetav_1} (\xv) \RB 
\end{align}
and the associated loss is 
\begin{align}
    \label{eq:bkg_loss_1}
    l = L(z, t) = L \LB f^{(2)}_{\theta_2} \LB f^{(1)}_{\thetav_1} (\xv) \RB , \frac{1}{2} mv^2 \RB.
\end{align}
We focus here on minimizing $L(z, t)$ \gls{wrt} the parameter of the input layer $\theta_{1,1}$, as the minimization \gls{wrt} other parameters would be similar.
To apply the gradient descent algorithm on $\theta_{1,1}$, the chain rule of derivation is a useful tool:
\begin{align}
    \frac{\partial l}{\partial \theta_{1, 1}} =  \frac{\partial l}{\partial z} \frac{\partial z}{\partial \theta_{1, 1}} =  \frac{\partial l}{\partial z} \frac{\partial z}{\partial y} \frac{\partial y}{\partial \theta_{1, 1}} 
\end{align}
From here, it can be seen that to compute the derivative of the loss \gls{wrt} a parameter in the first layer, gradients on the loss function and on the second layer needs to be computed as well.
In the following, we show that it is preferable to start by computing the gradient of the loss function $\frac{\partial l}{\partial z} $, then of the last layer $ \frac{\partial z}{\partial y}$, and finally of the first layer $\frac{\partial y}{\partial \theta_{1,1}} $.
This process, consisting in computing gradient from the loss function to the desired layer, is known as gradient $backpropagation$.

\subsubsection*{Performing backpropagation through a neural network}

\begin{figure}
    \centering
    \includegraphics[width=0.4\textwidth]{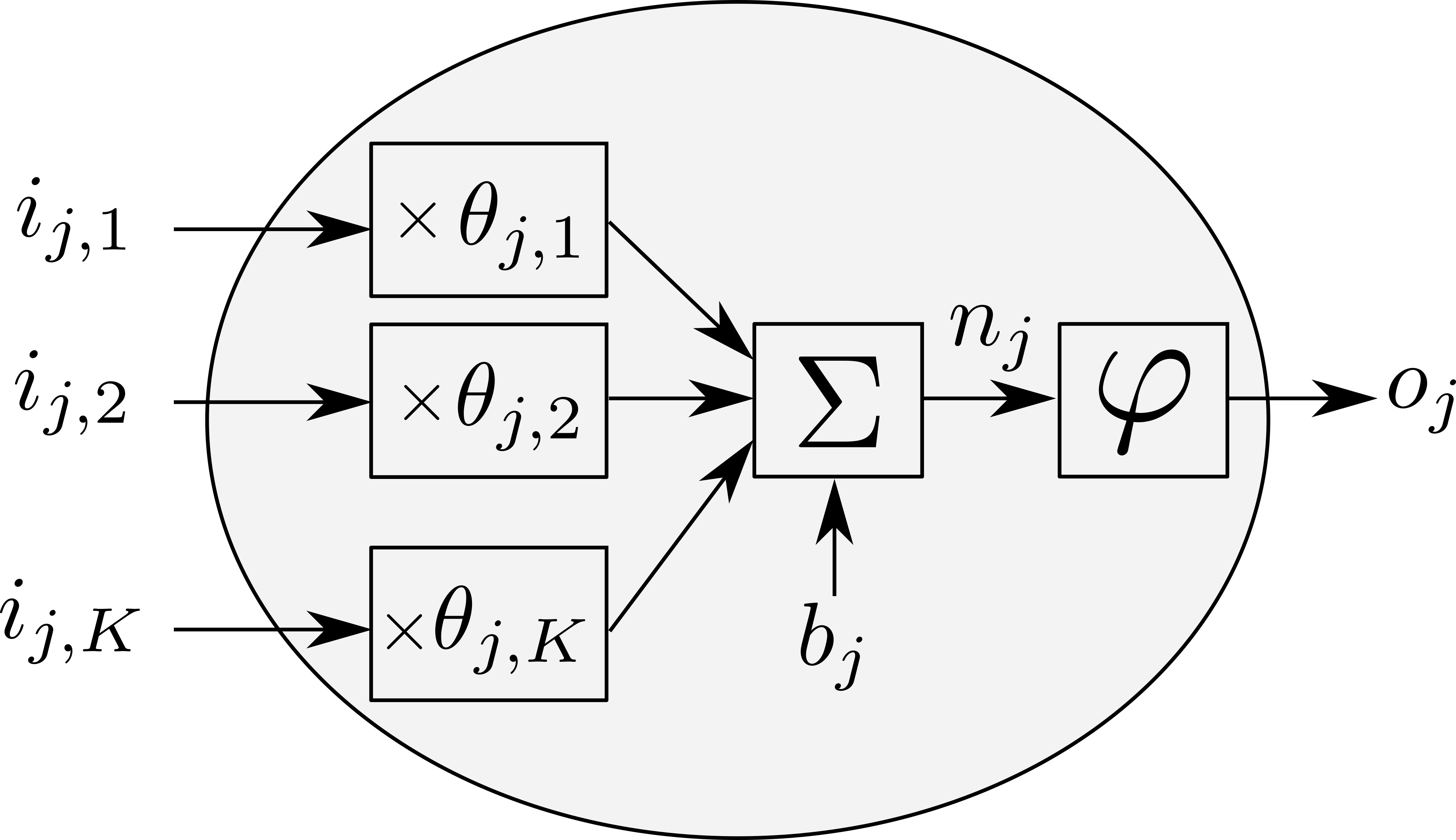}
    \caption{Representation of an artificial neuron.}
    \label{fig:bkg_neuron}
\end{figure}

The core elements of \emph{\glspl{NN}} are \emph{neurons}, each neuron $j$ performing (Fig.~\ref{fig:bkg_neuron})
\begin{align}
    \label{eq:bkg_neuron}
    o_j = \varphi(n_j) = \varphi \LB \sum_{k=1}^{K} \theta_{j,k} i_{j,k} + b_j \RB
\end{align}
where $i_{j,k} \in \RR$  is the $k^{\text{th}}$ input to the neuron (out of $K$), $o_j \in \RR$ is the unique neuron output,  $\varphi(\cdot)$ is an \emph{activation function}, $n_j$ is the neuron output prior to the activation function, $\theta_{j,k}$ is the trainable \emph{weight} corresponding to the $k^{\text{th}}$ input of the $j^{\text{th}}$ neuron, and $b_j$ is a trainable \emph{bias}.
In a neuron, the \emph{trainable parameters} refer to the set comprising all the trainable weights and biases.
For a neuron in the input layer, the vector $\iv_j = \LSB i_{j,1}, \cdots, i_{j,K} \RSB \tp$ corresponds to the input vector $\xv$, whereas $o_j =z$ for a neuron in the output layer (assuming a single output for simplicity).
Without activation function, every neuron would perform a linear transformation, and an \gls{NN} would simply be a composition of linear transformations, resulting in one large linear transformation.
The aim of the activation function is to add non-linearities to the \gls{NN} processing, and therefore to enable the handling of tasks that are more complex than  linear regression problems.

Let us consider an \gls{NN} with one input and one output layer, comprising one neuron each without bias ($b_1 = b_2 = 0$).
The first layer takes an input vector of dimension three $\xv \in \RR^3$ and outputs a scalar $y$, while both the input $y$ and the output $z$ of the second layer are scalars. 
This \gls{NN} can be expressed similarly to~\eqref{eq:bkg_double_func}, with the first and second neurons being respectively implemented by $f^{(1)}_{\thetav_1}$ and $f^{(2)}_{\theta_2}$:
\begin{align}
    \label{eq:bkg_nn_midle}
    z &= f^{(2)}_{\theta_2} \LB y \RB \nonumber \\
      &= f^{(2)}_{\theta_2} \LB f^{(1)}_{\thetav_1} (\xv) \RB \\
      &= \varphi \LB \theta_2 \LB \varphi \LB \sum_{k=1}^{3} \theta_{1,k} x_k \RB \RB \RB. \nonumber
\end{align}
In this example, the activation function is chosen to be the \emph{sigmoid} function, defined as:
\begin{align}
    \label{eq:bkg_sigmoid}
    \varphi(a) = \frac{1}{1+\exp{-a}}
\end{align}
for which the derivative is
\begin{align}
    \frac{d \varphi(a)}{da} = \varphi(a)(1-\varphi(a)).
\end{align}
Finally, the loss in~\eqref{eq:bkg_loss_1} is chosen to be the half of the squared error:
\begin{align}
    l = L(z, t) = \frac{1}{2} (z-t)^2
\end{align}
for which the derivative \gls{wrt} $z$ is
\begin{align}
    \label{eq:bkg_d_loss}
    \frac{\partial l}{\partial z} = z - t.
\end{align}

First, let us compute the derivative of $l$ \gls{wrt} to the parameter of the output layer $\theta_2$:
\begin{align}
    \label{eq:bkg_d_t1}
    \frac{\partial z}{\partial \theta_2} = \frac{\partial l}{\partial z} \frac{\partial z}{\partial \theta_2} =\frac{\partial l}{\partial z} \frac{\partial z}{\partial n_2} \frac{\partial n_2}{\partial \theta_2}
\end{align}
%
Each element of the right-hand side equation can be easily computed:
\begin{itemize}
    \item $\frac{\partial l}{\partial z} = z - t$ as per~\eqref{eq:bkg_d_loss}.
    \item $\frac{\partial z}{\partial n_2} = \frac{\partial \varphi(n_2)}{\partial n_2} =\varphi(n_2)(1-\varphi(n_2)) = z(1-z)$.
    \item $\frac{\partial n_2}{\partial \theta_2} = \frac{\partial \theta_2 y}{\partial \theta_2} = y$
\end{itemize}
Substituting each element in~\eqref{eq:bkg_d_t1}, we obtain
\begin{align}
    \frac{\partial z}{\partial \theta_2} = (z - t)z(1-z)y = \delta_2 y
\end{align}
with 
\begin{align}
    \delta_2 = \frac{\partial l}{\partial z} \frac{\partial z}{\partial n_2} = (z - t)z(1-z).
\end{align}

We can now compute the derivative of $l$ \gls{wrt} the parameter of the first layer $\theta_{1, 1}$:
\begin{align}
    \frac{\partial l}{\partial \theta_{1, 1}} = \frac{\partial l}{\partial z} \frac{\partial z}{\partial n_2} \frac{\partial n_2}{\partial y}\frac{\partial y}{\partial n_1} \frac{\partial n_1}{\partial \theta_{1, 1}}
\end{align}
\begin{itemize}
    \item The first two elements have been computed above:  $\frac{\partial l}{\partial z} \frac{\partial z}{\partial n_2} = \delta_2 $
    \item $\frac{\partial n_2}{\partial y} = \frac{\partial \theta_2 y}{\partial y} = \theta_2$
    \item $\frac{\partial y}{\partial n_1} = \frac{\partial \varphi(n_1)}{\partial n_1} = y(1-y)$ 
    \item  $\frac{\partial n_1}{\partial \theta_{1,1}} = \frac{\partial \theta_{1, 1} x_1 + \theta_{1, 2} x_2}{\partial \theta_{1, 1}} = x_1$
\end{itemize}
which leads to 
\begin{align}
    \frac{\partial l}{\partial \theta_{1, 1}} = \delta_2 \theta_2 y(1-y)  x_1 = \delta_1 x_1
\end{align}
with 
\begin{align}
    \delta_1 = \delta_2 \theta_2 y(1-y).
\end{align}
As one can see, the derivative needs to be computed starting from the last layer.
For an \gls{NN} with $J$ layers, the backpropagation algorithm successively computes $\delta_J,  \delta_{J-1},  \cdots, \delta_1$.
Using the notation of~\eqref{eq:bkg_neuron}, i.e., respectively denoting by $o_j$ and $i_{j,k}$ the output and inputs of a neuron $j$ and $\theta_{j,k}$ the parameter corresponding to the $k^{\text{th}}$ input of this same neuron, the backpropagation to any layer $j$ is given by:
\begin{align}
    \label{eq:bkg_backprop}
    \frac{\partial l}{\partial \theta_{j,k}} = i_{j,k} \delta_j, \text{ with }
    \delta_j = 
    \left\{
    \begin{array}{ll}
        (o_j-t)o_j(1-o_j) & \text{if } $j$ \text{ is an output neuron,} \\
        \LB \sum_{m=1}^{M} \theta_{m,j} \delta_m \RB o_j(1-o_j) & \text{otherwise}
    \end{array}
\right.
\end{align}
where $M$ denotes the number of neurons in the $(j+1)^{\text{th}}$ layer, and the summation accounts for the fact that each layer can have multiple neurons, which was not considered in the example above for clarity.
Please note that the expression~\eqref{eq:bkg_backprop} is only valid for an \gls{NN} that does not have any biases, uses logistic activation functions, and is evaluated using the loss $L(z, t) = \frac{1}{2}(z-t)^2$.

\begin{figure}
    \centering
    \includegraphics[width=0.4\textwidth]{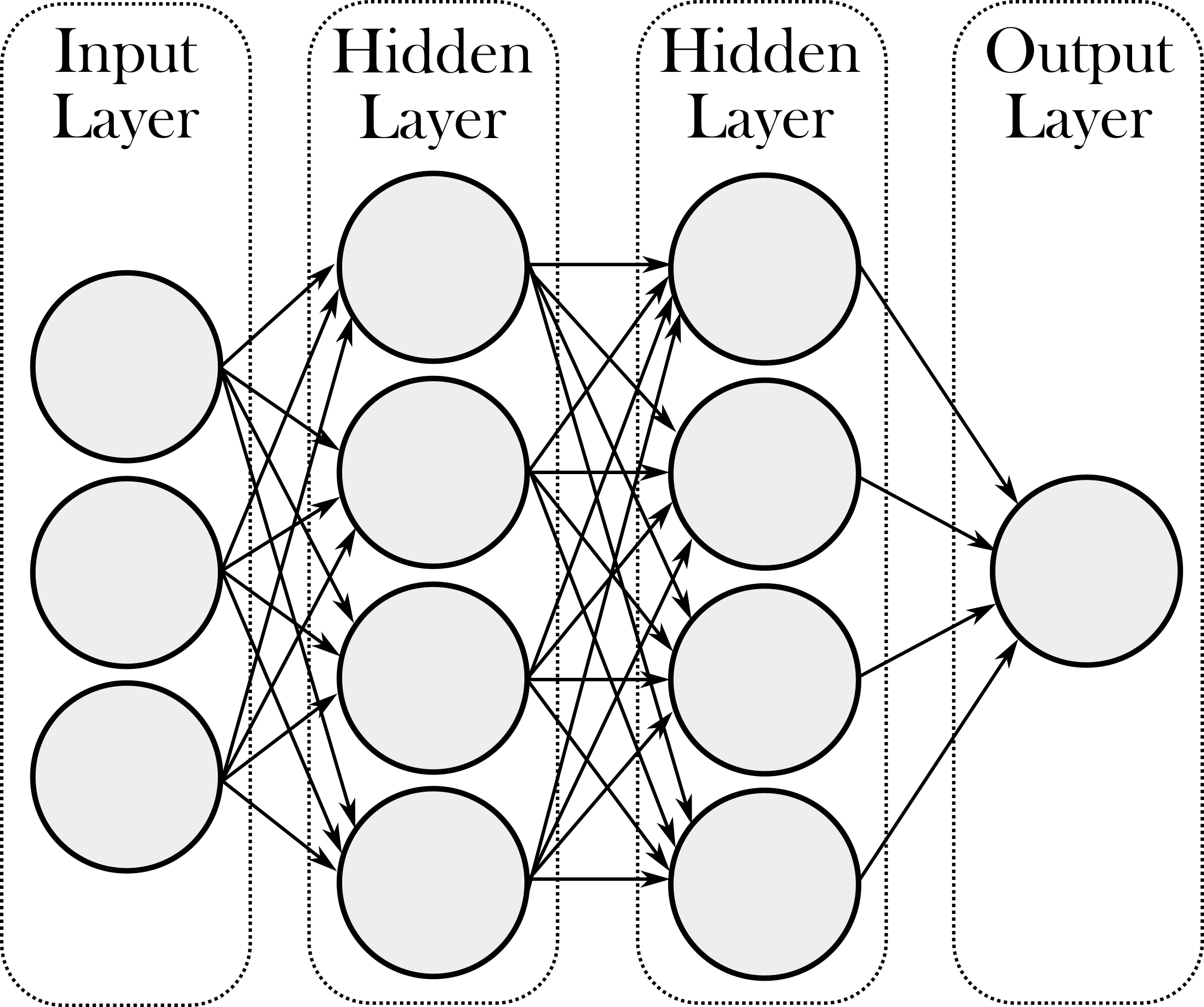}
    \caption{Representation of a neural network.}
    \label{fig:bkg_nn_neuron}
\end{figure}

One of the most used type of \glspl{NN} is the fully connected neural network (FCNN).
In such \glspl{NN}, each layer is \emph{dense}, i.e., composed of neurons for which the inputs are the outputs of all the neurons from the preceding layer.
A representation of an FCNN with 4 dense layers respectively containing three, four, four, and one neuron is presented in Fig.~\ref{fig:bkg_nn_neuron}.
Note that the \gls{NN} output dimension corresponds to the number of neuron in the output layer, and can be greater than one.

\subsubsection*{The stochastic gradient descent algorithm}

To train an \gls{NN} such as the one defined in~\eqref{eq:bkg_nn_midle}, a \emph{dataset} comprising \emph{features} and \emph{labels} is needed.
Let us denote by $D_S$ the size of this dataset, which typically contains thousands or millions of such samples.
In the example of a system that learns to predict the time of flight of a projectile, the features are vectors $\xv^{[s]} = \LSB v^{[s]}, \alpha^{[s]}, h^{[s]} \RSB\tp, s\in \{1, \cdots, D_S\}$ and the labels are the associated measured time of flights $t^{[s]}, s\in \{1, \cdots, D_S\}$.
If we denote by $\thetav$ the set containing all the trainable parameters and biases of the \gls{NN}, a gradient descent iteration on the dataset is
\begin{align}
    \thetav^{(i+1)} = \thetav^{(i)} - \frac{\eta}{D_S} \sum_{s=1}^{D_S}\nabla_{\thetav} L(z^{[s]}, t^{[s]})
\end{align}
where $z^{[s]}$ is the output of the \gls{NN} corresponding to the input $\xv^{[s]}$, and therefore depends on the parameters $\thetav$.
However, such iteration would be of prohibitive complexity due to the large amount of samples in the dataset.
The \glsreset{SGD}\emph{\gls{SGD}} algorithm therefore randomly select \emph{batches} of $B_S$ samples from the dataset, and at each iteration performs a gradient descent on a single batch only:
\begin{align}
    \thetav^{(i+1)} = \thetav^{(i)} - \frac{\eta}{B_S} \sum_{s=1}^{B_S}\nabla_{\thetav} L(z^{[s]}, t^{[s]}).
\end{align}
After each iteration, another random batch of samples is selected.
The algorithm can be stopped by multiple factors, including when the loss reaches a given threshold or after a fixed number of iterations have been performed.
Multiple enhancements of the \gls{SGD} algorithm have been proposed, an example of which being the Adam optimizer~\cite{Kingma15} which sets individual learning rates for each parameter and computes a moving average of the gradient magnitude.
A standard \gls{SGD} algorithm that is stopped after $I$ iteration is presented in algorithm~\ref{alg:bkg_SGD}.
\begin{algorithm}
    \SetAlgoLined
     Initialize $\thetav^{(0)}$ randomly \\
     \For{$I = 0, \cdots, I $}{
        $\triangleright$Select a random batch of $B_S$ sample from the dataset \\
        $\triangleright$Compute one \gls{NN} inference to obtain  $z^{[s]}, s\in\{1, \cdots, B_S\}$\\
        $\triangleright$Evaluate the losses $L(z^{[s]}, t^{[s]}), s\in\{1, \cdots, B_S\}$\\
        $\triangleright$Perform one gradient descent iteration on the batch:\\
         $\thetav^{(i+1)} = \thetav^{(i)} - \frac{\eta}{B_S} \sum_{s=1}^{B_S}\nabla_{\thetav} L(z^{[s]}, t^{[s]})$ \\
     }
     \caption{A standard \gls{SGD} algorithm}
     \label{alg:bkg_SGD}
\end{algorithm}

\subsection{Layers and Activation Functions}
\label{subsec:layers_act_func}

To simplify the notations in section, we recall that the vector $\mv_a$ and scalar $m_{a,b}$ formed by slicing the matrix $\Mm$ along its first and second dimensions can also be denoted by $[\Mm]_a$ and $[\Mm]_{a,b}$, respectively.

\subsubsection*{Dense layer}
A dense layer, as presented previously, is composed of $K$ neurons for which the inputs are the outputs of all $J$ neurons from the preceding layer.
If we respectively denote by $\xv \in \RR^{J}$ and $\yv \in \RR^{K}$ its input and output vectors of dimension $J$ and $K$, a dense layer can be expressed by 
\begin{align}
    \yv = f(\xv) = \varphi \LB \Thetam \xv + \bv \RB
\end{align}
where $\bv \in \RR^K$  and $\Thetam \in \RR^{K\times J}$ and are respectively the vector of trainable biases and the matrix of trainable weights, and $\varphi(\cdot)$ is the activation function.
One can see that each output $y_k = \varphi \LB  \LSB \Thetam \RSB_{k}\tp \xv + b_k \RB$ corresponds to the processing done by the $k^{\text{th}}$ neuron in the layer.

\subsubsection*{Convolutional layer}
\begin{figure}
    \centering
    \includegraphics[width=0.9\textwidth]{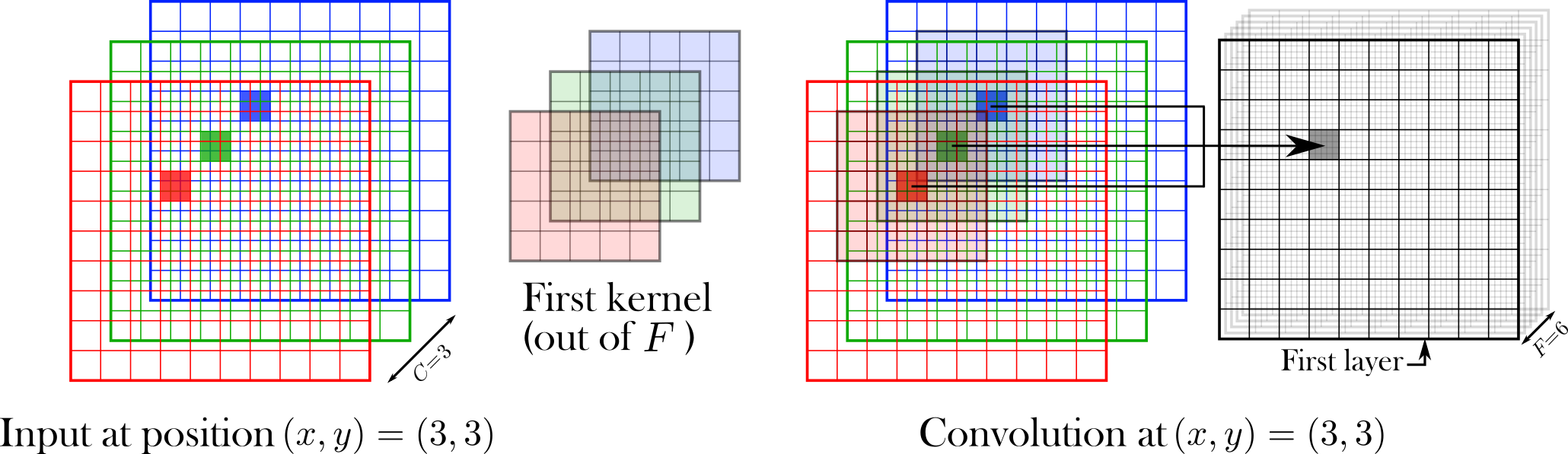}
    \caption{A convolution producing the first output layer (out of 6) at position $(x, y) = (3, 3)$.}
    \label{fig:bkg_conv}
\end{figure}

While dense layers are useful to process one-dimensional data, two- or three-dimensional inputs are usually processed by \emph{convolutional} layers.
Let us denote by $\Xm \in \RR^{V\times H \times  C}$ the 3D input of a convolutional layer.
This could correspond to an image, where $V$ and $H$ are respectively the vertical and horizontal dimensions of the image, and the last dimension $C=3$ corresponds to the red, green, and blue \emph{channels}.
Such convolutional layers have trainable \emph{kernels} $\Km \in \RR^{K_V \times K_H \times C}$, with which the input is convoluted.
$(K_V, K_H)$ is referred to as the \emph{kernel size}, and for simplicity let us assume that $K_V$ and $K_H$ are odd, i.e., $K_V = 2K_V' +1$ and $K_H = 2K_H' +1$.
The output of the convolution at any position $(x,y)$ is given by
\begin{align}
    [\text{conv}(\Xm, \Km)]_{x, y} = \sum_{v=0}^{K_V-1} \sum_{h=0}^{K_H-1}  \sum_{c=0}^{C-1} [\Km]_{v, h, c} [\Xm]_{x-K_V'+v, y-K_H'+h, c}.
\end{align}
For positions $x$ ($y$) lower than $K_V'$ ($K_H'$) or greater than $V-K_V'$ ($H-K_H'$), the indexes $x-K_V'+v$ ($y-K_H'+h$) are outside the dimension of $\Xm$.
This problem can be dealt with by assuming that $[\Xm]_{x-K_V'+v, y-K_H'+h, c} = 0$ at these indexes, which corresponds to a convolution with zero-padding.
Multiple kernels are usually defined for a given convolutional layer. 
Let us denote by $F$ the number of kernels, also known as number of \emph{filters}, and by $\Km_f$ the $f^{\text{th}}$ kernel, with $f\in\{0, \cdots, F-1\}$.
The convolution with each filter defines a new output layer $f$, such as the convolution can be written as
\begin{align}
    [\text{conv}(\Xm, \Km)]_{x, y, f} = \sum_{v=0}^{K_V-1} \sum_{h=0}^{K_H-1}  \sum_{c=0}^{C-1} [\Km_f]_{v, h, c} [\Xm]_{x-K_V'+v, y-K_H'+h, c}.
\end{align}
Such a convolution is depicted in Fig.~\ref{fig:bkg_conv}, where the input has $C=3$ channels of dimension $10 \times 10$ (possibly corresponding to the RGB values of an image), the kernels have dimension $5 \times 5 \times 3$, and the first output layer (out of $F=6$) is represented.
To obtain the final outputs of the convolutional layer, biases $b_f\in\RR$ are added for each convolution layer, and an activation function is applied:
\begin{align}
    [\Ym]_{x, y, f}  &= \varphi \LB [\text{conv}(\Xm, \Km)]_{x, y, f} + b_l \RB \\
                     &= \varphi \LB \sum_{v=0}^{K_V-1} \sum_{h=0}^{K_H-1}  \sum_{c=0}^{C-1} [\Km_f]_{v, h, c} [\Xm]_{x-K_V'+v, y-K_H'+h, c} + b_f \RB
\end{align}
with $\Ym \in \RR^{V \times H \times F}$ being the output matrix.

\begin{figure}
    \centering
    \includegraphics[width=0.8\textwidth]{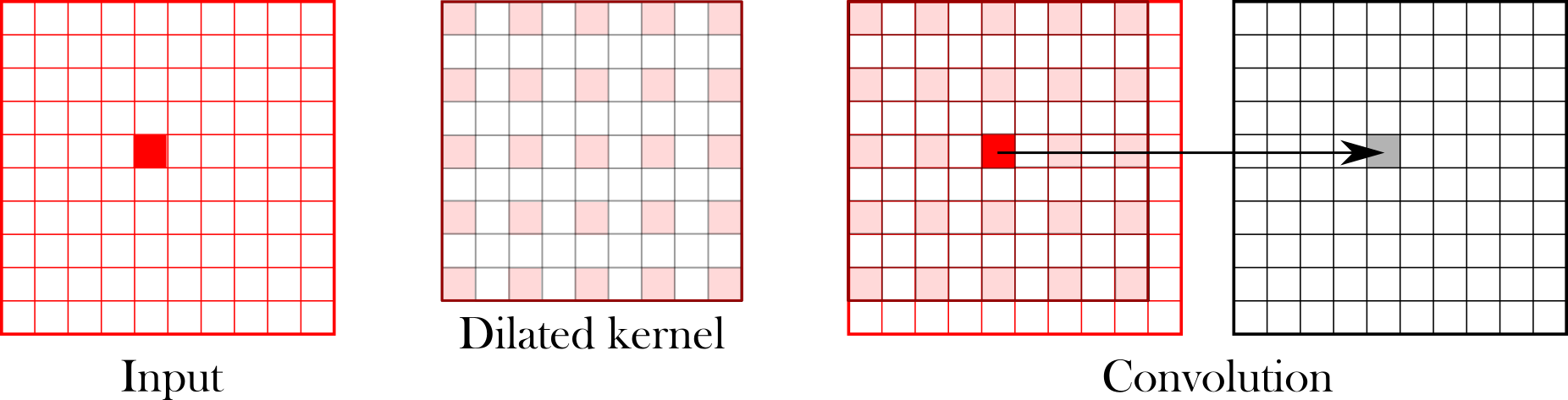}
    \caption{A dilated convolution at position $(x, y) = (5, 5)$, where $C=1$, $F=1$, $D=2$.}
    \label{fig:bkg_dilations}
\end{figure}

The \emph{receptive field} of an \gls{NN} is defined as the dimension of the set of inputs that affects a single output.
To increase the receptive field of a \gls{CNN}, it is common to use \emph{dilated} convolutions, in which the kernels are spread on the inputs:
\begin{align}
    [\text{dilated conv}_D(\Xm, \Km)]_{x, y, f} = \sum_{v=0}^{K_V-1} \sum_{h=0}^{K_H-1}  \sum_{c=0}^{C-1} [\Km_f]_{v, h,c} [\Xm]_{x-DK_V'+Dv, y-DK_H'+Dh, c} 
\end{align}
where $D$ is the dilation parameter.
A dilated convolution with $D=2$ is represented in Fig.~\ref{fig:bkg_dilations}.

\subsubsection*{Separable convolutional layer}
A \emph{separable} convolutional layer is composed of a depthwise convolution followed by a pointwise convolution.
In depthwise convolutions, each kernel $K_f^{\text{(d)}} \in \RR^{K_V \times K_H}$ has only two dimensions, and the number of kernels is the same as the number of input channels, i.e., $F=C$.
Each kernel therefore acts separately on each input channel, resulting in 
\begin{align}
   \Ym^{\text{(d)}}_{x, y, c} =  \sum_{v=0}^{K_V-1} \sum_{h=0}^{K_H-1} [\Km_c^{\text{(d)}}]_{v, h} [\Xm]_{x-K_V'+v, y-K_H'+h, c} + b_c^{\text{(d)}}
\end{align}
where $\Ym^{\text{(d)}}_{x, y, c} \in \RR^{V \times H \times C}$ is the output matrix and $\bv^{\text{(d)}} \in \RR^C$ is the vector of trainable biases.
Then, the pointwise convolution uses $F$ kernels $\Km_f^{\text{(p)}} \in \RR^{1 \times  1 \times C}$:
\begin{align}
    \Ym^{\text{(p)}}_{x, y, f} =  \sum_{c=0}^{C-1} [\Km_f^{\text{(p)}}]_{0, 0, c} [\Ym^{\text{(d)}}]_{x, y, c} + b_f^{\text{(p)}}
 \end{align}
resulting in an output of dimension of $V \times H \times F$.
Finally, the output of the separable convolution is given by
\begin{align}
    \Ym = \varphi \LB \Ym^{\text{(p)}} \RB.
\end{align}
It has been shown that performing a separable convolution instead of a traditional convolution drastically reduces the number of computations and of trainable parameters while maintaining a similar level of performance \cite{howard2017mobilenets}.

\subsubsection*{Batch normalization layer}

During training, the distribution of the outputs corresponding to each layer evolves as the trainable parameters are optimized.
Therefore, each hidden layer needs to constantly readjust its parameters to follow the changes in its input distribution.
This problem is amplified on deep \glspl{NN}, as a change in the output distribution of the first layer can have a significant effect on the input distribution of the last layers.
To tackle this problem, batch normalization layers normalize their inputs so that the corresponding distributions have optimized means and variances.
Let us denote by $\xv^{[s]} \in \RR^J, s \in \{1, \cdots, B_S \}$ the input vectors of a batch normalization layer, where $B_S$ is the batch size.
The mean and variance of each element $x_j$ can be estimated on the batch:
\begin{align}
    \mu_j = \frac{1}{B_S} \sum_{s=1}^{B_S} x_j^{[s]}, \text{ and } \sigma^2_j = \frac{1}{B_S} \sum_{s=1}^{B_S} \LB x_j^{[s]} - \mu_j \RB^2 .
\end{align}
Each dimension is then normalized separately to have zero-mean and unit variance:
\begin{align}
    \hat{x}_j^{[s]} = \frac{x_j^{[s]} - \mu_j}{\sqrt{\sigma^2_j + \epsilon}}
\end{align}
where $\epsilon$ is a small constant that is added to ensure numerical stability.
The means and variances of each dimension $j$ are then typically controlled by trainable parameters $\gamma_j$ and $\beta_j$, respectively:
\begin{align}
    y_j^{[s]} = \gamma_j \hat{x}_j^{[s]} + \beta_j.
\end{align}
This reparametrization of the input distribution enables faster training and improves both the performance and the generalization properties of the \gls{NN} \cite{Goodfellow-et-al-2016}.

\subsubsection*{Residual connections}

\begin{figure}
    \centering
    \includegraphics[width=0.4\textwidth]{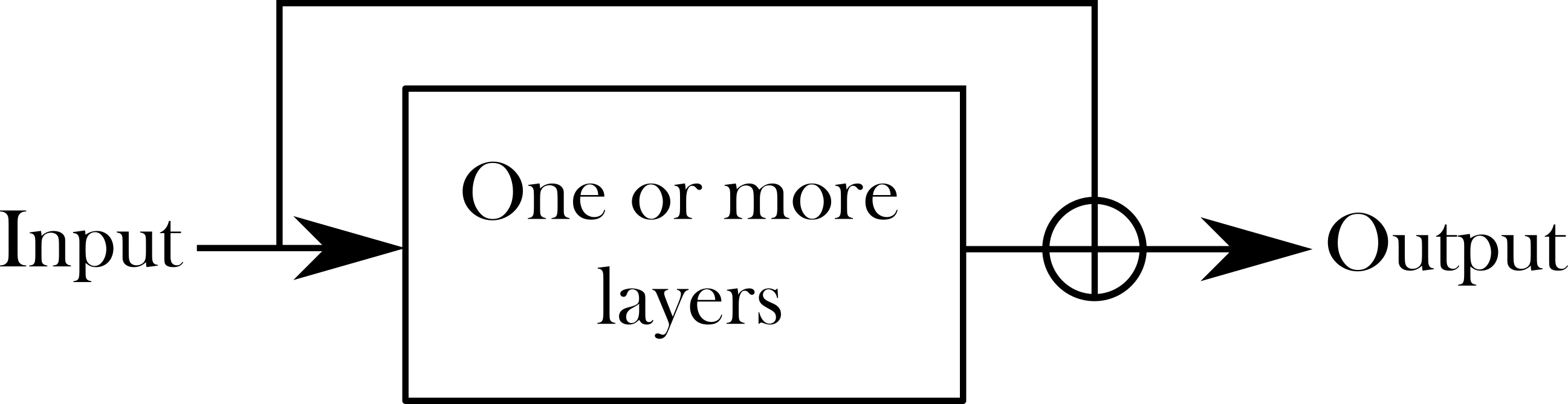}
    \caption{A residual connection.}
    \label{fig:bkg_resnet}
\end{figure}

NNs composed of many layers can be affected by the \emph{vanishing gradient} problem, where the gradients that are backpropagated to the first layers become increasingly small, thus restraining the optimization of their parameters.
Residual connections alleviate this effect by allowing the gradients to skip one or more layers.
A simple implementation consists in adding the input of a (suite of) layer(s) to its output, as depicted in Fig.~\ref{fig:bkg_resnet}.
\Glspl{NN} that contain residual connections are referred to residual networks, or \emph{resnets}.

\subsubsection*{The ReLU activation function}

One of the most used activation function is the \gls{ReLU}, depicted in Fig.~\ref{fig:bkg_act_func}:
\begin{align}
    y = \text{ReLU}(x)
    = \left\{
        \begin{array}{ll}
            0 & \mbox{if } x \leq 0 \\
            x & \mbox{if } x > 0
        \end{array}
        \right.
        = \max (x, 0).
\end{align}
The main advantage of the \gls{ReLU} is its simplicity, which allows for very efficient implementations.
But \glspl{NN} using this activation function might face the dying ReLU problem, where some neurons cannot output anything other than $0$.
Once a neuron is "dead", the gradient of its trainable parameters stays null as $\frac{d \; \text{ReLU}(x)}{dx}=0$ when $x < 0$, thus preventing further training.

\subsubsection*{The ELU activation function}

\begin{figure}[t]
	\centering
	\input{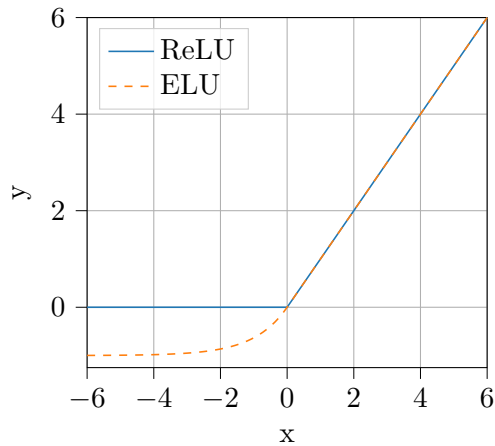}
	\caption{\gls{ReLU} and \gls{ELU} activation functions}
	\label{fig:bkg_act_func}
\end{figure}

To prevent the "dead" neuron problem, multiple variants of the \gls{ReLU} have been proposed.
One of them is the \gls{ELU} (Fig.~\ref{fig:bkg_act_func}), which mimics the \gls{ReLU} function for positive $x$, but maintains a non-constant output for $x<0$:
\begin{align}
    y = \text{ELU}(x)
    = \left\{
        \begin{array}{ll}
            \exp{x}-1 & \mbox{if } x \leq 0 \\
            x & \mbox{if } x > 0
        \end{array}.
        \right.
\end{align}
ELU has been shown to outperform many other ReLU variants \cite{clevert2016fast}, but the use of the exponential function leads to longer computation times. 

\subsection{Optimizing Communication Systems through SGD}

\subsection*{Modeling a communication system as an autoencoder}

\begin{figure}
    \centering
    \includegraphics[width=0.65\textwidth]{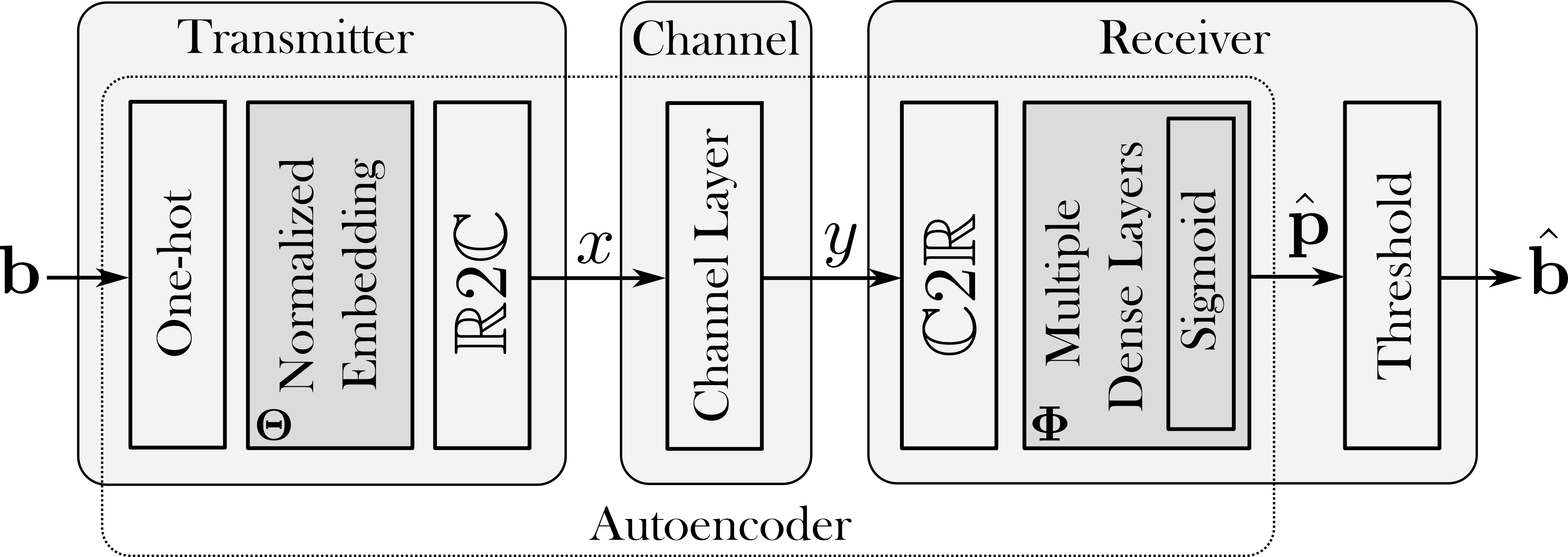}
    \caption{A communication system modeled as an autoencoder. The dark gray elements contain trainable parameters.}
    \label{fig:bkg_nn_comm}
\end{figure}

An autoencoder is a type of \gls{NN} that aims to reconstruct its input at its output. 
Such \glspl{NN} have one "bottleneck" layer of reduced dimensionality compared to its inputs and outputs, which means that an efficient, lower-dimensionality representation of the input data needs to be learned prior to this layer to enable a correct data reconstruction at the output.
The first half of the \gls{NN} is therefore usually referred to the encoder part, while the second half is the decoder part.
A communication system can be seen as an autoencoder, as recovering the transmitted bits involve implementing a bit mapping at the transmitter (encoder) and demapping at the receiver (decoder) that is robust the channel distortions (bottleneck layer).
A simple autoencoder-based communication system with simple \gls{AWGN} channel is detailed in the following (Fig.~\ref{fig:bkg_nn_comm}).

At the transmitter, the first layer is a one-hot layer.
This layer converts the vector of $Q$ bits $\bv\in \{0, 1\}^Q$ into a vector of dimension $2^Q$ containing only zeros except a one at the position corresponding to the decimal representation of $\bv$.
For example, if $Q=2$, $[0, 0]\tp$ is converted to $[1, 0, 0, 0]\tp$, $[0, 1]\tp $ to $[0, 1, 0, 0]\tp$, $[1, 0]\tp $ to $[0, 0, 1, 0]\tp$, and $[1, 1]\tp $ is converted to $[0, 0, 0, 1]\tp$.
The next layer is a normalized embedding layer, which implements
\begin{align}
    f^{(1)}(\av) = \frac{\sqrt{2^Q}}{||\boldsymbol{\Theta}||_F} \boldsymbol{\Theta} \av = \Wm \av 
\end{align}
where $\boldsymbol{\Theta}\in \RR^{2\times 2^Q}$ is a matrix of trainable coefficients, and $\Wm\in \RR^{2\times 2^Q}$ is normalized such that its column vectors have average energy of one.
Finally, the $\RR2\CC$ layers converts $2$ real numbers into a single complex number.
The combination of the one-hot encoding, normalized embedding and $\RR2\CC$ layers outputs symbol $x$ that have unit average energy, i.e., $\EE\LSB ||x||_2^2  \RSB = 1$.

The \gls{AWGN} channel is implemented as a channel layer that performs
\begin{align}
    y = x + n
\end{align}
where  $n \thicksim \Cc \Nc ( 0, \sigma^2)$ is a complex \gls{AWGN} with variance $\sigma^2$.
Such channel is differentiable and therefore allows the gradient to be backpropagated from the receiver to the transmitter.

The first receiver layer is a $\CC2\RR$ layer, which converts a complex number into two real numbers.
Then, the demapping is performed by multiple dense layers with trainable parameters denoted by $\boldsymbol{\Phi}$, the last layer being of dimension $2^Q$ and using the sigmoid activation function as defined in~\eqref{eq:bkg_sigmoid}.
The sigmoid has two advantages.
First, its outputs are in the range $[0, 1]$, and thus can be interpreted as a probability vector $\hat{\pv} = \LSB \hat{p}_0, \cdots,\hat{p}_{Q-1} \RSB \tp$, where each entry $\hat{p}_q$ corresponds to an estimated probability that a $q^{\text{th}}$ bit equals one given $y$:
\begin{align}
    \hat{p}_q = \widehat{P} \LB b_q = 1 | y \RB.
\end{align}
Second, if we denote by \emph{logits} the output of the last layers before the sigmoid activation function, we have
\begin{align}
    \widehat{P} \LB b_q = 1 | y\RB = \frac{1}{1+\exp{-\text{logits}}} \iff \text{logits} = \ln{\frac{\widehat{P} \LB b_q = 1 | y \RB}{\widehat{P} \LB b_q = 0 | y \RB}}.
\end{align}
These logits can thus be used as \glspl{LLR} by a channel decoder, as illustrated in fig.~\ref{fig:bkg_mimo}.

For a perfect transmission, the vector of estimated probabilities $\hat{\pv}$ matches the transmitted bit vector $\bv$, i.e., $\hat{\pv} = \bv$, and therefore the communication system can truly be seen as an autoencoder.
Finally, for each bit $q$, the threshold layer outputs 
\begin{align}
    \label{eq:bkg_threshold}
    \hat{b}_q = \left\{
    \begin{array}{ll}
        1 & \mbox{if } \hat{p}_q >0.5  \\
        0 & \mbox{otherwise}
    \end{array}.
\right.
\end{align}
\subsection*{From \gls{NN}-based systems to \gls{DL}-enhanced systems}

\begin{figure}
    \centering
    \begin{minipage}{.4\textwidth}
        \centering
        \includegraphics[width=.8\linewidth]{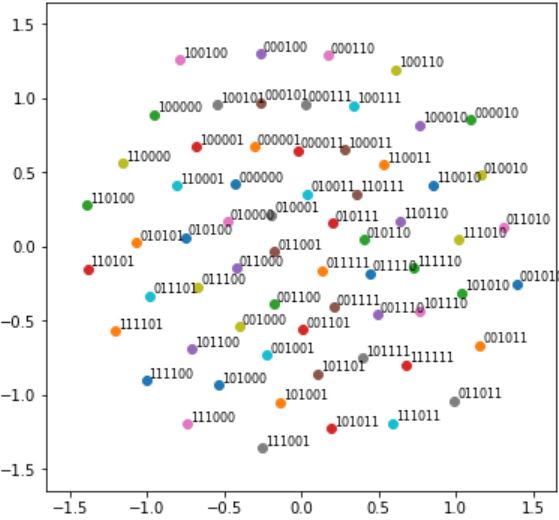}
        \captionof{figure}{A learned constellation with $Q=6$.}
        \label{fig:bkg_learned_64}
      \end{minipage}
      \hfill
    \begin{minipage}{.4\textwidth}
      \centering
      \includegraphics[width=.8\linewidth]{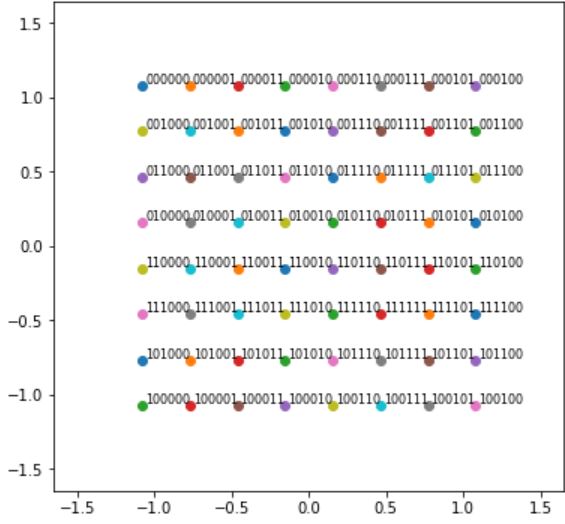}
      \captionof{figure}{Constellation corresponding to a 64-QAM.}
      \label{fig:bkg_64_qam}
    \end{minipage}%
\end{figure}

Estimating the bits that were transmitted is a binary classification problem, as for each bit the label is either $0$ or $1$.
The loss function associated with such problem is the binary cross-entropy, defined as
\begin{align}
\label{eq:bkg_cross_ent}
l =  \EE_y \LSB  L(\bv, \hat{\pv}) \RSB =  \EE_y \LSB - \frac{1}{Q} \sum_{q=0}^{Q-1}  b_q \cdot \text{log}_2 \LB \hat{p}_q \RB + (1- b_q) \cdot \text{log}_2 \LB 1-\hat{p}_q \RB \RSB
\end{align}
where the expected value reflects the fact that the metric should be independent of the noise realization $n$, and can be estimated through Monte-Carlo sampling with batches of size $B_S$:
\begin{align}
    l  \approx \frac{1}{B_S} \sum_{s=1}^{B_S} L \LB \bv^{[s]}, \hat{\pv}^{[s]} \RB.
\end{align}
Training such \gls{NN}-based communication systems does not require any dataset, as two identical infinite sequences of bits can be obtained at the transmitter and at the receiver by initializing a random number generator with the same seed.
These two sequences can be grouped into batches of bit vectors $\bv$, enabling an \gls{SGD}-based optimization:
\begin{align}
    \thetav^{(i+1)} = \thetav^{(i)} - \frac{\eta}{B_S} \sum_{s=1}^{B_S} \nabla_{\thetav} L\LB \bv^{[s]}, \hat{\pv}^{[s]} \RB
\end{align}
where $\thetav$ denotes the set of trainable parameters of the entire \gls{NN}, i.e., $\thetav=\{\boldsymbol{\Theta}, \boldsymbol{\Phi}\}$.
The constellation learned for a system trained with $Q=6$ is depicted in Fig.~\ref{fig:bkg_learned_64} and can be compared to a 64-\gls{QAM} modulation shown in Fig.~\ref{fig:bkg_64_qam}.
Performance evaluations will be carried out in Chapters 3, 4, and 5 for different \gls{DL}-enhanced communication systems.

The differentiability of every layer in the communication system is key to achieve \gls{SGD}-based optimization of every trainable parameters \footnote{When no channel model is available, resulting in a non-differentiable channel layer, one can leverage deep reinforcement learning techniques to train the transmitter and the receiver in an alternating fashion~\cite{8792076, 9144089} or use generative adversarial networks to learn a channel model from available data~\cite{ye2018channel, 8685573}.}.
For example the loss function can not be applied to the estimated bits $\hat{\bv}$, as the threshold layer~\eqref{eq:bkg_threshold} is not differentiable and therefore prevent the gradient to be backpropagated.
Moreover, the trainable communication system presented in Fig.~\ref{fig:bkg_nn_comm} is mostly composed of non-trainable layers, and the normalized embedding layer does not use any neurons.
That explains why such systems are usually referred to as ML or DL-enhanced communication systems, instead of \gls{NN}-based systems.
Overall, this paradigm shift is increasingly visible, as the difference between training \glspl{NN} to perform communication tasks and performing \gls{SGD} on trainable communication systems has never been so thin.

\subsubsection{An information theory perspective}
Let us denote by $\mathscr{b}_q$ the random variable associated with the bit $q$.
To simplify the notations, we denote by $P(b_q, y)$ and $\widehat{P}(b_q, y)$ the true and estimated probability that the $q^{\text{th}}$ bit was transmitted, i.e.
\begin{align}
    \label{eq:bkg_p_tilde}
    P(b_q, y) = \left\{
    \begin{array}{ll}
        P(\mathscr{b}_q = 1 | y) & \mbox{if } b_q = 1 \\
        P(\mathscr{b}_q = 0 | y) & \mbox{if }  b_q = 0
    \end{array}
\right.
\end{align}
and
\begin{align}
    \label{eq:bkg_p_tilde_hat}
    \widehat{P}(b_q, y) = \left\{
    \begin{array}{ll}
        \widehat{P}(\mathscr{b}_q = 1 | y) & \mbox{if } b_q = 1 \\
        \widehat{P}(\mathscr{b}_q = 0 | y) & \mbox{if }  b_q = 0
    \end{array}.
\right.
\end{align}
The \gls{CE} defined in~\eqref{eq:bkg_cross_ent} can be seen as an approximation through Monte Carlo sampling of the true \gls{CE}, defined as

\begin{align}
     l & = \sum_{q=0}^{Q-1} \EE_y \LSB H \LB P(b_q | y), \widehat{P}(b_q | y) \RB \RSB \\
       &= -\sum_{q=0}^{Q-1} \int_y P(y) \sum_{b_q\in \{0, 1 \}}  P (b_q | y) \text{log}_{2} \LB\widehat{P}(b_q | y) \RB dy \\
       &= - \sum_{q}  \sum_{b_q} \int_y  P (b_q, y) \text{log}_{2} \LB  P(b_q) \frac{ \widehat{P}( b_q | y) P (y)}{ P(y) P(b_q)} \RB dy \\
       &= \underbrace{-\sum_{q=0}^{Q-1} \sum_{b_q}  \int_y P (b_q, y) \text{log}_{2} \LB P(b_q) \RB dy}_{\sum_{q=0}^{Q-1}  H(b_q)}
        - \sum_{q=0}^{Q-1}  \sum_{b_q} \int_y P (b_q, y) \text{log}_{2} \LB  \frac{ \widehat{P}( b_q | y) P (y)}{ P(y) P(b_q)} \RB dy \\
        &= \sum_{q=0}^{Q-1}  H(b_q)
        - \sum_{q=0}^{Q-1}  \sum_{b_q} \int_y P (b_q, y) \text{log}_{2} \LB \frac{ P( b_q | y) P(y)  }{ P(y) P(b_q)}  \frac{\widehat{P}(b_q |y)}{P( b_q | y)} \RB dy \\
        &= Q
        - \underbrace{\sum_{q=0}^{Q-1}  \sum_{b_q} \int_y P (b_q, y) \text{log}_{2} \LB  \frac{ P( b_q , y) }{ P(y) P(b_q)} \RB dy }_{\sum_{q=0}^{Q-1} I(b_q, y)}
        -  \sum_{q=0}^{Q-1}  \sum_{b_q} \int_y P(y) P(b_q| y) \text{log}_{2} \LB  \frac{\widehat{P}(b_q |y)}{P( b_q | y)} \RB dy \\
        &= Q - \sum_{q=0}^{Q-1} I(b_q, y) + 
        \underbrace{\sum_{q=0}^{Q-1} \int_y P(y) \sum_{b_q}  P (b_q| y) \text{log}_{2} \LB  \frac{P(b_q |y)}{\widehat{P}(b_q |y)} \RB dy }_{\sum_{q=0}^{Q-1} \EE_y \LSB D_{\text{KL}} \LB P(b_q |y) || \widehat{P}(b_q |y) \RB \RSB}\\
        & = Q - \underbrace{\LB  \sum_{q=0}^{Q-1} I(b_q, y) - \sum_{q=0}^{Q-1} \EE_y \LSB D_{\text{KL}} \LB P(b_q |y) || \widehat{P}(b_q |y)  \RB \RSB  \RB }_{C}. \label{eq:bkg_link_ce_rate}
\end{align}

The first term of $C$ is mutual information between all $b_q$ and $y$, and corresponds to the maximum information rate that can be achieved assuming an ideal \gls{BMD} receiver~\cite{bocherer2018achievable}
This term both depends on the transmitter and on the channel model.
The second term of $C$ is the sum of the expected values of the KL-divergence between the true posterior probability $P(b_q |y)$ and the one estimated by the proposed receiver, and corresponds to a rate loss due to a suboptimal receiver.
It can be seen as a measure of distance between the probabilities that would be computed by an ideal receiver and the ones estimated by our NN-based implementation.
Minimizing $l$ therefore jointly optimizes the transmitter and the receiver to both maximize the information rate of the transmission and refine the estimated bit probabilities.
Finally, $C$ is an achievable rate assuming a mismatched BMD receiver \cite{pilotless20} meaning that improvements in $C$ directly translate to an improved \gls{BER} performance.

\clearpage

\chapter{HyperMIMO: a Deep HyperNetwork-Based MIMO Detector}
\label{ch:1}

\section{Motivation} 
\label{sec:chp1_introduction}


As introduced in Section~\ref{sec:intro_challenges}, \gls{MU-MIMO} is seen as a key technology to unlock the gains envisioned for beyond-5G systems.
But optimal detection in such \gls{MIMO} systems is known to be NP-hard~\cite{10.1007_s10107-016-1036-0}, and less complex approaches usually suffer from unsatisfying performance on correlated channels or become impractical with large number of receive antennas.
Examples of such approaches include the \gls{LMMSE} detector~\cite{tse2005fundamentals}, the \gls{AMP} algorithm~\cite{AMP}, and its extension to correlated channels~\cite{OAMP}.
Recently, advances in \gls{MIMO} detection have been made using \gls{DL} to improve the equalization block, detailed in Section~\ref{sec:OFDM_systems}, and which corresponds to a block-based optimization as depicted in Fig.~\ref{fig:intro_ml_syst_1}.
%
One technique consists in using an \gls{NN} to select a traditional detection algorithm from a predefined set~\cite{Samsung}.
According to available \gls{CSI}, the algorithm with lowest complexity that enables a \gls{BLER} lower than a predefined threshold is chosen.
Another technique is to design an \gls{NN} that directly performs the detection. 
One example is DetNet~\cite{DetNet}, which can be viewed as an unfolded \gls{RNN} where each iteration is made of three dense layers.
Although it achieves encouraging results on Rayleigh channels, DetNet's performance on correlated channels is not satisfactory and it suffers from a prohibitive complexity.
In~\cite{DetNet_small}, Mohammad et al. partially addressed this drawback by weights pruning.
A third promising approach is known as \emph{deep unfolding}, and consists in infusing existing iterative algorithms with \gls{DL} components~\cite{balatsoukasstimming2019deep}.
One possibility is to add trainable parameters to such algorithms and interpret the whole structure as an \gls{NN}~\cite{8642915, OAMPNet}, but most approaches still suffer from a performance drop on correlated channels.
This was mitigated by the MMNet detector~\cite{MMNet}, which effectively achieves state-of-the-art performance on such channels.
However, the need of retraining for each channel realization makes its practical implementation challenging.

\begin{figure}
\centering
	\includegraphics[width=0.4\linewidth]{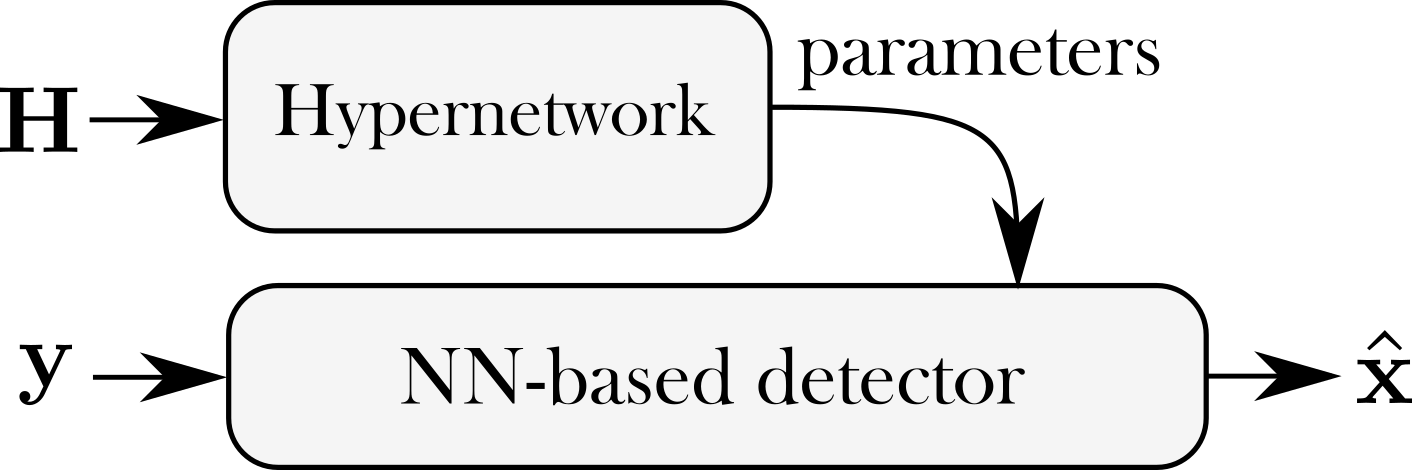}
	\caption{HyperMIMO: A hypernetwork generates the parameters of an NN-based detector.}
\label{fig:chp1_HG_small}
\end{figure}

In the following, we alleviate this issue using the emerging idea of \emph{hypernetworks}~\cite{learnet,talking_heads}.
Applied to our setup, it consists in having a secondary \gls{NN}, referred to as the hypernetwork, that generates for a given channel matrix an optimized set of weights for an \gls{NN}-based detector.
This scheme, which we referred to as \emph{HyperMIMO} in our introductory paper~\cite{goutay2020deep}, is illustrated in Fig.~\ref{fig:chp1_HG_small}.
Used with the MMNet detector from~\cite{MMNet}, HyperMIMO replaces the training procedure that would be required for each channel realization by a single inference of the hypernetwork.
We have evaluated the proposed approach using simulations on spatially correlated channels.
Our results show that HyperMIMO achieves performance close to that of MMNet trained for each channel realization, and outperforms the recently proposed OAMPNet~\cite{OAMPNet}.
They also reveal that HyperMIMO is robust to user mobility up to a certain point, which is encouraging for practical use.
However, HyperMIMO still suffers from performance drops when evaluated with channels that vary significantly from the ones it has been trained with, and is only able to handle a fixed number of users.
More recent works~\cite{zilberstein2021robust,pratik2020remimo} proposed to address these shortcomings and will be discussed in the closing section of this chapter.

\subsection*{Related literature}

Although most of the published literature regarding NN-based block optimization focuses exclusively on the equalization step~\cite{8642915, OAMPNet, MMNet, goutay2020deep, zilberstein2021robust, pratik2020remimo}, another line of research targets improved channel estimation.
In~\cite{8272484}, an \gls{NN} architecture derived from a conventional \gls{LMMSE} estimator is able to provide estimation gains on a wide range of channels with reduced complexity.
Two \gls{CNN}-based channel estimators are proposed in~\cite{8752012} for Massive MIMO mmWave transmissions, and either target lower computational complexity or reduced pilot overhead.
Pilot overhead reduction is also studied in~\cite{mashhadi2020pruning}, where the pilot insertion and channel estimation blocks are jointly optimized to reduce the number of pilots required to achieve satisfactory channel estimation.
However, these approaches require ground-truth of the channel realizations during training, which can only be approximated with costly measurement campaigns in practice.
Finally, it has been proposed to improve the estimation of the bit probabilities by replacing the demapper with an \gls{NN}, but this solution has only been studied for \gls{SISO} setups \cite{shental2020machine, pilotless20, oztekin2019efficient, 9345976}.


\clearpage
\section{Framework} 
\label{sec:chp1_Background}

\subsection{Problem Formulation and LMMSE Baseline}
\label{subsec:chp1_chp1_problem_definition}
We consider a conventional \gls{MIMO} uplink channel as presented in~\eqref{eq:_ofdm_mimo_model}.
In this section, a single \gls{RE} $(m,n)$ is considered, such that the channel transfer function can be simplified to
\begin{equation}
\label{eq:chp1_rayleigh}
\yv = \Hm \xv + \nv
\end{equation}
where $\xv \in \Cc^{K}$ is the vector of transmitted symbols, $\yv \in \CC^{L}$ is the vector of received distorted symbols, $\Hm \in \CC^{L \times K}$ is the channel matrix, and $\nv \thicksim \Cc\Nc(\mathbf{0}, \sigma^2 \Id_{L})$ is the \gls{iid} complex Gaussian noise with power $\sigma^2$ in each complex dimension. 
It is assumed that $\Hm$ and $\sigma$ are perfectly known to the receiver.
In the following, the problem of hard symbol detection is considered, in which the estimated symbol $\doublehat{\xv}$ must belong to the used constellation, i.e., $\doublehat{\xv} \in \Cc^{K}$.

The optimal maximum likelihood detector, as defined in~\eqref{eq:intro_ml_detector}, is known to be too complex for any practical implementation~\cite{10.1007_s10107-016-1036-0}. 
On the other hand, the zero-forcing equalizer~\eqref{eq:intro_zero_forcing} is known to perform poorly on ill-conditionned channels.
To tackle both issues, one well-known scheme is the \gls{LMMSE} estimator which aims to minimize the \gls{MSE} 
\begin{equation}
\label{eq:chp1_loss_mmse}
\hat{\xv} = \arg \min_{\xv' \in \CC^{K}; \xv' = \Wm \yv} \EE_{\xv, \nv} \left[ || \xv - \xv' ||_2^2 \right]
\end{equation}
by restricting to linear estimators, i.e., by left-multiplying $\yv$ with a matrix $\Wm \in \CC^{K\times L}$.
A derivation can be obtained by finding the matrix $\Wm$ that nulls the gradient 
\begin{align}
    \nabla_{\Wm}  \EE_{\xv, \nv} \left[ || \xv - \Wm \yv ||_2^2  \right]  = & \EE_{\xv, \nv}  \left[ -2 \xv \yv \htp + 2 \Wm \yv \yv\htp \right]  \stackrel{!}{=} 0 \\
     \iff & \Wm = \EE_{\xv, \nv} \left[ \xv \yv \htp \right] \LB  \EE_{\xv, \nv} \left[ \yv \yv\htp \right] \RB^{-1} \\
     \iff & \Wm = \Hm\htp \LB \Hm \Hm\htp + \sigma^2 \Id_K \RB^{-1}.
\end{align}
This allows for a closed-form expression of the solution to~(\ref{eq:chp1_loss_mmse})
\begin{equation}
\label{eq:chp1_mmse_1}
\hat{\xv} = (\Hm^H \Hm + \sigma^2 \Id_{K})^{-1} \Hm^H \yv.
\end{equation}
Because the transmitted symbols are known to belong to the finite alphabet $\Cc$, the closest symbol is typically selected for each user:
\begin{equation}
\label{eq:chp1_mmse_2}
\doublehat{x}_{k} = \arg \min_{x \in \Cc} || \hat{x}_{k} - x ||_2^2, \quad \quad \forall k \in \{1, \cdots, K\}.
\end{equation}
Although sub-optimal, this approach has the benefit of being computationally tractable. 
Multiple schemes have been proposed to achieve a better performance-complexity trade-off among which \gls{DL}-based algorithms form a particularly promising lead.

\subsection{Deep Learning-based MIMO Detectors}
\label{subsec:chp1_DL-based_algorithms}

As discussed previously, an interesting approach to design enhanced detectors is to add trainable parameters to existing schemes, and is often referred to as deep unfolding~\cite{balatsoukasstimming2019deep}.
Traditional iterative algorithms are particularly suitable since they can be viewed as \gls{NN} once unfolded.
Typically, each iteration aims to further reduce the \gls{MSE} and comprises a linear step followed by a non-linear denoising step.
The estimate $\hat{\xv}^{(i+1)}$ at the $(i+1)$\textsuperscript{th} iteration is 
\begin{equation} 
\label{eq:chp1_it_algo}
\begin{aligned}
\kappav^{(i)}  &= \hat{\xv}^{(i)} + \Am^{(i)} \left(\yv - \Hm \hat{\xv}^{(i)} + \cv^{(i)} \right)\\
\hat{\xv}^{(i+1)} &= \chi^{(i)}\left(\kappav^{(i)}, \tau^{(i)} \right)
\end{aligned}
\end{equation}
where the superscript $(i)$ is used to refer to the $i^{\text{th}}$ iteration and $\hat{\xv}^{(0)}$ is set to $\mathbf{0}$.
$\tau^{(i)}$ denotes the estimated variance of the components of the noise vector $\kappav^{(i)} -\xv^{(i)}$ at the input of the denoiser, which is assumed to be \gls{iid}.
Iterative algorithms differ by their choices of matrices $\Am^{(i)} \in \CC^{K \times L}$, bias vectors $\cv^{(i)} \in \CC^{K}$, and denoising functions $\chi^{(i)}(\cdot)$.
A limitation of most detection schemes is their poor performance on correlated channels. 
OAMP~\cite{OAMP} mitigates this issue by constraining both the linear step and the denoiser. 
OAMPNet~\cite{OAMPNet} improves the performance of OAMP by adding two trainable parameters per iteration, which respectively scales the matrix $\Am^{(i)}$ and the channel noise variance $\sigma^2$.
MMNet~\cite{MMNet} goes one step further by making all matrices $\Am^{(i)}$ trainable and by relaxing the constraint on $\kappav^{(i)} -\xv^{(i)}$ being identically distributed. 
Although MMNet achieves state-of-the-art performance on spatially-correlated channels, it needs to be re-trained for each channel matrix, which makes it unpractical.

\subsection{Hypernetworks}
\label{HyperNetworks}

Hypernetworks were introduced in~\cite{hypernetworks} as \glspl{NN} that generate the parameters of other \glspl{NN}.
The concept was first used in~\cite{learnet} in the context of image recognition.
The goal was to predict the parameters of an \gls{NN} given a new sample so that it could recognize other objects of the same class without the need for training.
This same idea was also leveraged to generate images of talking heads~\cite{talking_heads}.
In this later work, a picture of a person is fed to a hypernetwork that computes the weights of a second \gls{NN}.
This second \gls{NN} then generates realistic images of the same person with different facial expressions.
Motivated by these achievements, we propose to alleviate the need of MMNet to be retrained for each channel realization using hypernetworks. 
\clearpage
\section{HyperMIMO} 
\label{sec:chp1_HyperMIMO}

The key idea of our approach is to replace the training process required by MMNet for each channel realization by a single inference through a trained hypernetwork.
We first present a variation of MMNet which reduces its number of parameters, and then introduce the architecture of the hypernetwork, where a relaxed form of weight sharing is used to decrease its output dimension.
Both reducing the number of parameters of MMNet and using weight sharing in the hypernetwork are crucial to obtain a system of reasonable complexity.
The combination of the hypernetwork together with MMNet form the HyperMIMO system, schematically shown in Fig.~\ref{fig:chp1_HG_small}.

\subsection{MMNet with Less Parameters}
\label{subsec:chp1_mmnet}

To reduce the number of parameters of MMNet, we perform the QR-decomposition of the channel matrix, $\Hm = \Qm \Rm$, where $\Qm$ is an $L \times L$ orthogonal matrix and $\Rm$ an $L \times K$ upper triangular matrix.
It is assumed that $L > K$, and therefore $\Rm = \begin{bmatrix}\mathbf{R_A} \\ \boldsymbol{0}\end{bmatrix}$ where $\mathbf{R_A}$ is of size $K \times K$, and $\Qm = \left[\mathbf{Q_A} \boldsymbol{Q_B} \right]$ where $\mathbf{Q_A}$ has size $L \times K$.
We define $\bar{\yv} \coloneqq \mathbf{Q_A}^H \yv$ and $\bar{\nv} \coloneqq \mathbf{Q_A}^H\nv$, and rewrite~(\ref{eq:chp1_rayleigh}) as 
\begin{equation}
\label{eq:chp1_QR}
\bar{\yv} = \mathbf{R_A} \xv + \bar{\nv}.
\end{equation}
Note that $\bar{\nv} \thicksim \Cc\Nc(\mathbf{0}, \sigma^2 \Id_{K})$.
MMNet sets $\cv^{(i)}$ to $\zerov$ for all $i$ and uses the same denoiser for all iterations, which are defined by
\begin{equation} 
\label{eq:chp1_mmnet}
\begin{split}
\kappav^{(i)} & = \widehat{\xv}^{(i)} + \boldsymbol{\Theta}^{(i)} \left( \bar{\yv} - \mathbf{R_A} \hat{\xv}^{(i)} \right) \\
\widehat{\xv}^{(i+1)} & = \chi \left( \kappav^{(i)}, {\boldsymbol{\tau}^{(i)}} \right)
\end{split}
\end{equation}
where $\boldsymbol{\Theta}^{(i)}$ is a $K \times K$ complex matrix whose components need to be optimized for each channel realization.
The main benefit of leveraging the QR-decomposition is that the dimension of the matrices $\boldsymbol{\Theta}^{(i)}$ to be optimized is $K \times K$ instead of $K \times L$, which is the dimension of $\Am^{(i)}$ in~(\ref{eq:chp1_it_algo}).
This is significant since the number of active users $K$ is typically much smaller than the number of antennas $L$ of the \gls{BS}.

The noise at the input of the denoiser $\kappav^{(i)} -\xv^{(i)}$ is assumed to be independent but not identically distributed in MMNet.
The vector of estimated variances at the $i^{\text{th}}$ iteration is denoted by $\boldsymbol{\tau}^{(i)} \in \RR^{K}$ and computed by
\begin{equation} 
\label{eq:chp1_noise_std}
\begin{aligned} 
\boldsymbol{\tau}^{(i)} =\frac{\boldsymbol{\psi}^{(i)}}{K} & \left( \frac{||\Id_{K}-\boldsymbol{\Theta}^{(i)} \mathbf{R_A}||_{F}^{2}}{||\mathbf{R_A}||_{F}^{2}}  \left[||\bar{\yv}-\mathbf{R_A} \hat{\xv}^{(i)}||_{2}^{2}-L \sigma^{2}\right]^{+} + ||\boldsymbol{\Theta}^{(i)} ||_{F}^{2} \sigma^{2} \right) \end{aligned}
\end{equation}
where $[x]^{+} = \max(0,x)$, and $\boldsymbol{\psi}^{(i)} \in \RR^{K}$ needs to be optimized for each channel realization.
Further details on the origin of this equation can be found in~\cite{OAMP}.
The denoising function in MMNet is the same for all iterations, and is chosen to minimize the \gls{MSE} $\EE_{\xv} \left[||\hat{\xv}^{(I)} - \xv||_2^2  \right]$ assuming the noise is independent and Gaussian distributed.
This is achieved by applying element-wisely to $(\kappav^{(i)}, \boldsymbol{\tau}^{(i)})$
\begin{equation} 
\label{eq:chp1_denoiser}
\chi (\kappa, \tau) = \frac{1}{\sum_{x \in \Cc} \exp{- \frac{|\kappa - x|^2}{\tau}}} \sum_{x \in \Cc} x \exp{- \frac{|\kappa - x|^2}{\tau} }.
\end{equation}
MMNet consists of $I$ layers performing~(\ref{eq:chp1_mmnet}), and
a hard decision as in~(\ref{eq:chp1_mmse_2}) to predict the final estimate $\doublehat{\xv}$.
One could also use $\hat{\xv}^{(I)}$ to predict bit-wise \glspl{LLR}.

\subsection{HyperMIMO Architecture}
\label{subsec:chp1_implementation_details}

\begin{figure}
\centering
	\includegraphics[width=0.6\linewidth]{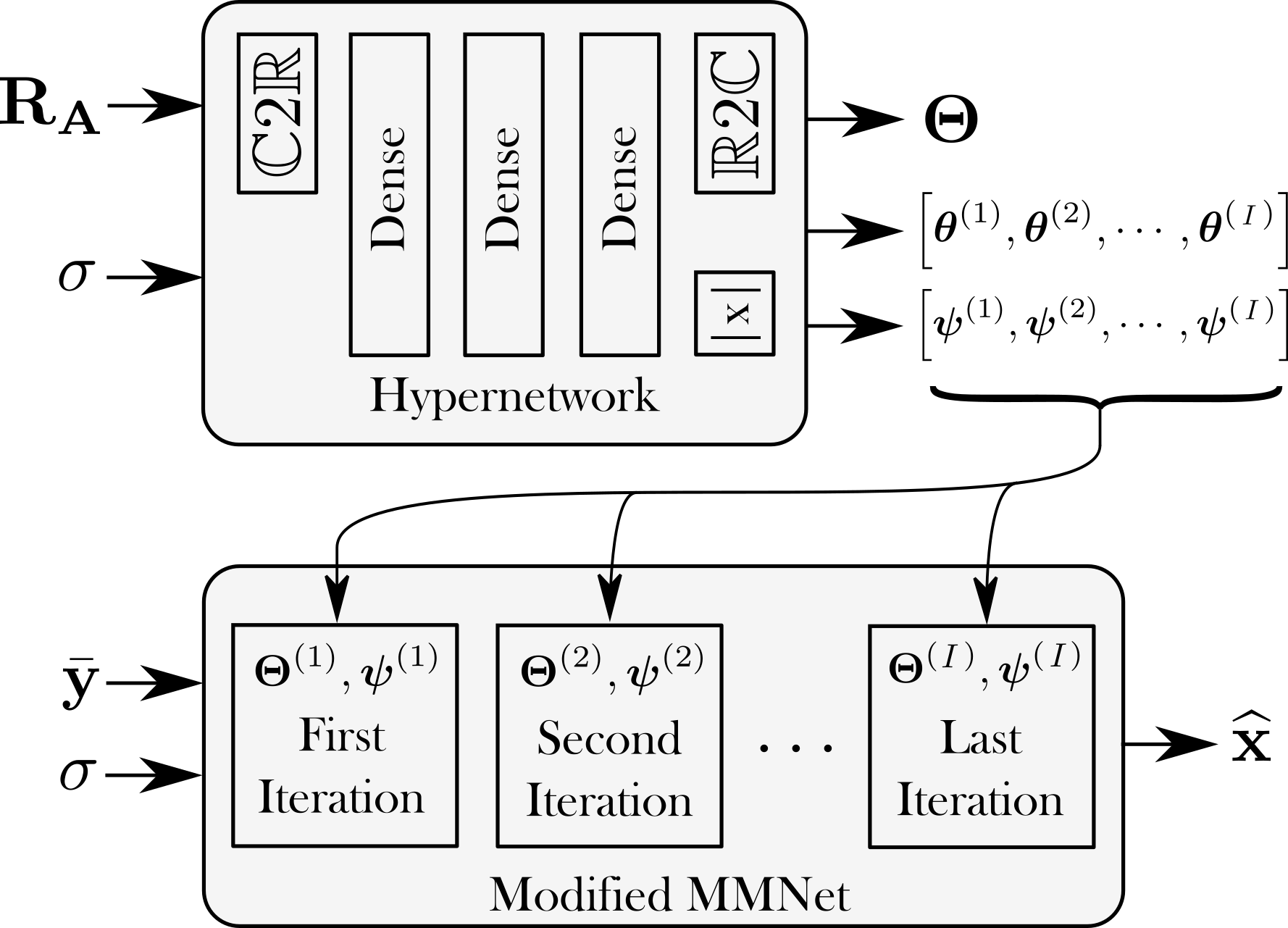}
	\caption{Detailed architecture of HyperMIMO }
\label{fig:chp1_HG}
\end{figure} 

Fig.~\ref{fig:chp1_HG} shows in details the architecture of HyperMIMO.
As our variant of MMNet operates on $\bar{\yv}$, the hypernetwork is fed with $\mathbf{R_A}$ and the channel noise standard deviation $\sigma$.
Note that because $\mathbf{R_A}$ is upper triangular, only $K(K+1)/2$ non-zero elements need to be fed to the hypernetwork. 
Moreover, using this matrix as input instead of $\Hm$ has been to found to be critical to achieve high performance.
As detailed previously, the number of parameters that need to be optimized in MMNet was reduced by leveraging the QR-decomposition.
To further decrease the number of outputs of the hypernetwork, we adopt a relaxed form of weight sharing inspired by~\cite{learnet}.
Instead of computing the elements of each $\boldsymbol{\Theta}^{(i)}, i = \{1,\dots,I \}$, the hypernetwork outputs a single matrix $\boldsymbol{\Theta}$ as well as $I$ vectors $\boldsymbol{\theta}^{(i)} \in \RR^{K}$.
For each iteration $i$, $\boldsymbol{\Theta}^{(i)}$ is computed by 
\begin{equation}
\label{eq:chp1_scaling}
\boldsymbol{\Theta}^{(i)} = \boldsymbol{\Theta} \left( \Id_{K} + \text{diag}\left(\boldsymbol{\theta}^{(i)}\right) \right).
\end{equation}
The idea is that all matrices $\boldsymbol{\Theta}^{(i)}$ differ by a per-column scaling different for each iteration.
We have experimentally observed that scaling of the rows leads to worse performance.

Because $\mathbf{R_A}$ is complex-valued, a $\CC2\RR$ layer maps the complex elements of $\mathbf{R_A}$ to real ones, by concatenating the real and imaginary parts of the complex scalar elements.
To generate a complex-valued matrix $\boldsymbol{\Theta}$, a $\RR2\CC$ layer does the reverse operation of $\CC2\RR$.
The hypernetwork also needs to compute the values of the $I$ vectors $\boldsymbol{\psi}^{(i)}$.
Because the elements of these vectors must be positive, an absolute-value activation function is used in the last layer.

HyperMIMO, which comprises the hypernetwork and MMNet, is trained by minimizing the \gls{MSE} between the transmitted and estimated symbols, denoted by $\widehat{\xv} = \widehat{\xv}^{(I)}$:
\begin{equation}
\label{eq:chp1_loss_hg}
\Lc = \EE_{\xv, \Hm, \nv} \left[ ||\widehat{\mathbf{x}}-\mathbf{x}||_{2}^{2} \right]
\end{equation}
Finally, the expected value can be approximated through Monte-Carlo sampling, by sending batches of $B_S$ samples:
\begin{align}
	\Lc \approx \frac{1}{B_S} \sum_{s=1}^{B_S} \left\Vert \widehat{\mathbf{x}}^{[s]}-\mathbf{x}^{[s]} \right\Vert_{2}^{2} 
\end{align}
Note that this loss differs from the one of~\cite{MMNet}, which is $\frac{1}{I} \sum_{i=1}^I \EE_{\xv, \Hm, \nv} \left[ ||\widehat{\mathbf{x}}^{(i)}-\mathbf{x}||_{2}^{2} \right]$.
When training HyperMIMO, the hypernetwork and MMNet form a single \gls{NN}, such that the output of the hypernetwork are the weights of MMNet.
The only trainable parameters are therefore the ones of the hypernetwork.
When performing gradient descent, their gradients are backpropagated through the parameters of MMNet.
\clearpage
\section{Experiments} 
\label{sec:chp1_Experiments}

HyperMIMO was evaluated by simulations.
This section starts by introducing the considered spatially correlated channel model.
Next, details on the simulation setting and training process are provided.
Finally, the obtained results are presented and discussed.

\subsection{Channel Model}
\label{subsec:chp1_channel_model}

\begin{figure}
\center
\includegraphics[width=0.6\linewidth]{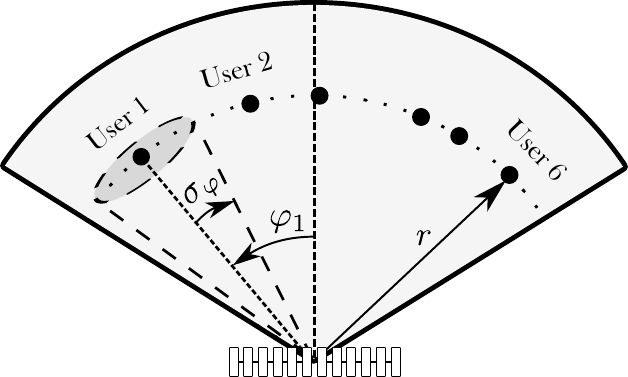}
\caption{Considered channel model. The \gls{BS} has no scatters in its near field, and scattering is only located near users.}
\label{fig:chp1_setup}
\end{figure}

The local scattering model with spatial correlation presented in~\cite[Ch.~2.6]{massivemimobook} and illustrated in Fig.~\ref{fig:chp1_setup} is considered.
The \gls{BS} is assumed to be equipped with a uniform linear array of $L$ antennas, located at the center of a 120$^{\circ}$-cell sector in which $K$ single-antenna users are dropped with random nominal angles $\varphi_{k},~k \in \{1, \cdots, K \}$.
Perfect power allocation is assumed, leading to all users appearing to be at the same distance $r$ from the \gls{BS} and an average gain of one.
The \gls{BS} is assumed to be elevated enough to have no scatterers in its near field, such that the scattering is only located around the users.
Given a user $k$, the multipath components reach the \gls{BS} with normally distributed angles with mean $\varphi_{k}$ and variance $\sigma_{\varphi}^2$.
For small enough $\sigma_{\varphi}$, a valid approximation of the channel covariance matrix is $\Cm_k \in \CC^{L \times L}$ with components
\begin{equation}
\label{eq:chp1_covariance}
\left[ C_k \right]_{a,b} = e^{2\pi j d(a-b)\sin(\varphi_k)} e^{-\frac{\sigma_{\varphi}^2}{2}(2\pi d(a-b)\cos(\varphi_k))^2}
\end{equation}
%
where $d$ is the antenna spacing measured in multiples of the wavelength.
For a given user $k$, a random channel vector $\hv_k \thicksim \Cc \Nc(\mathbf{0}, \Cm_k)$ is sampled by computing
\begin{equation}
\label{eq:chp1_h_vectors}
\hv_k = \Lm_k \mathbf{\Lambda}_k^{\frac{1}{2}} \Lm_k^H \ev
\end{equation}
where $\ev$ is sampled from $\Cc \Nc (\mathbf{0}, \Id_{L})$ and $\Lm_k \mathbf{\Lambda}_k \Lm_k^H$ is the eigenvalue decomposition of $\Cm_k$.
The \gls{SNR} of the transmission is defined by 
\begin{equation}
\label{eq:chp1_snr}
\mathrm{SNR}=\frac{\mathbb{E}\left[\frac{1}{N_{r}}\|\mathbf{y}\|_{2}^{2}\right]}{\sigma^{2}}=\frac{1}{\sigma^{2}}.
\end{equation}

\subsection{Simulation Setting}
\label{subsec:chp1_simulation_setting}

\begin{figure}[!t]
	\center
\begin{tikzpicture}
	  \begin{axis}[
	    grid=both,
	    grid style={line width=.01pt, draw=gray!10},
	    major grid style={line width=.2pt,draw=gray!50},
	    xlabel={$\varphi_k$ for $k = \{1,\dots,6\}$ (degrees)},
	    ytick={1,2,3,4,5,6,7,8,9,10},
	    ylabel={Drop number},
	    width = 239pt,
	    height = 140pt,
	  ]
	    \addplot[only marks, blue, mark=*] table [x=angle_1, y=idx_1, col sep=comma] {Chapter1/figs/set_angles.csv};
	    \addplot[only marks, blue, mark=*] table [x=angle_2, y=idx_2, col sep=comma] {Chapter1/figs/set_angles.csv};
	    \addplot[only marks, blue, mark=*] table [x=angle_3, y=idx_3, col sep=comma] {Chapter1/figs/set_angles.csv};
	    \addplot[only marks, blue, mark=*] table [x=angle_4, y=idx_4, col sep=comma] {Chapter1/figs/set_angles.csv};
	    \addplot[only marks, blue, mark=*] table [x=angle_5, y=idx_5, col sep=comma] {Chapter1/figs/set_angles.csv};
	    \addplot[only marks, blue, mark=*] table [x=angle_6, y=idx_6, col sep=comma] {Chapter1/figs/set_angles.csv};
	    \addplot[only marks, blue, mark=*] table [x=angle_7, y=idx_7, col sep=comma] {Chapter1/figs/set_angles.csv};
	    \addplot[only marks, blue, mark=*] table [x=angle_8, y=idx_8, col sep=comma] {Chapter1/figs/set_angles.csv};
	    \addplot[only marks, blue, mark=*] table [x=angle_9, y=idx_9, col sep=comma] {Chapter1/figs/set_angles.csv};
	    \addplot[only marks, blue, mark=*] table [x=angle_10, y=idx_10, col sep=comma] {Chapter1/figs/set_angles.csv};
	    
		\end{axis}
		\end{tikzpicture}
	    \caption{Ten randomly generated user drops}
	    \label{fig:chp1_users_drops}
\end{figure}
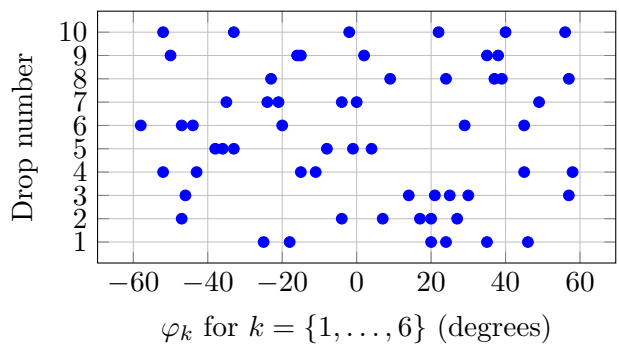

The number of antennas that equip the \gls{BS} was set to $L = 12$, and the number of users to $K = 6$.
\Gls{QPSK} modulation was used.
The standard deviation of the multipath angle distribution $\sigma_{\varphi}$ was set to $10^{\circ}$, which results in highly correlated channel matrices.
The number of layers of MMNet in the HyperMIMO detector was set to $I = 5$.
The hypernetwork was made of 3 dense layers (see Fig.~\ref{fig:chp1_HG}).
The first layer had a number of units matching the number of inputs, the second layer 75 units, and the last layer a number of units matching the number of parameters of the detector.
The first two dense layers used \gls{ELU} activation functions, and the last dense layer had no activation functions.

\pagebreak

Our simulations revealed that training with randomly sampled user drops leads to suboptimal results.
Therefore, HyperMIMO was trained with fixed channel statistics, i.e., fixed user positions.
If this might seem unpromising, our results show that HyperMIMO is still robust to user mobility (see Section~\ref{subsec:chp1_simulatioLesults}).
Moreover, our scheme only has $10 \times$ more parameters than MMNet as proposed in~\cite{MMNet}, which allows it to be quickly re-trained in the background when the channel statistics change significantly.
Note that this is different from MMNet that needs to be retrained every time the channel matrix changes, which is considerably more computationally demanding.
Moreover, it is possible that further investigations on the hypernetwork architecture may alleviate this issue.

Given a user drop, HyperMIMO was trained by randomly sampling channel matrices $\Hm$, \glspl{SNR} from the range [0,10]\si{dB}, and symbols from a \gls{QPSK} constellation for each user.
Training was performed using the Adam~\cite{Kingma15} optimizer with a batch size of 500 and a learning rate decaying from $10^{-3}$ to $10^{-4}$.

\subsection{Simulation Results}
\label{subsec:chp1_simulatioLesults}

\begin{figure}[!t]
	\center
\begin{tikzpicture}

	\pgfplotsset{
    width=.55\textwidth,
    height=0.5\textwidth
}
	  \begin{axis}[
	    ymode=log,
	    grid=both,
	    grid style={line width=.01pt, draw=gray!10},
	    major grid style={line width=.2pt,draw=gray!50},
	    minor tick num=1,
	    xtick={0, 2, 4, 6, 8, 10},
	    xlabel={SNR},
	    ylabel={SER},
	    legend style={at={(0.03, 0.03)},anchor=south west},
		ymax = 1e-1,	    
	    ymin = 2e-6,
	    legend cell align={left},
	  ]
	    \addplot[name path=mmse, blue, mark=diamond*] table [x=snr, y=mmse, col sep=comma] {Chapter1/figs/snr.csv};
	    \addplot[name path=oamp, violet, mark=triangle*] table [x=snr, y=oamp, col sep=comma] {Chapter1/figs/snr.csv};
	    \addplot[name path=hg, red, mark=*] table [x=snr, y=hg, col sep=comma] {Chapter1/figs/snr.csv};
	    \addplot[name path=mmnet, orange, mark=pentagon*] table [x=snr, y=mmnet, col sep=comma] {Chapter1/figs/snr.csv};
	    \addplot[name path=ml, black, mark=square*] table [x=snr, y=ml, col sep=comma] {Chapter1/figs/snr.csv};






	  \addlegendentry{LMMSE}
	  \addlegendentry{OAMPNet}
	  \addlegendentry{HyperMIMO}
	  \addlegendentry{MMNet}
	  \addlegendentry{Max. Likelihood}
		\end{axis}
		\end{tikzpicture}
	    \caption{\gls{SER} achieved by different schemes}
	    \label{fig:chp1_snr}
\end{figure}

All presented results were obtained by averaging over 10 randomly generated drops of 6 users, shown in Fig.~\ref{fig:chp1_users_drops}. 
Fig.~\ref{fig:chp1_snr} shows the \gls{SER} achieved by HyperMIMO, \gls{LMMSE}, OAMPNet with 10 iteration, MMNet with 10 iterations and trained for each channel realization, and the maximum likelihood detector.
As expected, MMNet when trained for each channel realization achieves a performance close to that of maximum likelihood.
One can see that the performance of OAMPNet are close to that of LMMSE on these highly correlated channels.
HyperMIMO achieves \gls{SER} slightly worse than MMNet, but outperforms OAMPNet and \gls{LMMSE}.
More precisely, to achieve a \gls{SER} of $10^{-3}$, HyperMIMO exhibits a loss of 0.65\si{dB} compared to MMNet, but a gain of 1.85\si{dB} over OAMPNet and 2.85\si{dB} over LMMSE.

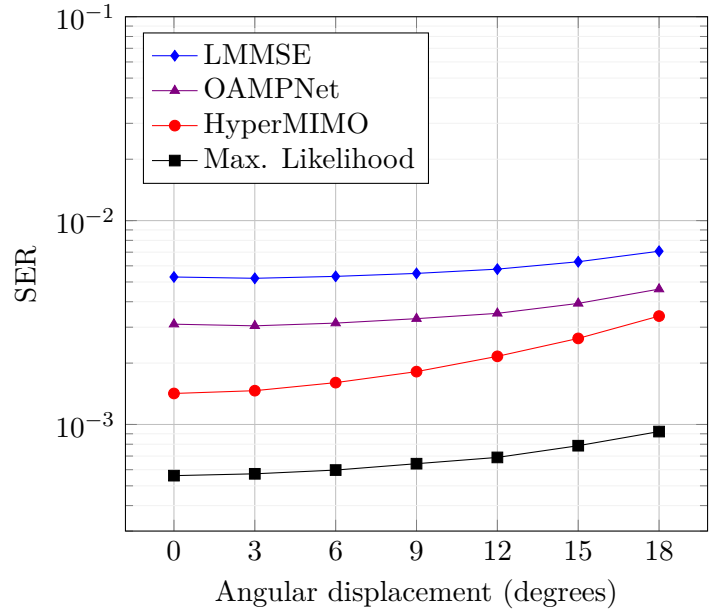
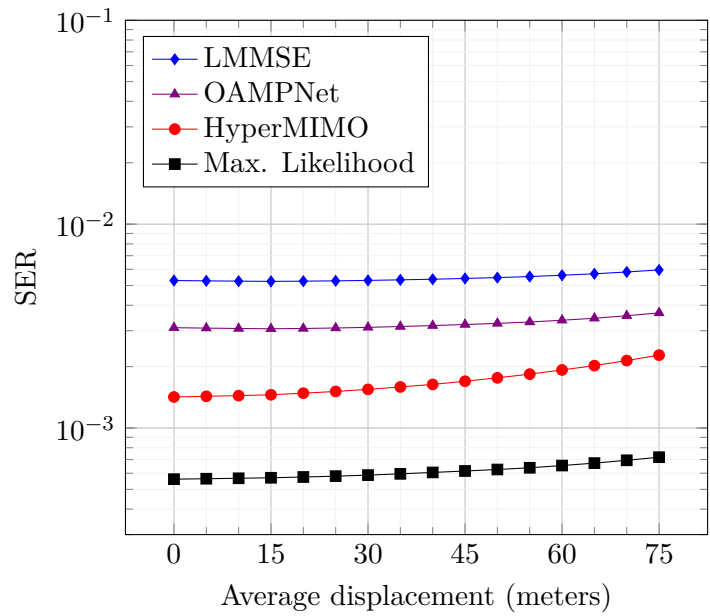
\begin{figure}
    \centering
    \begin{subfigure}{.6\textwidth}
		\begin{tikzpicture}

			\pgfplotsset{
				width=.99\textwidth,
				height=0.9\textwidth
			}

			\begin{axis}[
	    ymode=log,
	    grid=both,
	    grid style={line width=.01pt, draw=gray!10},
	    major grid style={line width=.2pt,draw=gray!50},
	    minor tick num=0,
	    xtick={0, 3, 6, 9, 12, 15, 18},
	    xlabel={Angular displacement (degrees)},
	    ylabel={SER},
	    legend style={at={(0.03, 0.675)},anchor=south west},
		ymax = 1e-1,	    
	    ymin = 3e-4,
	    legend cell align={left},
	  ]
	    \addplot[blue, mark=diamond*] table [x=angles, y=mmse, col sep=comma] {Chapter1/figs/angles_18.csv};
	    \addplot[violet, mark=triangle*] table [x=angles, y=oamp, col sep=comma] {Chapter1/figs/angles_18.csv};
	    \addplot[red, mark=*] table [x=angles, y=hg, col sep=comma] {Chapter1/figs/angles_18.csv};
	    \addplot[black, mark=square*] table [x=angles, y=ml, col sep=comma] {Chapter1/figs/angles_18.csv};

	  \addlegendentry{LMMSE}
	  \addlegendentry{OAMPNet}
	  \addlegendentry{HyperMIMO}
	  \addlegendentry{Max. Likelihood}
		\end{axis}
		\end{tikzpicture}
		\caption{Angular mobility}
		\label{fig:chp1_angles}
	 \end{subfigure} 
	 \hfill
    \begin{subfigure}{.6\textwidth}
	    \begin{tikzpicture}

			\pgfplotsset{
				width=.99\textwidth,
				height=0.9\textwidth
			}

		  \begin{axis}[
	    ymode=log,
	    grid=both,
	    grid style={line width=.01pt, draw=gray!10},
	    major grid style={line width=.2pt,draw=gray!50},
	    minor tick num=2,
	    xtick={0, 15, 30, 45, 60, 75},
	    xlabel={Average displacement (meters)},
	    ylabel={SER},
	    legend style={at={(0.03, 0.675)},anchor=south west},
		ymax = 1e-1,	    
	    ymin = 3e-4,
	    legend cell align={left},
	  ]
	    \addplot[blue, mark=diamond*] table [x=meters, y=mmse, col sep=comma] {Chapter1/figs/meters_75.csv};
	    \addplot[violet, mark=triangle*] table [x=meters, y=oamp, col sep=comma] {Chapter1/figs/meters_75.csv};
	    \addplot[red, mark=*] table [x=meters, y=hg, col sep=comma] {Chapter1/figs/meters_75.csv};
	    \addplot[black, mark=square*] table [x=meters, y=ml, col sep=comma] {Chapter1/figs/meters_75.csv};

	  \addlegendentry{LMMSE}
	  \addlegendentry{OAMPNet}
	  \addlegendentry{HyperMIMO}
	  \addlegendentry{Max. Likelihood}
		\end{axis}
		\end{tikzpicture}
	    \caption{Random 2D mobility}
	    \label{fig:chp1_distance}
	\end{subfigure}

	\caption{\gls{SER} achieved by the compared approaches under mobility}
	\label{fig:chp1_mobility}

\end{figure}

The robustness of HyperMIMO to user mobility was tested by evaluating the achieved \gls{SER} when users undergo angular mobility (Fig.~\ref{fig:chp1_angles}) or move in random 2D directions (Fig.~\ref{fig:chp1_distance}) from the positions for which the system was trained.
Fig.~\ref{fig:chp1_angles} was generated by moving all users by a given angle, and evaluating HyperMIMO for these new users positions (and therefore new channel spatial correlation matrices) without retraining.
Note that averaging was done over the two possible directions (clockwise or counterclockwise) for each user.
One can see that the \gls{SER} achieved by HyperMIMO gracefully degrades as the angular displacement increases, and never get worse than \gls{LMMSE} nor OAMPNet.

Fig.~\ref{fig:chp1_distance} was generated by randomly moving the users in random 2D directions.
Users were located at an initial distance of $r=250$\si{m}.
The \gls{SER} was computed by averaging over 100 randomly generated displacements.
As in Fig.~\ref{fig:chp1_angles}, the \gls{SER} achieved by HyperMIMO gracefully degrades as the displacement distance increases.
These results show that, despite having being trained for a particular set of user positions, HyperMIMO remains relatively robust to mobility.
 
\clearpage
\section{New Perspectives and Concluding Thoughts}
\label{sec:chp1_conclusion}

\begin{figure}
    \center
    \includegraphics[width=0.5\linewidth]{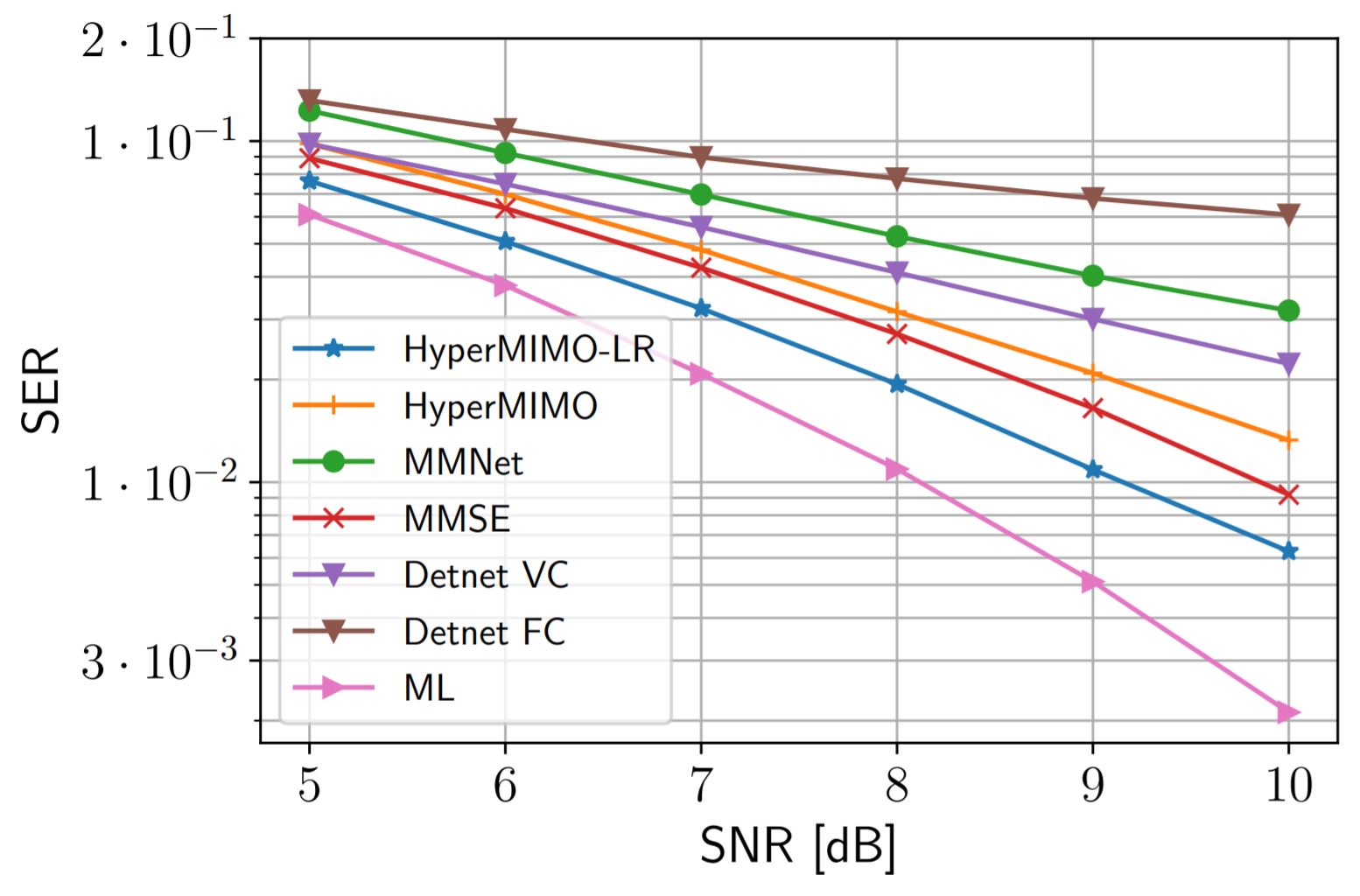}
    \caption{Performance comparison for different detectors, as can be found in~\cite{zilberstein2021robust}}
    \label{fig:chp1_hypermimo_lg_eval}
\end{figure} 
In order to enable performances that remain consistent across a wider set of channels, the authors in~\cite{zilberstein2021robust} proposed a variation of our HyperMIMO architecture, referred to HyperMIMO with learned regularizers, or HyperMIMO-LR.
The key idea is to regularize the hypernetwork outputs with a set of MMNet parameters optimized on a single channel realization.
Let us denote by $\Wm_s$ the set of MMNet parameters estimated by the hypernetwork for a given channel $\Hm_s$, i.e., $\Wm_s = \LP \Thetam, \LSB \thetav^{(1)}, \cdots, \thetav^{(I)} \RSB, \LSB \psiv^{(1)}, \cdots, \psiv^{(I)} \RSB \RP $ when the decomposition presented in Section~\ref{subsec:chp1_mmnet} is performed.
Prior to the hypernetwork training, multiple standalone MMNet detectors are trained on a set of $S$ channels $\Hc = \LP \Hm_1, \cdots, \Hm_S \RP$, resulting in a set of close-to-optimal parameters $\LP \Wm_1^*, \cdots, \Wm_S^* \RP$ that entail good detection performance for their corresponding channels.
Then, the hypernetwork is trained to both minimize the \gls{MSE}, as in~\eqref{eq:chp1_loss_hg}, and penalize the distance between parameters estimated for the set of channel $\Hc$ and the close-to-optimal ones:
\begin{equation}
    \label{eq:chp1_loss_hg_lr}
    \Lc = \EE_{\xv, \Hm, \nv} \left[ ||\widehat{\mathbf{x}}-\mathbf{x}||_{2}^{2} \right] + \beta \sum_{s=1}^{S} || \Wm_s - \Wm_s^* ||_1.
    \end{equation}
The parameter $\beta$ allows choosing a trade-off between optimizing for a given distribution of channels and achieving high performance on a set of pre-specified channels.
Moreover, the $l_1$ norm is leveraged in the right-hand term of~\eqref{eq:chp1_loss_hg_lr} to promote sparse differences between $\Wm_s$ and $\Wm_s^*$ so that some of their entries coincide.

HyperMIMO-LR is evaluated on channels generated following a more realistic Jakes model~\cite{8792117} with time sequences of lengh 4, but the number of user and receive antennas were reduced to $K=2$ and $L=4$, respectively.
The trade-off parameter was set to $\beta=1$ and the dataset $\Hc$ contained 561 channels.
Simulation results, as reported in Fig.~\ref{fig:chp1_hypermimo_lg_eval}, indicate that HyperMIMO-LG is able to provide gains compared to a standard HyperMIMO detector.
One can notice that the varying channels generated from the Jakes model pose significant difficulties to all other \gls{DL}-based methods, as the \gls{LMMSE} detector slightly outperforms HyperMIMO, which itself outperforms DetNet trained with varying channels (DetNet VC), MMNet, and DetNet trained with fixed channels (DetNet FC).

Although HyperMIMO-LG is reasonably close to the performance of the maximum likelihood equalizer (ML in Fig.~\ref{fig:chp1_hypermimo_lg_eval}), the fact that the detector only handles a specific number of users hinders its implementation on beyond-5G systems.
To tackle this problem, a recurrent equivariant MIMO (RE-MIMO) detector has been proposed in~\cite{pratik2020remimo}.
This detector capitalizes on the recent advances in transformer networks~\cite{NIPS2017_3f5ee243} and recurrent inference networks~\cite{putzky2017recurrent} to handle a variable number of users with a single \gls{NN}.
An additional advantage is that the RE-MIMO detector is permutation invariant, which means that the ordering of signals received from the users has no impact on the detection performance. 
RE-MIMO can also be seen as an unfolded \gls{RNN}, where each iterative unit consists of a module that computes the gradient of a likelihood model $P(\yv | \xv)$ and of an encoder and predictor modules that output an updated hidden state and an estimation of the transmitted signal.
Evaluations performed on correlated channels with $K=16$ users and $L=64$ receive antennas show tangible performance gains compared to the OAMPNet algorithm, indicating that RE-MIMO should be able to match the performance of HyperMIMO-based algorithms.
However, this performance and scalability come at a complexity cost, since the attention layers used at every iteration of the unfolded \gls{RNN} each contains multiple large trainable matrices.

Overall, this chapter has introduced a body of work dealing with NN-based optimization of the equalization block.
More specifically, we proposed to leverage the idea of hypernetworks to alleviate the need for retraining an MMNet detector for each channel realization, while still achieving competitive performance.
To reduce the complexity of the hypernetwork, MMNet was modified to decrease its number of trainable parameters, and a form of weights sharing was used.
Simulations revealed that the resulting HyperMIMO architecture achieves near state-of-the-art performance under highly correlated channels when trained and evaluated with the same number of users and with fixed channel statistics.
These weaknesses were subsequently addressed in more recent work, with the HyperMIMO-LR variant providing gains on a wider variety of channels, and more complex schemes such as the RE-MIMO detector being able to handle a varying number of users.
However, while these \gls{NN}-based detectors represent promising improvements compared to traditional algorithms, their block-based optimization still provides no guaranties on the overall receiver optimality.
An additional drawback is that all the presented NN-based detectors  require perfect channel estimates at training, which are usually not available in practice. 
For these reason, we focus in the next chapter on transceiver-based optimization, albeit for \gls{SISO} systems only.
 
\clearpage

\chapter{Learning OFDM Waveforms with PAPR and ACLR Constraints}
\label{ch:2}

\section{Motivation} 
\label{sec:chp2_introduction}

As discussed in Chapter~\ref{ch:1}, \gls{MU-MIMO} is an efficient technique to improve the spectral efficiency when the \gls{BS} is equipped with multiple antennas.
However, spatial multiplexing cannot be exploited on \gls{SISO} systems, and therefore other approaches must be considered.
One of them is the design of new waveforms that satisfy stronger requirements on the signal characteristics.
For example, a higher number of connected devices suggests that the available spectrum should be more efficiently shared among users, challenging the need for guard bands.
Moreover, the \gls{PA} nonlinearities lead to the use of large power back-offs that decrease its efficiency, and therefore the probability of high-amplitude signals should be reduced.
Among possible candidates, \gls{OFDM} is already used in most modern communication systems thanks to a very efficient hardware implementation and a single-tap equalization at the receiver.
However, conventional \gls{OFDM} suffers from multiple drawbacks, including the need for pilots, a high sensitivity to Doppler spread, and both a high \gls{PAPR} and \gls{ACLR}, which might hinder its use in beyond-5G systems.

Future base stations and user equipments are expected to be equipped with dedicated \gls{DL} accelerators~\cite{hoydis20216g}, which should be eventually be able to support fully NN-based transceivers.
In this chapter, we take advantage of such capabilities and review the learning-based approach to OFDM waveform design that we proposed in~\cite{goutay2021learning}.
More specifically, this approach is based on a \gls{CNN} transmitter that implements a high-dimension modulation scheme and a CNN-based receiver that computes \glspl{LLR} on the transmitted bits.
Both transceivers operate on top of OFDM to benefit from its efficient hardware implementation and process OFDM symbols instead of individual \glspl{RE}.
We derive a training procedure which allows to both maximize an achievable information rate and to offset OFDM drawbacks by defining specific optimization constraints.
The end-to-end system training is performed through \gls{SGD}, and therefore the achievable rate and the constraints need to be expressed as functions that can be evaluated and differentiated during training.

In the following, we focus on designing OFDM-based waveforms that enable pilotless transmissions and satisfy \gls{PAPR} and \gls{ACLR} constraints.
The end-to-end system is benchmarked against a close to ideal implementation of a \gls{TR} baseline, in which a number of subcarriers are used to generate peak-reduction signals.
Both systems are evaluated on a \gls{3GPP}-compliant channel model, and the baseline uses a pilot configuration supported by the 5G \gls{NR} specifications.
The end-to-end system, on the contrary, does not use any pilots and learns a high-dimensional modulation that enables accurate detection at the receiver.
Evaluation results show that the learning-based system allows meeting \gls{PAPR} and \gls{ACLR} targets and enables throughput gains ranging from 3\% to 30\% compared to a baseline with similar characteristics.
To get insight into how the proposed system reduces the \gls{PAPR} and \gls{ACLR} while maintaining high rates, we have carried out a detailed study of the learned high-dimensional modulation scheme.
We have found out that the \gls{ACLR} and \gls{PAPR} reductions are achieved through a combination of spectral filtering, uneven energy allocation across subcarriers, and positional adjustments of constellation symbols.
To the best of our knowledge, this method is the first \gls{DL}-based approach that jointly maximizes an information rate of OFDM transmissions and allows setting \gls{PAPR} and \gls{ACLR} targets.

\subsection*{Related literature}

To counteract the drawbacks of OFDM-based waveforms, previous works suggested to filter the analog signal, to modify the constellation used for modulation, or to inject custom signals on reserved subcarriers.
The first approach is known as iterative clipping and filtering, and iterates between clipping in the time-domain  and filtering in the frequency domain to constrain the signal amplitude and possibly the spectral leakage~\cite{wang2010optimized}.
The second scheme, named active constellation extension, extends the outer symbols of a constellation to reduce the signal \gls{PAPR} at the cost of an increased power consumption~\cite{krongold2003reduction}.
Finally, the third technique reserves a subset of available subcarriers to create a peak-cancelling signal, and is referred to as tone reservation (TR)~\cite{1261335}.

Motivated by the success of \gls{DL} in other physical layer tasks, multiple works suggested replacing existing \gls{PAPR} reduction algorithms with \gls{DL} components. 
For example, an \gls{NN} was used to generate the constellation extension vectors in~\cite{8928056}.
Similarly, a \gls{TR} algorithm was unfolded as \gls{NN} layers in~\cite{wang2020novel, wang2021model, 8928103}.
It has also been proposed to model the communication system as an autoencoder that is trained to both minimize the \gls{PAPR} and symbol error rate~\cite{8240644}.
Although closer to our approach, the proposed scheme operates on symbols, meaning that the bit mapping and demapping have to be implemented separately, and has only been evaluated on a simple Rayleigh fading channel.
Moreover, the time-domain signal is not oversampled, which is required to obtain an accurate representation of the underlying waveform for \gls{PAPR} calculations~\cite{922754}.
Finally, none of these works allow setting precise \gls{PAPR} and \gls{ACLR} targets, which means that the trade-off between the \gls{PAPR}, \gls{ACLR}, and spectral efficiency can not be accurately controlled.

Regarding the end-to-end training of communication systems, numerous works proposed to use this technique to design transceivers aimed at different channels.
For example, the learning of constellation geometries to achieve pilotless and \gls{CP}-less communication over OFDM channels was done in~\cite{pilotless20}, and the design of NN-based systems for multicarrier fading channels and optical fiber communications were respectively studied in~\cite{9271932} and~\cite{8433895}.
Finally, the design of transmit and receive filters for single-carrier systems under spectral and \gls{PAPR} constraints was presented in~\cite{aoudia2021waveform}.

\clearpage
\section{Problem Statement} 
\label{sec:chp2_problem_positioning}

A \gls{SISO} system using OFDM is considered ($K=L=1$). 
In this section, the OFDM channel model and the expressions of the signal waveform and spectrum are recalled.
The \gls{ACLR} and \gls{PAPR} metrics typically used to characterize the analogue signal are then detailed. 
Finally, a close to ideal implementation of a \gls{TR} baseline is introduced, where a subset of subcarriers are reserved to minimize the signal \gls{PAPR} and pilots are transmitted to estimate the channel.

\subsection{System Model}

\subsubsection{Channel model}
 
\sloppy The \gls{OFDM} channel model as derived in Section~\ref{sec:the_wireless_channel} is considered, with $N$ subcarriers and one time slot, which consists of $M=14$ OFDM symbols.
In this chapter, the subcarriers are indexed by the set $\mathcal{N}= \LP -\frac{N -1}{2}, \cdots, \frac{N -1}{2} \RP $, with $N$ assumed odd for convenience.
When considering the entire \gls{RG}, the OFDM channel of~\eqref{eq:intro_mimo_vectors} can be expressed as
\begin{align}
    \Ym = \Hm \odot \Xm + \Nm
    \label{eq:chp2_OFDM_channel}
\end{align}
where $\Xm \in \mathbb{C}^{M \times N}$ and $\Ym \in \mathbb{C}^{M \times N}$ respectively represent the matrix of transmitted and received \glspl{FBS}, $\Hm \in \mathbb{C}^{M \times N}$ is the matrix of channel coefficients, and {$\Nm~\in~\mathbb{C}^{M \times N}$} is the additive Gaussian noise matrix such that each element has a variance $\sigma^2$.
We consider a slow-varying environment so that the channel can be assumed constant over the duration of a slot.
The matrix of bits to be transmitted on the OFDM symbol $m$ is denoted $\Bm_{m} = \left[  \bv_{m,1}, \cdots, \bv_{m, N} \right]\tp$, where $\bv_{m, n}\in \{0,1\}^{Q}, m \in \{1, \cdots, M\}, n \in \Nc$,  is a vector of bits to be transmitted and $Q$ is the number of bits per channel use. 
The transmitter modulates each $\Bm_{m}$ onto the \glspl{FBS} $\xv_{m} \in \CC^{N}$, which are mapped on the orthogonal subcarriers to form the spectrum
\begin{align}
    S_{m}(f) = \sum_{n \in \mathcal{N}} x_{m, n} \frac{1}{\sqrt{\Delta_f}}\text{sinc} \left( \frac{f}{\Delta_f} - n \right)
\end{align}
where $\Delta_f$ is the subcarrier spacing.
When \glspl{CP} on duration $T^{\text{CP}}$ are prepended to the \gls{OFDM} symbols of duration $T$, the signal spectrum becomes 
\begin{align}
    \label{eq:chp2_s_cp}
    S_{m}^{\text{CP}}(f) = \sum_{n \in \mathcal{N}} x_{m, n} \frac{1}{\sqrt{\Delta_f^{\text{CP}}}}\text{sinc} \left( \frac{f-n\Delta_f}{\Delta_f^{\text{CP}}} \right)
\end{align}
where $\Delta_f^{\text{CP}} = \frac{1}{T+T^{\text{CP}}}$.
We recall from~\eqref{eq:intro_time_signal} that the corresponding signal is expressed as
\begin{align}
    s(t)=\sum_{m=0}^{M-1} s_{m}(t)=\sum_{m=0}^{M-1} \sum_{n _in \mathcal{N}} x_{m,n} \phi_{n}(t-mT^{\text{tot}})
\end{align}
where $T^{\text{tot}} = T+T^{\text{CP}}$ and the transmit filters $\phi_{n}(t), n\in \Nc $ are defined as
\begin{align}
    \phi_{n}(t) = \frac{1}{\sqrt{T^{\text{tot}}}} \text{rect}\LB \frac{t}{T^{\text{tot}}} -\frac{1}{2} \RB e^{j 2 \pi n \frac{ t-T^{\text{CP}}}{T}}.
\end{align}

\subsubsection{Relevant metrics}
\Gls{OFDM} waveforms have, inter alia, two major drawbacks.
The first one is their high amplitude peaks, which create distortions in the output signal due to the saturation of the \gls{PA}.
Such distortions are usually reduced by operating the PA with a large power back-off or by leveraging complex digital pre-distortion, thus reducing the power efficiency.
Let us denote by $\nu(t) = \frac{|s(t)|^2}{\EE \left[|s(t)|^2 \right]} $ the ratio between the instantaneous and average power of a signal.
We define the $\text{PAPR}_{\epsilon}$ as the smallest $e\geq 0$, such that the probability of $\nu(t)$ being larger than $e$ is smaller than a threshold $\epsilon \in \left( 0, 1 \right)$:
\begin{align}
    \label{eq:chp2_papr}
    \text{PAPR}_{\epsilon}  \coloneqq \mathrm{min} \; e, \;\; \text{s. t.} \;\;  P\left(\nu(t) > e  \right) \leq \epsilon.
\end{align}
Setting $\epsilon=0$ leads to the more conventional \gls{PAPR} definition $\frac{\max{|s(t)|^2}}{\EE \left[|s(t)|^2 \right]}$.
However, the maximum signal power occurs with very low probability, and therefore such a definition of the \gls{PAPR} only has a limited practical interest. 
Relaxing $\epsilon$ to values greater than 0 allows considering more frequent, and therefore more practically relevant, power peaks.

\begin{figure}
    \centering
\input{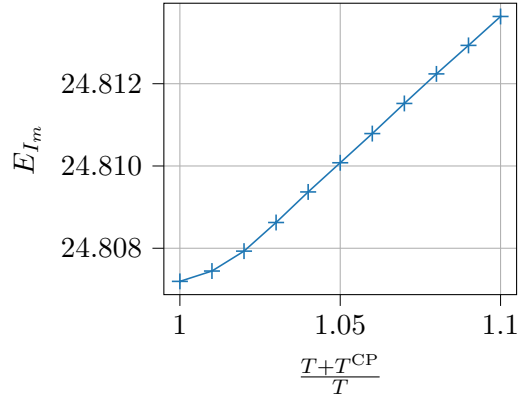}
\caption{Effect of the \gls{CP} length on the in-band energy.}
\label{fig:chp2_cp}
\end{figure}

The second drawback of OFDM is its low spectral containment. 
This characteristic is typically measured with the \gls{ACLR}, which is the ratio between the expected out-of-band energy $\EE_{\xv_m}\left[ E_{O_m} \right]$ and the expected in-band energy $\EE_{\xv_m}\left[ E_{I_m}\right]$:
\begin{align}
    \label{eq:chp2_aclr}
    \text{ACLR} \coloneqq  \frac{\EE_{\xv_m} \left[ E_{O_m} \right]}{\EE_{\xv_m} \left[ E_{I_ m}\right]} 
     =   \frac{\EE_{\xv_m} \left[ E_{A_m} \right]}{\EE_{\xv_m} \left[ E_{I_m} \right]}-1 
\end{align}
where $E_{O_m}$, $E_{I_m}$, and $E_{A_m} = E_{O_m} + E_{I_m}$ are respectively the out-of-band, in-band, and total energy of the OFDM symbol $m$.
The in-band energy $E_{I_m}$ is given by
\begin{equation}
\begin{split}
E_{I_m} & \coloneqq \int_{-\frac{N \Delta_f}{2}}^{\frac{N \Delta_f}{2}} \left| S_m (f)\right|^2 df = \xv_m ^H \Jm \xv_m\\
\end{split} 
\end{equation}
where each element $j_{a, b}$ of the matrix $\Jm \in \RR^{N \times N}$ is 
\begin{equation}
j_{a, b} = \frac{1}{\Delta_f^{\text{CP}}} \int_{-\frac{N \Delta_f}{2}}^{\frac{N \Delta_f}{2}}  \text{sinc} \left(\frac{f - a \Delta_f}{\Delta_f^{\text{CP}}} \right) \text{sinc} \left(\frac{f - b \Delta_f}{\Delta_f^{\text{CP}}} \right) df, \quad a, b \in \Nc.
\end{equation}
The effect on the \gls{CP} length on the in-band energy is shown in~Fig.\ref{fig:chp2_cp}, which have been obtained by sending $10^6$ random \glspl{FBS} $\xv_m \thicksim \Cc \Nc (\mathbf{0}, \frac{1}{\sqrt{2}}\Id)$ on $N=25$ subcarriers with \gls{CP} lengths of  $T^{\text{CP}} \in [0, 0.1T]$.
It can be seen that the in-band energy increases with the CP length, which is to be expected as increasing  $T^{\text{CP}}$ amounts to modulating the subcarriers with sinc for which the ripples are brought closer together, thus containing more energy in $\LSB \Delta_f \LB n-\frac{1}{2} \RB, \Delta_f \LB n+\frac{1}{2} \RB \RSB$.
This in-band energy increase can be directly mapped to an \gls{ACLR} decrease, as the total energy does not depend on CP length.
In the following, we therefore consider that $T^{\text{CP}} = 0$ when computing the time-domain representation and spectrum of the signal, as it corresponds to the worst-case scenario in terms of spectral energy leakage. 

Finally, the total energy can be more conveniently computed in the time domain:
\begin{align}
E_{A_m} \coloneqq \int_{ -\frac{T}{2}}^{\frac{T}{2}} \left| s (t)\right|^2 dt = \xv_m ^H \Km \xv_m
\end{align}
where $\Km \in \RR^{N \times N}$ has elements
\begin{equation}
k_{a, b} = \frac{1}{T} \int_{-\frac{T}{2}}^{\frac{T}{2}}  e^{i 2 \pi (a-b) t /T} dt, \quad a, b \in \Nc .
\end{equation}

\subsection{Baseline}
One technique to reduce the \gls{PAPR} of OFDM signal is \gls{TR}, in which a subset of $R$ tones (subcarriers) is used to create peak-reduction signals.
These subcarriers are referred to as \glspl{PRT}, and the remaining $D$ subcarriers are used for data and pilot transmission.
The sets containing the \glspl{PRT} and the data-carrying subcarriers are respectively denoted by $\mathcal{R}$ and $\mathcal{D}$, and are such that $\mathcal{R} \cup \mathcal{D} = \mathcal{N}$.

\subsubsection{Transmitter}

\begin{figure}[t]
    \centering
    \includegraphics[height=115pt]{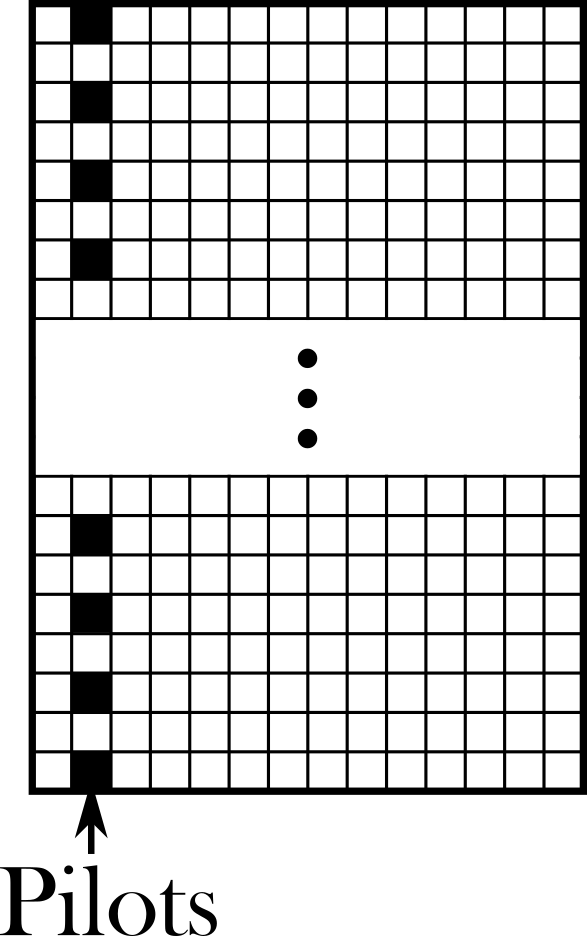}
    \caption{Pilot pattern used by the baseline.} 
    \label{fig:chp2_pilots}
\end{figure}

The \gls{TR}-based transmitter sends three types of signals: data signals, peak reduction signals, and pilot signals which are used by the receiver to estimate the channel.
Such pilots are inserted in the \gls{RG} following the 5G \gls{NR} pattern illustrated in Fig.~\ref{fig:chp2_pilots}, i.e., every two \glspl{RE} on the second \gls{OFDM} symbol, and the value of each pilot is chosen randomly on the unit circle. 
For clarity, only data and peak-reduction signals are considered when describing the transmitter, as transmitting pilots is achieved by simply replacing some \glspl{RE} carrying data by reference signals.
We denote by $d_{m,n\in\mathcal{D}}$ and $r_{m,n\in\mathcal{R}}$ the \glspl{FBS} carrying data and peak-reduction signals, respectively.
An \gls{FBS} $ d_{m, n\in\mathcal{D}}$ corresponds to the mapping of a vector of bits $\bv_{m, n\in\mathcal{D}}$ following a $2^Q$-\gls{QAM}, the constellation of which is denoted by $\mathcal{C}\in \CC^{2^{Q}}$.
The vector of \glspl{FBS} $\dv_{m}\in\CC^{N}$ is composed of all $ d_{m, n\in\mathcal{D}}$ and of zeros at positions that correspond to  \glspl{PRT}, i.e., $d_{m, n\in\mathcal{R}} = 0$.
The reduction vector $\rv_{m}\in\CC^{N}$ is formed by the signals $r_{m,n\in\mathcal{R}}$ mapped to the \glspl{PRT}, and is conversely such that $r_{m, n\in\mathcal{D}} = 0$.
As an example, if three subcarriers are used and only the last one is used as a \gls{PRT}, these vectors are expressed as $\dv_m = \left[ d_{m, -1}, d_{m, 0}, 0 \right]\tp$ and $\rv_m = \left[ 0, 0, r_{m, 1} \right]\tp$.
The vector of discrete baseband signals to be transmitted and the corresponding continuous-time waveform are finally denoted by $ \xv_m = \dv_m + \rv_m$ and $s_m(t)$, respectively.
\gls{TR} aims at finding $\rv_m$ that minimizes the maximum squared signal amplitude: 
\begin{align}
    \arg\min_{\rv_m} \max_{t} |s_m (t)|^2.
    \label{eq:chp2_papr_1}
\end{align}
As minimization over the time-continuous signal leads to intractable calculations, $s_m (t)$ is first discretized. 
Many previous studies considered a discrete vector $\zv_m \in \CC^{N}$, sampled with a period $\frac{T}{N}$, as a substitute for the underlying signal~\cite{wang2021model, 8928103, 8240644}.
However, it has been shown that using a vector $\underline{\zv}_m \in \CC^{NO_S}$, oversampled by a factor $O_S$, is necessary to correctly represent the analog waveform~\cite{922754}.
\begin{figure}
    \centering
\input{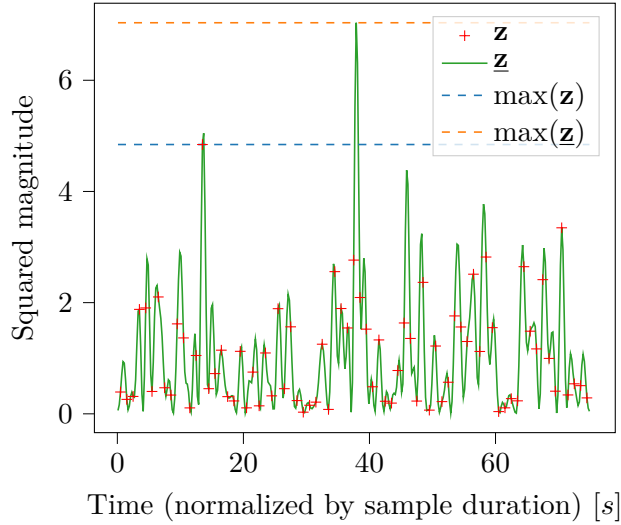}
\caption{OFDM signal generated from $N=75$ subcarriers.}
\label{fig:chp2_waveform}
\end{figure}
The difference between the two discretized vectors are visible in Fig.~\ref{fig:chp2_waveform}, where the squared amplitude of $\zv_m$ and $\underline{\zv}_m$ are plotted for an arbitrary OFDM symbol, with $N=75$ subcarriers and an oversampling factor of $O_S=5$.
It can be seen that the oversampled signal exhibits a different maximum peak with a higher amplitude as well as numerous secondary peaks that are not present in the non-oversampled signal.
These peaks might lie in the \gls{PA} saturation region, causing distortions of the transmitted waveform.
Let us define the \gls{IDFT} matrix $\Fm\htp \in \CC^{NO_S \times N}$, where each element is expressed as $f_{a, b} = \frac{1}{\sqrt{N}O_S}e^{\frac{j2\pi a b}{NO_S}}$.
The oversampled vector can be obtained with
\begin{align}
    \underline{\zv}_m = \Fm\htp \xv_m = \Fm \left( \dv_m + \rv_m \right).
\end{align}

\pagebreak
The value of $\rv_m$ that minimizes the \gls{PAPR} can now be approximately found by minimizing the oversampled signal:
\begin{align}
    \arg\min_{\cv_m} \left\Vert g \left( \Fm\htp (\dv_m + \rv_m) \right) \right\Vert_{\infty}
\label{eq:chp2_papr_2}
\end{align}
where $g(\cdot)$ denotes the element-wise squared magnitude $| \cdot |^{2}$ and $ \left\Vert \cdot \right\Vert_{\infty}$ denotes the infinity norm.
\sloppy The convexity of $\left\Vert g \left( \Fm (\dv_m + \rv_m) \right) \right\Vert_{\infty}$ theoretically allows to find the optimal value of $\rv_m$, but the associated complexity leads to the development of algorithms that approximate this value in a limited number of iterations~\cite{1261335}.
In this thesis, however, we use a convex solver~\cite{diamond2016cvxpy} to find the exact solution of \eqref{eq:chp2_papr_2} for each symbol $\xv_m$.
Although such a scheme would be prohibitively complex in practice, it is considered here to provide a close to ideal implementation of a TR-based baseline.
Moreover, for fairness with conventional \gls{QAM} systems, we add the convex constraint $\rv_m \htp \rv_m \leq R$ so that the average energy per OFDM symbol equals at most $N$. 
We experimentally verified that the average energy of the peak reduction signals $\EE_{\rv_m}\left[ \rv_m\htp \rv_m \right] $ was always close to $R$, leading to $ \EE_{\xv_m}\left[ \xv_m\htp \xv_m \right] \approx N$.
Finally, it was shown that placing the \glspl{PRT} at random locations at every transmission leads to the lowest \gls{PAPR} among other placement techniques~\cite{fcd29097-44f6-450e-8fc9-981cffc60574}.
The baseline therefore implements such a random positioning scheme for all OFDM symbols, except for the one carrying pilots for which peak-reduction signals cannot be inserted on pilot-carrying subcarriers.
On this specific OFDM symbol, the number of \glspl{PRT} is also reduced to $\frac{R}{2}$ in order to always maintain a significant number of subcarriers carrying data.

\subsubsection{Receiver}
On the receiver side, channel estimation is performed first, using the pilot signals received in the pilot-carrying OFDM symbol $m^{(p)}\in\mathcal{M}$.
The pattern depicted in Fig.~\ref{fig:chp2_pilots} allocates $\frac{N+1}{2}$ \glspl{RE} to pilot transmissions. 
Let us denote by $\pv_{m^{(p)}}\in\CC^{\frac{N+1}{2}}$ the vector of received pilot signals, extracted from $\yv_{m^{(p)}}$.
The channel covariance matrix, providing the spectral correlations between all \glspl{RE} carrying pilots, is denoted by $\Sigmam\in\CC^{\frac{N+1}{2} \times \frac{N+1}{2}}$.
This matrix can be empirically estimated by constructing a large dataset of received pilot signals and computing the statistics over the entire dataset.
The channel coefficients at \glspl{RE} carrying pilots are estimated using an \gls{LMMSE} channel estimator:
\begin{align}
    \widehat{\hv}_{m^{(p)}}^{(p)} = \Sigmam \left( \Sigmam + \sigma^2 \mathbf{I}_{\frac{N+1}{2}} \right)^{-1} \pv_{m^{(p)}} \quad \in \CC^{\frac{N+1}{2}} .
\end{align}
Channel estimation at the remaining $N$ \glspl{RE} of the OFDM symbol $m^{(p)}$ is achieved through linear interpolation.
As the channel is assumed to be invariant over the duration of a slot, the so obtained vector $\widehat{\hv}_{m^{(p)}}  \in \CC^{N}$ is also used for all other OFDM symbols, forming the channel estimate matrix $\widehat{\Hm} \in \CC^{M \times N}$ where all columns are equal.
On fast changing channels, pilots could be inserted in other OFDM symbols to better track the evolution of the channel.

The transmitted \glspl{FBS} are estimated through equalization:
\begin{align}
    \widehat{\Xm} = \Ym \oslash \widehat{\Hm}.
\end{align}
Finally, the \gls{LLR} of the $q^{\text{th}}$ bit corresponding to the RE $(m,n)$ is computed with a conventional \gls{AWGN} demapper:
\begin{align}
    \text{LLR}_{m,n}(q) = \text{ln} \left( \frac{
        \sum_{c \in \mathcal{C}_{q, 1}} \text{exp} \left( - \frac{|\hat{h}_{m,n}|^2 }{\sigma^2 } | \hat{x}_{m,n} - c |^2 \right)}
        {\sum_{c \in \mathcal{C}_{q, 0}} \text{exp} \left( - \frac{|\hat{h}_{m,n}|^2 }{\sigma^2 } | \hat{x}_{m,n} - c |^2 \right)} 
        \right)
\end{align}
where $\mathcal{C}_{q, 1}$ ($\mathcal{C}_{q, 0}$) is the subset of $\mathcal{C}$ containing the symbols that have the $q^{\text{th}}$ bit set to 1 (0), and $\frac{|\hat{h}_{m,n}|^2 }{\sigma^2 }$ is the post-equalization noise variance.

\clearpage
\section{Learning a High Dimensional Modulation} 
\label{sec:chp2_high_dimensional_modulation}

\begin{figure}[t]
    \centering
    \includegraphics[width=0.55\textwidth]{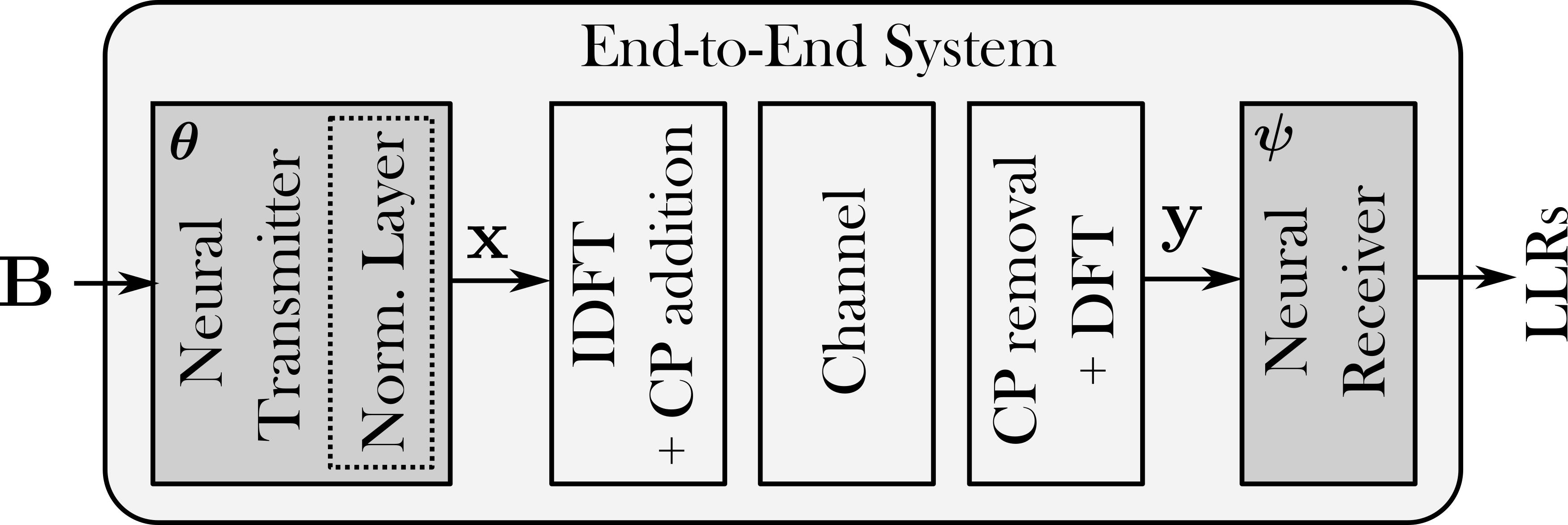}
    \caption{Trainable system, where grayed blocks represent trainable components.}
    \label{fig:chp2_E2E_system}
\end{figure}

In the following, we train an \gls{NN}-based transmitter and receiver to maximize an achievable rate under \gls{ACLR} and \gls{PAPR} constraints.
This end-to-end system is referred to as "E2E" system for brevity, and is schematically shown in Fig.~\ref{fig:chp2_E2E_system}.
An optimization procedure is derived to handle the constrained optimization problem, in which the loss function is expressed as a differentiable augmented Lagrangian. 
Next, we detail the transmitter and receiver architectures, both implemented as \glspl{CNN}.

\subsection{Optimization Procedure}

\subsubsection{Problem formulation}
We aim at finding a high-dimensional modulation and associated detector that both maximize an information rate for the OFDM transmission and satisfy constraints on the signal \gls{PAPR} and \gls{ACLR}.
The transmitter and receiver of the E2E system operate on top of OFDM, i.e., \gls{IDFT} \glsunset{DFT} (\gls{DFT}) is performed and a cyclic prefix is added (removed) before (after) transmission (see Fig.~\ref{fig:chp2_E2E_system}).
The considered rate \cite{pilotless20} is an extension of the one derived in~\eqref{eq:bkg_link_ce_rate} for an entire \gls{RG}, and depends on the transmitter and receiver trainable parameters respectively denoted by $\thetav$ and $\psiv$:
\begin{align}
    \label{eq:chp2_rate0}
    C(\thetav, \psiv) & = \frac{1}{MN}\sum_{m \in \mathcal{M}}\sum_{n \in \mathcal{N}}\sum_{q=0}^{Q-1}I\left( b_{m,n,q}; \yv_m \right | \thetav)\\
    & - \frac{1}{MN}\sum_{m\in\mathcal{M}}\sum_{n \in \mathcal{N}}\sum_{q=0}^{Q-1}\EE_{\yv_m}\left[\text{D}_\text{KL} \left( P (b_{m,n,q}|\yv_m)|| \widehat{P}_{\psiv}(b_{m,n,q}| \yv_m ) \right)\right]. \nonumber
\end{align}
As the E2E system outputs \glspl{LLR}, the estimated posterior probabilities can be obtained from
\begin{equation}
    \text{LLR}_{m,n}(q) \coloneqq \text{ln} \left( \frac{\widehat{P}_{\psiv}(b_{m,n,q} = 1| \yv_m ) }{\widehat{P}_{\psiv}(b_{m,n,q} = 0| \yv_m ) } \right).
\end{equation}

To ensure a unit average energy per \gls{RE}, a normalization layer is added to the transmitter (see  Fig.~\ref{fig:chp2_E2E_system}).
Perfect normalization would perform
\begin{align}
    \label{eq:chp2_layer_norm*}
    l_{\text{norm}}^*(\xv_m) = \frac{\xv_m}{ \left( \frac{1}{N} \EE_{\xv_m}[E_{A_m}] \right)^{\frac{1}{2}} }
\end{align}
but the $2^{MNQ}$ different combinations of bits would make the computation of the expected value too complex for any practical system.
Batch normalization is therefore preferred, ensuring that the average energy per RE in the batch is one:
\begin{align}
    \label{eq:chp2_layer_norm}
    l_{\text{norm}}(\xv_m^{[j]}) = \frac{\xv_m^{[j]}}{  \left( \frac{1}{M N B_S}  \sum_{m\in\mathcal{M}}\sum_{i=1}^{B_S}  \xv_m^{[i]^{\mathsf{H}}} \Km \xv_m^{[i]} \right)^{\frac{1}{2}} }
\end{align}
where the superscript $[j]$ denotes the $j^{\text{th}}$ element in the batch and $B_S$ is the batch size.
This expression is slightly different from the one typically used in related works, since it accounts for the correlation that can appear between the \glspl{FBS} generated by the transmitter.
Conventional bit-interleaved modulation systems produces \gls{iid} symbols, and therefore does not need to take such correlation into account.

We can now formulate the constrained optimization problem we aim to solve: 
\begin{subequations}
    \label{eq:chp2_rate}
     \begin{align}
    \underset{\thetav, \psiv}{\text{maximize}} & \quad\quad C(\thetav, \psiv) \label{eq:chp2_rate1} \\
   \text{subject to}  & \quad\quad \text{PAPR}_{\epsilon}(\thetav) = \gamma_{\text{peak}} \label{eq:chp2_rate3} \\
    &  \quad\quad \text{ACLR}(\thetav) \leq \beta_{\text{leak}} \label{eq:chp2_rate4} 
     \end{align}
\end{subequations}
where $\gpeak$ and $\bleak$ respectively denote the target \gls{PAPR} and \gls{ACLR}.
Note that the \gls{PAPR} and \gls{ACLR} depend on the transmitter parameters $\thetav$.

\subsubsection{System training}
One of the main advantages of implementing the transmitter-receiver pair as an E2E system is that it enables optimization of the trainable parameters through \gls{SGD}.
As seen in Chapter~\ref{ch:background}, this requires a differentiable loss function so that the gradients can be computed and backpropagated through the E2E system.
In the following, the augmented Lagrangian method is leveraged to convert the problem \eqref{eq:chp2_rate} into its augmented Lagrangian, which acts a differentiable loss function that can be minimized with respect to $\thetav$ and $\psiv$~\cite{bertsekas2014constrained}.
The key idea is to relax the constrained optimization problem into a sequence of unconstrained problems that are solved iteratively.
This method is known to be more effective than the quadratic penalty method, enabling a faster and more stable convergence~\cite{nocedal2006numerical}.
In the following, we express the objective \eqref{eq:chp2_rate1} and the constraints \eqref{eq:chp2_rate3} and \eqref{eq:chp2_rate4} as differentiable functions that can be evaluated during training and minimized with \gls{SGD}.

First, the achievable rate \eqref{eq:chp2_rate1}  can be equivalently expressed using the system \gls{BCE}~\cite{9118963}, which is widely used in binary classification problems:
\begin{align}
    \label{eq:chp2_CE}
    L_C(\thetav, \psiv) &:= - \frac{1}{MN}\sum_{m\in\mathcal{M}}\sum_{n\in\mathcal{N}}\sum_{q=0}^{Q-1} \EE_{\yv_m} \left[ \text{log}_2 \left(\widehat{P}_{\psiv} (b_{m,n,q}| \yv_m ) \right) \right]  \\
    & = Q -  C(\thetav, \psiv). 
\end{align}
To overcome the complexity associated with the computation of the expected value, an approximation is typically obtained through Monte Carlo sampling:
\begin{align}
    \label{eq:chp2_CE_batch}
    L_C(\thetav, \psiv) \approx & - \frac{1}{MNB_S}\sum_{m\in\mathcal{M}}\sum_{n\in\mathcal{N}}\sum_{q=0}^{Q-1}\sum_{i=0}^{B_S-1}  \text{log}_2 \left( \widehat{P}_{\psiv} \left( b_{m,n,q}^{[i]}| \yv_m ^{[i]} \right) \right).
\end{align}

Second, evaluating the constraint \eqref{eq:chp2_rate3} requires the computation of the probability $P(\frac{|s(t)|^2}{\EE \left[|s(t)|^2 \right]}  > e)$, where $e$ is the energy threshold defined in \eqref{eq:chp2_papr}.
However, computing such probability would be prohibitively complex due to the sheer amount of possible OFDM symbols.
During training, we therefore enforce the constraint by setting $\epsilon=0$ and penalizing all signals whose squared amplitude exceed $\gpeak$. 
With $\epsilon=0$, the constraint \eqref{eq:chp2_rate3} is equivalent to enforcing $L_{\gamma_{\text{peak}}}(\thetav) =0$, with
\begin{align}
    \label{eq:chp2_loss_papr}
    L_{\gamma_{\text{peak}}}(\thetav)  = \EE_m \left[ \int_{-\frac{T}{2}}^{\frac{T}{2}} \left(|s_m(t)|^2-\gamma_{\text{peak}} \right)^+ dt \right]
\end{align}
where $(x)^+$ denotes the positive part of $x$, i.e., $ (x)^+= \text{max}(0, x)$.
To evaluate $L_{\gamma_{\text{peak}}}(\thetav)$ during training, the value of the expectation can be obtained through Monte Carlo sampling, and the integral can be approximated using a Riemann sum:
\begin{align}
    L_{\gamma_{\text{peak}}}(\thetav)  \approx \frac{T}{B_S N O_S} \sum_{i=0}^{B_S-1} \sum_{t=-\frac{NO_S-1}{2}}^{\frac{NO_S-1}{2}}  \left(\left| \underline{z}_{m,t}^{[i]} \right| ^2 - \gamma_{\text{peak}} \right)^+
    \label{eq:chp2_loss_papr2}
\end{align}
where $\underline{\zv}_m = \Fm\htp \xv_m \in \CC^{NO_S}$ is the vector of the oversampled time signal corresponding to the neural transmitter output  $\xv_m$.

Third, the inequality constraint \eqref{eq:chp2_rate4} can be converted to the  equality constraint $\text{ACLR}(\thetav) - \beta_{\text{leak}}  = -v$, where $v\in\RR_+$ is a positive slack variable.
This equality constraint is then enforced by minimizing $L_{\beta_{\text{leak}}}(\thetav) + v$, with
\begin{align}
L_{\beta_{\text{leak}}}(\thetav)  & =  \frac{ \EE \left[ E_A \right]}{ \EE \left[ E_I \right]}-1   - \beta_{\text{leak}} \\
& \approx \frac{  \frac{1}{B_S} \sum_{i=0}^{B_S-1}  \xv^{[i]^{\mathsf{H}}} \Wm \xv^{[i]}}{ \frac{1}{B_S} \sum_{i=0}^{B_S-1}  \xv^{[i]^{\mathsf{H}}} \Vm \xv^{[i]}} -1   - \beta_{\text{leak}} .
\end{align}

Finally, for $\epsilon=0$, the problem \eqref{eq:chp2_rate} can be reformulated as 
\begin{subequations}
    \label{eq:chp2_pb}
     \begin{align}
    \underset{\thetav, \psiv}{\text{minimize}} & \quad\quad L_C(\thetav, \psiv) \label{eq:chp2_pb1} \\
    \text{subject to} & \quad\quad L_{\gamma_{\text{peak}}}(\thetav)  = 0 \label{eq:chp2_pb2} \\
    &  \quad\quad L_{\beta_{\text{leak}}}(\thetav) + v = 0 \label{eq:chp2_pb3} 
     \end{align}
\end{subequations}
where the objective and the constraints are differentiable and can be estimated at training.
The augmented Lagrangian method introduces two types of hyperparameters that are iteratively updated during training.
The first one corresponds to the penalty parameters which are slowly increased to penalize the constraint with increasing severity. 
The second one corresponds to estimates of the Lagrange multipliers, as defined in~\cite{bertsekas2014constrained}. 
Let us denote by  $\mu_p > 0$ and $\mu_l>0$ the penalty parameters and by $\lambda_p$ and $ \lambda_l$ the Lagrange multipliers for the constraint functions $L_{\gamma_{\text{peak}}}(\thetav)$ and $L_{\beta_{\text{leak}}}(\thetav)$, respectively.
The corresponding augmented Lagrangian is~\cite{bertsekas2014constrained}
\begin{align}
    \label{eq:chp2_lagrange_slack}
    \overline{L}^* (\thetav, \psiv, \lambda_p, \lambda_l, & \mu_p, \mu_l, q)  = L_C(\thetav, \psiv) \nonumber \\
    & + \lambda_p L_{\gamma_{\text{peak}}}(\thetav) + \frac{1}{2} \mu_p |L_{\gamma_{\text{peak}}}(\thetav)|^2 \\
    & + \lambda_l  \left( L_{\beta_{\text{leak}}}(\thetav) + v \right) + \frac{1}{2} \mu_l \left\lvert L_{\beta_{\text{leak}}}(\thetav) + v \right\rvert^2. \nonumber
\end{align}
As derived in~\cite{bertsekas2014constrained}, the minimization of~\eqref{eq:chp2_lagrange_slack} with respect to $v$ can be carried out explicitly for each fixed pair of $\{\thetav, \psiv\}$ so that the augmented Lagrangian can be expressed as
\begin{align}
    \label{eq:chp2_lagrange}
    \overline{L} (\thetav, \psiv, \lambda_p, \lambda_l, & \mu_p, \mu_l)  = L_C(\thetav, \psiv) \nonumber \\
    & + \lambda_p L_{\gamma_{\text{peak}}}(\thetav) + \frac{1}{2} \mu_p |L_{\gamma_{\text{peak}}}(\thetav)|^2 \\
    & + \frac{1}{2\mu_l} \left( \text{max}(0, \lambda_l + \mu_l L_{\beta_{\text{leak}}}(\thetav) )^2 - \lambda_l^2 \right). \nonumber
\end{align}
Each training iteration comprises multiples steps of SGD on the augmented Lagrangian \eqref{eq:chp2_lagrange} followed by an update of the hyperparameters.
The optimization procedure is detailed in Algorithm \ref{alg:chp2_lagrangian}, where $\tau \in \RR^+$ controls the evolution of the penalty parameters and the superscript $(u)$ refers to the $u^{\text{th}}$ iteration of the algorithm.

\begin{algorithm}
    \SetAlgoLined
     Initialize $\thetav, \psiv, \lambda_p^{(0)}, \lambda_l^{(0)}, \mu_p^{(0)}, \mu_l^{(0)}$ \\
     \For{$u = 0, \cdots $}{
      $\triangleright$ Perform multiple steps of SGD \\
      on $\overline{L} (\thetav, \psiv, \lambda, \lambda_l, \mu_p, \mu_l)$ w.r.t. $\thetav$ and $ \psiv$ \\
      $\triangleright$ Update optimization hyperparameters: \\
      $\lambda_p^{(u+1)} = \lambda_p^{(u)} + \mu_p^{(u)} L_{\gamma_{\text{peak}}}(\thetav) $\\
      $\lambda_l^{(u+1)} = \text{max} \left(0, \lambda_l^{(u)}  + \mu_l^{(u)} L_{\beta_{\text{leak}}}(\thetav) \right)$\\
      $\mu_p^{(u+1)} = (1+\tau) \mu_p^{(u)}$ \\
      $\mu_l^{(u+1)} = (1+\tau) \mu_l^{(u)}$ 
     }
     \caption{Training procedure}
     \label{alg:chp2_lagrangian}
\end{algorithm}

\subsection{System Architecture}

\begin{figure}
    \centering
    \begin{subfigure}{.45\textwidth}
        \centering
        \includegraphics[height=170pt]{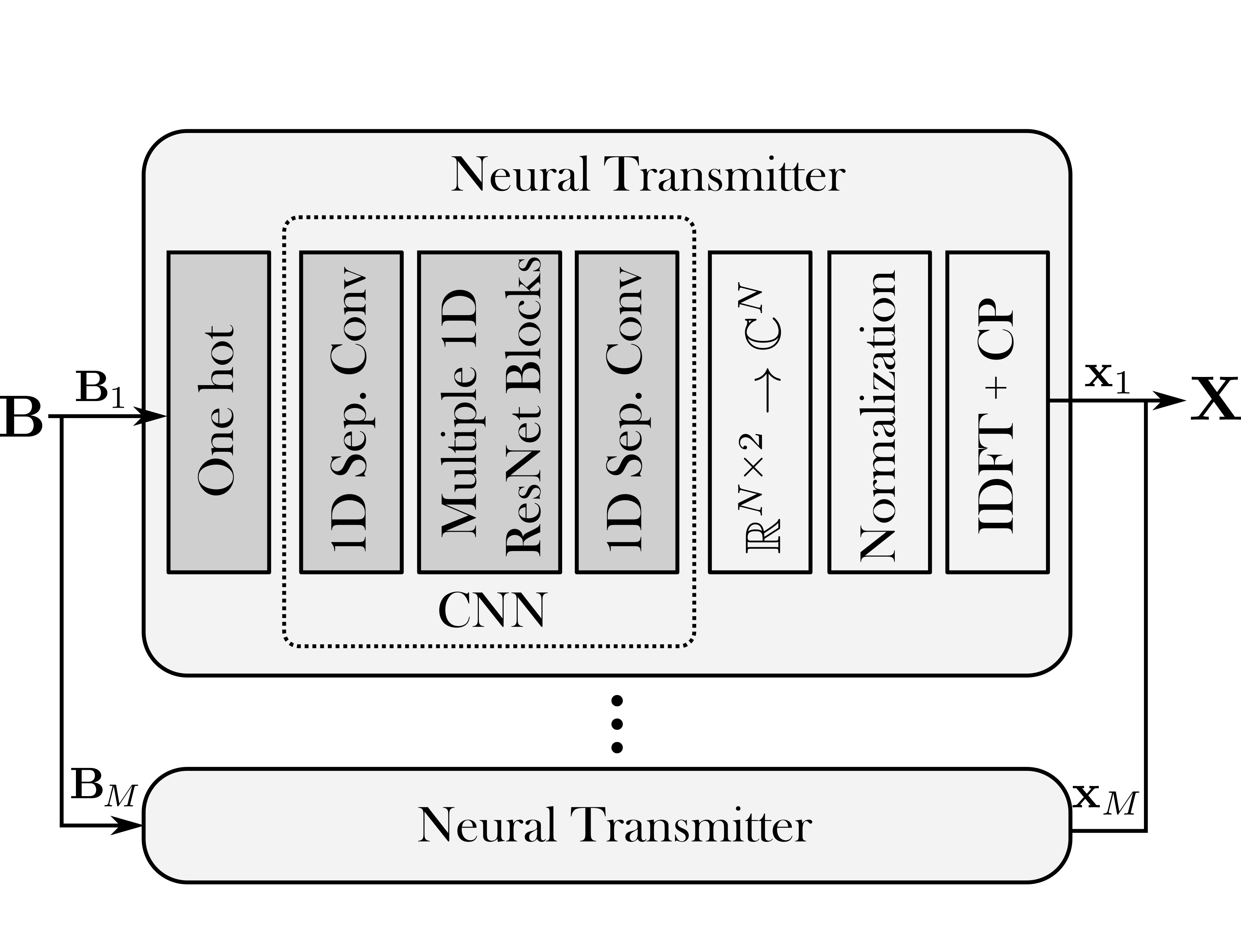}
        \caption{Neural Transmitter.}
        \label{fig:chp2_nn_tx}
      \end{subfigure}
      \hfill
      \begin{subfigure}{.2\textwidth}
        \centering
        \includegraphics[height=170pt]{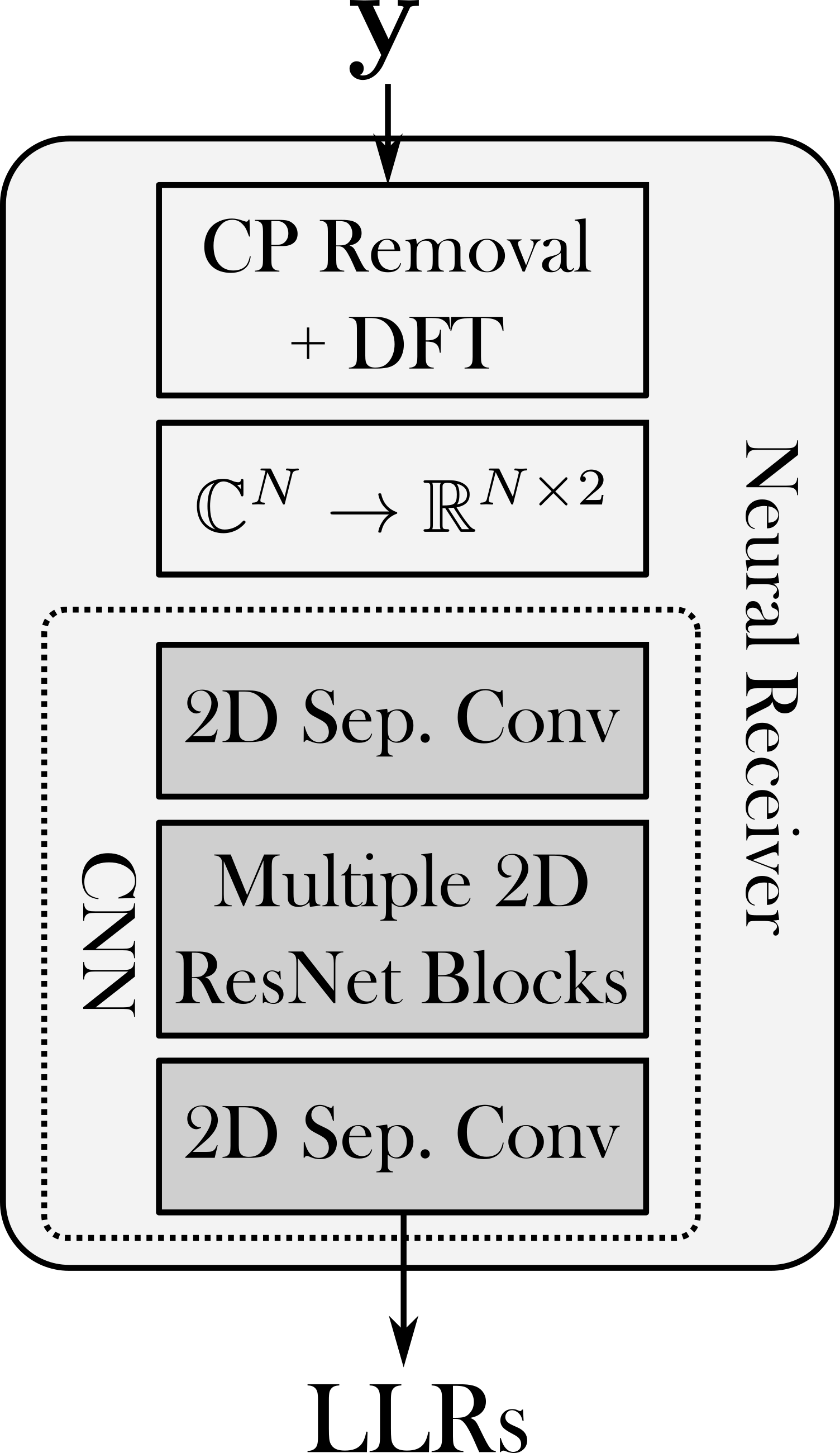}
        \caption{Neural Receiver.}
        \label{fig:chp2_nn_rx}
      \end{subfigure}%
      \hfill
    \begin{subfigure}{.2\textwidth}
      \centering
      \includegraphics[height=170pt]{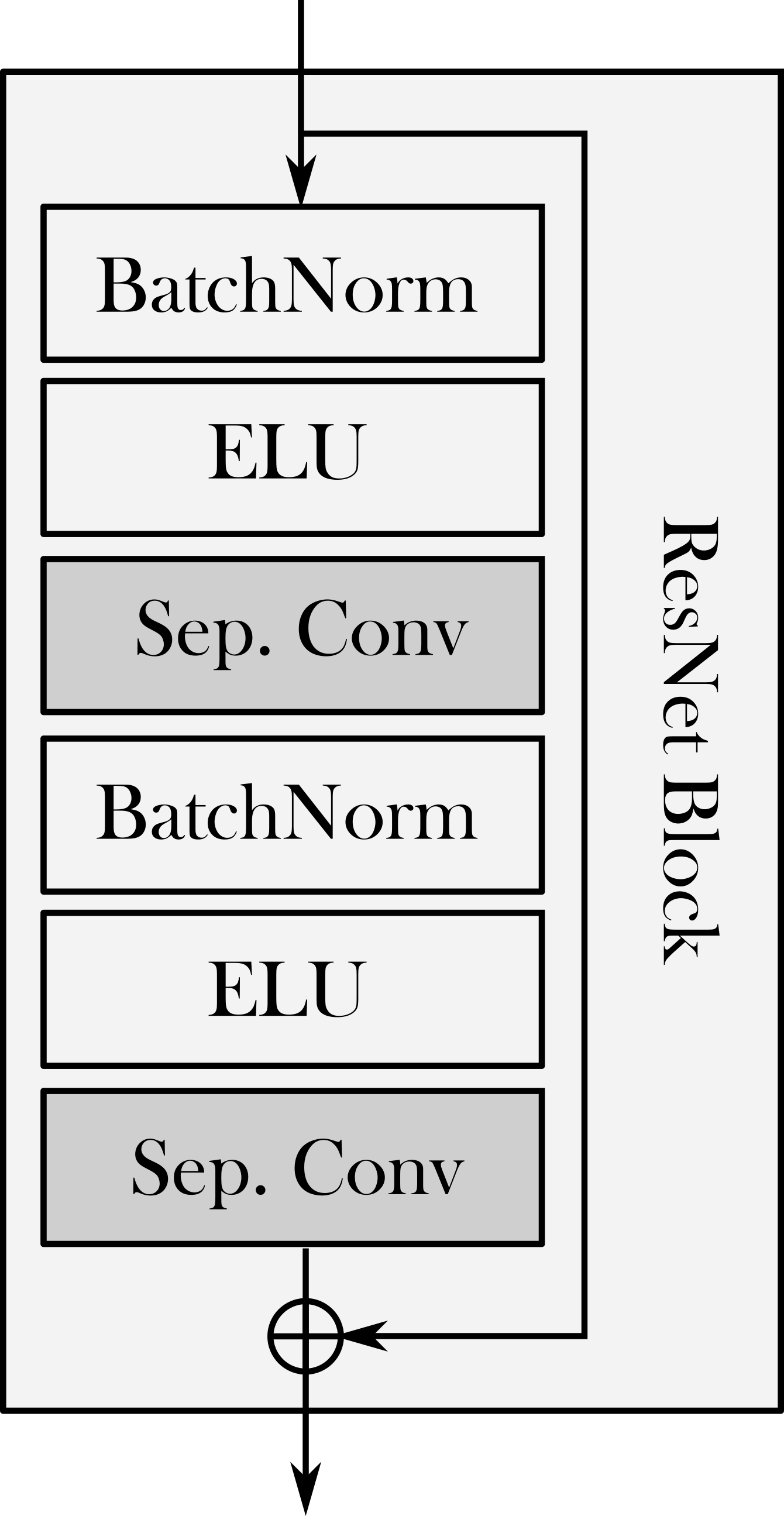}
      \caption{ResNet block.}
      \label{fig:chp2_resnet}
    \end{subfigure}%
    
    \caption{Different parts of the end-to-end system, where grayed blocks are trainable components.}
    \label{fig:chp2_components}
\end{figure}

The neural transmitter and receiver are based on similar architectures, schematically shown in Fig.~\ref{fig:chp2_components}.
The core element is a \emph{ResNet block}, which was introduced in the physical layer to implement a fully \gls{NN}-based radio receiver \cite{honkala2020deeprx}, and whose effectiveness has been demonstrated in other related works~\cite{pilotless20, korpi2020deeprx, goutay2020machine}.
A ResNet block is made of two identical sequences of layers followed by the addition of the input, as depicted in Fig.~\ref{fig:chp2_resnet}.
In the original ResNet block~\cite{he2016identity}, each sequence was composed of a batch normalization layer, a rectified linear unit (ReLU) activation function, and a convolution. 
Our architecture differs from the original one in that ReLUs are replaced by exponential linear units (ELUs) to alleviate the vanishing gradient problem~\cite{clevert2016fast} and separable convolutions are used since they enable similar performance than conventional ones but at a fraction of the computational cost~\cite{howard2017mobilenets}.
Finally, we use zero-padding on all 1D (2D) convolutions to maintain constant the size of the first (and possibly second) dimension(s).
In the following, the architectures of the neural transmitter and receiver are detailed, although the exact numbers of parameters for each layer are given in Section~\ref{sec:evaluations}.

The transmitter processes all OFDM symbols in parallel (see Fig.~\ref{fig:chp2_nn_tx}). 
Each instance of the CNN implemented at the transmitter takes as input the matrix of bits $\Bm_m$, corresponding to the OFDM symbol $m$, and outputs the OFDM symbol $\xv_m$.
The vector of bits is first converted into its one hot representation, i.e., into vectors of $\{0, 1\}^{2^Q}$ where all elements but one are set to zero.
Then, a \gls{CNN} comprises one 1D separable convolution, multiple 1D ResNet blocks, and another 1D separable convolution.
This CNN is fed with the one-hot matrix of dimension $N \times 2^Q$, were $N$ corresponds to the dimension of the 1D convolution and $2^Q$ to different convolution channels, and outputs $N\times 2$ elements.
The next layers convert these $N\times 2$ real numbers into $N$ complex symbols and normalize them, as in \eqref{eq:chp2_layer_norm}, to have a unit average energy per \gls{RE}.
Finally, an \gls{IDFT} is performed on the symbols and a \gls{CP} is added before transmission.
We experimentally verified that independent processing of all OFDM symbols, resulting in the use of 1D convolutions, leads to better performance than 2D convolutions that would process all OFDM symbols at once.
This could be explained by the 1D nature of \gls{PAPR} and \gls{ACLR} measurements, which are computed for all OFDM symbols separately.

The neural receiver, on the contrary, performs a 2D processing on all OFDM symbols since it enables more accurate channel estimation and equalization~\cite{pilotless20, goutay2020machine}.
At reception, the \gls{CP} is first removed and an DFT is applied to the received signals $\Ym$.
The $M \times N$ symbols are then converted into $M \times N \times 2$ real numbers that are fed, along with the transmission SNR of size $M \times N \times 1$, into a 2D CNN.
The architecture of the receiver CNN is similar to the one of the transmitter, except that 2D separable convolutions are used.
The last 2D separable layer outputs $M \times N \times Q$ real numbers that correspond to the \glspl{LLR} of all transmitted symbols, as shown in Fig.~\ref{fig:chp2_nn_rx}.
Note that no pilots are used as it was shown in~\cite{pilotless20} that pilotless communication is possible over OFDM channels when  neural receivers are used.

\clearpage
\section{Simulations Results and Insights} 
\label{sec:chp2_evaluations_insights}

In this section, the E2E system is benchmarked against the TR baseline.
We first describe the training and evaluation setup, followed by rate, \gls{BER}, and goodput comparisons.

\begin{table}

    \centering
    
    \newcolumntype{M}[1]{>{\centering\arraybackslash}m{#1}}
    \newcolumntype{N}{@{}m{0pt}@{}}
    
    \renewcommand{\arraystretch}{1.3}
    
    \begin{tabular}{ |M{1.4cm}||M{1.9cm}|M{1.3cm}|M{1.3cm}|M{1.5cm}|M{1.3cm}|M{1.3cm}|M{1.9cm}| N }
        \hline
            & Sep. Conv. 1D (2D) & \multicolumn{5}{c|}{ResNet blocks 1D (2D)} & Sep. Conv 1D (2D) \\
		\hhline{|=|=|=====|=|}
        Kernel size & 1 (1,1) & 3 (3,2)& 9 (9,4)& 15 (15,6)& 9 (9,4)& 3 (3,2)& 1 (1,1)\\
        \hline
        Dilation rate & 1 (1,1)& 1 (1,1)& 2 (2,1) & 4 (4,1)& 2 (2,1)& 1 (1,1)& 1 (1,1)\\
        \hline
        Filters & \multicolumn{6}{c|}{128} & $2$ ($Q$) \\
        \hline
    \end{tabular}
    
    \caption{Parameters for the neural transmitter (receiver).}
    \label{table:chp2_arch}
    \renewcommand{\arraystretch}{1}
    
\end{table}

\subsection{Evaluations}
\label{sec:evaluations}

\subsubsection{Training and evaluation setup}
Separate datasets were used for training and testing, both generated using a mixture of \gls{3GPP}-compliant \gls{UMi} line-of-sight (LOS) and non-LOS models. The channel responses were generated using  QuaDRiGa 2.4.0~\cite{quadriga}, and perfect power control was assumed such that the average channel energy per \gls{RE} was one, i.e., $\EE \left[ |h_{m, n}|^2 \right]= 1$.
$N=75$ subcarriers were considered with $M=14$ OFDM symbols using \glspl{CP} of sufficient lengths so that the channel can be represented by \eqref{eq:chp2_OFDM_channel} in the frequency domain.
The center carrier frequency and subcarrier spacing were respectively set to 3.5~\si{GHz} and 30~\si{kHz}, the number of bits per channel use was set to $Q=4$, and 16-QAM modulation was used by the baseline.
Coded \gls{BER} comparisons were performed using a standard 5G-compliant low-density parity-check (LDPC) code with length 1024 and rate $\eta = \frac{1}{2}$.
The baseline was evaluated for $R\in\{0, 2, 4, 8, 16, 32\}$ \glspl{PRT} and the E2E system was trained to achieve \gls{ACLR} targets of $\bleak \in \{-20, -30, -40\}$~\si{dB} and \gls{PAPR} targets of $\gpeak \in \{ 4, 5, 6, 7, 8, 9 \}$~\si{dB}.

The transmitter and receiver architectures that were used are detailed in Table~\ref{table:chp2_arch}.
Inspired by~\cite{honkala2020deeprx}, the receptive fields of the \glspl{CNN} were increased using dilations, and the kernel sizes had an increasing, then decreasing number of parameters.
The Lagrange multipliers were initialized to $\lambda^{(0)}_p = \lambda^{(0)}_l = 0$ and $\tau$ was set to $0.004$.
The penalty parameters were initialized to $\mu^{(0)}_p = 10^{-1}$ and $\mu^{(0)}_l = 10^{-3}$, which mirrors the fact that $L_{\gamma_{\text{peak}}}(\thetav)$ is usually two orders of magnitude lower than $L_{\beta_{\text{leak}}}(\thetav)$.
The optimization procedure was composed of 2500 iterations in which 20 SGD steps were performed with a learning rate of $10^{-3}$ and a batch size of $B_S = 100$.
The oversampling factor used to compute $\underline{\zv}$ in \eqref{eq:chp2_loss_papr2} was set to $O_S = 5$, as it was shown to be sufficient to correctly represent the analog signal~\cite{1261335}.
Finally, the \gls{SNR} 
\begin{align}
	\text{SNR} =  \frac{\EE_{h_{m,n}} \left[ |h_{m,n}|^2 \right]}{\sigma^2}  =  \frac{1}{\sigma^2}
\end{align}
was chosen randomly in the interval $[10, 30]$~\si{dB} for each \gls{RG} in the batch during training.
The simulations parameters are listed in Table~\ref{table:chp2_params}.

\begin{table}[h!]
	\centering
	\renewcommand{\arraystretch}{1.3}
	\begin{tabular}{|p{5.5cm}|c|c|c|c|}
	  \hline
	  Parameters  & Symbol (if any) & Value  \\ 
	  \hhline{|=|=|=|}
	  Number of OFDM symbols & $M$ & 14 \\   \hline
	  Number subcarriers & $N$ & 75   \\   \hline
	  Number of bits per channel use & $Q$ & 4 \\   \hline
	  \Gls{PAPR} targets & $\gpeak$ & $\{4, 5, 6, 7, 8, 9\}$~\si{dB} \\   \hline
	  \Gls{ACLR} targets & $\bleak$ & $\{-20, -30, -40\}$~\si{dB} \\   \hline
	  \Glspl{PRT} used by the baseline & R & $\{0, 2, 4, 8, 16, 32\}$\\   \hline
	  Batch size & $B_S$ & 100 \glspl{RG}  \\   \hline
	  Oversampling factor & $O_S$ & 5   \\   \hline
	  Code rate & $\eta $ & $\frac{1}{2}$ \\   \hline
	  Code length & - & \SI{1024}{\bit} \\   \hline
	  Center frequency (subcarrier spacing) & - & \SI{3.5}{\GHz} (\SI{30}{\kHz}) \\   \hline
	  Scenario & - & 3GPP 38.901 UMi LOS + NLOS \\   \hline
	  SNR & - & $[10, 30]$~\si{dB} \\   \hline
	  Learning rate & - & $10^{-3}$ \\   \hline
	  
	\end{tabular}
	
	\vspace{20pt}
	\caption{Training and evaluation parameters.}
	\label{table:chp2_params}
	\renewcommand{\arraystretch}{1}
  \end{table}

\subsubsection{Evaluation results}
\begin{figure}[t]
	\centering
	\input{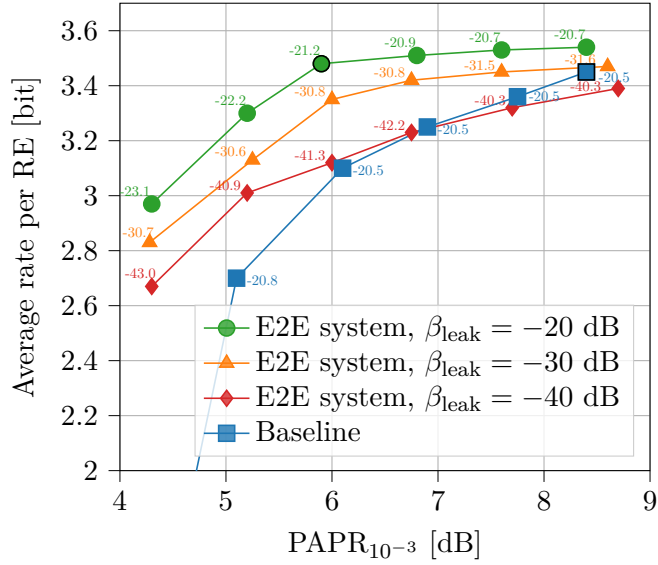}
	\caption{Rates achieved by the compared systems. Numbers near scatter plots indicate the \glspl{ACLR}.}
	\label{fig:chp2_rates}
\end{figure}

In the following evaluations, the \gls{PAPR} probability threshold was set to $\epsilon = 10^{-3}$.
Note that setting $\epsilon=0$ would only take into account the maximum signal peak, which is achieved by a single possible waveform that has probability $\frac{1}{2^{NQ}} \approx 10^{-90}$.
The average rates per \gls{RE} achieved by the baseline and the E2E systems are shown in Fig.~\ref{fig:chp2_rates}, where the numbers next to the data points are the corresponding \glspl{ACLR}.
First, it can be seen that at the maximum \gls{PAPR} of approximately $8.5$~\si{dB}, the E2E system trained with $\bleak = -20$ ~\si{dB} achieves a $3\%$ higher throughput than the baseline with no \gls{PRT}. 
This can be explained by the rate loss due to the presence of pilots in the baseline, which do not carry data and account for approximately $4\%$ of the total number of \glspl{RE}.
Second, at lower \glspl{PAPR}, the rates achieved by the E2E system trained with $\bleak\in\{-20, -30\}$ are significantly higher than the ones achieved by the baseline. 
For example, the E2E system trained with $\bleak=-20$ and $\gpeak=5$ achieves an average rate $22\%$ higher than the baseline for the same \gls{PAPR}.  
Finally, the E2E systems are able to meet their respective \gls{PAPR} and \gls{ACLR} targets.

\begin{figure}
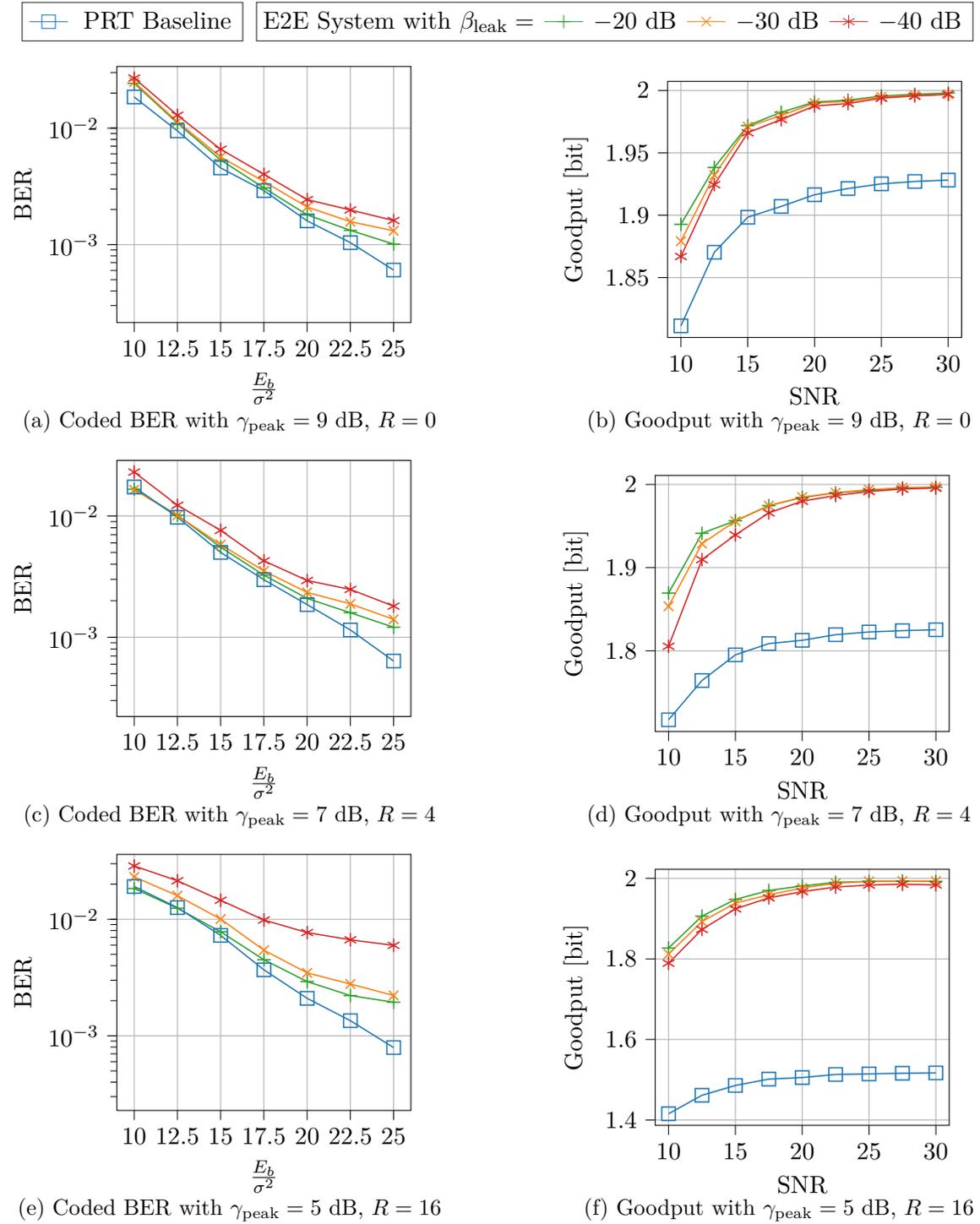


	\centering

	\begin{subfigure}[b]{0.2\textwidth}
		\begin{tikzpicture} 
		
			\definecolor{color0}{rgb}{0.12156862745098,0.466666666666667,0.705882352941177}
			\definecolor{color1}{rgb}{1,0.498039215686275,0.0549019607843137}
			\definecolor{color2}{rgb}{0.172549019607843,0.627450980392157,0.172549019607843}
			\definecolor{color3}{rgb}{0.83921568627451,0.152941176470588,0.156862745098039}
		
				\begin{axis}[%
				hide axis,
				xmin=10,
				xmax=50,
				ymin=0,
				ymax=0.4,
				legend columns=4, 
				legend style={draw=white!15!black,legend cell align=left,column sep=1.0ex}
				]
				\addlegendimage{color0, mark=square, mark size=3}
				\addlegendentry{PRT Baseline};
				\end{axis}
		\end{tikzpicture}
	\end{subfigure}
	\hspace{1pt}
	\begin{subfigure}[b]{0.75\textwidth}
	\centering
	\begin{tikzpicture} 
	
	\definecolor{color0}{rgb}{0.12156862745098,0.466666666666667,0.705882352941177}
	\definecolor{color1}{rgb}{1,0.498039215686275,0.0549019607843137}
	\definecolor{color2}{rgb}{0.172549019607843,0.627450980392157,0.172549019607843}
	\definecolor{color3}{rgb}{0.83921568627451,0.152941176470588,0.156862745098039}

    	\begin{axis}[%
    	hide axis,
    	xmin=10,
   	 xmax=50,
    	ymin=0,
    	ymax=0.4,
    	legend columns=4, 
    	legend style={draw=white!15!black,legend cell align=left,column sep=.5ex}
    	]
    	\addlegendimage{white ,mark=*, mark size=2}
    	\addlegendentry{\hspace{-0.9cm} E2E System with $\bleak=$};
    	\addlegendimage{color2, mark=+, mark size=3}
    	\addlegendentry{$-20$~\si{dB}};
    	\addlegendimage{color1, mark=x, mark size=3}
    	\addlegendentry{$-30$~\si{dB}};
    	\addlegendimage{color3, mark=asterisk, mark size=3}
    	\addlegendentry{$-40$~\si{dB}};
    	\end{axis}
	\end{tikzpicture}
\end{subfigure}

	\vspace{10pt}

	  	\begin{subfigure}[b]{0.45\textwidth}
	  		\input{Chapter2/eval/ber_20.tex}
			\vspace{-8pt}
	  		\caption{Coded BER with $\gpeak=9$~\si{dB}, $R=0$}
	  		\label{fig:chp2_ber_1}
		\end{subfigure}%
		\hfill
		\begin{subfigure}[b]{0.45\textwidth}
			\input{Chapter2/eval/gp_20.tex}
			\vspace{-8pt}
			\caption{Goodput with $\gpeak=9$~\si{dB}, $R=0$}
			\label{fig:chp2_goodput_1}
		\end{subfigure}

		\vspace{10pt}

		\begin{subfigure}[b]{0.45\textwidth}
			\input{Chapter2/eval/ber_30.tex}
			\vspace{-8pt}
			\caption{Coded BER with $\gpeak=7$~\si{dB}, $R=4$}
			\label{fig:chp2_ber_2}
		\end{subfigure}%
		\hfill
	  	\begin{subfigure}[b]{0.45\textwidth}
	  		\input{Chapter2/eval/gp_30.tex}
			\vspace{-8pt}
			\caption{Goodput with $\gpeak=7$~\si{dB}, $R=4$}
	  		\label{fig:chp2_goodput_2}
		\end{subfigure}%

		\vspace{10pt}

		\begin{subfigure}[b]{0.45\textwidth}
			\input{Chapter2/eval/ber_40.tex}
			\vspace{-8pt}
			\caption{Coded BER with $\gpeak=5$~\si{dB}, $R=16$}
			\label{fig:chp2_ber_3}
		\end{subfigure}
		\hfill
		\begin{subfigure}[b]{0.45\textwidth}
			\input{Chapter2/eval/gp_40.tex}
			\vspace{-8pt}
			\caption{Goodput with $\gpeak=5$~\si{dB}, $R=16$}
			\label{fig:chp2_goodput_3}
		\end{subfigure}%
	
	\caption{BER and goodput achieved by the different schemes.}
	\label{fig:chp2_ber_goodput}
	\end{figure}

The coded \glspl{BER} of the baseline and of the three E2E systems are shown in the left column of Fig.~\ref{fig:chp2_ber_goodput}.
As the baseline transmits pilots and reduction signals in addition to data signals, the energy per transmitted bit is higher than that of the of E2E systems.
To reflect this characteristic, we define the energy-per-bit-to-noise-spectral-density ratio as
\begin{align}
	\frac{E_b}{\sigma^2} = \frac{\EE_{x_{m,n}} \left[ |x_{m,n}|^2 \right]}{\rho K \sigma^2} = \frac{1}{\rho Q \sigma^2}
\end{align}
where $\rho$ is the ratio of \glspl{RE} carrying data signals over the total number of \glspl{RE} in the \gls{RG}.
Fig.~\ref{fig:chp2_ber_1},~\ref{fig:chp2_ber_2}, and~\ref{fig:chp2_ber_3} correspond to systems achieving \glspl{PAPR} of approximately $8.5$, $6.8$, and $5.2~\si{dB}$, respectively.
Such \glspl{PAPR} were obtained using $R=\{0, 4, 16\}$ \gls{PRT} for the baseline, and $\gpeak = \{9, 7, 5\}~\si{dB}$ for the trained systems.
Note that evaluation results corresponding to systems trained with $\bleak = \{-30, -40\}$~\si{dB} are provided for completeness, but a fair comparison is only possible between the baseline and the system trained with $\bleak = -20$~\si{dB} as this value corresponds to the baseline \gls{ACLR}.
Overall, one can see that the baseline consistently achieves slightly lower \gls{BER} than the E2E systems.
However, the baseline also transmits fewer bits per \gls{RG}, due to some \glspl{RE} being used to transmit pilots or reduction signals.

To understand the benefits provided by the E2E approach, the second column of Fig.~\ref{fig:chp2_ber_goodput} presents the \emph{goodputs} achieved by each compared system.
The goodput is defined as the average number of information bits that have been successfully received in a RE, i.e., 
\begin{align}
	\text{Goodput} = \eta \rho Q (1-\text{BER}),
\end{align}
and is plotted with respect to the SNR as it already accounts for the different number of \glspl{RE} transmitting information bits through the parameter $\rho$.
Evaluation results show that the goodputs achieved by all trained systems, including those trained for lower \glspl{ACLR}, are significantly higher than the ones of the baseline. 
Indeed, at an SNR of $30~\si{dB}$, all E2E systems are able to successfully transmit close to two bits per RE, while the baseline saturates at $1.93$, $1.82$, and $1.53$ bits for \glspl{PAPR} of $8.5$, $6.8$, and $5.2$ \si{dB}, respectively.
The BER improvements range from a $3\%$ increase at low SNR with $\text{PAPR}_{\epsilon} \approx 8.5~\si{dB}$ to a $30\%$ increase at high SNR with $\text{PAPR}_{\epsilon} \approx 5.2~\si{dB}$, indicating that the E2E approach is particularly effective when the \gls{PAPR} reduction is high.
These gains are jointly enabled by the pilotless nature of the E2E transmission, the effective \gls{ACLR} reduction scheme learned through the proposed optimization procedure, and the fact that every \glspl{RE} can be used to transmit data.

\subsection{Insights on the Learned ACLR and PAPR Reduction Techniques}

\subsubsection{Interpreting the \gls{ACLR} minimization process}

In order to get insight into the processing done by the neural transmitter, we first study the \gls{ACLR} reduction technique learned by the E2E systems.
Three covariance matrices $\EE_{\xv_m} \left[\xv_m \xv_m \htp \right]$ are shown in Fig.~\ref{fig:chp2_cov_subcarriers} for systems trained with $\gpeak = 9$.
The first covariance matrix is from a system trained with an \gls{ACLR} target of $\bleak=-20~\si{dB}$, while the second and third matrices were respectively obtained with $\bleak=-30~\si{dB}$ and $\bleak=-40~\si{dB}$.
It can be observed that the correlation between the elements close to the diagonal increase inversely to the \gls{ACLR} target. 
Moreover, the correlation increases at subcarriers located near the edges of the spectrum, indicating that a subcarrier-dependent filtering is learned.
Fig.~\ref{fig:chp2_energy_subcarriers} depicts the distribution of the energy across the subcarriers. 
On the one hand, the system trained with a lax \gls{ACLR} constraint ($\bleak=-20~\si{dB}$) equally distributes the available energy on all subcarriers, which can be shown to maximize the information rate of the transmission.
On the other hand, the systems trained with lower \gls{ACLR} targets learn to reduce the energy of the subcarriers located at the \gls{RG} edges, also contributing the out-of-band emissions reduction. 
The joint effect of the filtering and of the uneven energy distribution across subcarriers is visible in Fig.~\ref{fig:chp2_psd}, where the \glspl{PSD} of the compared systems are shown.
One can see that the system trained with $\bleak=-20~\si{dB}$ and the baseline have similar \glspl{PSD}, while the \glspl{PSD} of the systems trained with lower \gls{ACLR} targets present significantly lower out-of-band emissions.
It therefore appears that the improved \gls{ACLR} allowed by the E2E system is the result of both reducing the power of the subcarriers located near the \gls{RG} edges and the introduction of correlations between the transmitted symbols.

\begin{figure} 
	\centering 
	  \begin{subfigure}[b]{0.20\linewidth}
		\input{Chapter2/eval/cov_20.tex}
		\vspace{-10pt}
		\caption{$\bleak=-20~\si{dB}$} \label{fig:chp2_M1}  
	  \end{subfigure}
	  \hspace{20pt}
	\begin{subfigure}[b]{0.20\linewidth}
	  \input{Chapter2/eval/cov_30.tex}
	  \vspace{-10pt}
	\caption{$\bleak=-30~\si{dB}$} \label{fig:chp2_M2}  
	\end{subfigure}
	\hspace{20pt}
	\begin{subfigure}[b]{0.20\linewidth}
		\input{Chapter2/eval/cov_40.tex}
		\vspace{-10pt}
	  \caption{$\bleak=-40~\si{dB}$} \label{fig:chp2_M3}  
	  \end{subfigure}
	  \hspace{30pt}
	  \begin{subfigure}[b]{0.15\linewidth}
		\input{Chapter2/eval/cov_colorbar.tex}
		\vspace{19pt}
	  \end{subfigure}
	\caption{Covariance matrices $\EE_m \left[\xv_m \xv_m \htp \right]$ for systems trained with different target ACLRs.}
	\label{fig:chp2_cov_subcarriers}
\end{figure}

\begin{figure}
	\centering
	\begin{minipage}{.45\textwidth}
	  \centering
	  \input{Chapter2/eval/power_subcarriers.tex}
	  \captionof{figure}{Mean energy of the symbols transmitted on each subcarrier.}
		\vspace{-20pt}
	  \label{fig:chp2_energy_subcarriers}
	\end{minipage}%
	\hfill
	\begin{minipage}{.45\textwidth}
	  \centering
	  \input{Chapter2/eval/PSD_smooth.tex}
	  \captionof{figure}{PSD of four systems having no \gls{PAPR} constraint.}
		\vspace{-20pt}	
	  \label{fig:chp2_psd}
	\end{minipage}
\end{figure}

\subsubsection{Interpreting the \gls{PAPR} minimization process}
\begin{figure}[t]
	\centering
	\includegraphics[width=1\textwidth]{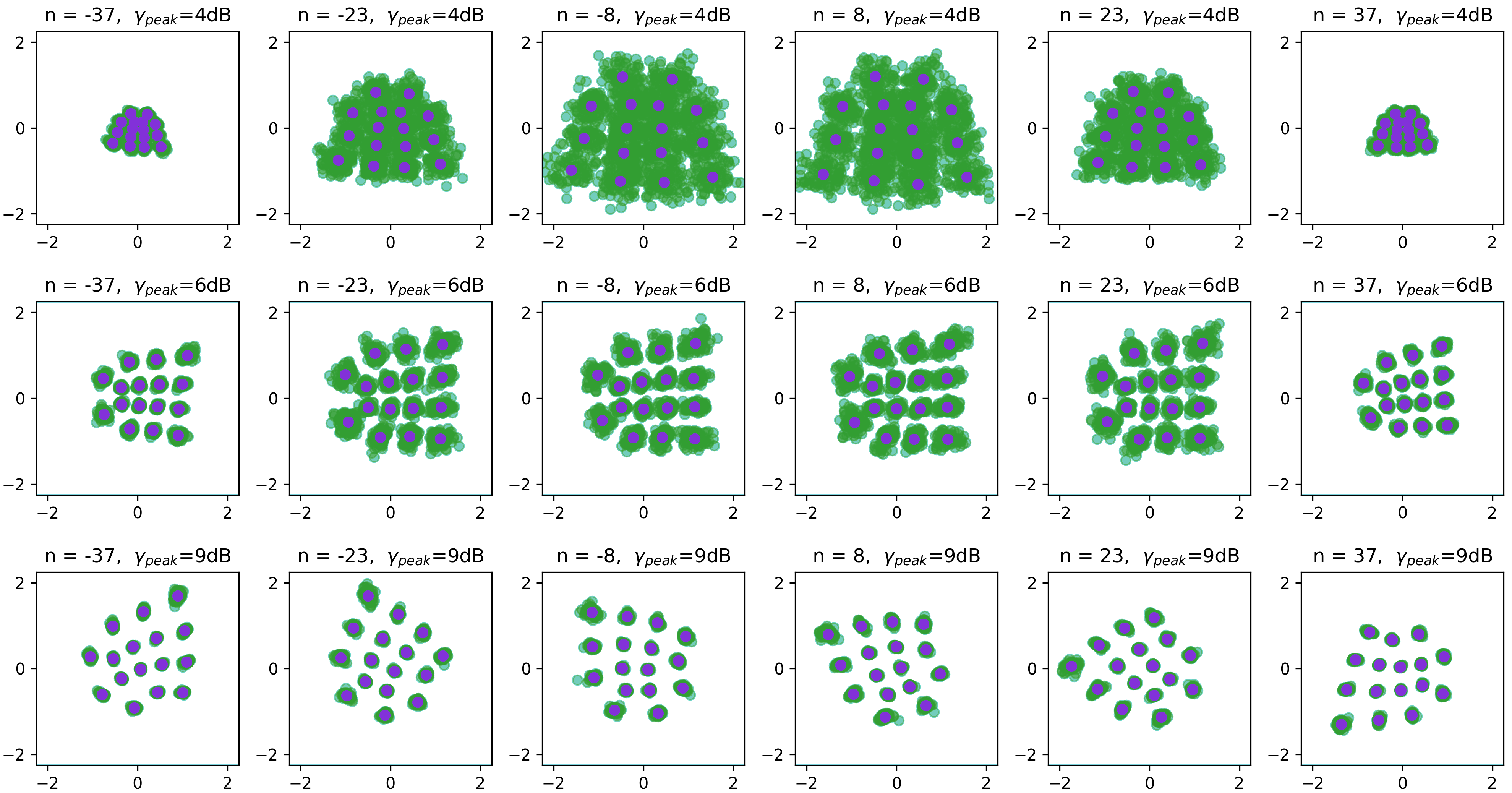}
	\caption{Transmitted signals on six subcarriers for E2E systems trained with different \gls{PAPR} targets but a lax \gls{ACLR} constraint. The purple dots represent the center of each cluster.}
	\label{fig:chp2_clusters}
\end{figure}

To understand how the neural transmitter and the optimization procedure enable significant \gls{PAPR} reduction, it is insightful to look at the symbols transmitted for different \gls{PAPR} targets, as illustrated in Fig.~\ref{fig:chp2_clusters}.
The three rows represent E2E systems trained for \gls{PAPR} targets of $\gpeak\in\{4, 6, 9\}~\si{dB}$, and the six columns show the signals transmitted on the subcarriers $n\in\{-37, -23, -8, 8, 23, 37\}$.
All systems where trained with a lax \gls{ACLR} constraint \mbox{($\bleak=-20\si{dB}$)}.
This figure was obtained by sending 25 \glspl{RG}, and each green point represents one signal transmitted on the corresponding subcarrier. 
One can see that the signals are gathered into $2^Q$ groups, that will be referred to as \emph{clusters} in the following.
It can be observed that for low \gls{PAPR} targets ($\gpeak=4\si{dB}$), the clusters at the subcarriers located at the center of the \gls{RG} exhibit more dispersion than the ones located near the \gls{RG} edges. 
Moreover, the average energy of the transmitted signals also appears higher on the central subcarriers.
On the contrary, for high \gls{PAPR} targets ($\gpeak=9\si{dB}$), the clusters seem equally dispersed and the energy is evenly spread across the subcarriers.
Finally, it can be noted that the positions of the clusters are not rotationally symmetrical, which should help the neural receiver in estimating and equalizing the channel.

\begin{figure}
	\centering
	\begin{minipage}{.45\textwidth}
	  \centering
	  \input{Chapter2/eval/energy_const_pates.tex}
	  \captionof{figure}{Mean energy of the clusters centers per subcarrier.}
	  \label{fig:chp2_energy_clusters}
	\end{minipage}%
	\hspace{20pt}
	\begin{minipage}{.45\textwidth}
	  \centering
	  \input{Chapter2/eval/var_pates.tex}
	  \vspace{-20pt}
	  \captionof{figure}{Mean variance of the clusters per subcarrier.}
	  \label{fig:chp2_var_clusters}
	\end{minipage}
\end{figure}

To better understand the neural transmitter behavior, we index each cluster by a tuple $(n,k)$, where $n$ is the subcarrier index and $k\in\{1, \cdots, 2^Q\}$ is the index of the cluster for the $n^{\text{th}}$ subcarrier.
Let us denote by $\left\{ \bv^{(k)} \right\}_{1 \leq k \leq 2^Q}$ the set of all possible vectors of bits indexed by their decimal representation, i.e., $\bv^{(0)} = [0, 0]\tp, \bv^{(1)} = [0, 1]\tp, \bv^{(2)} = [1, 0]\tp$, and $\bv^{(3)} = [1, 1]\tp$ for $Q=2$.
We verified that each cluster corresponds to a unique vector of bits $\bv^{(k)}$, and the center of these clusters are represented by purple dots in Fig.~\ref{fig:chp2_clusters}.
For each E2E system, we define a new high-dimensional constellation $\bar{\mathcal{C}} \in \CC^{N \times 2^Q}$ that comprises the centers of the $2^Q$ clusters on all $N$ subcarriers.
We denote by $\bar{\mathcal{C}}(n, k) = \EE_{\xv_m} \left[ x_{m,n}^{(k)}\right]$ the $k^{\text{th}}$ constellation point in the $n^{\text{th}}$ subcarrier, where $x_{m,n}^{(k)}$ denotes the output of the neural transmitter for the RE $(m,n)$ when $\bv^{(k)}$ was given as input.
Similarly, we define $\bar{\mathcal{V}}\in \CC^{N \times 2^Q}$, such that $\bar{\mathcal{V}}(k) = \EE_{\xv_m} \left[ x_{m,n}^{(k)} {x_{m,n}^{(k)}}^\star \right]$ represents the variance of the cluster $(n,k)$.

Fig.~\ref{fig:chp2_energy_clusters} shows the mean energy of the constellation on each subcarrier for E2E systems trained with $\gpeak\in\{4, 6, 9\}~\si{dB}$ and a lax \gls{ACLR} constraint, corresponding to the three rows of Fig.~\ref{fig:chp2_clusters}.
We can verify that when the \gls{PAPR} constraint is lax ($\gpeak=9\si{dB}$), the energy is evenly distributed across the subcarriers.
On the contrary, the border subcarriers are given less energy when a harsh constraint is applied ($\gpeak=4\si{dB}$).
One explanation could be that the center subcarriers have longer wavelength, and therefore contribute less than their counterparts with shorter wavelengths. 
By focusing the available energy on these subcarriers, the neural transmitter jointly minimizes the probability of peaks in the analog waveform and decreases the number of \glspl{FBS} that have a significant impact on them.
Note that since the carrier frequency is typically much higher than the bandwidth of each subcarrier, the average power of the passband and baseband signals differs by a factor two but their maximum are approximately equal, which amounts to a roughly 3~\si{dB} difference between the passband and baseband \gls{PAPR}~\cite{eras2019analysis}.
Finally, the mean variance of the clusters on each subcarrier are plotted in Fig.~\ref{fig:chp2_var_clusters} for the same three systems.
One can observe that the clusters exhibit almost no variance with $\gpeak=9\si{dB}$, but a high variance in the center frequencies with $\gpeak=4\si{dB}$.
These high variances observed with harsh \gls{PAPR} constraints tend to indicate that the neural transmitter is able to slightly relocate the relevant \glspl{FBS} in order to minimize the waveform \gls{PAPR}.
Overall, the transmitter seems to focus its energy on central subcarriers to reduce the probability of peaks and the number of relevant \glspl{FBS}, and to adjust the positions of these \glspl{FBS}  to minimize the peak amplitudes.

\begin{figure}
	\centering
	\begin{minipage}{.45\textwidth}
	  \centering
	  \input{Chapter2/eval/ccdf_const.tex}
	  \captionof{figure}{CCDF of the power of the E2E system trained for $\gpeak=4\si{dB}$ \mbox{($\bleak=-20\si{dB}$)} and of a conventional system using its extracted constellation.} 
	  \label{fig:chp2_ccdf_const}
	\end{minipage}%
	\hspace{20pt}
	\begin{minipage}{.45\textwidth}
	  \centering
	  \input{Chapter2/eval/ccdf.tex}
	  \captionof{figure}{CCDF of the power of the E2E system trained with $\gpeak=6\si{dB}$ \mbox{($\bleak=-20\si{dB}$)} and of the baseline using 16-QAM modulation and no \gls{PRT}.}
	  \label{fig:chp2_ccdf}
	\end{minipage}
\vspace{-10pt}
\end{figure}

In order to verify this claim, the  \gls{CCDF}  of the ratio between the instantaneous and average power of the transmitted signal is
\begin{align}
	\text{CCDF}_{|\nu(t)|^2}(e) = P \left( |\nu(t)|^2 > e \right) \approx P \left( \left| \frac{\underline{z}_{m,t}}{\EE\LSB \underline{z}_{m,t} \RSB } \right| ^2 > e \right)
\end{align}
and is approximated by sending $10000$ batches of $1000$ \glspl{RG}, each \gls{RG} $i$ being composed of $N=75$ subcarriers and $M=14$ OFDM symbols $\underline{z}_{m,t}^{[i]} $ oversampled with a factor $O_S = 5$.
The \gls{CCDF} of two systems are presented in Fig.~\ref{fig:chp2_ccdf_const}, the first one corresponds to the first row of Fig.~\ref{fig:chp2_clusters}, i.e., it is trained with the harshest \gls{PAPR} constraint ($\gpeak=9\si{dB}$) and a lax \gls{ACLR} constraint ($\bleak=-20\si{dB}$).
The second one is a conventional system that uses the constellation $\bar{\mathcal{C}}$ extracted from the aforementioned E2E system (and represented by purple dots in the first row of Fig.~\ref{fig:chp2_clusters}).
The goal of this comparison is to evaluate the effect of the adjustment of the \glspl{FBS}  positions from the clusters centers operated by the transmitter.
It can be seen that the \gls{CCDF} of the E2E system is drastically lower than the one of the conventional system using the extracted constellation, indicating that the \gls{PAPR} reduction is indeed performed through the \glspl{FBS}  position adjustments.

Finally, the \glspl{CCDF} of the E2E system trained with $\gpeak=6\si{dB}$ (and a lax \gls{ACLR} constraint) and of a conventional \gls{QAM} system are compared in Fig.~\ref{fig:chp2_ccdf}.
The E2E system corresponds to the second row of Fig.~\ref{fig:chp2_clusters}, and the two compared systems are respectively highlighted by a black circle and a black square in Fig.~\ref{fig:chp2_rates}.
On can see that the \gls{PAPR} minimization process operated by the neural transmitter is particularly effective, as the \gls{CCDF} of the E2E system is significantly lower despite the two systems enabling similar rates.
Finally, it is also interesting to note that the \gls{CCDF} of the system using the extracted constellation in Fig.~\ref{fig:chp2_ccdf_const} is higher than the one of the 16-QAM system in Fig.~\ref{fig:chp2_ccdf}.
This indicates that the underlying learned constellation alone has worse \gls{PAPR} characteristics than a standard \gls{QAM}, demonstrating the efficiency of the positional adjustments enabled by the neural transmitter.

\clearpage
\section{Concluding Thoughts} 
\label{sec:chp2_conclusion}

In this chapter, we proposed a learning approach to design OFDM waveforms that meet specific constrains on the envelope and spectral characteristics.
We leveraged the end-to-end strategy to model the transmitter and receiver as two \glspl{CNN} that perform high-dimensional modulation and demodulation, respectively.
The associated training procedure first requires all optimization constraints to be expressed as differentiable functions that can be minimized through \gls{SGD}.
Then, a constrained optimization problem is formulated and solved using the augmented Lagrangian method.
We evaluated the proposed approach on the learning of OFDM waveforms that maximize an information rate of the transmission while satisfying \gls{PAPR} and \gls{ACLR} constraints.
Simulations were performed using \gls{3GPP}-compliant channel models, and results show that the optimization procedure is able to design waveforms that satisfy the \gls{PAPR} and \gls{ACLR} constraints.
Moreover, the end-to-end system enables up to 30\% higher throughput than a close to optimal implementation of a \gls{TR} baseline with similar \gls{ACLR} and \gls{PAPR}.

Evaluation insights revealed that the neural transmitter achieves \gls{PAPR} and \gls{ACLR} reduction through a subcarrier-dependent filtering, an uneven energy distribution across subcarrier, and a positional readjustment of each constellation point.
On the other side, the neural receiver is able to equalize the channel without any pilot transmitted thanks to the learned asymmetrical constellations.
This directly translates into throughput gains as the associated overhead is removed. 
It is interesting to note that all these improvements are possible because the communication system is entirely \emph{designed by DL}.
Current standards however require the use of well-known modulations and pilot patterns to ensure compatibility across a wide range of hardware.
Moreover, commercially available devices are not yet optimized to perform \gls{NN} inferences at speeds that coincide with the processing time of physical layer tasks.
Short-term implementations of end-to-end systems are therefore unlikely, and more practical solutions must be derived to unlock the potential of \gls{DL} in the near future.

\clearpage

\chapter[DL-enhanced Receive Processing for MU-MIMO Systems]{DL-enhanced \\Receive Processing \\for MU-MIMO Systems}
\label{ch:ml_mu_mimo}
\label{ch:3}

\section{Motivations} 
\label{sec:chp3_introduction}

Both the block-based and end-to-end optimization strategies present advantages and drawbacks that were respectively discussed in Chapter~\ref{ch:1} and~\ref{ch:2}.
On the one hand, the block-based strategy for MU-MIMO detection has the advantages of remaining interpretable and scalable to any number of users, but is not trained to optimize the end-to-end performance and requires perfect \gls{CSI} that is not available in practice.
On the other hand, the end-to-end strategy allows the emergence of new waveforms that can be tailored for specific needs, but NN-based transceivers are computationally expensive and acts as black-boxes that are not suited for \gls{MU-MIMO} transmissions.
The quest for practical, standard-compliant, and reasonably complex designs for \gls{DL}-based \gls{MU-MIMO} systems therefore remains an active research topic.

In this context, we introduced a new hybrid approach in~\cite{goutay2020machine}, in which multiple \gls{DL} components are trained in an end-to-end manner to enhance a conventional \gls{MU}-\gls{MIMO} receiver.
The goal is to combine the interpretability of the first strategy, the efficiency of the second one, and the flexibility of traditional receivers.
The resulting architecture is easily scalable to any number of users and is composed of components that are individually interpretable and of reasonable complexity.
Moreover, the end-to-end training alleviates the need for perfect channel estimates.
Our solution exploits the \gls{OFDM} structure  to counteract two known drawbacks of conventional receivers that are overlooked in the current literature.
First, it improves the prediction of the channel estimation error statistics, that can only be obtained for the pilot signals in a standard receiver, resulting in largely sub-optimal detection accuracy. 
Motivated by the recent successes of \glspl{CNN} in physical layer tasks~\cite{8752012, honkala2020deeprx}, we use \glspl{CNN} to predict such error statistics for every \gls{RE} in the  \gls{OFDM} grid.
In contrast to the traditional approach that is based on mathematical models, the proposed \glspl{CNN} learn these statistics from the data during training.
The second improvement is a \gls{CNN}-based demapper that operates on the entire \gls{OFDM} grid, unlike traditional demappers that operate on individual \glspl{RE}. 
In doing so, our demapper is able to better cope with the residual distortions of the equalized signal.
The resulting \gls{DL}-enhanced receiver is optimized such that all \gls{DL} components are jointly trained to maximize the information rate of the transmission \cite{9118963}.

The proposed architecture is evaluated on \gls{3GPP}-compliant channel models with two different pilot configurations supported by the 5G \gls{NR} specifications. 
Both uplink and downlink transmissions are studied using \gls{TDD}.
We compare the coded \gls{BER} achieved by different schemes for user speeds ranging from \SIrange[range-units = single]{0}{130}{\km\per\hour}.
Two baselines were implemented, the first one being a traditional receiver implementing a \gls{LMMSE} channel estimation and an \gls{LMMSE} equalizer. 
The second baseline also uses \gls{LMMSE} equalization, but is provided with perfect channel knowledge at pilot positions and the exact second order statistics of the channel estimation errors.
Our results show that the gains provided by the \gls{DL}-enhanced receiver increase with the user speed, with small \gls{BER} improvements at low speeds and significant ones at high speeds.
We have also observed that the gains in the uplink are more pronounced than in the downlink, due to channel aging which significantly penalizes the downlink precoding scheme.



\subsection*{Related literature}

Multiple recent papers proposed to use one large \gls{NN} to jointly perform multiple processing steps.
This idea has first been proposed in \cite{8052521} to perform joint channel estimation and equalization in a \gls{SISO} setup.
This work has then been extended by additionally learning the demapper and operating directly on time-domain samples~\cite{zhao2021deepwaveform, korpi2020deeprx}.
The DeepRX architecture presented in \cite{honkala2020deeprx} shows impressive results on \gls{SIMO} channels while being 5G compliant.
Another standard-compliant receiver has been proposed for Wi-Fi communications using both synthetic and real-world data \cite{zhang2020deepwiphy}.
Regarding \gls{MIMO} transmissions, a special form of \gls{RNN} called reservoir computing has been leveraged in \cite{zhou2019learning} to process time-domain \gls{OFDM} signals.
The DeepRX receiver has also been extended with a so-called transformation layer to handle \gls{MIMO} transmissions \cite{korpi2020deeprx}. 
Their \gls{CNN}-based solution is fed with frequency-domain signals and outputs \glspl{LLR} for the transmitted bits. 
DeepRX MIMO shows important gains on \gls{SU}-\gls{MIMO} channels, but remains very computationally expensive.
Compared to our solution, the main disadvantages of these \gls{NN}-based \gls{MIMO} receivers are their lack of scalability and interpretability. 
They are tailored for a specific number of users or transmit antennas, and the \gls{CSI} needed for downlink precoding can not be easily extracted.

\clearpage
\section{System Model} 
\label{sec:chp3_system_model}

\begin{figure*}[t!]
  	\begin{subfigure}{0.20\textwidth}
    	\includegraphics[height=140pt]{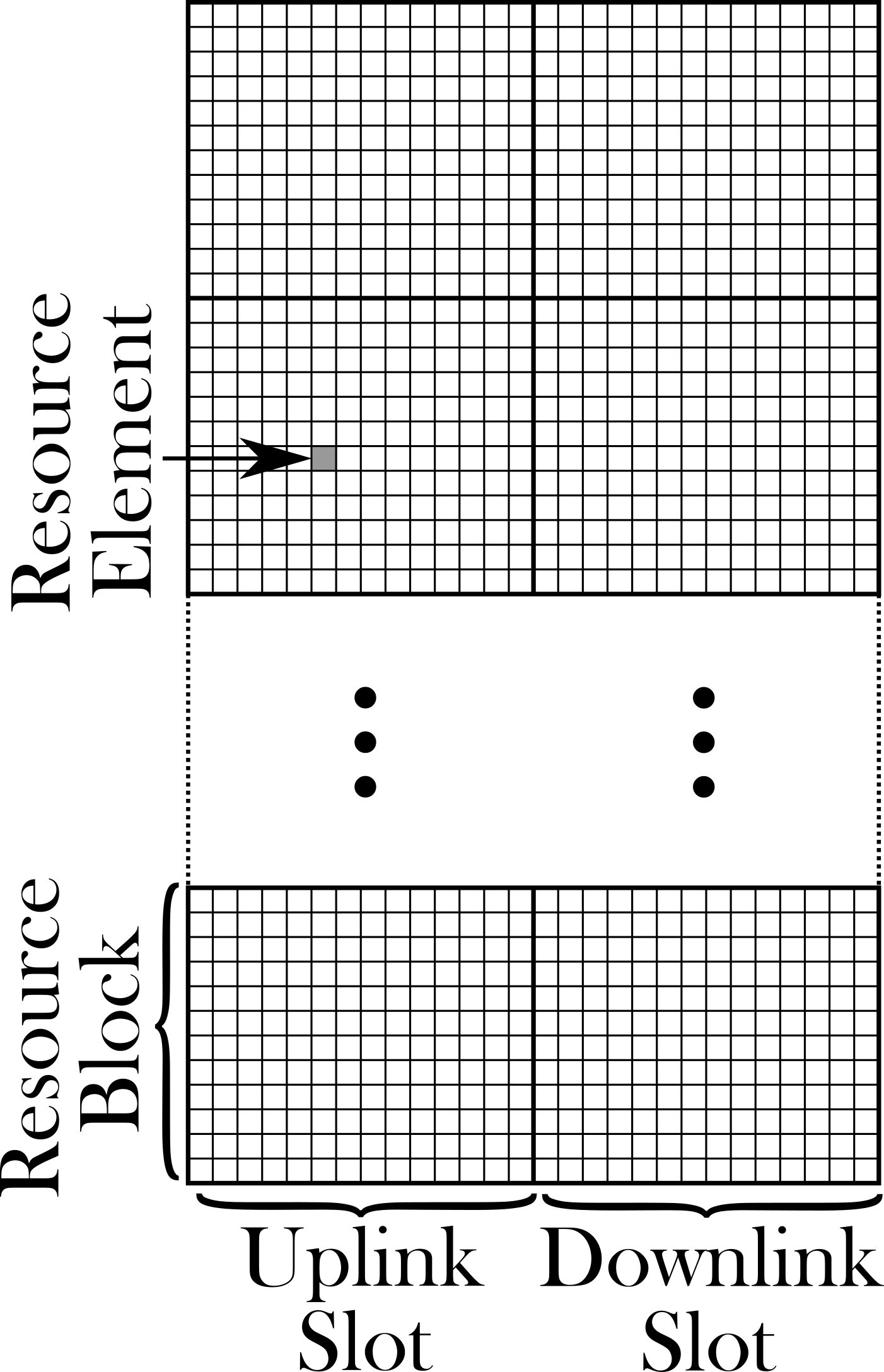} 
  	\caption{Resource grid.}
  	\label{fig:chp3_resource_grid}
	\end{subfigure}%
\hspace{13pt}
  	\begin{subfigure}{0.33\textwidth}
  	\includegraphics[height=140pt]{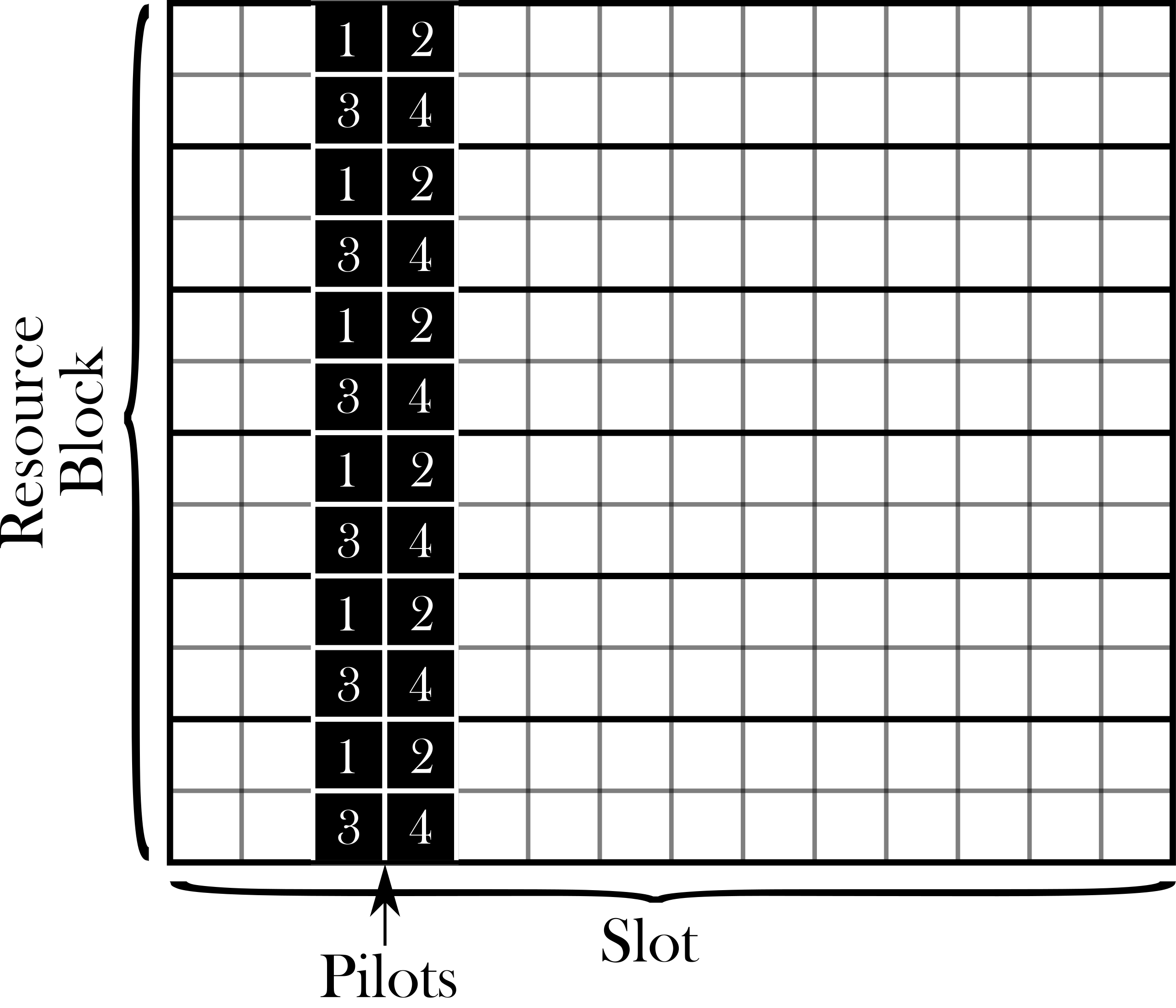}
  	\caption{1P pilot pattern.}
  	\label{fig:chp3_1P_pattern}
	\end{subfigure}%
\hspace{23pt}
  	\begin{subfigure}{0.33\textwidth}
  	\includegraphics[height=140pt]{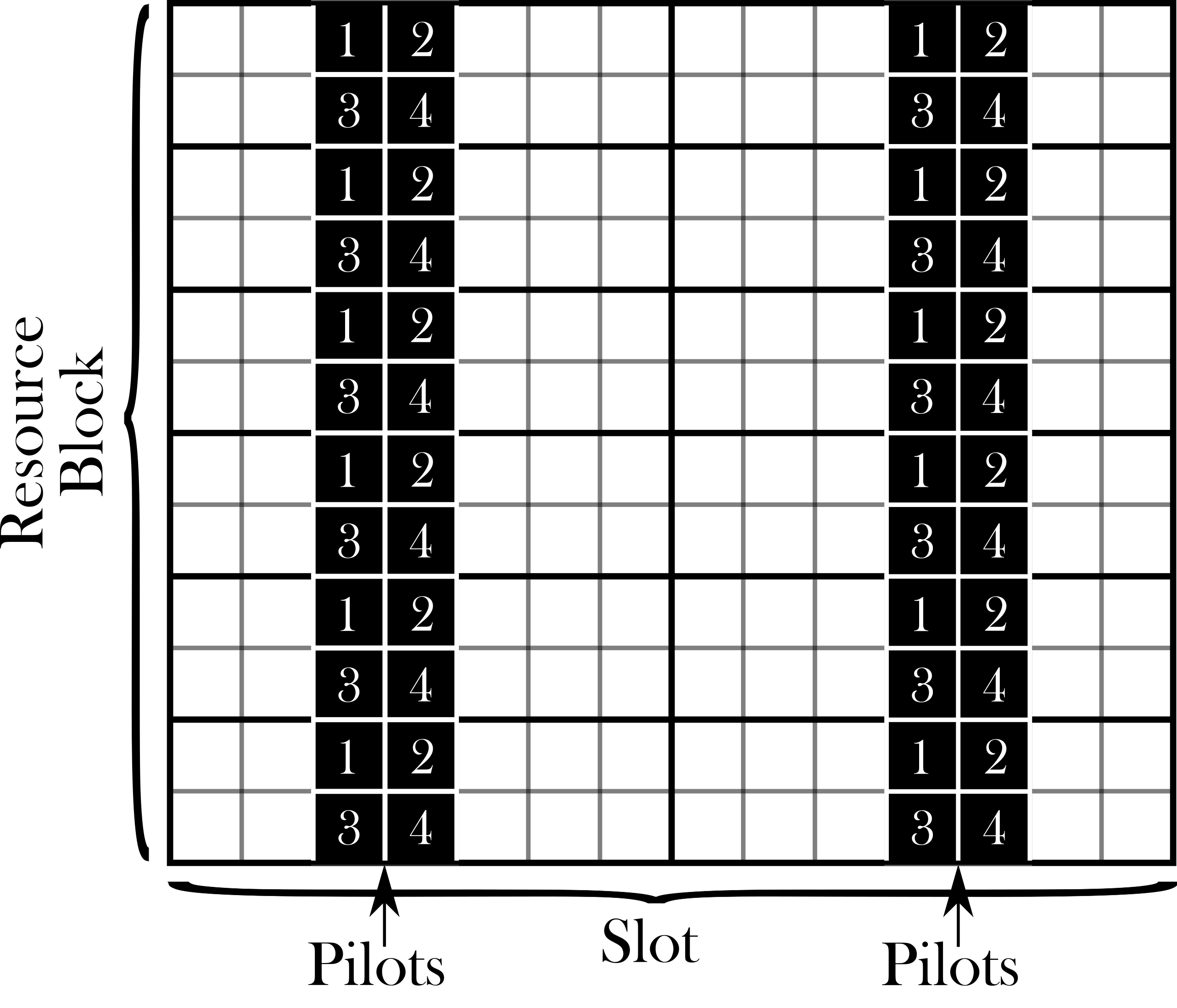}
  	\caption{2P pilot pattern.}
  	\label{fig:chp3_2P_pattern}
	\end{subfigure}%
\caption{Pilots are arranged on the \gls{RG} according to two different patterns, where each number corresponds to a different transmitter.}
\label{fig:chp3_channel_model}
\end{figure*}

We consider a \gls{MU-MIMO} system, as presented in Section~\ref{sec:mimo_systems}, in which $K$ single-antenna users communicate with a \gls{BS} equipped with $L$ antennas in the uplink and downlink.
\gls{OFDM} transmissions are considered, and the \gls{RG} is divided into resource blocks consisting of twelve adjacent subcarriers (Fig.~\ref{fig:chp3_resource_grid}). 
$2^Q$-\gls{QAM} modulations are used to transmit data.
This section introduces the channel model and the baselines against which the proposed approach is benchmarked.

\subsection{Channel Model}

Following the notations of Section~\ref{sec:mimo_systems}, the channel coefficients form a 4-dimensional tensor denoted by $\Hm \in \CC^{2 M \times N \times L \times K}$, such that $\Hm_{m, n} \in \CC^{L \times K}$ is the channel matrix at \gls{RE} $(m,n)$, and $\hv_{m, n, k}\in\CC^{L}$ is the channel vector at \gls{RE} $(m,n)$ and for user $k$.
Duplexing is achieved through \gls{TDD}, such that a slot is either assigned to the  uplink or downlink in an alternating fashion, as illustrated in Fig.~\ref{fig:chp3_resource_grid}.
More precisely, the first slot is assigned to the uplink and the second slot is assigned to the downlink.
It is assumed that channel reciprocity holds, i.e., $\Hm_{m,n}$ refers as well to the uplink or the downlink channel.
To enable channel estimation, a transmitter sends pilot signals on dedicated \glspl{RE} according to a predefined pilot pattern.
We assume, without loss of generality, that all pilots are equal to one.
Two different pilot patterns are considered, referred to as the 1P and 2P pilot patterns, which respectively contain pilots on two or four symbols within a slot.
Fig.~\ref{fig:chp3_1P_pattern} and \ref{fig:chp3_2P_pattern} respectively show the 1P and 2P pilot patterns over a resource block assuming 4 users.
The set of \glspl{RE} carrying pilots for a user $k \in \LP 1, \dots, K \RP$ is denoted by $\Pc^{(k)}$ and the numbers of symbols and subcarrier carrying pilots are respectively denoted by $|\Pc_M|$ and $|\Pc_N|$. 
As an example, if the 1P pattern shown in Fig.~\ref{fig:chp3_1P_pattern} is used with $N = 12$, the positions $(\text{symbol}, \text{subcarrier})$ of all \glspl{RE} carrying pilots for user 1 are denoted by $\mathcal{P}^{(1)} = \{(3, 1), (3, 3), (3, 5), (3, 73), (3, 93), (3, 11)\}$, resulting in $|\Pc_M| = 1$ and $|\Pc_N| = 6$.
Note that when a \gls{RE} is allocated to a user for the transmission of a pilot, other users do not transmit any signal (data nor pilot) on that \gls{RE}.
As a consequence, pilots do not experience any interference.
The noise power is denoted by $\sigma^2$ and assumed equal for all users and all \glspl{RE}.
In the following, perfect power control is assumed over the \gls{RG} such that the mean energy corresponding to a single \gls{BS} antenna and a single user is one, i.e., $\EE\LSB |h_{m,n,l,k}|^2 \RSB = 1$.
The \gls{SNR} of the transmission is defined as 
\begin{equation}
\label{eq:chp3_snr}
\text{SNR} = 10 \log{\frac{\EE\LSB |h_{m,n,l,k}|^2 \RSB}{\sigma^2}} = 10 \log{\frac{1}{\sigma^2}} \, [\si{dB}].
\end{equation}

\subsection{Uplink Baseline}
\label{sec:chp3_baseline_ul}
\begin{figure*}[!t]
    \centering
    \includegraphics[width=1\textwidth]{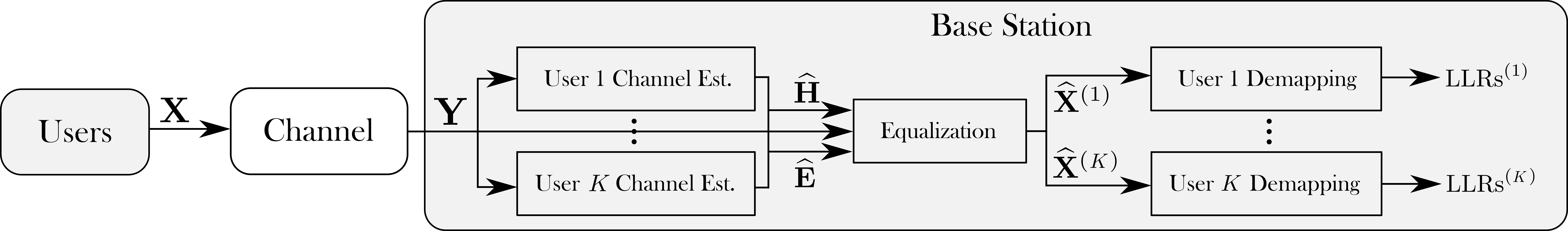}
    \caption{Architecture of the uplink communication system.}
    \label{fig:chp3_uplink}
\end{figure*}

In uplink, the \gls{BS} aims to recover the bits transmitted simultaneously by the $K$ users on the \glspl{RE} carrying data. 
The tensors of transmitted and received signals of all users are respectively denoted by $\Xm \in \CC^{2M \times N \times K}$ and $\Ym \in \CC^{2M \times N \times  L}$, and the transfer function on the uplink is $\yv_{m,n} = \Hm_{m,n} \xv_{m,n} + \nv_{m,n}$, where $\nv_{m,n} \sim \Cc\Nc \LB \zerov, \sigma^2\Id_{L} \RB$ is the noise vector.
In this scenario, only the uplink slot is used and therefore all signals with indices $m > M$ are ignored, i.e., the  corresponding values are set to $0$.
The architecture of the uplink system is shown in Fig.~\ref{fig:chp3_uplink}, where the IDFT (DFT) operation and the addition (removal) of the cyclic prefix before (after) the channel are not shown for clarity.
The channel estimation, equalization, and demapping stages of the baseline will be explained in the following.

\subsubsection{Channel estimation}
\label{sec:chp3_ch_est_1}

As the pilots are assumed to be orthogonal, \gls{LMMSE} channel estimation can be carried out for each user independently.
The channel covariance matrix providing the spatial, temporal, and spectral correlations between all \glspl{RE} carrying pilots is denoted by $\Sigmam \in \CC^{|\Pc_M| \cdot |\Pc_N|\cdot  L \times |\Pc_m| \cdot |\Pc_n| \cdot L}$.
In the following, it is assumed that the precise local statistics of the receivers are not available.
The channel and receiver statistics are therefore averaged over the entire cell, resulting in a zero-mean channel, and a discussion on how these statistics can be obtained is provided in Section~\ref{sec:chp3_stats}.
For a user $k \in \{1,\dots,K\}$, the \gls{LMMSE} channel estimate at \glspl{RE} carrying pilots is denoted by $\widehat{\Hm}^{(k)}_{\mathcal{P}^{(k)}} \in \CC^{|\Pc_M| \times |\Pc_N|  \times L}$ and given by
\begin{equation}
\text{vec}\LB\widehat{\Hm}^{(k)}_{\mathcal{P}^{(k)}}\RB = \Sigmam \LB \Sigmam  + \sigma^2\Id_{|\Pc_M| |\Pc_N| L } \RB^{-1} \text{vec}\LB\Ym^{(k)}_{\mathcal{P}^{(k)}}\RB
\end{equation}
where $\Ym^{(k)}_{\mathcal{P}^{(k)}} \in \CC^{|\Pc_M| \times |\Pc_N| \times L}$ is the tensor of received pilots for user $k$.
Channel estimation could also be performed at \glspl{RE} carrying data \cite{pilotless20}, but this would require knowledge of the channel statistics at those \glspl{RE}, which are typically not available in practice.

Inspired by the \gls{3GPP} guidelines \cite{std3gpp}, the channel estimates for all \glspl{RE} are computed by first linearly interpolating the estimates from \glspl{RE} carrying pilots in the frequency dimension and then using the estimate at the \gls{NIRE} on the neighboring \glspl{RE}.
It is also possible to leverage temporal linear interpolation between the \gls{OFDM} symbols carrying pilots when the 2P pilot pattern is used.
The so-obtained tensor of channel estimates is denoted by $\widehat{\Hm}^{(k)} \in \CC^{2M \times  N \times L}$. 
The overall channel estimation for all users $\widehat{\Hm} \in \CC^{2M \times N \times  L \times K} $ is obtained by stacking the channel estimates of all users.
Since only the uplink slot is considered here, the channel estimates for the last $M$ symbols (downlink slot) are set to be null.
The channel estimation error is denoted by $\widetilde{\Hm}$ and is such that $\Hm = \widehat{\Hm} + \widetilde{\Hm}$.
For a \gls{RE} $(m,n)$, we define
\begin{equation}
\label{eq:chp3_E}
\Em_{m,n} \coloneqq \EE \LSB \widetilde{\Hm}_{m, n} \widetilde{\Hm}_{m, n}\htp \RSB
\end{equation}
as the sum of the \emph{spatial} channel estimation error covariance matrices from all users.

\begin{figure}
	\center
\includegraphics[width=0.9\textwidth]{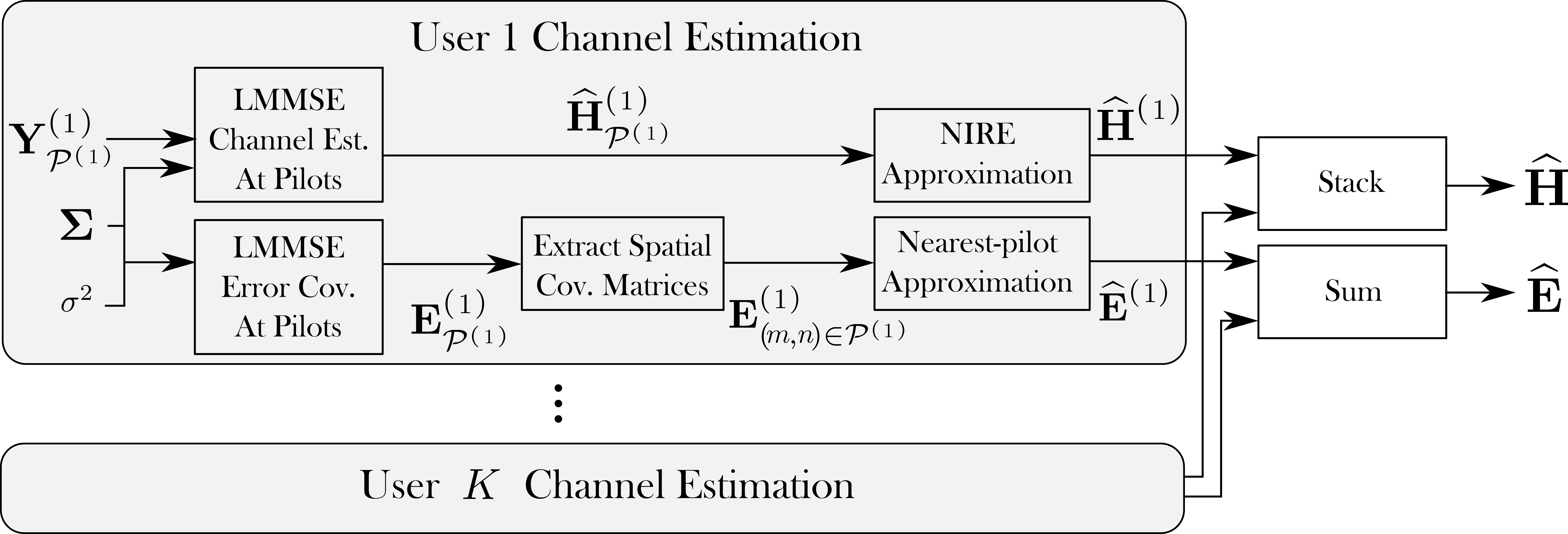}
  \caption{Uplink channel estimation.}
    \label{fig:chp3_ch_est_tradi}
\end{figure}

\begin{figure}
	\center
\includegraphics[width=0.55\textwidth]{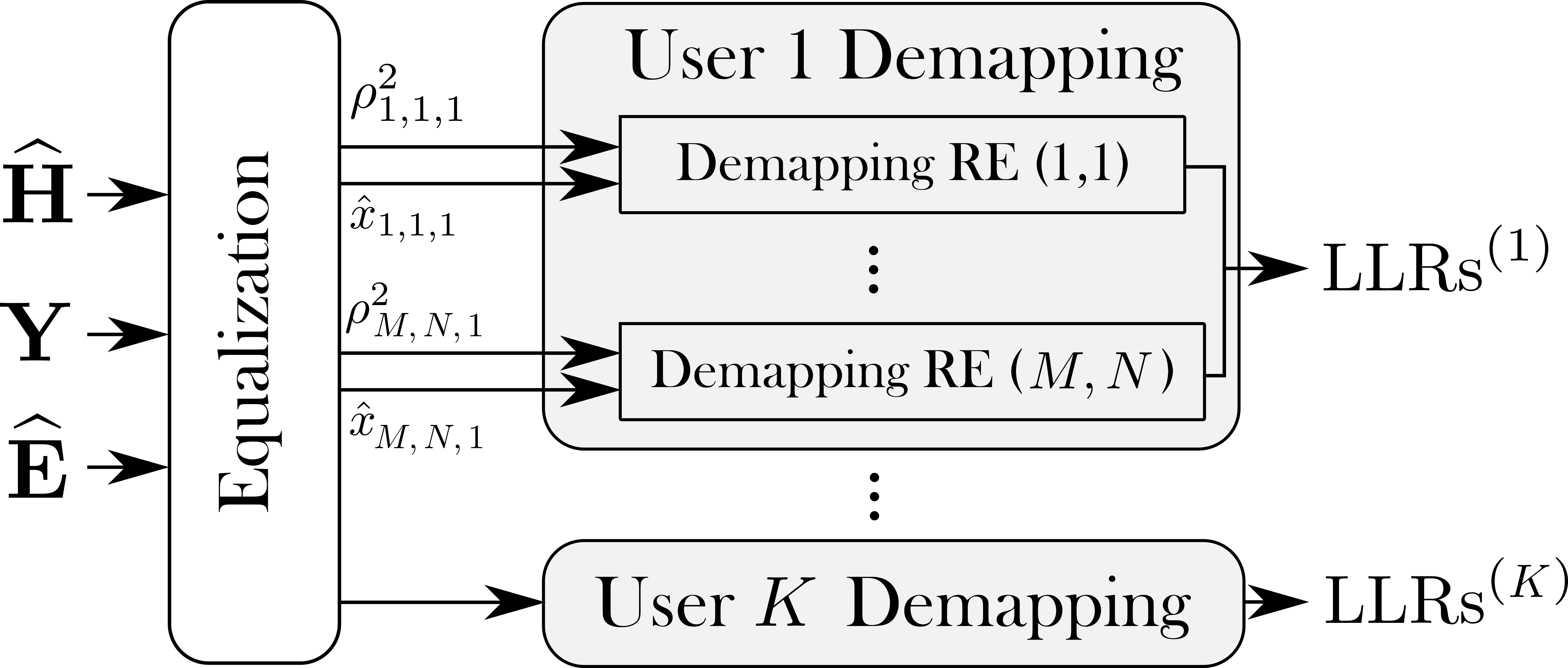}
\vspace{8pt}
  \caption{Uplink demapping}
  \label{fig:chp3_eq_dmp_tradi}
\end{figure}

\subsubsection{Equalization}

The \gls{LMMSE} equalizer derived in Chapter 3 is used, except that perfect channel estimation is not assumed anymore.
Moreover, as computing a dedicated \gls{LMMSE} operator for each \gls{RE} is infeasible in practice due to prohibitive complexity, we resort to a grouped-\gls{LMMSE} equalizer, i.e., a single \gls{LMMSE} operator is applied to a group of adjacent \glspl{RE} spanning multiple symbols $m \in \{M_b,\dots,M_e\}$ and subcarriers $n \in \{N_b,\dots,N_e\}$.
Under this assumption, the \gls{LMMSE} operator for the group of \glspl{RE} spanning $\{M_b,\dots, M_e\}\times\{N_b,\dots,N_e\}$ is 
\begin{align}
\label{eq:chp3_W}
\Wm_{m,n} = & \LB \sum_{m'=M_b}^{M_e} \sum_{n'=N_b}^{N_e}  \widehat{\Hm}_{m', n'}\htp \RB 
\LB \sum_{m'=M_b}^{M_e} \sum_{n'=N_b}^{N_e}  \widehat{\Hm}_{m', n'} \widehat{\Hm}_{m',n'}\htp + \Em_{m',n'} + \sigma^2 \Id_{L}  \RB^{-1} 
\end{align}
where $\Wm_{m,n} \in \CC^{K \times L}$ is constant over $\{M_b,\dots, M_e\}\times\{N_b,\dots,N_e\}$ (see Appendix~\ref{app:LMMSE}).

The post-equalization channel is expected to be an additive noise channel. More precisely, for any \gls{RE} $(m,n)$, the demapper expects the output of the equalizer $\hat{\xv}_{m, n} \in \CC^{K}$ to be such that $\hat{\xv}_{m,n} = \xv_{m,n} + \nv'_{m,n}$ where $\nv'_{m,n}$ is an additive noise term.
However, this decomposition is not achieved by the \gls{LMMSE} equalizer (see Section~1.6.1 of \cite{heath2018foundations} for a more detailed discussion).
To obtain such a post-equalized channel, the following diagonal matrix is applied to the output of the \gls{LMMSE} equalizer
\begin{equation}
\Dm_{m,n} = \LB  \LB \Wm_{m,n} \hat{\Hm}_{m, n} \RB \odot \Id_{K} \RB^{-1}
\end{equation}
which re-scales the equalizer output so that the post-equalization \gls{SNR} remains maximized.
For a \gls{RE} $(m,n) \in \{M_b,\dots, M_e\}\times\{N_b,\dots,N_e\}$, the equalized vector $\hat{\xv}_{m, n}$ is computed by
\begin{equation}
\hat{\xv}_{m, n} = \Dm_{m,n} \Wm_{m,n} \yv_{m,n}.
\end{equation}
The equalized symbols of user $k$ are denoted by $\widehat{\Xm}^{(k)} \in \mathbb{C}^{M \times N}$, as shown in Fig.~\ref{fig:chp3_uplink}.

\subsubsection{Demapping} 
\label{sec:chp3_demapping_ul}

For a \gls{RE} $(m,n)$, let us denote by $\wv_{m,n,k}$ the column vector made of the $k^{\text{th}}$ line of the matrix $\Wm_{m,n}$ and by $\Hm_{m, n, -k}$ the tensor made of the channel coefficients of all users except user $k$. 
After equalization, the uplink channel can be viewed as $M N K$ parallel additive noise channels that can be demodulated independently for every \gls{RE} and every user.
For a \gls{RE} $(m,n)$ and user $k$, the post-equalization channel is expressed as

\begin{equation}
\hat{x}_{m, n, k}= x_{m, n, k} + \underbrace{\frac{\wv_{m, n, k}\tp \LB \widehat{\Hm}_{m, n, \shortminus k} \xv_{m, n, \shortminus k} + \widetilde{\Hm}_{m, n} \xv_{m, n} +  \nv_{m, n} \RB}{\wv_{m, n, k}\tp \widehat{\hv}_{m, n, k} }}_{\zeta_{m, n, k}}
\end{equation}

where the noise $\zeta_{m, n, k}$ includes both the interference and the noise experienced by user $k$.
Its variance is given by
\begin{align}
\label{eq:chp3_eq_snr_ul}
\rho_{m, n, k}^2 & =  \EE\LSB \zeta_{m, n, k}^* \zeta_{m, n, k} \RSB\\
& = \frac{\wv_{m, n, k}\htp \LB\widehat{\Hm}_{m, n, \shortminus k} \widehat{\Hm}_{m, n, \shortminus k}\htp + \Em_{m,n} + \sigma^2 \Id_{L} \RB \wv_{m, n, k}}{\wv_{m, n, k}\htp \widehat{\hv}_{m, n, k} \widehat{\hv}_{m, n, k}\htp \wv_{m, n, k}} \nonumber .
\end{align}
We denote by $\mathcal{C}_{q,0}$~($\mathcal{C}_{b,1}$) the subset of $\mathcal{C}$ which contains all symbols with the $q^{\text{th}}$ bit set to 0~(1).
Assuming that the noise $\zeta_{m,n,k}$ is Gaussian\footnote{This is not true in general as the interference and channel estimation errors are not Gaussian distributed.}, the \glspl{LLR} of the $q^{\text{th}}$ bit transmitted by user $k$ on the \gls{RE} $(m, n)$ is given by

\begin{equation}
\text{LLR}_{m, n, k}^{\text{UL}}(q) = \ln{\frac{
\sum_{c\in\mathcal{C}_{q,1}} \exp{  - \frac{1}{\rho_{m, n, k}^2} \abs{ \hat{x}_{m, n, k} - c}^2 }
}{
\sum_{c\in\mathcal{C}_{q,0}} \exp{  - \frac{1}{\rho_{m, n, k}^2} \abs{\hat{x}_{m, n, k} - c}^2 }
} } .
\end{equation}

The equalization and demapping process is schematically shown in Fig.~\ref{fig:chp3_eq_dmp_tradi}.

\subsection{Downlink Baseline}
\label{sec:chp3_baseline_dl}

\begin{figure*}
    \centering
    \includegraphics[width=1\textwidth]{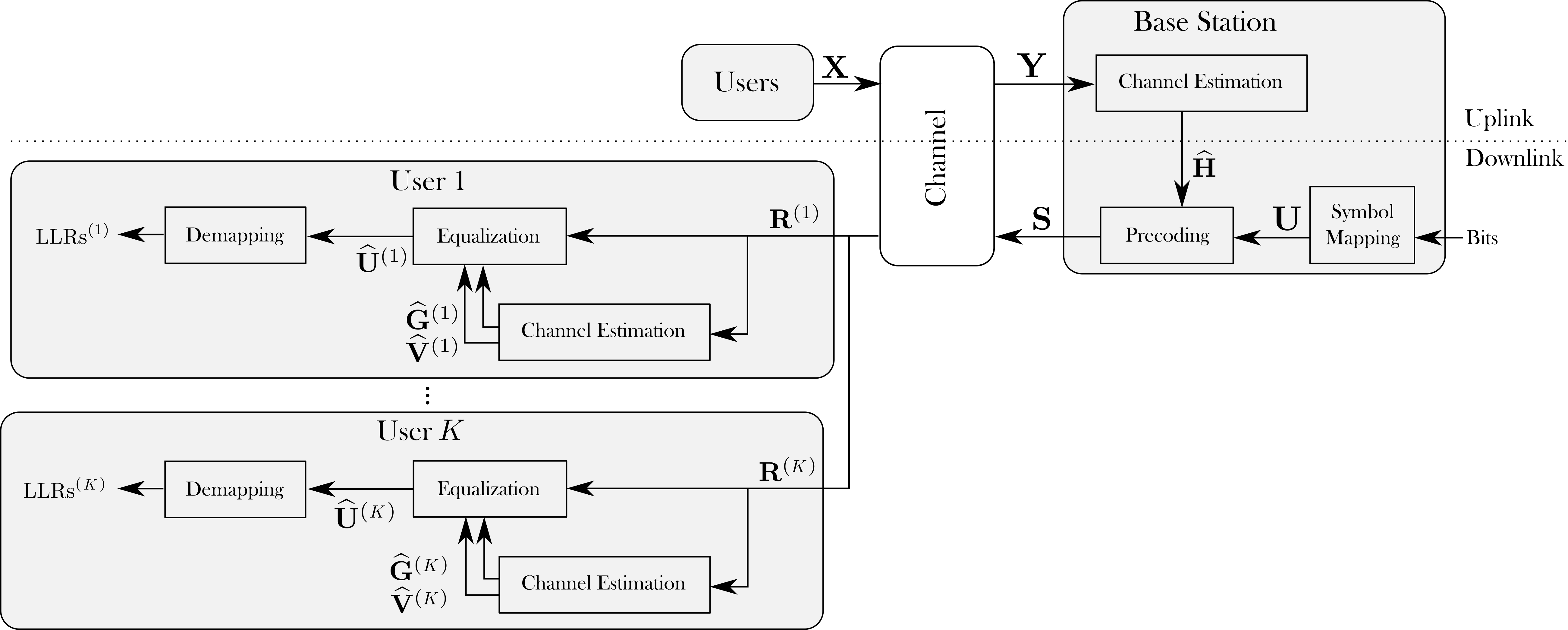}
    \caption{Architecture of the downlink communication system.}
    \label{fig:chp3_downlink}
\end{figure*}

The \gls{BS} aims to simultaneously transmit to  $K$ users on all \glspl{RE} of the downlink slot. 
The signal transmitted by the \gls{BS} is precoded to mitigate interference.
We remind that downlink transmissions occur after the uplink slot, as shown in Fig.~\ref{fig:chp3_resource_grid}.
Let us denote by $\Sm \in \mathcal{C}^{2M \times N \times  K}$ and by $\Tm \in \CC^{2M \times N \times  L}$ the tensors of unprecoded and precoded symbols, respectively.
We denote by $\Um \in \CC^{ 2 M \times N \times K}$ the tensor of symbols received by the $K$ users.
Those quantities are only relevant on the downlink slot and therefore are considered null on the first $M$ symbols, i.e., $\sv_{m,n} = \tv_{m,n} = \uv_{m,n} =  \mathbf{0} \quad  \forall (m,n) \in \{1, \dots, M\} \times \{1, \dots, N\}$.
The downlink transfer function of the channel  for a \gls{RE} $(m,n)$ is
\begin{equation}
\label{eq:chp3_down_chan}
\uv_{m,n} = \Hm_{m,n}\htp \tv_{m,n} + \qv_{m,n}
\end{equation}
where $\qv_{m,n} \sim \Cc\Nc(\zerov,\sigma^2 \Id_{K})$ is the noise vector,  considered null in the first $M$ symbols.
For convenience, the noise variance $\sigma^2$ is assumed to be the same as in the uplink.
Fig.~\ref{fig:chp3_downlink} shows the architecture of the downlink system, where the IFFT (FFT) operation and the addition (removal) of the cyclic prefix before (after) the channel are again not shown for clarity.
In the rest of this section, we detail the downlink precoding, channel estimation and equalization, and demapping steps.

\subsubsection{Precoding}

Precoding requires estimation of the downlink channel. 
As \gls{TDD} is used, we can exploit channel reciprocity so that the downlink channel can be estimated using the nearest-pilot approach, i.e., $\widehat{\Hm}_{M + 1 \leq m \leq 2M, n} = \widehat{\Hm}_{M, n}$.
Precoding is achieved by exploiting the uplink-downlink duality~\cite{viswanath2003sum}, which results in using $\Wm_{m,n}\htp$ as precoding matrix that can be computed using~\eqref{eq:chp3_W}.
Normalization is performed to ensure that the average energy per transmitted symbol equals one, by applying the diagonal matrix 
\begin{equation}
\Cm_{m, n} = \LB \LB \Wm_{m,n}\Wm_{m,n}\htp \RB \odot \Id_{K} \RB^{-\frac{1}{2}}
\end{equation}
leading to the precoded signal
\begin{equation}
\tv_{m,n}  = \Cm_{m,n} \Wm_{m, n}\htp  \sv_{m,n}.
\end{equation}
The channel transfer function~\eqref{eq:chp3_down_chan} can be rewritten as
\begin{align}
\uv_{m,n}& = \Hm_{m,n}\htp \tv_{m,n}  + \qv_{m,n} \\
&= \underbrace{\Hm_{m,n}\htp \Cm_{m,n} \Wm_{m, n}\htp }_{\Gm_{m,n}} \sv_{m,n} + \qv_{m,n} \nonumber
\end{align}
where $\Gm_{m, n}  \in \CC^{K \times K}$ is referred to as the \emph{equivalent} downlink channel for the \gls{RE} $(m,n)$.
Each user $k$ receives its signal $u^{(k)}_{m,n}$  and the corresponding channel, i.e., the $k^{\text{th}}$ row of $\Gm_{m, n}$, is denoted by ${\gv^{(k)}_{m,n}}\tp \in \mathbb{C}^{K}$.
Finally, the equivalent channel experienced by user $k$ for the entire \gls{RG} is denoted  by $\Gm^{(k)} \in \mathbb{C}^{2M \times N \times  K}$.

\subsubsection{Channel estimation and equalization}
\label{sec:chp3_dl_ch_est}

To enable estimation of the equivalent  downlink channel by the users, pilot signals are transmitted by the \gls{BS} using the same pilot patterns as in the uplink (Fig.~\ref{fig:chp3_channel_model}).
Each user $k$ estimates its equivalent channel $\widehat{\Gm}^{(k)} \in \mathbb{C}^{2M \times N \times  K}$, where for a given \gls{RE} $(m,n)$, the element $\hat{g}^{(k)}_{m,n,k}$ corresponds to the main channel coefficient, whereas the elements $\hat{g}^{(k)}_{m,n,i}, i \neq k$ correspond to the interference channel coefficients.
As in the uplink, \gls{LMMSE} estimation, followed by spectral and possibly temporal interpolation, is used, but it is assumed that the elements of ${\hat{\gv}^{(k)}_{m,n}}$ are uncorrelated.
Therefore, channel estimation is performed independently for the main channel and each interference channel, enabling easy scalability to any number of interferers.
The covariance matrices used to estimate the main channel and one of the interfering channels of a given user are denoted by $\Omegam \in \CC^{|\Pc_M|\cdot|\Pc_N| \times |\Pc_M|\cdot |\Pc_N|} $ and by $\Psim \in \CC^{|\Pc_M|\cdot|\Pc_N| \times |\Pc_M|\cdot |\Pc_N|}$, respectively, and are equal for all users and interfering channels.
The tensor of the equivalent channel estimation error for user $k$ is denoted by $\widetilde{\Gm}^{(k)} \in \CC^{2M \times N \times  K}$, and is such that $\Gm^{(k)} = \widehat{\Gm}^{(k)} + \widetilde{\Gm}^{(k)}$.
The estimation error variances for the main and the $i^{\text{th}}$ interfering channel of user $k$ are respectively denoted by $v^{(k)}_{m,n,k} \coloneqq \EE \LSB \abs{\tilde{g}^{(k)}_{m,n,k}}^2 \RSB$ and $v^{(k)}_{m,n,i} \coloneqq \EE \LSB \abs{\tilde{g}^{(k)}_{m,n,i}}^2 \RSB$.
Similarly, we denote by $\Vm^{(k)} \in\mathbb{R}^{N, \times M \times K}$ the tensor of estimated error variances for user $k$, as shown in Fig.~\ref{fig:chp3_downlink}.
An estimation of the transmitted unprecoded symbol for user $k$ is computed by equalizing the received signal as follows

\begin{equation}
\hat{s}^{(k)}_{m, n} = \frac{u^{(k)}_{m, n}}{\hat{g}^{(k)}_{m,n,k}}.
\end{equation}

\subsubsection{Demapping}

The post-equalization downlink channel can be seen as $ M \times N \times K$ parallel additive noise channels.
More precisely, for a user $k \in \{1,\dots,K\}$ and \gls{RE} $(m,n)$,
\begin{equation}
\label{eq:chp3_eq_eq_ch_dl}
\hat{s}^{(k)}_{m, n}
= s_{m, n, k}
+ \underbrace{\frac{\tilde{g}^{(k)}_{m, n, k} s_{m, n, k} 
+ {\gv^{(k)}_{m,n, \shortminus k}}\tp ~ \sv_{m, n, \shortminus k}
+ q_{m, n, k}}{\hat{g}^{(k)}_{m, n, k}}}_{\xi_{m, n, k}}
\end{equation}
where $\xi_{m,n,k}$ comprises the channel noise and interference and has variance
\begin{align}
\label{eq:chp3_eq_snr_dl}
\tau_{m, n, k}^2
& =  \EE\LSB \xi_{m, n, k}^* \xi_{m, n, k} \RSB  \\
& = \frac{v^{(k)}_{m,n,k} + \hat{\gv}^{(k)~~^{\scriptstyle \mathsf{H}}}_{m, n,\shortminus k} ~ \hat{\gv}^{(k)}_{m,n,\shortminus k}
+ \sum_{i=1, i\neq k}^{K} v^{(k)}_{m,n,i} + \sigma^2 }{\abs{\hat{g}^{(k)}_{m, n, k}}^2} \nonumber .
\end{align}
Assuming $\xi_{m,n,k}$ is Gaussian distributed\footnote{Similarly to the uplink scenario, this is not true in general.}, the \gls{LLR} for the $q^{\text{th}}$ bit transmitted to user $k$ on \gls{RE} $(m, n)$ is given by

\begin{equation}
\text{LLR}_{m, n, k}^{\text{DL}}(q) = \ln{\frac{
\sum_{c\in\mathcal{C}_{q,1}} \exp{  - \frac{1}{\tau_{m, n, k}^2} \abs{\hat{s}_{m, n, k} - c}^2 }
}{
\sum_{c\in\mathcal{C}_{q,0}} \exp{  - \frac{1}{\tau_{m, n, k}^2} \abs{\hat{s}_{m, n, k} - c}^2 }
} } .
\end{equation}

\subsection{Estimation of the Required Statistics}
\label{sec:chp3_stats}

The baselines described above require the knowledge of the covariance matrices $\Sigmam$, $\Omegam$, and $\Psim$ which provide the spatial, time, and spectral correlations between the \glspl{RE} carrying pilots.
These matrices can be set based on models or can be empirically estimated by constructing large datasets of uplink, downlink, and interfering pilot signals, as was done in this Chapter.
The channel estimation error covariances $\Em_{m,n}$, defined in~\eqref{eq:chp3_E}, also need to be estimated to compute both the uplink equalization matrices $\Wm_{m,n}$ and the downlink precoding matrices $\Wm_{m,n}\htp$.
Focusing on \glspl{RE} carrying pilots, the estimation error covariance for a user $k$ is 
\begin{align}
\label{eq:chp3_cov_err}
\begin{split}
\Em^{(k)}_{\mathcal{P}^{(k)}}
& =   \EE\LSB \text{vec}\LB \widetilde{\Hm}^{(k)}_{\mathcal{P}^{(k)}}\RB  \text{vec}\LB \widetilde{\Hm}^{(k)}_{\mathcal{P}^{(k)}}\RB\htp \RSB \\
& = \Sigmam - \Sigmam \LB \Sigmam  + \sigma^2 \RB^{-1}\Sigmam 
\end{split}
\end{align}
where $\widetilde{\Hm}^{(k)}_{\mathcal{P}^{(k)}}$ is the channel estimation error at \glspl{RE} carrying pilots.
However, we are only interested in the \emph{spatial} channel estimation error correlations, whereas~\eqref{eq:chp3_cov_err} provides the correlations of channel estimation errors between all the receive antennas, subcarriers, and symbols.
For a single pilot position $(m, n) \in \mathcal{P}^{(k)}$, this spatial correlation matrix is defined by
\begin{equation}
\Em^{(k)}_{(m,n) \in \mathcal{P}^{(k)}} = \EE \LSB  \LB \widetilde{\hv}^{(k)}_{(m,n) \in \mathcal{P}^{(k)}} \RB \LB \widetilde{\hv}_{(m,n) \in \mathcal{P}^{(k)}}^{(k)}\RB ^\mathsf{H} \RSB \in  \CC^{L \times L}
\end{equation}
and can be extracted from $\Em^{(k)}_{\mathcal{P}^{(k)}}$.
To estimate $\Em_{m,n}$ for \glspl{RE} carrying data, a nearest-pilot approach is used, which sets the value $\Em_{m,n}$ for a \gls{RE} carrying a data signal to the one of the nearest \gls{RE} carrying a pilot signal.
The so-obtained estimation is denoted by $\widehat{\Em}_{m,n}^{(k)}$ for a \gls{RE} $(m,n)$.
The overall spatial estimation error covariance matrix for any \gls{RE} $(m,n)$ is obtained by summing the estimations for all users:
\begin{equation}
\widehat{\Em}_{m,n}
= \sum_{k=1}^{K} \widehat{\Em}^{(k)}_{m,n} \quad \in \CC^{L \times L}.
\end{equation}
The uplink channel and error covariance estimations are depicted in Fig.~\ref{fig:chp3_ch_est_tradi}.

In the downlink, the estimation error variances $v^{(k)}_{m,n,k}$ and $v^{(k)}_{m,n,i}, i \neq k$ for user $k$ are estimated following a similar procedure, but with only one receive antenna and using the downlink covariances matrices $\Omegam$ and $\Psim$. 
The resulting quantity is denoted by $\widehat{\Vm}^{(k)}$, as shown in Fig.~\ref{fig:chp3_downlink}.
\clearpage
\section{DL-enhanced Receiver Architecture}
\label{sec:chp3_ml-receiver}

The baselines presented in the previous section have several limitations. 
Especially, the \gls{NIRE} approximation leads to high channel estimation errors for \glspl{RE} that are far from pilots.
Similarly, the grouped equalization can be inaccurate at those \glspl{RE}.
This section details the architecture of a receiver that builds on the presented baseline and uses  multiple \glspl{CNN} to improve its performance.

\subsection{Receiver Training}

\begin{figure}
    \centering
    \includegraphics[width=1\textwidth]{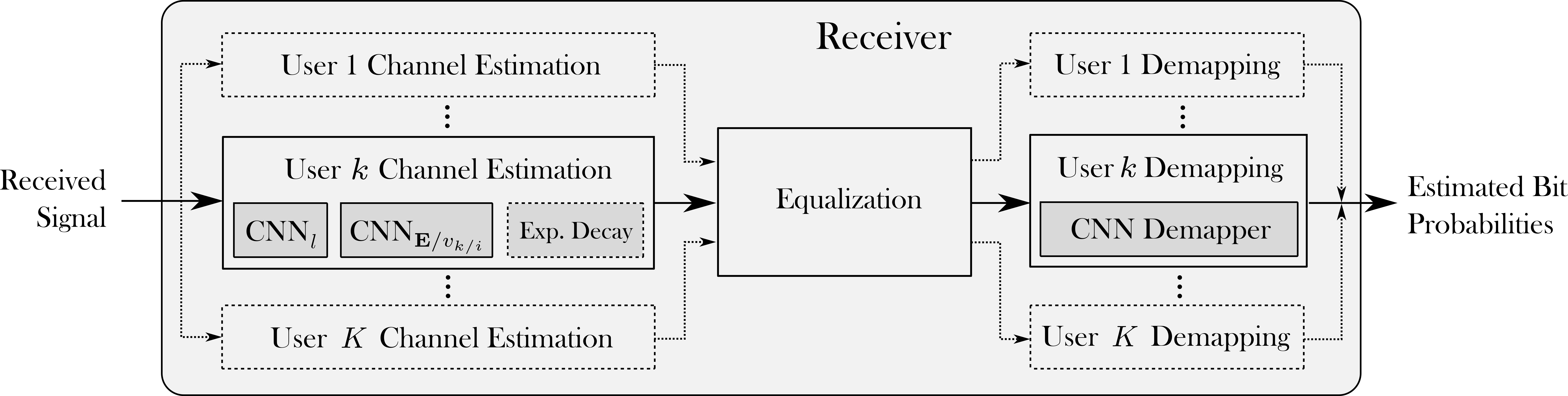}
    \caption{DL-enhanced receiver architecture. The dotted elements are only present in the uplink, where the \gls{BS} jointly processes all users. The dark gray elements are trainable components.}
    \label{fig:chp3_ml_receiver}
\end{figure}

The \gls{DL}-enhanced receiver architecture is shown in Fig.~\ref{fig:chp3_ml_receiver}, where the trainable components are represented in dark gray.
In the downlink, each user $k$ only performs the channel estimation, equalization, and demapping of its own signal, and the corresponding components are illustrated with continuous outlines.
However, in the uplink, the \gls{BS} processes all users in parallel, and the additional components are delimited with dotted lines.
Although the internal processing of the trainable components might not be interpretable, using multiple \gls{DL}-based blocks to perform precise and relatively simple signal processing tasks allows to precisely control which parts of the receiver are enhanced.
Moreover, this approach makes the output of each \gls{DL} components easier to interpret, as discussed in Section~\ref{sec:chp3_E_comp_speed} for the error statistics $\widehat{\Em}$.

Scalability is achieved by using different copies of the same \gls{DL} components for every user, where all copies share the same set of trainable parameters.
We propose to jointly optimize all these components based only on the estimated bit probabilities, and not by training each of them individually. 
This approach is practical as it does not assume knowledge of the channel coefficients at training, that can only be estimated through extensive measurement campaigns for practical channels.
Let's denote by $\thetav$ the set of trainable parameters of the \gls{DL}-enhanced receiver, and by  $b_{m,n,k,q}$ the transmitted bit $(m,n,k,q)$.
In the uplink, those parameters are optimized to minimize the total \gls{BCE}:
%
%
\begin{align}
  \label{eq:chp3_bce}
  \mathcal{L} \triangleq & - \sum_{k=1}^{K} \sum_{(m,n)\in \mathcal{D}} \sum_{q=0}^{Q-1} \EE_{b, \Ym} \left[  \text{log}_2 \LB 
  \widehat{P}_{\thetav} \LB b_{m,n,k,q} | \Ym \RB \RB   \right]
  \end{align}
where $\mathcal{D}$ denotes the set of \glspl{RE} carrying data and $\widehat{P}_{\thetav} \LB \cdot | \Ym \RB $ is the receiver estimate of the posterior
distribution on the bits given $\Ym$. 
In the downlink, the receiver parameters are optimized in a similar manner, except that the signal received by the users is $\Rm$.
The expectation in \eqref{eq:chp3_bce} is estimated through Monte Carlo sampling using batches of $B_S$ samples:
%
\begin{align}
\label{eq:chp3_loss_mc}
\mathcal{L} \approx & - \frac{1}{B_S} \sum_{i=1}^{B_S} \sum_{k=1}^{K} \sum_{(m,n)\in \mathcal{D}} \sum_{q=0}^{Q-1} \LB  \text{log}_2 \LB 
\widehat{P}_{\thetav} \LB b_{m,n,k,q}^{[i]}| \Ym^{[i]} \RB \RB \RB
\end{align}
where the superscript $[i]$ refers to the $s^{\text{th}}$ sample in the batch.
Following a derivation similar to~\eqref{eq:bkg_link_ce_rate}, the loss~\eqref{eq:chp3_bce} can be redefined as
\begin{equation}
\mathcal{L} = \sum_{k=1}^{K} \LB \text{Card}(\mathcal{D}) Q - C_k \RB
\end{equation}
where $\text{Card}(\mathcal{D})$ is the number of \glspl{RE} carrying data, $\text{Card}(\mathcal{D}) Q $ is the total number of bits transmitted by one user, and $C_k$ is an achievable rate for user $k$:
\begin{align}
\label{eq:chp3_Ck}
C_k =  &\sum_{(m,n)\in \mathcal{D}} \sum_{q=0}^{Q-1} I\LB b_{m,n,k,q}; \Ym \RB  \\
&-  \sum_{(m,n)\in \mathcal{D}} \sum_{q=0}^{Q-1}  \EE_\Ym \LSB D_{KL} \LB P \LB  b_{m,n,k,q} | \Ym \RB ||  \widehat{P}_{\thetav} \LB  b_{m,n,k,q} | \Ym \RB \RB \RSB \nonumber.
\end{align}
Minimizing $\Lc$ therefore maximizes $C_k$, which directly translates to improved \gls{BER} performances.

\begin{figure}
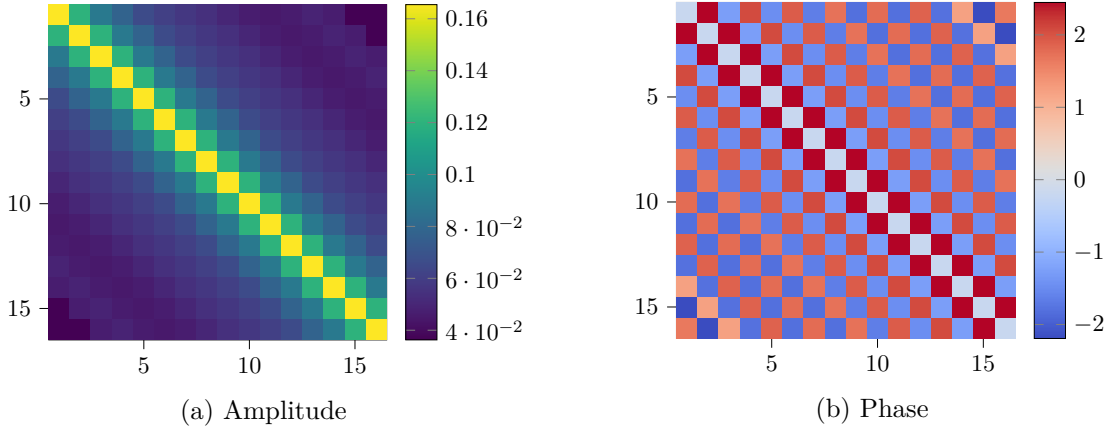


\centering
	
  \centering
  	\begin{subfigure}{0.45\textwidth}
      \center
    	\begin{adjustbox}{height=5cm} 
   		 \input{Chapter3/eval/E_single_abs.tex}
  	\end{adjustbox}    
  	\caption{Amplitude}
	\end{subfigure}%
\hspace{25pt}
  	\begin{subfigure}{0.45\textwidth}
      \center
  	\begin{adjustbox}{height=5cm} 
  		\input{Chapter3/eval/E_single_phase.tex}
  	\end{adjustbox} 
  	\caption{Phase}
	\end{subfigure}%

\caption{Example of amplitude and phase for $\Em_{m,n}^{(k)}$.}
\label{fig:chp3_E}
\end{figure}

\subsection{DL-enhanced Channel Estimator}

As seen in Section~\ref{sec:chp3_stats}, the channel estimation error statistics can only be obtained for \glspl{RE} carrying pilots.
However, the estimation accuracy decreases as we move away from them.
In the following, we present \glspl{CNN} that estimate the channel estimation error covariance matrices in the uplink and the estimation error variances in the downlink.

\subsubsection{Uplink scenario}
\label{sec:chp3_ml_ch_est_ul}

\begin{figure}
    \centering
    \includegraphics[width=0.9\textwidth]{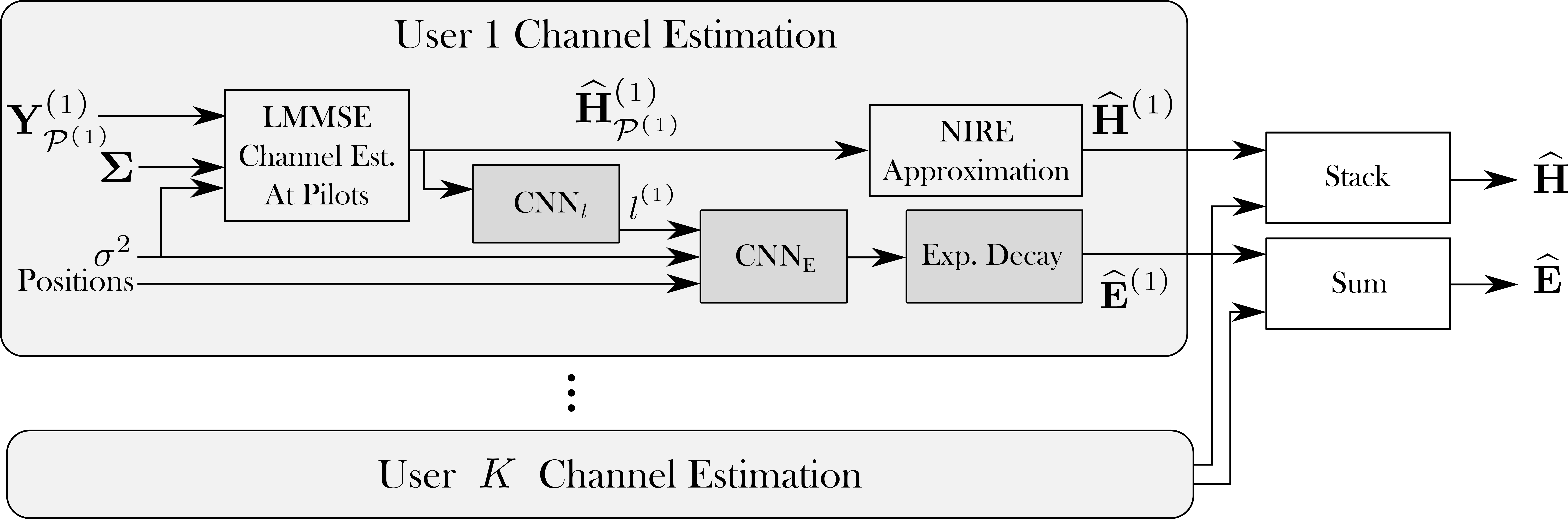}
    \caption{DL-enhanced uplink channel estimation.}
    \label{fig:chp3_ch_est_ml}
\end{figure}

\begin{figure*}
    \centering
    \includegraphics[width=1\textwidth]{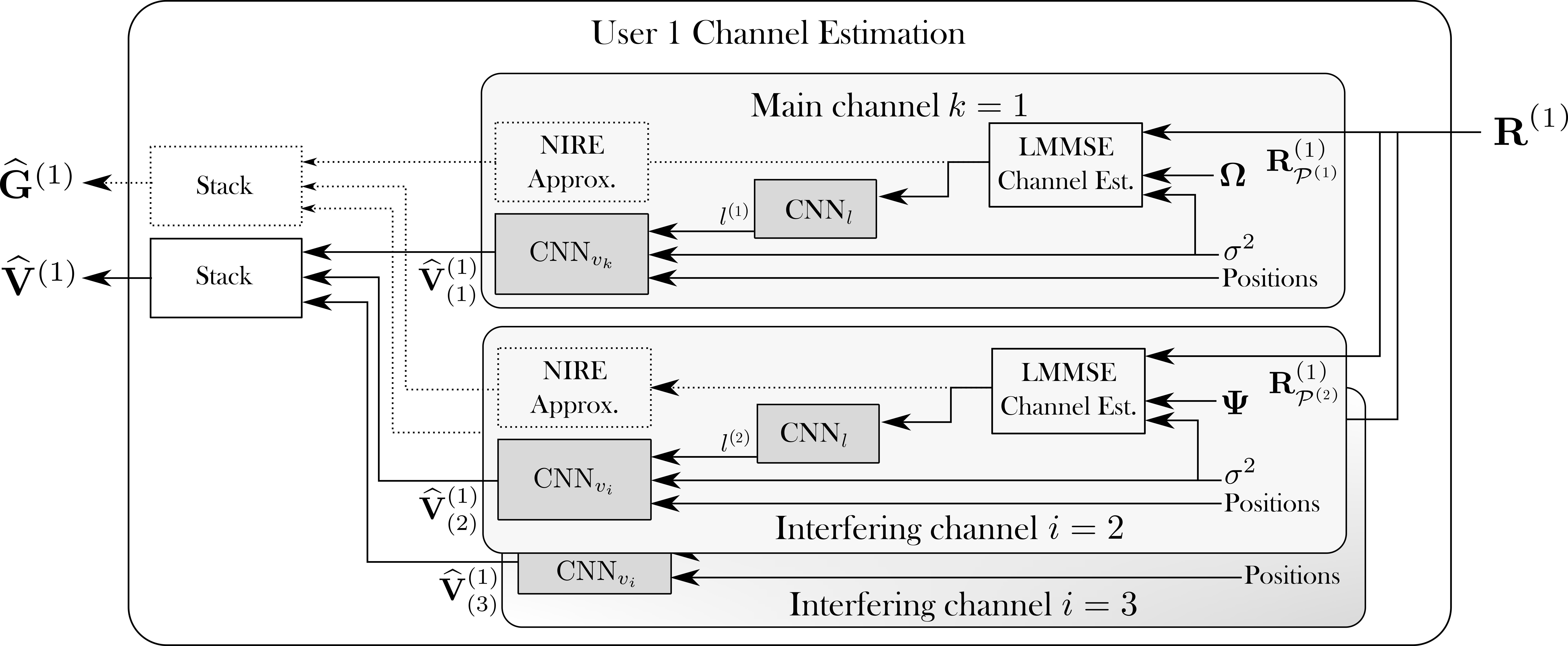}
    \caption{Detailed view of the downlink \emph{Channel Estimation} component of user $1$ out of $K = 3$, as depicted in Fig~\ref{fig:chp3_downlink}.}
    \label{fig:chp3_ch_est_dl}
\end{figure*}

The \gls{DL}-enhanced uplink channel estimation architecture is depicted in Fig.~\ref{fig:chp3_ch_est_ml}.
In this scenario, the spatial channel estimation error covariance matrices $\Em_{m,n}$ is needed to compute both the equalized symbols \eqref{eq:chp3_W} and the uplink post-equalization noise variance \eqref{eq:chp3_eq_snr_ul}.
An example of the amplitude and phase of an estimate of a covariance matrix $\widehat{\Em}_{1,1}^{(1)}$ is shown in Fig.~\ref{fig:chp3_E}  for a \gls{ULA} of antennas at the \gls{BS}.
One can see that the amplitude of the coefficients of $\widehat{\Em}^{(1)}_{1,1}$ decays rapidly when moving away from the diagonal.
The phase, on the other hand, exhibits a more surprising pattern, with a phase difference of roughly $\pi$ between two adjacent antennas.
To predict every element of $\widehat{\Em}^{(k)}$ for user $k$, a naively designed \gls{CNN} would need to output $M  N  L ^2$ complex parameters.
This would be of prohibitive complexity for any large number of subcarriers, symbols, or receiving antennas.
For this reason, we propose to approximate every element $(x,y)$ of $\widehat{\Em}_{m,n}^{(k)}$ with a complex power decay model:
\begin{equation}
\label{eq:chp3_exp_decay}
\hat{e}^{(k)}_{m,n,a,b}= \alpha_{m,n} \beta_{m,n}^{|b-a|} \exp{j \gamma (b-a)}
\end{equation}
where $b-a$ is the horizontal position difference between that element and the diagonal, and $\alpha_{m,n}$, $\beta_{m,n}$, and $\gamma$ are parameters of this model.
For a planar array, one could use such a model for each dimension, and take their Kronecker product to obtain the spatial channel estimation error covariance matrices.
A constant phase offset between two adjacent \glspl{RE} is assumed, which matches our experimental observations.
The parameters $\alpha_{m,n}$ and $\beta_{m,n}$ respectively control the scale and the decay of the model, and depend on the \gls{RE} $(m,n)$.

To estimate those two parameters for each \gls{RE}, we use a $\text{CNN}$, denoted by $\text{CNN}_{\mathbf{E}}$, which takes four inputs, each of size $M \times N$, for a total input dimension of $M \times N \times 4$.
$\text{CNN}_{\mathbf{E}}$ outputs $\alpha_{m,n}$ and $\beta_{m,n}$ for every \glspl{RE}, resulting in an output dimension of $M \times N \times 2$.
The first two inputs provide the location of every \glspl{RE} in the \gls{RG}.
More precisely, the first input matrix has all columns equal to $[-\frac{M}{2}, \dots, -1, 1, \dots, \frac{M}{2}]\tp$, whereas the second one has all rows equal to $[-\frac{N}{2}, \dots, -1, 1, \dots, \frac{N}{2}]$.
The third input provides the \gls{SNR} of the transmission and is given as a matrix $\text{SNR} \cdot \mathds{1}_{M \times N}$.
Finally, the fourth input is a feature $f^{(k)} \in \RR$ provided by another \gls{CNN}, denoted by $\text{CNN}_{f}$, which was designed with the intuition to predict the time-variability of the channel experienced by user $k$.
To do so, $\text{CNN}_{f}$ uses the channel estimates at \glspl{RE} carrying pilots to estimate the Doppler and delay spread.
Although we cannot be certain that $\text{CNN}_f$ effectively learns to extract such information, the evaluations presented in Section~\ref{sec:chp3_E_comp_speed} tend to support this hypothesis.
$\text{CNN}_{f}$ takes an input of dimension $|\Pc_M| \times |\Pc_N| \times 2L$, which corresponds to the stacking of the real and imaginary parts of $\widehat{\Hm}^{(k)}_{\mathcal{P}^{(k)}}$ along the last dimension.
It outputs the scalar $f^{(k)}$, which is fed to $\text{CNN}_{\mathbf{E}}$ as the matrix $f^{(k)}\cdot \mathds{1}_{M \times N}$.

\subsubsection{Downlink scenario}
\label{sec:chp3_ml_ch_est_dl}
In the downlink, the equivalent channel estimation error variances $\Vm^{(k)}$ are needed to compute $\tau^2_{m,n,k}$ in \eqref{eq:chp3_eq_snr_dl}.
To estimate those variances, we take inspiration from the architecture presented in Section~\ref{sec:chp3_system_model} which uses two different downlink covariance matrices $\Omegam$ and $\Psim$ to estimate the error variances of the main and interfering channels.
Similarly, we use two separate \glspl{CNN}, denoted by $\text{CNN}_{v_k}$ and by $\text{CNN}_{v_i}$, to respectively predict the estimation error variances of the main channel $\hat{v}^{(k)}_{m,n,k}$ and of an interfering channel $\hat{v}^{(k)}_{m,n,i}, i \neq k$.
Both \glspl{CNN} take the same inputs as $\text{CNN}_{\mathbf{E}}$ but their outputs are of dimension $M \times N$ as variances are predicted for all \glspl{RE} $(m,n)$. 
To preserve the scalability of the conventional architecture, $K-1$ copies of $\text{CNN}_{v_i}$ are used to estimate the error variances $\hat{v}^{(k)}_{m,n,i}$ of all $K - 1$ interferers.
The downlink channel estimation is schematically shown in Fig.~\ref{fig:chp3_ch_est_dl}, where $\widehat{\Vm}^{(k)}_{(a)} = \{ \hat{v}^{(k)}_{m,n,a} \}_{ (m,n) \in \{0, \dots, M \} \times \{0, \dots, N \}  }$ denotes the estimation error variances seen by user $k$ on its main or interfering channel $a$.

\begin{figure*}[t!]
  \centering
  \begin{subfigure}{1\textwidth}
    \includegraphics[width=1\textwidth]{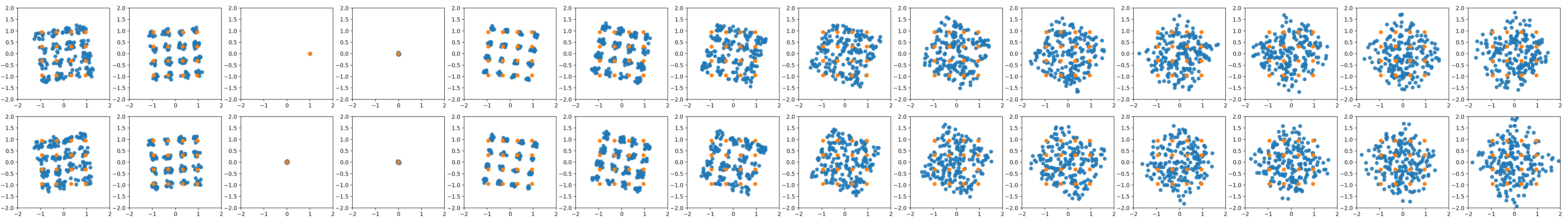}
    \caption{1P pilot pattern.}
    \label{fig:chp3_mismatch_1P}
    \vspace{10pt}
  \end{subfigure}
   \begin{subfigure}{1\textwidth}
    \includegraphics[width=1\textwidth]{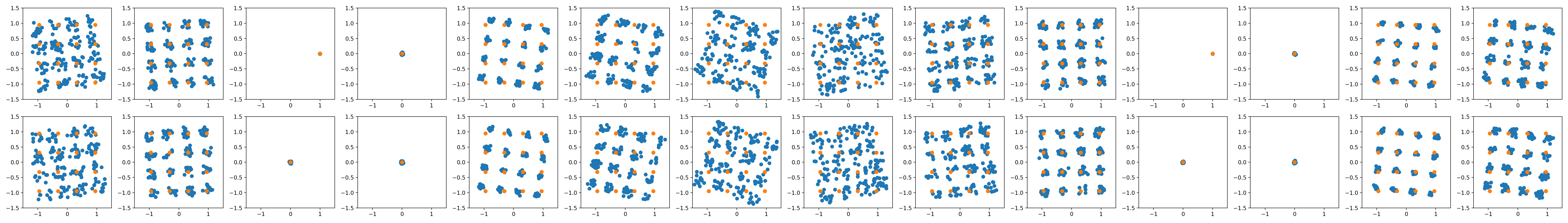}
    \caption{2P pilot pattern.}
    \label{fig:chp3_mismatch_2P}
  \end{subfigure}
  \caption{Mismatch between the transmitted signals (orange) and the equalized received ones (blue) for a single user out of four and using 16-QAM modulation.}
  \label{fig:chp3_mismatch}
\end{figure*}

\begin{figure*}[b!]
\centering
\begin{minipage}[b]{0.6\textwidth}
\centering
\includegraphics[height=110pt]{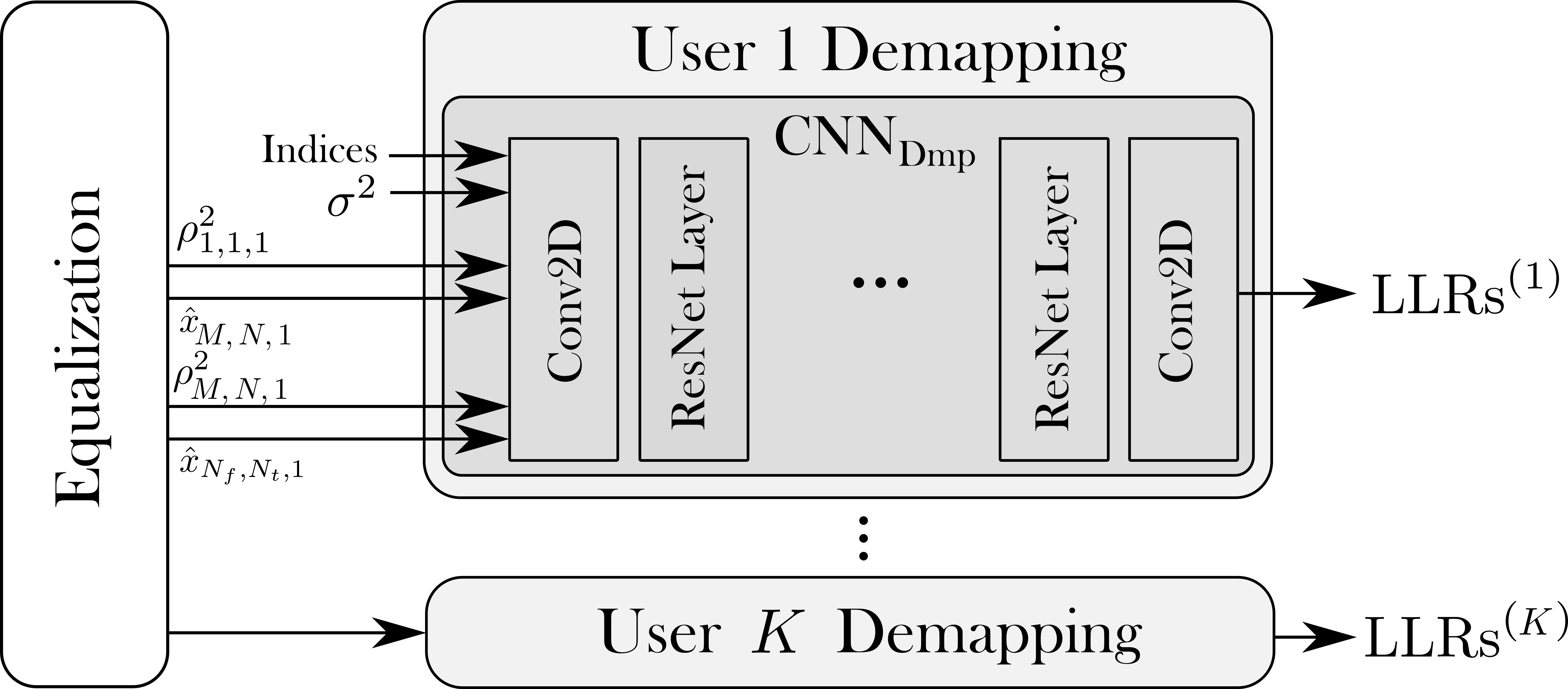}
  \captionof{figure}{\gls{CNN}-based uplink demapper.}
  \label{fig:chp3_demapper_ml}
\end{minipage}
\hfill
\begin{minipage}[b]{0.35\textwidth}
\centering
\includegraphics[height=110pt]{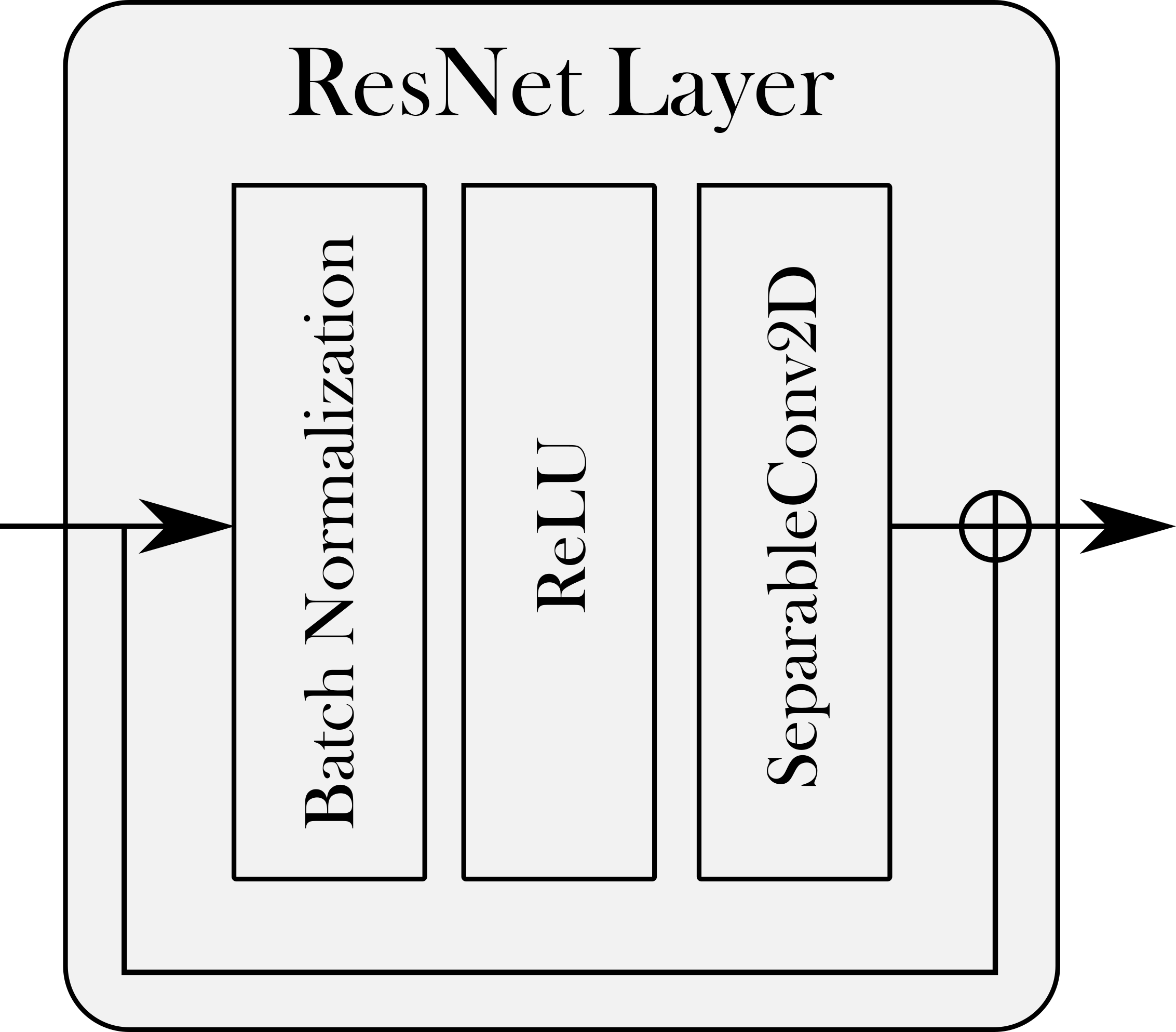}
\captionof{figure}{ResNet layer.}
  \label{fig:chp3_resnet}
\end{minipage}
\end{figure*}

\subsection{DL-enhanced Demapper}
\label{sec:chp3_ml_demapper}

A consequence of unperfect channel estimation and equalization is channel aging, which leads to residual distortion on the equalized signals, as illustrated in Fig.~\ref{fig:chp3_mismatch}. 
The signals transmitted by a single user out of four are represented in orange, while the corresponding equalized received signals are shown in blue, assuming spectral channel interpolation only. 
This figure has been obtained by sending a large batch of signals using a fixed realization of a fast-varying channel, and only displays the first two subcarriers of the uplink slot.
Moreover, an infinite \gls{SNR} is assumed so that only the effects of channel aging and interference are visible.
One can see that the equalized symbols suffer from little distortion and interference at \glspl{RE} close to the pilots, but these unwanted effects become increasingly stronger at \glspl{RE} that are away from them.

A traditional demapper, as presented in Section~\ref{sec:chp3_demapping_ul}, operates independently on each \gls{RE} and therefore only sees one equalized symbol at a time.
In contrast, we propose to use a \gls{CNN}, called $\text{CNN}_{\text{Dmp}}$, to perform a joint demapping of the entire \gls{RG}. 
By jointly processing all equalized symbols, the \gls{CNN} can estimate and correct the effects of channel aging to compute better \glspl{LLR}.
The input of $\text{CNN}_{\text{Dmp}}$ is of dimension $M \times N \times 6$ and carries the subcarriers and symbol indices for each RE, the SNR, the real and imaginary parts of the equalized symbols, and the post-equalization channel noise variances.
The output of $\text{CNN}_{\text{Dmp}}$ has dimensions $M \times N \times Q$ and corresponds to the predicted $\text{LLRs}^{(k)}$ over the \gls{RG} for a user $k$. 
As with a conventional receiver, the demapping is performed independently for each user to make the architecture easily scalable.
The $\text{CNN}_{\text{Dmp}}$ demapper is shown in Fig.~\ref{fig:chp3_demapper_ml}, which depicts the uplink demapping process.

\begin{table*}
  \centering
  \renewcommand{\arraystretch}{1.3}
  \begin{tabular}{|p{1.8cm}||c|c|c|c||c|c|c||c|c|c|}
    \hline
     & \multicolumn{4}{c||}{$\text{CNN}_{\mathbf{E}} / \text{CNN}_{v_{k,i}}$} & \multicolumn{3}{c||}{$\text{CNN}_{l}$} & \multicolumn{3}{c|}{$\text{CNN}_{\text{Dmp}}$} \\ \hhline{|=||====||===||===|}
    Input size & \multicolumn{4}{c||}{$M \times K \times 4 $} & \multicolumn{3}{c||}{$N_{P_f} \times N_{P_t} \times 2L$} & \multicolumn{3}{c|}{$M \times K \times 6$} \\  \hline

    Parameters  & filt. & kern. & dilat. & act. & filt. & kern. & dilat. & filt. & kern. & dilat.  \\

    \hline
    Conv2D  & 32 & (5,3) & (1,1) & ReLU      & 32 & (1,1) & (1,1)  & 128 & (1,1) & (1,1)  \\ \hline
    ResNet Layer  & \multicolumn{4}{c||}{-}  & 32 & (3,2) & (1,1)  & 128 & (3,3) & (1,1)  \\ \hline
    ResNet Layer  & \multicolumn{4}{c||}{-}  & 32 & (5,2) & (2,1)  & 128 & (5,3) & (2,1)  \\ \hline
    ResNet Layer  & \multicolumn{4}{c||}{-}  & 32 & (7,2) & (3,1)  & 128 & (7,3) & (3,2)  \\ \hline
    ResNet Layer  & \multicolumn{4}{c||}{-}  & 32 & (5,2) & (2,1)  & 128 & (9,3) & (4,3)  \\ \hline
    ResNet Layer  & \multicolumn{4}{c||}{-}  & 32 & (3,2) & (1,1)  & 128 & (7,3) & (3,2)  \\ \hline
    ResNet Layer  & \multicolumn{4}{c||}{-}  & \multicolumn{3}{c||}{-} & 128 & (5,3) & (2,1)    \\ \hline
    ResNet Layer  & \multicolumn{4}{c||}{-}  & \multicolumn{3}{c||}{-} & 128 & (3,3) & (1,1)   \\ \hline
    Conv2D   & 32 & (5,3) & (1,1)& ReLU  & 1 &(3,2) & (1,1)  & $Q$  & (1,1) & (1,1)  \\ \hline
    Conv2D   & 2 / 1 & (1,1) & (1,1)& Sigm.  & \multicolumn{3}{c||}{-}  & \multicolumn{3}{c|}{-}\\ \hline
    
    Output Layer & \multicolumn{4}{c||}{-} & \multicolumn{3}{c||}{Dense, units = 1}  & \multicolumn{3}{c|}{-} \\  \hline
    
    Output size  & \multicolumn{4}{c||}{$M \times K \times 2 / M \times K \times 1 $} & \multicolumn{3}{c||}{$1$} & \multicolumn{3}{c|}{$M \times K \times Q$} \\  \hline
    
  \end{tabular}
  
  \caption{Parameters of the different CNNs.}
  \label{table:chp3_CNNs}
\end{table*}

\begin{figure*}[h]
\centering
\begin{minipage}[b]{0.4\textwidth}
\centering
\includegraphics[height=80pt]{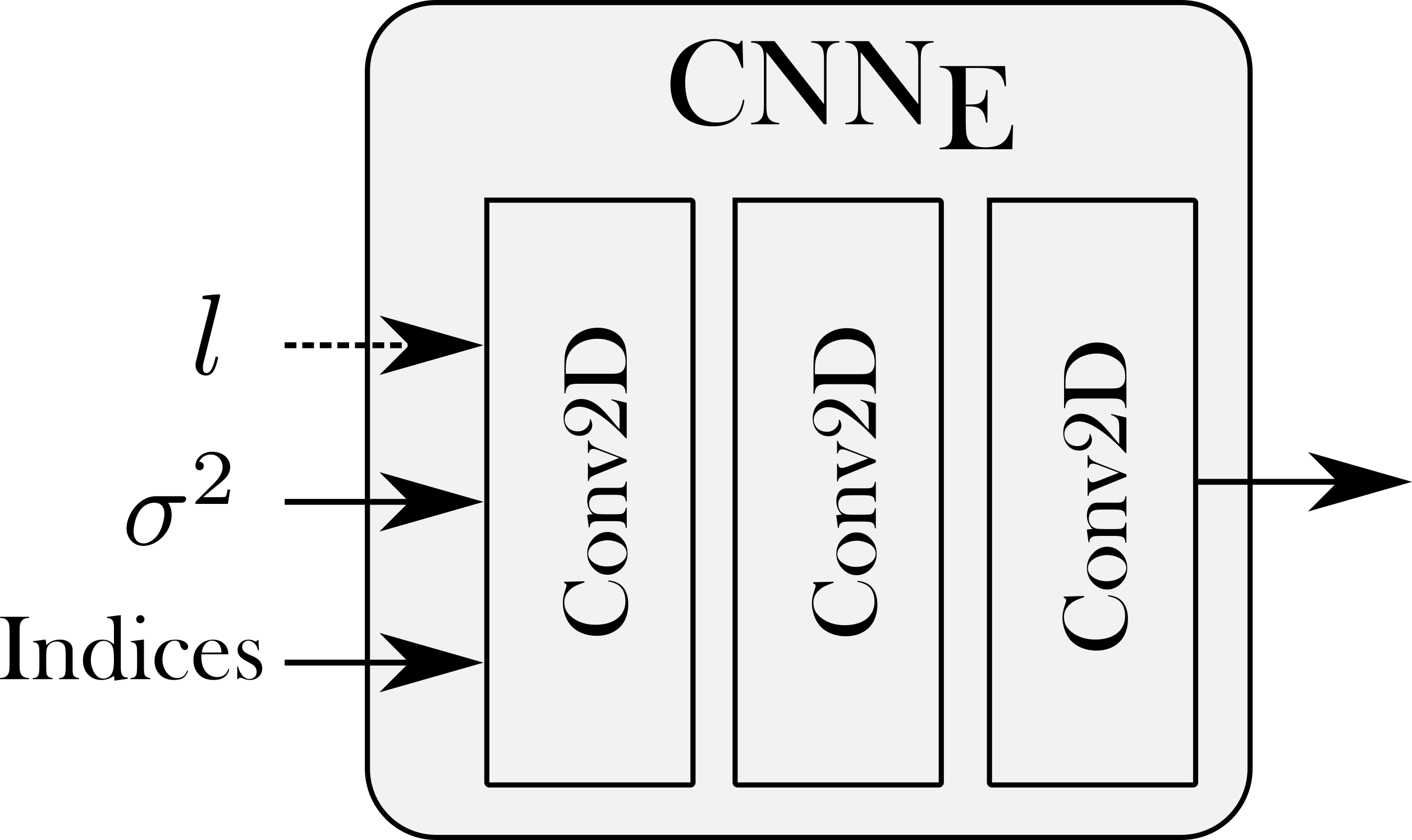}
  \captionof{figure}{Architecture of $\text{CNN}_{\mathbf{E}}$.}
  \label{fig:chp3_CNN_E}
\end{minipage}
\hfill
\begin{minipage}[b]{0.55\textwidth}
\centering
\includegraphics[height=80pt]{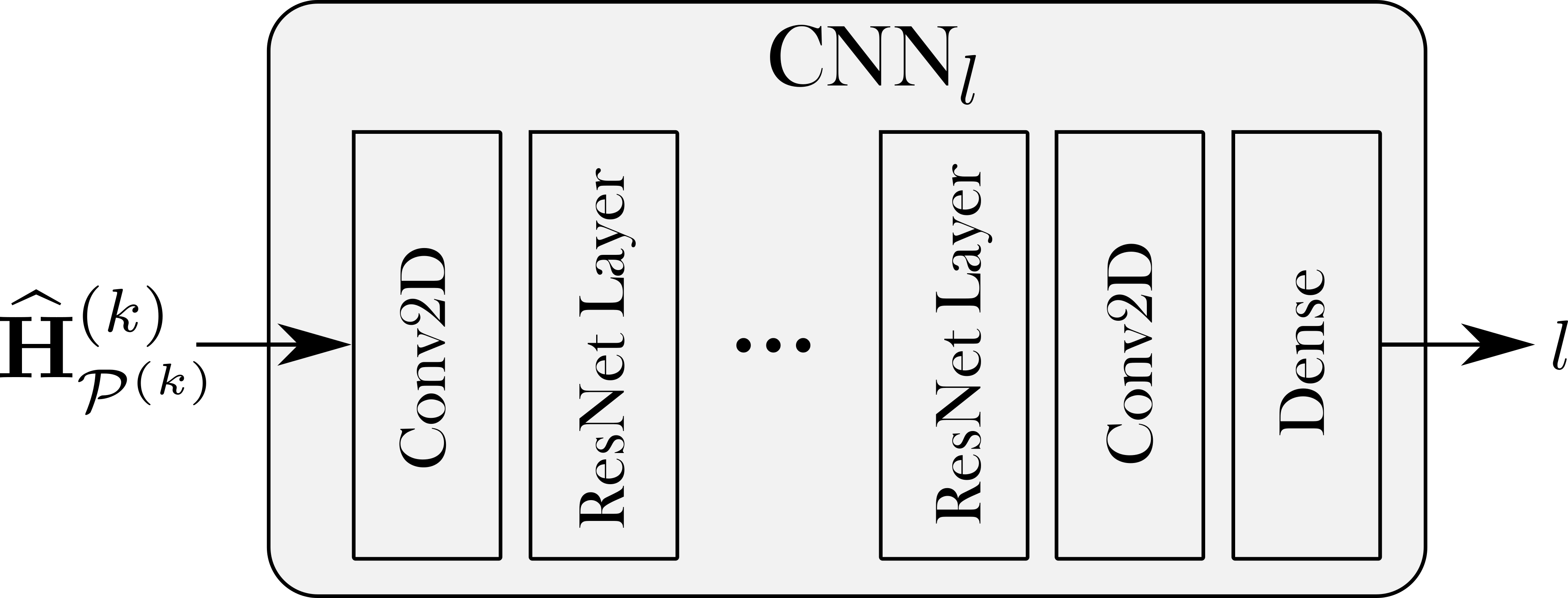}
  \captionof{figure}{Architecture of $\text{CNN}_{l}$.}
  \label{fig:chp3_CNN_l}
\end{minipage}
\end{figure*}

\subsection{CNN Architectures}

All \glspl{CNN} presented above share the same building blocks: convolutional 2D layers, dense layers, and custom ResNet layers. 
The custom ResNet layers, inspired by \cite{he2016identity}, consist of a batch normalization layer, a \gls{ReLU}, a 2D separable convolutional layer, and finally the addition of the input, as depicted in Fig.~\ref{fig:chp3_resnet}.
Separable convolutions are less computationally expensive while maintaining similar performance as regular convolutional layers \cite{howard2017mobilenets}.
$\text{CNN}_{\mathbf{E}}$, $\text{CNN}_{v_k}$, and $\text{CNN}_{v_i}$ are all made of three 2D convolutional layers, as shown in Fig.~\ref{fig:chp3_CNN_E} for $\text{CNN}_{\mathbf{E}}$. 
The first two layers are followed by a \gls{ReLU} activation function, while the last layer is followed by a sigmoid activation function.
$\text{CNN}_{\text{Dmp}}$ and $\text{CNN}_{f}$ also share a similar architecture, depicted in Fig.~\ref{fig:chp3_demapper_ml}  and Fig~\ref{fig:chp3_CNN_l}, respectively.
Both are composed of a 2D convolutional layer, followed by multiple ResNet layers and a 2D convolutional layer. 
$\text{CNN}_{f}$ outputs a single scalar, which is ensured by using a dense layer with a single unit and no activation function as output layer.
Inspired by~\cite{honkala2020deeprx}, we used increasing followed by decreasing kernel sizes and dilation rates to increase the receptive field of the \gls{CNN}. 
All convolutional and separable convolutional layers use zero-padding so that the output dimensions matches the input dimensions.
Details of the architectures of all \glspl{CNN} are given in  Table~\ref{table:chp3_CNNs}.
\clearpage
\section{Evaluations} 
\label{sec:chp3_evaluations}

In this section, the proposed \gls{DL}-enhanced receiver is evaluated and compared against two baselines: the one presented in Section~\ref{sec:chp3_system_model} as well as a perfect \gls{CSI} baseline that will be detailed later on.
The training and evaluation setups are first introduced, followed by evaluations of the uplink and downlink performance for the 1P and 2P pilot patterns from Fig.~\ref{fig:chp3_channel_model}.

\begin{table*}[t!]
  \centering
  \renewcommand{\arraystretch}{1.5}
  \begin{tabular}{|p{5.5cm}|c|c|c|c|}
    \hline
    Parameters  & Symbol (if any)& \multicolumn{3}{c|}{Value}  \\ \hhline{|=|=|===|}
    Number of users & $K$ & \multicolumn{3}{c|}{4 (2 in high speed downlink)}   \\   \hline
    Number of antennas at the \gls{BS}& $L$ & \multicolumn{3}{c|}{16}  \\   \hline
    Number of subcarrier  & $N$ & \multicolumn{3}{c|}{72  \, = 6 resource blocks} \\   \hline
    Number of \gls{OFDM} symbols & $M$ & \multicolumn{3}{c|}{14 (uplink) + 14 (downlink)} \\   \hline
    Bit per channel use & $Q$ & \multicolumn{3}{c|}{\SI{4}{\bit} (uplink), \SI{2}{\bit}(downlink)}  \\   \hline
    Center frequency & - & \multicolumn{3}{c|}{\SI{3.5}{\GHz}}  \\   \hline
    Subcarrier spacing & - & \multicolumn{3}{c|}{\SI{15}{\kHz}} \\   \hline
    Scenario & - & \multicolumn{3}{c|}{3GPP 38.901 UMi NLOS}  \\   \hline
    Code length & - & \multicolumn{3}{c|}{\SI{1296}{\bit}} \\   \hline
    Code rate & $\eta$ & \multicolumn{3}{c|}{$\frac{1}{2}$ (uplink), $\frac{1}{3}$ (downlink)} \\   \hline

    Learning rate & - & \multicolumn{3}{c|}{$10^{-3}$}  \\   \hline
    Batch size & $B_S$ & \multicolumn{3}{c|}{27 \glspl{RG}}   \\   \hline

  \end{tabular}
  
  \caption{Training and evaluation parameters.}
  \label{table:chp3_parameters}
\end{table*}

\subsection{Training and Evaluation Setup}

For realistic training and evaluation, the channel realizations were generated with QuaDRiGa version 2.0.0 \cite{quadriga}.
 It has been experimentally observed in \cite{honkala2020deeprx} that a receiver trained on certain scenarios was able to generalize to other channel models.
For this reason, we only focus on the \gls{3GPP} \gls{NLOS} \gls{UMi} scenario \cite{3gpp.38.901}.
The number of users was set to $K = 4$, except for the downlink at high speeds where it was reduced to $K=2$, and the number of antennas at the \gls{BS} was set to $L = 16$. 
All users were randomly placed within a \SI{120}{\degree} cell sector, with a minimum distance of \SI{15}{\meter} and a maximum distance of \SI{150}{\meter} from the \gls{BS}.
The user and \gls{BS} heights were respectively set to \SI{1.5}{\meter} and \SI{10}{\meter}.
The \glspl{RG} were composed of six resource blocks for a total of $N=72$ subarriers, with a center frequency of \SI{3.5}{\GHz} and a subcarrier spacing of \SI{15}{kHz}.
Both the uplink and downlink slots contained $M=14$ \gls{OFDM} symbols.
A Gray-labeled \gls{QAM} was used with $Q= 4$ bits per channel use on the uplink and $Q=2$ bits per channel use on the downlink.
The receivers were trained and evaluated on users moving at independent random speeds.
 Three ranges of low speeds were considered with the 1P pilot pattern: \SIrange[range-units=single]{0}{15}{\km\per\hour}, \SIrange[range-units=single]{15}{30}{\km\per\hour}, and \SIrange[range-units=single]{30}{45}{\km\per\hour}.
Similarly, three high speed ranges were considered with the 2P pilot pattern: \SIrange[range-units=single]{50}{70}{\km\per\hour}, \SIrange[range-units=single]{80}{100}{\km\per\hour}, and  \SIrange[range-units=single]{110}{130}{\km\per\hour}.
We noticed that $\text{CNN}_f$ was not able to extract useful information when using the 1P pattern, and therefore was not used in the corresponding trainings and evaluations.
For equalization, the grouped-\gls{LMMSE} equalizer operated on groups of $2\times 7$ \glspl{RE}, following the segmentation delimited by the thick black line of the 2P pattern in Fig.~\ref{fig:chp3_2P_pattern}.
The training and evaluation parameters are given in Table \ref{table:chp3_parameters}.

Separate training sets were constructed for the 1P and 2P pilot pattern, both made of channel realizations corresponding to 1000 \glspl{RG} of each respective user speed range, for a total of 3000 \glspl{RG} per training dataset.
Evaluations were performed separately for each user speed range with datasets containing 3000 \glspl{RG}.
A standard IEEE 802.11n \gls{LDPC} code of length \SI{1296}{\bit} \cite{wlan} was used, with code rates of $\eta = \frac{1}{2}$ and $\eta = \frac{1}{3}$ for uplink and downlink transmissions, respectively.
Decoding was done with 40 iterations using a conventional belief-propagation decoder.
To satisfy the perfect power assumption in \eqref{eq:chp3_snr}, each \gls{RG} was normalized to have an average energy of one per antenna and per user, i.e., 
\begin{equation}
\sum_{m=1}^{2M} \sum_{n=1}^{N} ||\hv_{m,m}^{(k)}||_2^2 = 2MN L.
\end{equation} 

Each training was carried out using batches of size $B_S=27$ so that the total number of bits transmitted for each user was a multiple of the code length.
The trainable parameters $\thetav$ were all initialized randomly except for $\gamma$ in \eqref{eq:chp3_exp_decay} that was initialized with $\pi$.
Training was done through \gls{SGD} using the Adam \cite{Kingma15} optimizer and a learning rate of $10^{-3}$.
The best performing random initialization out of ten was selected.

\subsection{Uplink Simulation Results}

\begin{figure}
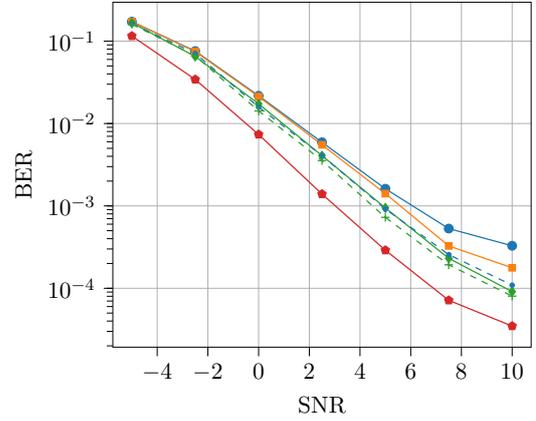
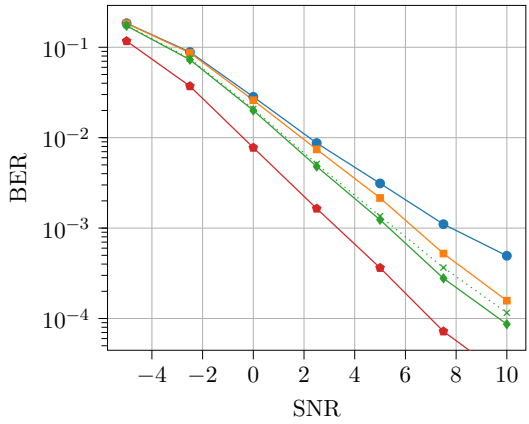
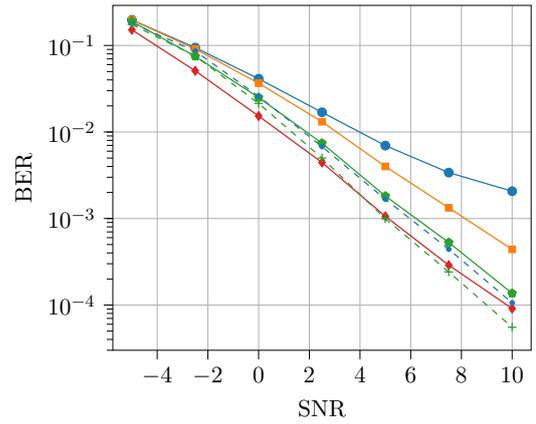
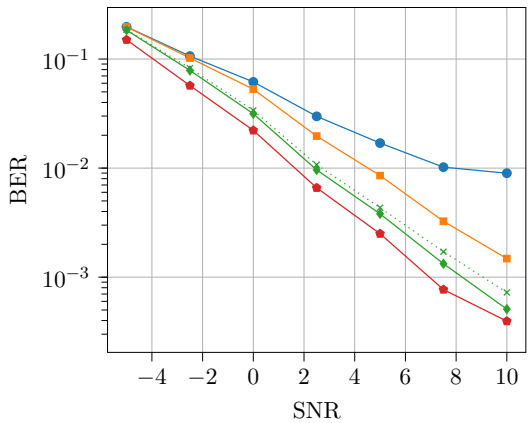
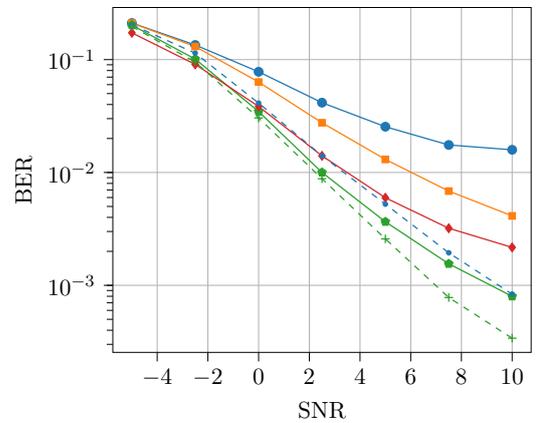

	\begin{subfigure}[l]{1\textwidth}
	\begin{tikzpicture} 
	
	\definecolor{color0}{rgb}{0.12156862745098,0.466666666666667,0.705882352941177}
	\definecolor{color1}{rgb}{1,0.498039215686275,0.0549019607843137}
	\definecolor{color2}{rgb}{0.172549019607843,0.627450980392157,0.172549019607843}
	\definecolor{color3}{rgb}{0.83921568627451,0.152941176470588,0.156862745098039}

	\begin{axis}[%
		hide axis,
		xmin=10,
		xmax=50,
		ymin=0,
		ymax=0.4,
		legend columns=5, 
		legend style={draw=white!15!black,legend cell align=left,column sep=2.022ex}
		]
		\addlegendimage{white,mark=square*}
		\addlegendentry{\hspace{-1.15cm} Spectral interp.:};
		\addlegendimage{color0,mark=square*}
		\addlegendentry{\hspace{-0.25cm} Baseline};
		\addlegendimage{color1, mark=*}
		\addlegendentry{\hspace{-0.25cm} DL ch. est. };
		\addlegendimage{color2, mark=pentagon*}
		\addlegendentry{\hspace{-0.25cm} DL receiver};
		\addlegendimage{color3, mark=diamond*}
		\addlegendentry{\hspace{-0.25cm} Perfect CSI};
		
		\end{axis}
	\end{tikzpicture}
	\end{subfigure}%
	\vspace{1pt}
	
	  \begin{subfigure}[c]{1\textwidth}
	  
	\begin{tikzpicture} 
	
	\definecolor{color2}{rgb}{0.172549019607843,0.627450980392157,0.172549019607843}

		\begin{axis}[%
		hide axis,
		xmin=10,
	   xmax=50,
		ymin=0,
		ymax=0.4,
		legend columns=2, 
		legend style={draw=white!15!black,legend cell align=right,column sep=1.9ex}
		]
		\addlegendimage{white,mark=square*, mark size=2}
		\addlegendentry{\hspace{-1.15cm} Spectral interp. (1P pattern only):};
		\addlegendimage{color2, dotted, mark=x, mark size=2, mark options={solid}}
		\addlegendentry{DL receiver trained with $K=2$ \hspace{2.29cm} \quad};
		\end{axis}
	\end{tikzpicture}
	
	\end{subfigure}
	
	\vspace{1pt}

	  \begin{subfigure}[c]{1\textwidth}
	  
	\begin{tikzpicture} 
	
	\definecolor{color0}{rgb}{0.12156862745098,0.466666666666667,0.705882352941177}
	\definecolor{color2}{rgb}{0.172549019607843,0.627450980392157,0.172549019607843}

		\begin{axis}[%
		hide axis,
		xmin=10,
		xmax=50,
		ymin=0,
		ymax=0.4,
		legend columns=3, 
		legend style={draw=white!15!black,legend cell align=left, column sep=1.9ex}
		]
		\addlegendimage{white,mark=square*, mark size=2}
		\addlegendentry{\hspace{-1.15cm} Spectral + temporal interp. (2P pattern only):};
		\addlegendimage{color0, dashed, mark=*, mark size=1, mark options={solid}}
		\addlegendentry{\hspace{-0.15cm}  Baseline};
		\addlegendimage{color2, dashed, mark=+, mark size=2, mark options={solid}}
		\addlegendentry{\hspace{-0.15cm}  DL receiver  \hspace{0.89cm} \quad };
		\end{axis}
	\end{tikzpicture}
	\end{subfigure}%
	
	\vspace{10pt}

	\centering
	\begin{subfigure}[b]{0.45\textwidth}
	  \begin{adjustbox}{width=\linewidth} 
		  \input{Chapter3/eval/UL_1P_S.tex}
	\end{adjustbox}    
	\caption{1P pilot pattern at \SIrange[range-units=single]{0}{5}{\km\per\hour}.}
	\label{fig:chp3_UL_1P_L}
  \end{subfigure}%
 \hfill
	\begin{subfigure}[b]{0.45\textwidth}
	\begin{adjustbox}{width=\linewidth} 
		\input{Chapter3/eval/UL_2P_S.tex}
	\end{adjustbox} 
	\caption{2P pilot pattern at \SIrange[range-units=single]{40}{60}{\km\per\hour}.}
	\label{fig:chp3_UL_2P_L}
  \end{subfigure}%
  
\vspace{15pt}
\centering
	\begin{subfigure}[b]{0.45\textwidth}
	  \begin{adjustbox}{width=\linewidth} 
		  \input{Chapter3/eval/UL_1P_M.tex}
	\end{adjustbox}     
	\caption{1P pilot pattern at \SIrange[range-units=single]{10}{20}{\km\per\hour}.}  %
	\label{fig:chp3_UL_1P_M}
  \end{subfigure}%
 \hfill
	\begin{subfigure}[b]{0.45\textwidth}
	\begin{adjustbox}{width=\linewidth} 
		\input{Chapter3/eval/UL_2P_M.tex}  		
	\end{adjustbox} 
	\caption{2P pilot pattern at \SIrange[range-units=single]{70}{90}{\km\per\hour}.}
	\label{fig:chp3_UL_2P_M}
  \end{subfigure}%
  
\vspace{15pt}
\centering

	\begin{subfigure}[b]{0.45\textwidth}
	  \begin{adjustbox}{width=\linewidth} 
		  \input{Chapter3/eval/UL_1P_H.tex}
	\end{adjustbox}    
	\caption{1P pilot pattern at \SIrange[range-units=single]{25}{35}{\km\per\hour}.}   %
	\label{fig:chp3_UL_1P_H}
  \end{subfigure}%
 \hfill
	\begin{subfigure}[b]{0.45\textwidth}
	\begin{adjustbox}{width=\linewidth} 
		\input{Chapter3/eval/UL_2P_H.tex}
	\end{adjustbox} 
	\caption{2P pilot pattern at \SIrange[range-units=single]{110}{130}{\km\per\hour}.}
	\label{fig:chp3_UL_2P_H}
  \end{subfigure}%

\vspace{-5pt}
\caption{Uplink BER achieved by the different receivers with the 1P and 2P pilot patterns.}
\label{fig:chp3_eval_UL}
\end{figure}

Different schemes were benchmarked in the uplink simulations.
The first one is the uplink baseline presented in  Sections~\ref{sec:chp3_baseline_ul} and~\ref{sec:chp3_stats}.
The second one, named `DL channel estimator', uses the DL-enhanced channel estimator presented in \ref{sec:chp3_ml_ch_est_ul} but  is trained and tested with a standard demapper.
The third one is the \gls{DL}-enhanced receiver, leveraging both the enhanced channel estimator and  demapper.
We refer to it as `\gls{DL} receiver'.
Evaluating the \gls{DL} channel estimator separately allows us to better understand the role both components play in the observed gains.
An ideal baseline with perfect knowledge of the channel at the \glspl{RE} carrying pilots and of $\Em$ is also considered, and referred to as `Perfect \gls{CSI}'.
All schemes used spectral interpolation followed by \gls{NIRE} approximation for channel estimation.
Additional simulations were conducted for the 2P pilot pattern using spectral and temporal interpolation for both the baseline and the DL receiver.
Finally, a DL receiver trained with only $K=2$ users was also evaluated to study the scalability of the system.

Simulation results for the 1P pattern are shown in the first column of Fig.~\ref{fig:chp3_eval_UL} for the three different speed ranges.
At speeds ranging from \SIrange[range-units=single]{0}{15}{\km\per\hour}, one can see that the \gls{DL} channel estimator does not bring any improvements, whereas the \gls{DL} receiver achieves a \SI{1}{\dB} gain at a coded \gls{BER} of $10^{-3}$.
The benefit of having a better estimation of $\Em$  becomes more significant as the speed increases.
At the highest speed range (Fig.~\ref{fig:chp3_UL_1P_H}), the \gls{DL} receiver enables gains of \SI{3}{\dB} over the baseline at a coded \gls{BER} of $10^{-2}$.
It can be observed that the DL receiver trained with only two users nearly matches the performance of the one trained with four users on all considered speeds, demonstrating the scalability of the proposed scheme with respect to the number of users.

Results for the 2P pilot pattern are shown in the second column of Fig.~\ref{fig:chp3_eval_UL}.
The gains provided by the \gls{DL} channel estimator alone and by the entire \gls{DL} receiver follow the same trend as with the 1P pattern, with moderate gains at \SI{50}{}-\SI{70}{\km\per\hour}, but significant improvements at \SI{110}{}-\SI{130}{\km\per\hour}.
More precisely, the \gls{DL} receiver provides a \SI{1}{\dB} gain over the baseline at a coded \gls{BER} of $10^{-3}$ in the \SI{50}{}-\SI{70}{\km\per\hour} range, and is the only scheme that achieves a coded \gls{BER} of $10^{-3}$ for the highest speeds with spectral interpolation only.
Indeed, at high speeds, the learned demapper is still able to mitigate the effects of channel aging, whereas even the perfect \gls{CSI} baseline suffers from strongly distorted equalized signals.
Using both spectral and temporal interpolations reduces the gains provided by the DL receiver, which can be explained by the better channel estimates leading to less channel aging, but they still amount to a \SI{2.2}{\dB} gap at a \gls{BER} of $10^{-3}$ for the highest speeds.
We have also experimentally verified that a DL receiver trained with only $K=2$ users was able to closely match the performance of the one trained with $K=4$, but decided not to include the corresponding curves for clarity reasons.
Overall, one can see that only the combination of a \gls{CNN}-based estimation of the channel estimation error statistics and \gls{CNN}-based demapper enables gains for both pilot patterns and all speed ranges.

\subsection{Visualizing the Channel Estimation Error Statistics}
\label{sec:chp3_E_comp_speed}

\begin{figure*}[t!]
\centering
\begin{minipage}[b]{0.55\textwidth}
  	\begin{subfigure}[b]{0.4\textwidth}
  	\begin{adjustbox}{height=6cm} 
  		\input{Chapter3/eval/E_60.tex}
  	\end{adjustbox} 
  	\caption{\SI{60}{\km\per\hour}}
	\end{subfigure}%
\hfill
  	\begin{subfigure}[b]{0.50\textwidth}
  	\begin{adjustbox}{height=6cm} 
  		\input{Chapter3/eval/E_120.tex}
  	\end{adjustbox} 
  	\caption{\SI{120}{\km\per\hour}}  	
  	\label{fig:chp3_E_RG_120}
	\end{subfigure}%
\caption{Predicted $||\widehat{\Em}_{m,n}||_{\text{F}}$ vs true $||\Em_{m,n}||_{\text{F}}$ for different user speeds.}
\label{fig:chp3_E_RG}
\end{minipage}
\hspace{35pt}
\begin{minipage}[b]{0.35\textwidth}
\hfill
  	\begin{subfigure}[b]{0.45\textwidth}
  	\begin{adjustbox}{height=6cm} 
  		\input{Chapter3/eval/E_error_60.tex}
  	\end{adjustbox} 
  	\caption{\SI{60}{\km\per\hour}}
	\end{subfigure}%
\hfill
  	\begin{subfigure}[b]{0.50\textwidth}
  	\begin{adjustbox}{height=6cm} 
  		\input{Chapter3/eval/E_error_120.tex}
  	\end{adjustbox} 
  	\caption{\SI{120}{\km\per\hour}}  	
	\end{subfigure}%
\caption{Normalized error for different user speeds.}
\label{fig:chp3_E_errors}
\end{minipage}
\end{figure*}

\begin{figure*}[b!]
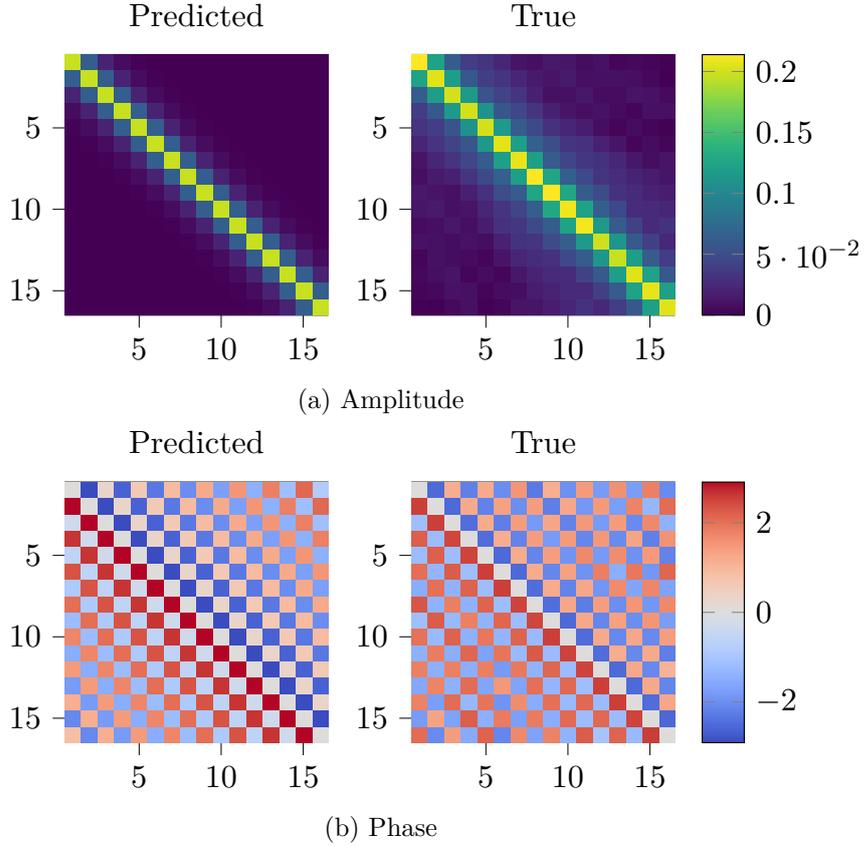


\centering
  	\begin{subfigure}[b]{0.65\textwidth}
  	\centering
    	\begin{adjustbox}{height=5cm} 
   		 \input{Chapter3/eval/E_abs.tex}
  	\end{adjustbox}    
  	\caption{Amplitude}
  	\label{fig:chp3_E_RE_abs}
	\end{subfigure}%
\hfill
  	\begin{subfigure}[b]{0.65\textwidth}
  	\centering
  	\begin{adjustbox}{height=5cm} 
  		\input{Chapter3/eval/E_angle.tex}
  	\end{adjustbox} 
  	\caption{Phase}
  	\label{fig:chp3_E_RE_phase}
	\end{subfigure}%

\caption{Amplitude and phase of $\widehat{\Em}_{7, 36}$ and $\Em_{7, 36}$ for a single realization of the channel at \SI{120}{\km\per\hour}.}
\label{fig:chp3_E_RE}
\end{figure*}

In order to get insights into the \gls{DL} channel estimator abilities, we visualize the channel estimation error covariance matrices $\Em$ for different user speeds.
Two batches of uplink signals were sent, with users respectively moving at \SI{60}{\km\per\hour} and \SI{120}{\km\per\hour}.
All batches comprised 90 \glspl{RG}, and used the 2P pilot pattern with an \gls{SNR} of \SI{5}{\dB}.
The Frobenius norms $||\Em_{m,n}||_{\text{F}}$ corresponding to all \glspl{RE} are shown in Fig.~\ref{fig:chp3_E_RG}, and the normalized errors $\left| \frac{||\widehat{\Em}_{m,n}||_{\text{F}} - ||\Em_{m,n}||_{\text{F}}}{||\Em_{m,n}||_{\text{F}}} \right|$ are shown in Fig.~\ref{fig:chp3_E_errors}.
The figures labeled as `Predicted' refers to the estimations  $\widehat{\Em}_{m,n}$ produced by the \gls{DL} channel estimator and averaged over the corresponding batches, using spectral interpolation only.
As expected, the Frobenius norms of the \glspl{RE} carrying data strongly depend on their distance to their closest pilot, reflecting the distortions visible in  Fig.~\ref{fig:chp3_mismatch_2P}.
The figures labeled as `True' are presented for reference, and are computed by Monte Carlo simulations assuming knowledge of the true channel realizations.
One can see that the normalized errors are low on \glspl{RE} carrying data, confirming that $\text{CNN}_\Em$ is able to learn the channel estimation error statistics from the data during training.
The higher errors on \glspl{RE} carrying pilots are due to the loss \eqref{eq:chp3_loss_mc} taking solely into account the positions $(m,n)\in\mathcal{D}$, giving no opportunities for $\text{CNN}_\Em$ to learn the statistics at pilot positions.
Experiments conducted using both spectral and temporal interpolation yielded higher normalized errors, probably because the increased accuracy of the channel estimates results in a more difficult learning of the error statistics.
Additionally, $\text{CNN}_l$ seems to correctly estimate the time-variability of the channels since the Frobenius norms of the predicted covariances increase with the user speed, matching the behavior of the true covariances.

It is also insightful to look at the predicted and true spatial covariance matrix for a single \gls{RE}.
We chose to focus on the \gls{RE} $(7, 36)$, positioned at the center of the \gls{OFDM} grid, and on the \SI{120}{\km\per\hour} scenario (Fig.~\ref{fig:chp3_E_RG_120}).
The amplitudes and phases of all elements of $\widehat{\Em}_{7, 36}$ and $\Em_{7, 36}$ are shown in Fig.~\ref{fig:chp3_E_RE_abs} and~\ref{fig:chp3_E_RE_phase}.
One can see that the predicted amplitudes and phases are close to the true ones, thus supporting the proposed power decay model.

\subsection{Downlink Simulation Results}

\begin{figure}
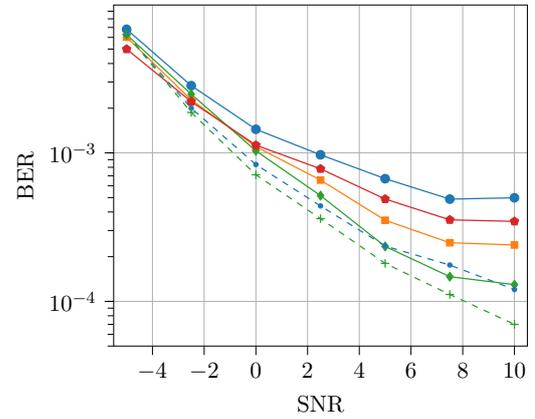
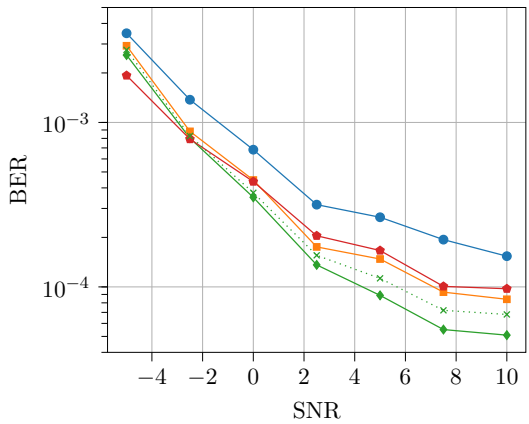
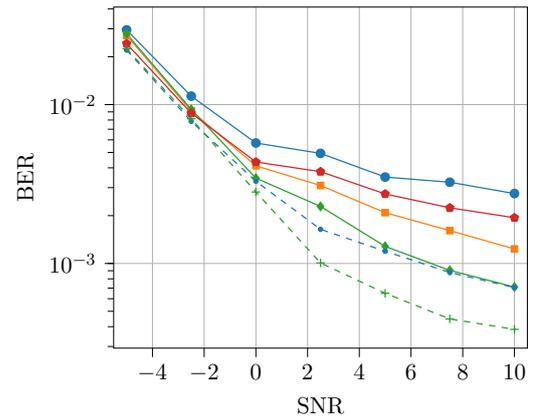
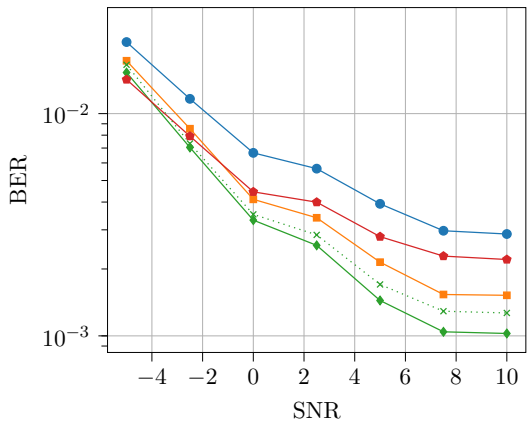
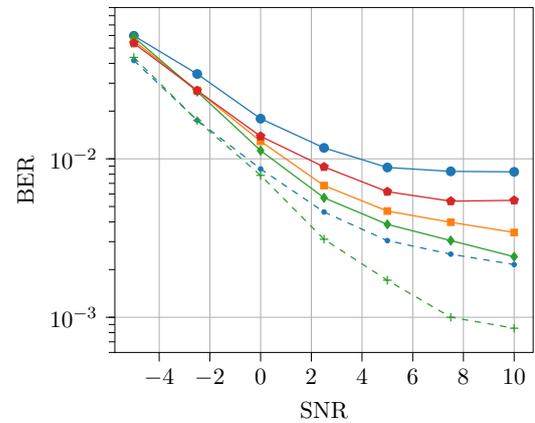

		\begin{subfigure}[l]{1\textwidth}
		\begin{tikzpicture} 
		
		\definecolor{color0}{rgb}{0.12156862745098,0.466666666666667,0.705882352941177}
		\definecolor{color1}{rgb}{1,0.498039215686275,0.0549019607843137}
		\definecolor{color2}{rgb}{0.172549019607843,0.627450980392157,0.172549019607843}
		\definecolor{color3}{rgb}{0.83921568627451,0.152941176470588,0.156862745098039}
	
		\begin{axis}[%
			hide axis,
			xmin=10,
			xmax=50,
			ymin=0,
			ymax=0.4,
			legend columns=5, 
			legend style={draw=white!15!black,legend cell align=left,column sep=2.022ex}
			]
			\addlegendimage{white,mark=square*}
			\addlegendentry{\hspace{-1.15cm} Spectral interp.:};
			\addlegendimage{color0,mark=square*}
			\addlegendentry{\hspace{-0.25cm} Baseline};
			\addlegendimage{color1, mark=*}
			\addlegendentry{\hspace{-0.25cm} DL ch. est. };
			\addlegendimage{color2, mark=pentagon*}
			\addlegendentry{\hspace{-0.25cm} DL receiver};
			\addlegendimage{color3, mark=diamond*}
			\addlegendentry{\hspace{-0.25cm} Perfect CSI};
			
			\end{axis}
		\end{tikzpicture}
		\end{subfigure}%
		\vspace{1pt}
		
		  \begin{subfigure}[c]{1\textwidth}
		  
		\begin{tikzpicture} 
		
		\definecolor{color2}{rgb}{0.172549019607843,0.627450980392157,0.172549019607843}
	
			\begin{axis}[%
			hide axis,
			xmin=10,
		   xmax=50,
			ymin=0,
			ymax=0.4,
			legend columns=2, 
			legend style={draw=white!15!black,legend cell align=right,column sep=1.9ex}
			]
			\addlegendimage{white,mark=square*, mark size=2}
			\addlegendentry{\hspace{-1.15cm} Spectral interp. (1P pattern only):};
			\addlegendimage{color2, dotted, mark=x, mark size=2, mark options={solid}}
			\addlegendentry{DL receiver trained with $K=2$ \hspace{2.29cm} \quad};
			\end{axis}
		\end{tikzpicture}
		
		\end{subfigure}
		
		\vspace{1pt}
	
		  \begin{subfigure}[c]{1\textwidth}
		  
		\begin{tikzpicture} 
		
		\definecolor{color0}{rgb}{0.12156862745098,0.466666666666667,0.705882352941177}
		\definecolor{color2}{rgb}{0.172549019607843,0.627450980392157,0.172549019607843}
	
			\begin{axis}[%
			hide axis,
			xmin=10,
			xmax=50,
			ymin=0,
			ymax=0.4,
			legend columns=3, 
			legend style={draw=white!15!black,legend cell align=left, column sep=1.9ex}
			]
			\addlegendimage{white,mark=square*, mark size=2}
			\addlegendentry{\hspace{-1.15cm} Spectral + temporal interp. (2P pattern only):};
			\addlegendimage{color0, dashed, mark=*, mark size=1, mark options={solid}}
			\addlegendentry{\hspace{-0.15cm}  Baseline};
			\addlegendimage{color2, dashed, mark=+, mark size=2, mark options={solid}}
			\addlegendentry{\hspace{-0.15cm}  DL receiver  \hspace{0.89cm} \quad };
			\end{axis}
		\end{tikzpicture}
		\end{subfigure}%
		
		\vspace{10pt}
		\centering
		\begin{subfigure}[b]{0.45\textwidth}
		  \begin{adjustbox}{width=\linewidth} 
			  \input{Chapter3/eval/DL_1P_S.tex}
		\end{adjustbox}    
		\caption{1P pilot pattern at \SIrange[range-units=single]{0}{5}{\km\per\hour}.}
		\label{fig:chp3_DL_1P_L}
	  \end{subfigure}%
	 \hfill
		\begin{subfigure}[b]{0.45\textwidth}
		\begin{adjustbox}{width=\linewidth} 
			\input{Chapter3/eval/DL_2P_S.tex}
		\end{adjustbox} 
		\caption{2P pilot pattern at \SIrange[range-units=single]{40}{60}{\km\per\hour}.}
		\label{fig:chp3_DL_2P_L}
	  \end{subfigure}%
	  
  \vspace{15pt}
  \centering
		\begin{subfigure}[b]{0.45\textwidth}
		  \begin{adjustbox}{width=\linewidth} 
			  \input{Chapter3/eval/DL_1P_M.tex}
		\end{adjustbox}     
		\caption{1P pilot pattern at \SIrange[range-units=single]{10}{20}{\km\per\hour}.}  %
		\label{fig:chp3_DL_1P_M}
	  \end{subfigure}%
	 \hfill
		\begin{subfigure}[b]{0.45\textwidth}
		\begin{adjustbox}{width=\linewidth} 
			\input{Chapter3/eval/DL_2P_M.tex}  		
		\end{adjustbox} 
		\caption{2P pilot pattern at \SIrange[range-units=single]{70}{90}{\km\per\hour}.}
		\label{fig:chp3_DL_2P_M}
	  \end{subfigure}%
	  
  \vspace{15pt}
  \centering
  
		\begin{subfigure}[b]{0.45\textwidth}
		  \begin{adjustbox}{width=\linewidth} 
			  \input{Chapter3/eval/DL_1P_H.tex}
		\end{adjustbox}    
		\caption{1P pilot pattern at \SIrange[range-units=single]{25}{35}{\km\per\hour}.}   %
		\label{fig:chp3_DL_1P_H}
	  \end{subfigure}%
	 \hfill
		\begin{subfigure}[b]{0.45\textwidth}
		\begin{adjustbox}{width=\linewidth} 
			\input{Chapter3/eval/DL_2P_H.tex}
		\end{adjustbox} 
		\caption{2P pilot pattern at \SIrange[range-units=single]{110}{130}{\km\per\hour}.}
		\label{fig:chp3_DL_2P_H}
	  \end{subfigure}%
  
  \vspace{-5pt}
  \caption{Downlink BER achieved by the receivers with the 1P and 2P pilot patterns.}
  \label{fig:chp3_eval_DL}
  \end{figure}

Four downlink schemes were evaluated on the considered speed ranges.
The first one is the baseline presented in Section~\ref{sec:chp3_system_model}.
In the second one, referred to as `\gls{DL} channel estimator', each user uses the enhanced channel estimator of Section~\ref{sec:chp3_ml_ch_est_dl} and  is trained and tested with a standard demapper.
The third one is the \gls{DL}-enhanced receiver using both the enhanced channel estimator and \gls{DL} demapper for each user, and is referred to as `\gls{DL} receiver'.
The last scheme has perfect \gls{CSI}, i.e., all users perfectly know both the channel at \glspl{RE} carrying pilots and the estimation error variances $v^{(k)}_{m,n,k}$ and $v^{(k)}_{m,n,i}$ everywhere.
All schemes use the 2P pilot pattern to perform the uplink channel estimation, but are evaluated on both patterns in the downlink.
Similarly to the uplink, all schemes leveraged spectral interpolation with \gls{NIRE} approximation for channel estimation, but additional simulations were performed with the 2P pilot pattern using spectral and temporal interpolation for both the baseline and the DL receiver.
A DL receiver trained with only $K=2$ users was also evaluated on the three slowest speed ranges to study the scalability of the system.

The 1P pilot pattern downlink evaluations are shown in the first row of Fig.~\ref{fig:chp3_eval_DL}.
Between \SI{0}{} and \SI{15}{\km\per\hour}, all schemes achieve good results but the DL receivers are the only ones to not saturate at high \glspl{SNR}.
As expected, the gains allowed by the learned channel estimator and demapper increase with the speed, and the \gls{DL} receiver is the only scheme to reach a \gls{BER} of approximately $10^{-3}$ in the \SI{30}{}-\SI{45}{\km\per\hour} speed range.
It can also be observed that the DL receiver trained with only $K=2$ users is able to closely match the performance of the DL receiver trained with four users.
The 2P pilot pattern evaluations are shown in the second row of Fig.~\ref{fig:chp3_eval_DL} for higher speeds.
In this new setup, the \gls{DL} receiver outperforms the baseline by \SI{1.1}{\dB} at a \gls{BER} of $10^{-3}$ in the speed range \SI{50}{}-\SI{70}{\km\per\hour}, and is the only one with spectral interpolation only to achieve a \gls{BER} of $10^{-3}$ in the speed range  \SI{70}{}-\SI{90}{\km\per\hour}.
Using both spectral and temporal interpolations, the DL receiver still provides significant gains at high speeds, being the only one to achieve a \gls{BER} of $10^{-3}$ in the \SI{110}{}-\SI{130}{\km\per\hour} speed range.

In all downlink scenarios except for the slowest speed range, the \gls{DL} channel estimator outperforms the perfect \gls{CSI} baseline.
This can be surprising since this baseline uses the exact noise variances $v^{(k)}_{m,n,k}$ and $v^{(k)}_{m,n,i}$ while the \gls{DL} schemes can only estimate them.
We suppose that this is because the baselines assume that the post-equalization downlink noise $\xi_{m,n,k}$ in \eqref{eq:chp3_eq_eq_ch_dl} is Gaussian distributed and uncorrelated to the transmitted signal, which is typically untrue.
The \gls{DL} channel estimator seems able to learn this model mismatch during training and predict variances that counteract it.

\clearpage
\section{Concluding Thoughts}
\label{sec:chp3_conclusion}

A new hybrid strategy to the design of \gls{DL}-enhaced \gls{MU-MIMO} receivers was proposed in this chapter.
Our architecture builds on top of a traditional MU-MIMO receiver and enhances it with \gls{DL} components that are trained to maximize the end-to-end performance of the system.
More precisely, \glspl{CNN} are used to improve both the demapping and the computation of the channel estimation error statistics.
All components of the proposed architecture are jointly optimized to refine the estimation of the \glspl{LLR}.
This approach does not require any knowledge of the channel for training, is interpretable, and easily scalable to any number of users.
Uplink and downlink evaluations were performed with multiple user speeds and two different pilot configurations on \gls{3GPP}-compliant channel models.
The results reveal that the proposed architecture effectively exploits the \gls{OFDM} structure to achieve tangible gains at low speeds and significant ones at high speeds compared to a traditional receiver.
On the one hand, the enhanced demapper jointly processes all \glspl{RE} of the \gls{OFDM} grid to counteract the effects of channel aging.
On the other hand, we demonstrated that the enhanced channel estimator is able to learn its error statistics during training.
In order to get insights into the improvements enabled by each of the trainable components, we also evaluated a conventional structure where only the channel estimator was enhanced.
The results indicate that the combined use of both the \gls{DL} channel estimator and demapper is key to achieve a substantial reduction of the coded \glspl{BER} across all scenarios.

We believe that such architectures, enabling performance improvements while remaining standard-compliant and reasonably complex, could be deployed in \glspl{BS} for the next generation of wireless communication systems.
However, it has be shown that additional gains can be still be achieved with a fully NN-based receiver that jointly processes all users~\cite{korpi2020deeprx}, in contrast to our approach in which they are processed independently.
Unlocking these gains while preserving the flexibility and interpretability of conventional architectures remains a significant challenge that is yet to be solved.

\clearpage

\chapter{Conclusion and Future Directions}

\bigskip

\subsection*{Summary of contributions}

Over the course of this manuscript, we studied three strategies that aim at bringing the benefits of DL to the physical layer of communication systems.
The first strategy, that we referred to as the NN-based block optimization, was used to design an NN that could replace current MU-MIMO detectors.
Available works often exploit the deep unfolding approach, in which a traditional iterative receiver algorithm is unfolded and trainable parameters are inserted.
Among these works, the MMNet detector showed impressive gains, but had to be retrained for each new channel realization.
We therefore proposed in Chapter 3 to use a second NN, the hypernetwork, that takes as input a channel realization and generates a set of optimized parameters for an MMNet-based detector.
To reduce the complexity of the system and the number of parameters that must be estimated, we both leveraged the QR-decomposition of the channel matrix and adopted a relaxed form of weight sharing.
The hypernetwork training was carried out by backpropagating the gradients through the MMNet detector up to the hypernetwork trainable parameters.
Simulation result demonstrated that the resulting architecture, referred to as HyperMIMO, achieves near state-of-the-art performance as long as the channel statistics does not change significantly.
Other works subsequently reduced or eliminated some HyperMIMO downsides, for example by enabling scalability to any number of users and performance improvements on a wider range of channels.
But these detectors are still trained to optimize their own performance only, which does not guaranty any system-level optimality.

The second strategy, in which the system is optimized end-to-end by implementing the transceivers as NNs, was discussed in Chapter 4.
It allows for a deeper optimization of the transmit and receive processing since it does not have to comply with existing protocols.
We applied this approach to the design of OFDM waveforms tailored for specific needs.
Specifically, we derived an NN architecture and a corresponding training algorithm that maximize an achievable rate while satisfying PAPR and ACLR constraints.
This was achieved by expressing the objective and the constraints as differentiable functions that can be minimized using SGD.
The constrained optimization problem was then solved using the augmented Lagrangian method.
Numerical results demonstrated the effectiveness of the proposed approach, with trained systems that meet ACLR and PAPR targets, does not require pilot signals, and achieve throughputs comparatively higher than a TR baseline.
These gains are driven by the emergence of new high-dimensional and asymmetrical constellations, in which the position and the energy associated with each symbol is a function of the subcarrier index and of all other symbols transmitted simultaneously.
Although promising, the corresponding transmitter and receiver are too complex for any short-term implementation, does not satisfy current standards, and are limited to SISO transmissions.
These limits pave the way for future research.

Finally, a third hybrid strategy was proposed in Chapter 5, and consists in inserting DL components into a traditional architecture that is trained in an end-to-end manner.
We leveraged this strategy to enhance a conventional MO-MIMO receiver with three CNNs.
The first and second ones learn the channel estimation error statistics from the data, which improve the symbol detection accuracy.
The third one performs the bit demapping on the entire RG so that it can estimate and correct the distortions present on the equalized symbols.
The end-to-end training also allows the system to be optimized on an achievable transmission rate.
Simulations were performed with 3GPP-compliant channels with different speed ranges and two pilot patterns.
The results indicate that our receiver achieves gains across all scenarios, both in uplink and downlink, and especially at high speeds.
More importantly, it preserves the scalability of conventional architectures, and is composed of components that are individually interpretable and of reasonable complexity.
Such systems, trained to handle multiple modulations and pilot patterns, could become standard-compliant as the DL components are only inserted in the receiver.
Finally, and in contrast to many comparable works, perfect channel estimates are not needed during training.
Overall, this strategy seems more suitable for practical use in the near future.

\subsection*{Future directions}

The deployment of DL in the physical layer will probably be conducted in several phases.
The first one corresponds to the integration of DL-based blocks into traditional BS receivers, trained following either the block-based or the hybrid strategy.
The success of this integration is first tied to the availability of dedicated DL accelerators that can run at sufficient speeds.
Fortunately, such processing units are already being designed to be implemented on BSs, where the constraints on the chip size and energy consumption are less stringent than on mobile devices.
But DL is first and foremost about data, so that the gains provided by its integration are also tied to our capacity to build sufficient datasets.
Indeed, most published works focus on off-line training using channel simulators, which might not be representative of every scenario that will be encountered in real life.
Collecting representative datasets from many diverse environments is therefore essential to ensure the reliability of these DL components.

These challenges are exacerbated on mobile devices, in which the integration of DL might correspond to a second deployment phase.
To limit the amount of on-device computations, DL components will probably be trained in a centralized manner.
This both requires that data be regularly collected from the devices and that updated parameters be transferred to them.
On the one hand, such periodic data transfers will generate energy and bandwidth consumption, in addition to raising confidentiality and privacy concerns.
On the other hand, they will enable more flexible communications, as the transceivers could be adapted to any specific hardware and environment.
This should lead to improved gains since the DL models will be able to embrace distortions and channel specificities that are not present in traditional models.
Such flexibility and adaptability are especially interesting for upcoming 6G networks, which are expected to support a wide range of use-case such as vehicular communications or small-scale sub-networks.

The increasing availability of DL accelerators at both ends of the transmission should eventually allow the emergence of fully NN-based transceivers, trained using the end-to-end strategy.
Many predict that this third phase will simplify the set of different options and parameters used in current standards, as each transmitter-receiver pair could be continuously optimized to maximize its performance.
However, the benefits provided by NN-based systems are jointly tied to the chosen NN architectures, training procedures, and to the quality of the available data.
In industry sectors in which AI already plays an important role, such as voice assistants or autonomous driving, companies are usually reluctant to give any details about their specific NNs since they are at the core of their technical advances.
The joint training of transmitters and receivers designed by different manufacturers therefore represents another major challenge for the telecommunication industry.
Future standards need to shift from regulating the behavior of communication algorithms to allowing the end-to-end optimization of NN-based systems.
Shared datasets and procedures will also be needed to enable proper testing, as the black-box nature of NNs prevents the derivation of the performance guarantees traditionally offered by model-based systems.

In its Ph.D.dissertation published in 2000, Joseph Mitola III described the concept of cognitive radio as “\emph{the point in which wireless personal digital assistants (PDAs) and the related networks are sufficiently computationally intelligent about radio resources and related computer-to-computer communications to detect user communications needs as a function of use context, and to provide radio resources and wireless services most appropriate to those needs}”~\cite{mitola2000cognitive}.
The next phases of DL deployment in communication systems could therefore be focused on the joint training of NNs for the physical and medium access control (MAC) layers, finally giving rise to the first fully AI-based cognitive radios.
 
\clearpage

\appendix
\chapter{Grouped-LMMSE Equalizer}
\label{app:LMMSE}

We aim to find the \gls{LMMSE} matrix to equalize a group of \glspl{RE} that spans multiple symbols $m \in [M_b, M_e]$ and subcarriers $n \in [N_b, N_e]$, with $1 \leq M_b \leq M_e \leq M$ and $1 \leq N_b \leq N_e \leq N$.
Let us denote by $\widehat{\Hm}_{m,n}$ the channel estimated at a \gls{RE} $(m,n)$ and by $\widetilde{\Hm}_{m,n}$ the corresponding estimation errors.
It is assumed that $\widehat{\Hm}$ is known, but that $\widetilde{\Hm}$, the symbols, and the noise, conditioned on the channel estimates, are random and uncorrelated.
The channel transfer function for that group is
\begin{equation}
\yv_{m,n} = \LB \widehat{\Hm}_{m,n} + \widetilde{\Hm}_{m,n} \RB \xv_{m,n} + \nv_{m,n}, \quad \forall  m\in \LSB M_b, M_e \RSB, n \in \LSB N_b, N_e \RSB.
\end{equation}
We denote by $\Wm_{m,n}, \;m \in [M_b, M_e], n \in [N_b, N_e]$, the \gls{LMMSE} matrix that minimizes

\begin{align}
 \mathcal{L}(\Wm_{m,n}) 
 &=\sum_{m'=M_b}^{M_e} \sum_{n'=N_b}^{N_e} \text{MSE}\LB  \xv_{m',n'}, \Wm_{m,n} \yv_{m',n'} \RB \\
&= \EE \LSB \sum_{m'=M_b}^{M_e} \sum_{n'=N_b}^{N_e} 
\LB\xv_{m', n'} - \Wm_{m,n} \yv_{m', n'} \RB
\LB\xv_{m', n'} - \Wm_{m,n} \yv_{m', n'} \RB\htp
 \RSB \nonumber.
\end{align}
Therefore, $\Wm_{m,n}$ nulls the gradient 
\begin{align}
\nabla_{\Wm_{m,n}} \mathcal{L}(\Wm_{m,n})
 =  2\Wm  \EE \LSB
\sum_{m'=M_b}^{M_e} \sum_{n'=N_b}^{N_e} 
 \yv_{m', n'} \yv_{m', n'}\htp  \RSB - 2 \EE \LSB
\sum_{m'=M_b}^{M_e} \sum_{m'=N_b}^{N_e} 
\xv_{m', n'} \yv_{m', n'}\htp \RSB
\stackrel{!}{=} 0 
\end{align}

which leads to
\begin{align}
\Wm_{m,n} 
&= \LB \EE \LSB
\sum_{m'=M_b}^{M_e} \sum_{n'=N_b}^{N_e} 
 \xv_{m',n'} \yv_{m',n'}\htp  \RSB \RB \LB \EE \LSB
\sum_{m'=M_b}^{M_e} \sum_{n'=N_b}^{N_e} 
 \yv_{m',n'} \yv_{m',n'}\htp  \RSB \RB^{-1} \\
&=\LB 
\sum_{m'=M_b}^{M_e} \sum_{n'=N_b}^{N_e} 
 \widehat{\Hm}_{m',n'}\htp  \RB \LB 
\sum_{m'=M_b}^{M_e} \sum_{n'=N_b}^{N_e} \LB
 \widehat{\Hm}_{m',n'} \widehat{\Hm}_{m',n'}\htp  
 + \EE \LSB \widetilde{\Hm}_{m',n'} \widetilde{\Hm}_{m',n'}\htp \RSB
 + \sigma^2\Id_{N_m}
 \RB \RB^{-1}. \nonumber
\end{align}

\begin{otherlanguage}{french}
\chapter{Résumé en Français}
\label{chp:french}

\tocless\section{Introduction}

\tocless\subsection{Quand l'Apprentissage Profond Rencontre le Traitement du Signal}
\begin{figure}[h]
    \centering
    \includegraphics[width=0.70\textwidth]{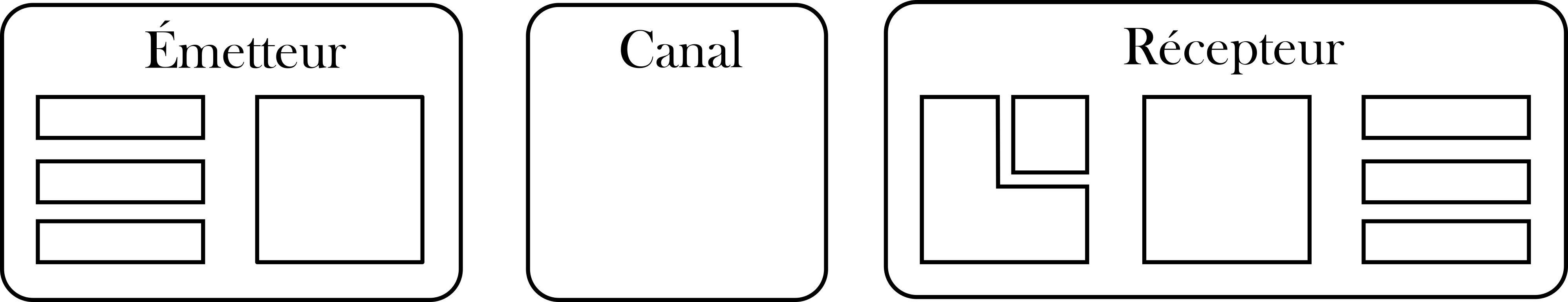}
    \caption{Un système de communication traditionnel basé sur des blocs.}
    \label{fig:bkg_regular_system_fr}
\end{figure}

Le premier modèle de réseau de neurones (neural network, NN) a été inventé en 1943~\cite{mcculloch1943logical}, mais soixante années de recherche et une importante augmentation de la puissance de calcul ont été nécessaires pour permettre une adoption massive  de l'apprentissage profond (deep learning, DL) par l'industrie~\cite{Bennett07thenetflix}.
Dans le même temps, de nouvelles générations de systèmes de communication sont apparues tous les dix ans, à partir de 1979~\cite{rappaport1996wireless}. 
Au fur et à mesure que les émetteurs et les récepteurs devenaient de plus en plus complexes, la tractabilité a été obtenue en divisant les chaînes de traitement d'émission et de réception en petits composants, généralement appelés blocs de traitement et illustrés sur la Figure~B.1. 
Ces systèmes de communication basés sur des blocs souffrent de multiples inconvénients.
D'un côté, les modèles de canal simplistes ne parviennent pas à capturer toutes les spécificités du matériel et des phénomènes de propagation sous-jacents. 
D'un autre côté, l'optimisation conjointe de l'émetteur et du récepteur devient rapidement intractable lorsque des modèles de canaux plus réalistes sont utilisés, et donc l'optimisation de chaque bloc est généralement réalisée indépendamment.
Cela ne garantit pas l'optimalité du système résultant, comme démontré pour les blocs de codage de canal et de modulation~\cite{141453}.
Enfin, de la signalisation est souvent nécessaire entre l'émetteur et le récepteur, ce qui introduit une surcharge qui réduit le débit du système.

\begin{figure}[h]
    \centering
       \begin{subfigure}{1\textwidth}
        \centering
       \includegraphics[width=0.70\textwidth]{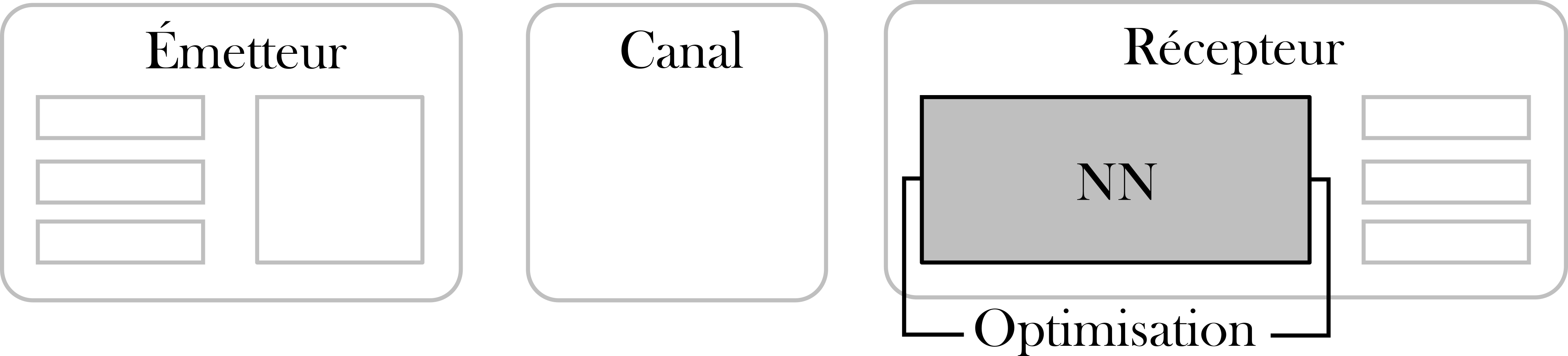}
       \caption{Optimisation des blocs basée sur un NN : un NN est optimisé pour remplacer un ou plusieurs blocs dans un système de communication.}
       \label{fig:intro_ml_syst_1_fr} 
       \vspace{10pt}
    \end{subfigure}
    \begin{subfigure}{1\textwidth}
        \centering
       \includegraphics[width=0.7\textwidth]{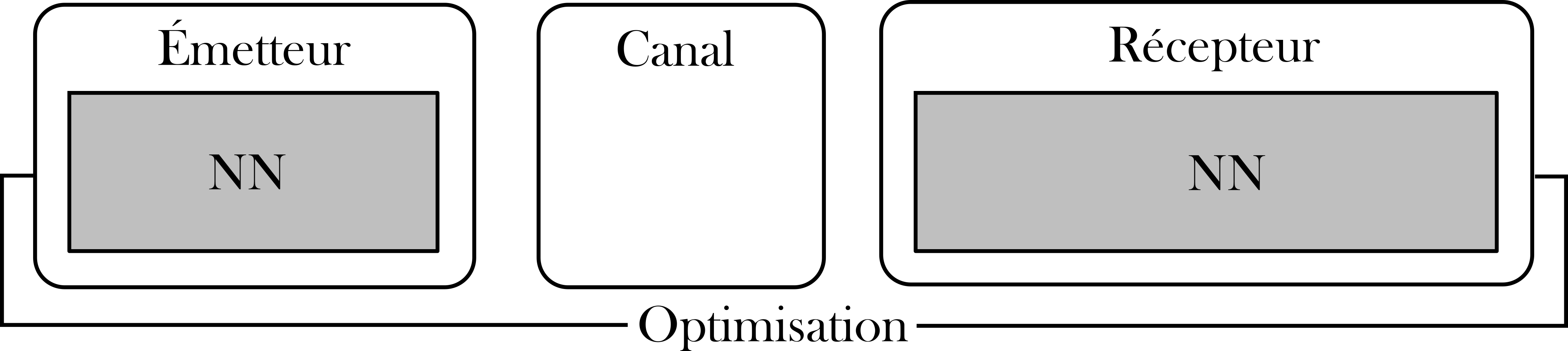}
       \caption{Optimisation de bout en bout : des émetteurs-récepteurs basés sur les NN sont optimisés pour maximiser les performances de bout en bout d'un système.}
       \label{fig:intro_ml_syst_3_fr}
    \end{subfigure}
    \caption{Différents niveaux d'intégration des NNs dans les systèmes de communication.}
    \label{fig:intro_ml_syst_fr}
    \end{figure}

Deux stratégies principales se dessinent concernant le DL dans la couche physique. 
La première approche correspond à une stratégie d'optimisation des blocs basée sur un NN, comme le montre la Figure~B.2a, dans laquelle un ou plusieurs blocs de traitement consécutifs sont remplacés par un NN qui est entraîné indépendamment des autres blocs.
La deuxième approche consiste à implémenter l'émetteur et le récepteur sous forme de NNs qui sont entraînés conjointement afin de maximiser les performances du système de bout en bout~\cite{8054694, 8335670}.
Cette approche est souvent appelée stratégie d'optimisation de bout en bout, comme le montre la Figure~B.2b. 
Cette stratégie permet aux systèmes d'être entièrement optimisés à partir de données récoltées sur le terrain, ce qui permet de traiter efficacement les dégradations matérielles et autres distorsions du canal sans nécessiter de modèle mathématique~\cite{8792076}. 
Cependant, chaque stratégie a ses défauts : les NNs basés sur des blocs (Figure~B.2a) ne sont pas entraînés pour maximiser la performance globale du système.
De plus les émetteurs-récepteurs entièrement appris (Figure~B.2b) manquent d'interprétabilité et d'adaptabilité à un nombre variable d'utilisateurs. 
Cette thèse vise donc à apporter une réponse à la question de l'intégration optimale des composants DL dans les systèmes de communication sans fil.

\tocless\subsection{Défis Actuels et Contributions de ce Travail}

La prochaine génération de réseaux cellulaires devra prendre en charge un nombre croissant de services et de dispositifs différents~\cite{9040431}.
À cette fin, les systèmes multi-utilisateurs à entrées et sorties multiples (multi-user multiple-input multiple-output, MU-MIMO) permettent un partage plus efficace des ressources disponibles.
Les deux principaux défis liés au déploiement de ces systèmes sont la complexité de l'algorithme de détection de symboles et la dégradation des performances sur des canaux mal conditionnés.
Ces dernières années, plusieurs approches ont tenté de relever ces défis en implémentant le détecteur comme un NN, ce qui correspond à la stratégie d'optimisation par blocs basée sur des NNs. 
Cependant, elles obtiennent toujours des performances insatisfaisantes sur les canaux spatialement corrélés, ou sont exigeantes en termes de calcul puisqu'elles nécessitent un réentraînement pour chaque réalisation de canal.
Dans ce travail, nous abordons ces deux problèmes en utilisant un réseau de neurones supplémentaire, appelé hyper-réseau, qui prend en entrée la matrice de canal et génère des poids optimisés pour un détecteur basé sur un NN.
Une autre direction de recherche est l'amélioration de la forme d'onde liée au multiplexage par répartition en fréquences orthogonales (orthogonal frequency-division multiplexing, OFDM), qui souffre de multiples inconvénients, tels qu'un rapport élevé de puissance de crête à puissance moyenne (peak-to-averge power ratio, PAPR) et un rapport de fuite dans les cannaux adjacents (adjacent channel leakage ratio, ACLR). 
Pour résoudre ces problèmes, nous tirons parti de la stratégie d'optimisation de bout en bout et modélisons l'émetteur et le récepteur comme deux NNs qui implémentent un schéma de modulation à haute dimension et estiment les bits transmis.

\begin{figure}[h]
    \centering
    \includegraphics[width=0.65\textwidth]{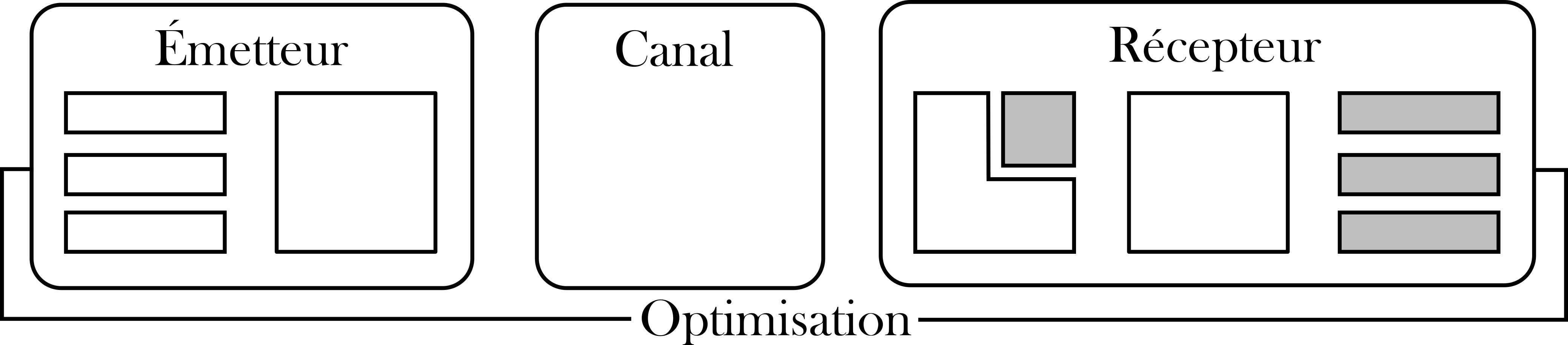}
    \caption{Une stratégie d'optimisation hybride.}
    \label{fig:into_ml_syst_4_fr}
\end{figure}

Cependant, ces systèmes de bout en bout manquent d'interprétabilité et d'adaptabilité, ce qui est particulièrement important dans les transmissions MU-MIMO puisque l'algorithme de réception doit permettre une adaptation facile à un nombre variable d'utilisateurs. 
Pour cette raison, nous proposons un récepteur MU-MIMO amélioré par DL qui s'appuie sur une architecture conventionnelle pour préserver son interprétabilité et son adaptabilité.
Cette approche peut être considérée comme une stratégie hybride, dans laquelle plusieurs composants basés sur du DL sont insérés dans une architecture traditionnelle basée sur des blocs, mais sont optimisés pour maximiser les performances du système de bout en bout (Figure~B.3).

\subsubsection{\textit{\textmd{\selectfont{Notations}}}}

Les tenseurs et les matrices sont désignés par des lettres majuscules en gras et les vecteurs par des lettres minuscules en gras.
Nous désignons respectivement par $\mv_a$ et $m_{a,b}$  le vecteur et le scalaire formés par le découpage de la matrice $\Mm$  le long de sa première et de ses deux premières dimensions.
Similairement, nous désignons par $\Tm_{a, b} \in \CC^{N_c \times N_d}$ ($\tv_{a, b, c} \in \CC^{N_d}$, $t_{a, b, c, d} \in \CC$) la matrice (le vecteur, le scalaire) formé par le découpage du tenseur $\Tm \in \CC^{N_a \times N_b \times N_c \times N_d}$ le long de ses deux (trois, quatre) premières dimensions.
La notation $ \Tm^{(k)}$ indique que la quantité en question n'est considérée que pour le  $k^{\text{ième}}$ utilisateur.
$\odot$, $(\cdot)\tp$ et $(\cdot)\htp$ désignent respectivement le produit par éléments, la transposition, et la transposition conjuguée.

\tocless\section{HyperMIMO : un Détecteur MIMO Basé sur un Hyper-réseau}

\tocless\subsection{Motivations}

Comme nous l'avons vu précédemment, la détection optimale dans les systèmes MIMO est connue pour être NP-difficile~\cite{10.1007_s10107-016-1036-0}, et les approches moins complexes souffrent généralement de performances insatisfaisantes sur des canaux corrélés.
Récemment, des progrès dans la détection MIMO ont été réalisés en utilisant le DL pour améliorer le bloc d'égalisation, ce qui correspond à une optimisation basée sur un bloc.
Une possibilité est d'ajouter des paramètres entraînables à ces algorithmes itératifs et d'interpréter l'ensemble de la structure comme un NN~\cite{8642915, OAMPNet}, mais la plupart des approches souffrent toujours d'une baisse de performance sur des canaux corrélés.
Ce problème a été atténué par le détecteur MMNet~\cite{MMNet}, qui atteint de bonnes performances sur ces canaux. 
Cependant, le besoin de réapprentissage pour chaque réalisation de canal rend son utilisation difficile.

\begin{figure}
	\centering
		\includegraphics[width=0.45\linewidth]{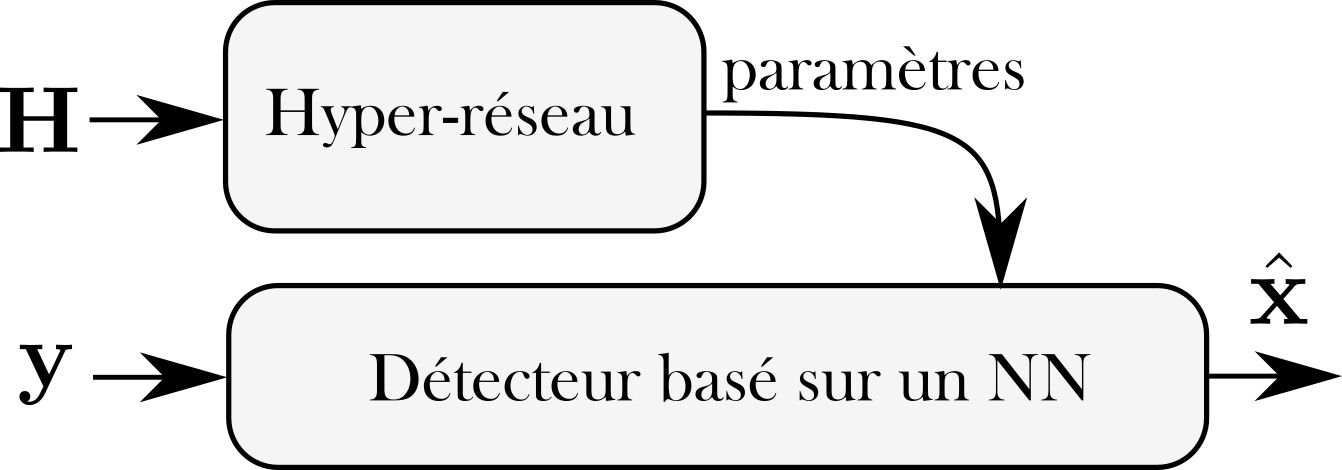}
		\caption{HyperMIMO : un hyper-réseau génère les poids d'un détecteur basé sur un NN.}
	\label{fig:chp1_HG_small_fr}
\end{figure} 

Dans ce qui suit, nous atténuons ce problème en exploitant l'idée émergente de l'hyper-réseaux~\cite{learnet,talking_heads}. 
Appliqué à notre configuration, elle consiste à avoir un NN secondaire, appelé hyper-réseau, qui génère pour une matrice de canal donnée un ensemble optimisé de poids pour un détecteur basé sur un NN.
Ce schéma, que nous avons appelé HyperMIMO dans notre article d'introduction~\cite{goutay2020deep}, est illustré dans la Figure~\ref{fig:chp1_HG_small_fr}.
Nous avons évalué l'approche proposée en utilisant des simulations sur des canaux spatialement corrélés. 
Nos résultats montrent qu'HyperMIMO atteint une performance proche de celle d'MMNet entraîné pour chaque réalisation de canal, et surpasse l'OAMPNet récemment proposé~\cite{OAMPNet}.

\bigskip

\tocless\subsection{Cadre de Travail} 

Nous considérons un canal de liaison montante MU-MIMO classique, où $K$ utilisateurs à antenne unique communiquent avec une station de base (base station, BS) ayant $L$ antennes de réception.
La fonction de transfert du canal est
\begin{equation}
\label{eq:chp1_rayleigh_fr}
\yv = \Hm \xv + \nv
\end{equation}
où $\xv \in \Cc^{K}$ est le vecteur des symboles transmis, $\yv \in \CC^{L}$ est le vecteur des symboles déformés reçus, $\Hm \in \CC^{L \times K}$ est la matrice de canal, et $\nv \thicksim \Cc\Nc(\mathbf{0}, \sigma^2 \Id_{L})$ est le bruit gaussien complexe indépendant et identiquement distribué (i.i.d.) avec une puissance $\sigma^2$ dans chaque dimension complexe.
On suppose que $\Hm$ et $\sigma$ sont parfaitement connus du récepteur. 
Dans la suite, nous considérons le problème de la détection dure de symboles, dans lequel le symbole estimé $\doublehat{\xv}$ doit appartenir à la constellation utilisée, c'est-à-dire, $\doublehat{\xv} \in \Cc^{K}$.

De multiples schémas DL ont été proposés pour obtenir un meilleur compromis performance-complexité que les détecteurs à maximum de vraisemblance ou à erreur quadratique moyenne minimale linéaire (linear minimum mean square error, LMMSE). 
Une technique prometteuse, appelée \emph{deep unfolding}, consiste à améliorer les schémas itératifs existants en ajoutant des paramètres entraînables, et en entraînant le tout comme un NN.
Généralement, chaque itération comprend une étape linéaire suivie d'une étape de débruitage non linéaire :
\begin{equation} 
\label{eq:chp1_it_algo_fr}
\begin{aligned}
\kappav^{(i)}  &= \hat{\xv}^{(i)} + \Am^{(i)} \left(\yv - \Hm \hat{\xv}^{(i)} + \cv^{(i)} \right)\\
\hat{\xv}^{(i+1)} &= \chi^{(i)}\left(\kappav^{(i)}, \tau^{(i)} \right)
\end{aligned}
\end{equation}
où l'exposant $(i)$ est utilisé pour faire référence à la  $i^{\text{ème}}$ itération et $\hat{\xv}^{(0)}$ est fixé à $\mathbf{0}$. 
$\tau^{(i)}$ désigne la variance estimée des composantes du vecteur bruit $\kappav^{(i)} -\xv^{(i)}$ à l'entrée du débruiteur, qui est supposé être i.i.d..
Les algorithmes itératifs diffèrent par leurs choix de matrices $\Am^{(i)} \in \CC^{K \times L}$, de vecteurs de biais $\cv^{(i)} \in \CC^{K}$, et de fonctions de débruitage $\chi^{(i)}(\cdot)$.
Une limitation de la plupart des schémas de détection est leur faible performance sur les canaux corrélés. 
OAMPNet~\cite{OAMPNet} apporte des améliorations en ajoutant deux paramètres entraînables par itération.
MMNet~\cite{MMNet} va plus loin en rendant toutes les matrices $\Am^{(i)}$ entraînables et en relâchant la contrainte selon laquelles les $\kappav^{(i)} -\xv^{(i)}$ doivent être identiquement distribué. 
Bien qu'MMNet atteigne des performances de pointe sur les canaux spatialement corrélés, il doit être réentraîné pour chaque matrice de canal, ce qui le rend peu pratique. 
L'idée clé de notre approche est donc de remplacer le processus d'entraînement requis par MMNet pour chaque réalisation de canal par une seule inférence à travers un hyper-réseau entraîné.

\bigskip
\tocless\subsection{HyperMIMO}


Pour réduire le nombre de paramètres d'MMNet, nous réalisons la décomposition QR de la matrice de canal, $\Hm = \Qm \Rm$, où $\Qm$ est une matrice orthogonale de dimension $L \times L$ et $\Rm$ est une matrice triangulaire supérieure de dimension $L \times K$.
Nous supposons que $L > K$, et par conséquent $\Rm = \begin{bmatrix}\mathbf{R_A} \\ \boldsymbol{0}\end{bmatrix}$ où $\mathbf{R_A}$ a pour dimension $K \times K$, et $\Qm = \left[\mathbf{Q_A} \boldsymbol{Q_B} \right]$ où $\mathbf{Q_A}$ a pour dimension $L \times K$.
Nous définissons $\bar{\yv} \coloneqq \mathbf{Q_A}^H \yv$ et $\bar{\nv} \coloneqq \mathbf{Q_A}^H\nv$, et réécrivons~(\ref{eq:chp1_rayleigh_fr}) comme suit
\begin{equation}
\label{eq:chp1_QR_fr}
\bar{\yv} = \mathbf{R_A} \xv + \bar{\nv}.
\end{equation}
Notez que $\bar{\nv} \thicksim \Cc\Nc(\mathbf{0}, \sigma^2 \Id_{K})$.
MMNet définit $\cv^{(i)}$ à $\zerov$ pour tous les $i$ et utilise le même débruiteur pour toutes les itérations, qui sont définies par
\begin{equation} 
\label{eq:chp1_mmnet_fr}
\begin{split}
\kappav^{(i)} & = \widehat{\xv}^{(i)} + \boldsymbol{\Theta}^{(i)} \left( \bar{\yv} - \mathbf{R_A} \hat{\xv}^{(i)} \right) \\
\widehat{\xv}^{(i+1)} & = \chi \left( \kappav^{(i)}, {\boldsymbol{\tau}^{(i)}} \right)
\end{split}
\end{equation}
où $\boldsymbol{\Theta}^{(i)}$ est une matrice complexe de dimension $K \times K$ dont les éléments doivent être optimisées pour chaque réalisation de canal.
Le principal avantage de l'utilisation de la décomposition QR est que la dimension des matrices $\boldsymbol{\Theta}^{(i)}$ à optimiser est de $K \times K$ au lieu de $K \times L$, qui est la dimension $\Am^{(i)}$ dans~(\ref{eq:chp1_it_algo_fr}).
Ceci est significatif puisque le nombre d'utilisateurs actifs $K$ est généralement beaucoup plus petit que le nombre d'antennes $L$ de la BS. 
MMNet se compose de $I$ couches exécutant~(\ref{eq:chp1_mmnet_fr}), et une décision dure est prise pour prédire l'estimation finale $\doublehat{\xv}$.


\begin{figure}
\centering
	\includegraphics[width=0.6\linewidth]{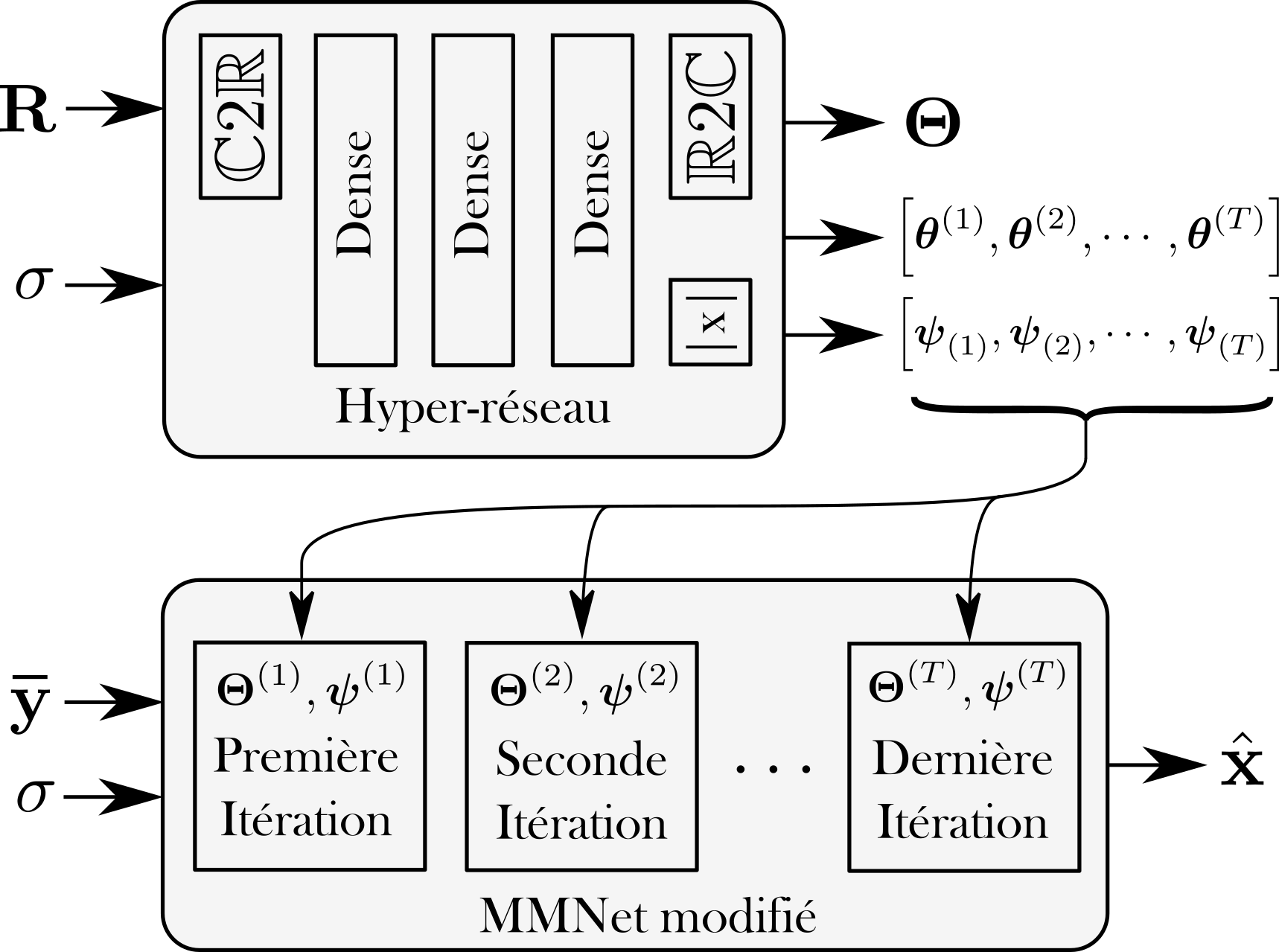}
	\caption{Architecture détaillée d'HyperMIMO.}
\label{fig:chp1_HG_fr}
\vspace{-10pt}
\end{figure} 

La Figure~\ref{fig:chp1_HG_fr} montre en détail l'architecture d'HyperMIMO.
Comme notre variante d'MMNet opère sur $\bar{\yv}$, l'hyper-réseau est alimenté par $\mathbf{R_A}$ et l'écart type du bruit du canal $\sigma$. 
Notez que, comme $\mathbf{R_A}$ est triangulaire supérieur, seuls $K(K+1)/2$ éléments non nuls doivent être fournis à l'hyper-réseau. 
Pour diminuer encore le nombre de sorties de l'hyper-réseau, nous adoptons une forme relaxée de partage de poids inspirée de~\cite{learnet}.
Au lieu de calculer les éléments de chaque $\boldsymbol{\Theta}^{(i)}, i = 1,\dots,I$, l'hyper-réseau estime une seule matrice $\boldsymbol{\Theta}$ ainsi que $I$ vecteurs $\boldsymbol{\theta}^{(i)} \in \RR^{K}$.
Pour chaque itération $i$, $\boldsymbol{\Theta}^{(i)}$ est calculé par
\begin{equation}
\label{eq:chp1_scaling_fr}
\boldsymbol{\Theta}^{(i)} = \boldsymbol{\Theta} \left( \Id_{K} + \text{diag}\left(\boldsymbol{\theta}^{(i)}\right) \right).
\end{equation}

Comme $\mathbf{R_A}$ est à valeur complexe, une couche $\CC2\RR$ transforme les éléments complexes de $\mathbf{R_A}$ en éléments réels, en concaténant les parties réelles et imaginaires des éléments complexes.
Pour générer une matrice $\boldsymbol{\Theta}$ à valeur complexe, une couche $\RR2\CC$  effectue l'opération inverse de $\CC2\RR$.
L'hyper-réseau doit également calculer les valeurs des $I$ vecteurs $\boldsymbol{\psi}^{(i)}$.
Comme les éléments de ces vecteurs doivent être positifs, une fonction d'activation à valeur absolue est utilisée dans la dernière couche, comme le montre la Figure~\ref{fig:chp1_HG_fr}. 
HyperMIMO, qui comprend l'hyper-réseau et MMNet, est entraîné en minimisant l'erreur quadratique moyenne (mean squared error, MSE) entre les symboles transmis et estimés.
Lors de l'apprentissage d'HyperMIMO, l'hyper-réseau et MMNet forment un seul NN, de sorte que la sortie de l'hyper-réseau constitue les poids du détecteur MMNet. 
Les seuls paramètres entraînables sont donc ceux de l'hyper-réseau.

\pagebreak

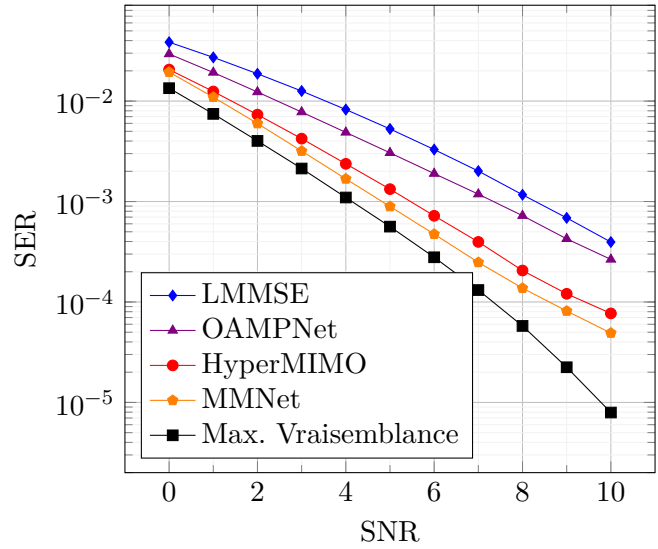
\begin{figure}[h]
	\center
\begin{tikzpicture}

	\pgfplotsset{
    width=.55\textwidth,
    height=0.5\textwidth
}
	  \begin{axis}[
	    ymode=log,
	    grid=both,
	    grid style={line width=.01pt, draw=gray!10},
	    major grid style={line width=.2pt,draw=gray!50},
	    minor tick num=1,
	    xtick={0, 2, 4, 6, 8, 10},
	    xlabel={SNR},
	    ylabel={SER},
	    legend style={at={(0.03, 0.03)},anchor=south west},
		ymax = 9e-2,	    
	    ymin = 2e-6,
	    legend cell align={left},
	  ]
	    \addplot[name path=mmse, blue, mark=diamond*] table [x=snr, y=mmse, col sep=comma] {Chapter1/figs_fr/snr.csv};
	    \addplot[name path=oamp, violet, mark=triangle*] table [x=snr, y=oamp, col sep=comma] {Chapter1/figs_fr/snr.csv};
	    \addplot[name path=hg, red, mark=*] table [x=snr, y=hg, col sep=comma] {Chapter1/figs_fr/snr.csv};
	    \addplot[name path=mmnet, orange, mark=pentagon*] table [x=snr, y=mmnet, col sep=comma] {Chapter1/figs_fr/snr.csv};
	    \addplot[name path=ml, black, mark=square*] table [x=snr, y=ml, col sep=comma] {Chapter1/figs_fr/snr.csv};

	  \addlegendentry{LMMSE}
	  \addlegendentry{OAMPNet}
	  \addlegendentry{HyperMIMO}
	  \addlegendentry{MMNet}
	  \addlegendentry{Max. Vraisemblance}
		\end{axis}
		\end{tikzpicture}
	    \caption{SER atteint par différents systèmes.}
	    \label{fig:chp1_snr_fr}
\end{figure}

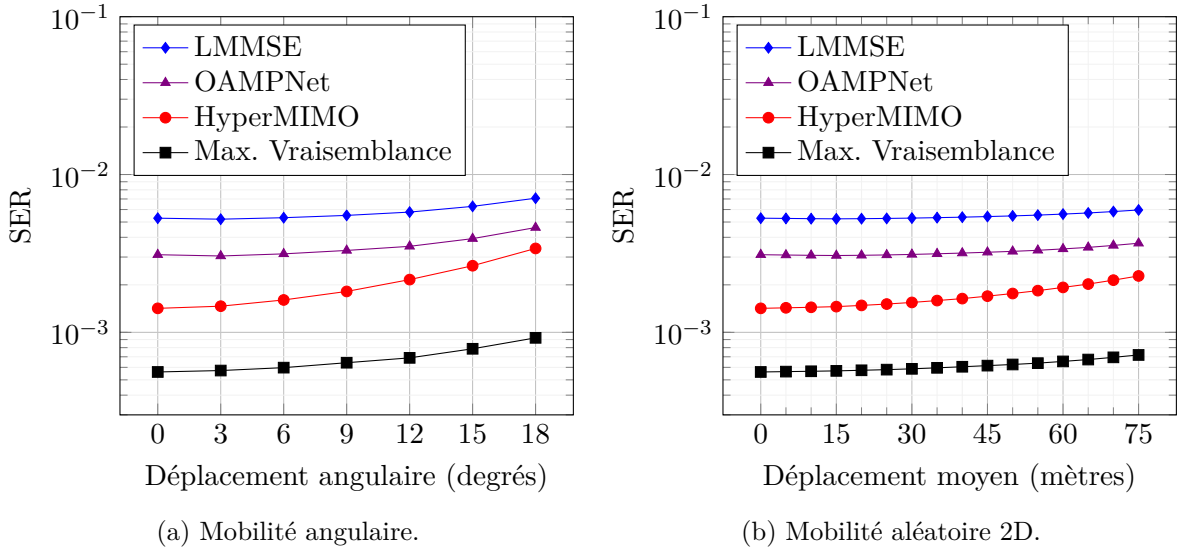
\begin{figure}[h]
    \centering
    \begin{subfigure}{.49\textwidth}
		\begin{tikzpicture}

			\pgfplotsset{
				width=.99\textwidth,
				height=0.9\textwidth
			}

			\begin{axis}[
	    ymode=log,
	    grid=both,
	    grid style={line width=.01pt, draw=gray!10},
	    major grid style={line width=.2pt,draw=gray!50},
	    minor tick num=0,
	    xtick={0, 3, 6, 9, 12, 15, 18},
	    xlabel={Déplacement angulaire (degrés)},
	    ylabel={SER},
	    legend style={at={(0.03, 0.6)},anchor=south west},
		ymax = 1e-1,	    
	    ymin = 3e-4,
	    legend cell align={left},
	  ]
	    \addplot[blue, mark=diamond*] table [x=angles, y=mmse, col sep=comma] {Chapter1/figs_fr/angles_18.csv};
	    \addplot[violet, mark=triangle*] table [x=angles, y=oamp, col sep=comma] {Chapter1/figs_fr/angles_18.csv};
	    \addplot[red, mark=*] table [x=angles, y=hg, col sep=comma] {Chapter1/figs_fr/angles_18.csv};
	    \addplot[black, mark=square*] table [x=angles, y=ml, col sep=comma] {Chapter1/figs_fr/angles_18.csv};

	  \addlegendentry{LMMSE}
	  \addlegendentry{OAMPNet}
	  \addlegendentry{HyperMIMO}
	  \addlegendentry{Max. Vraisemblance}
		\end{axis}
		\end{tikzpicture}
		\caption{Mobilité angulaire.}
		\label{fig:chp1_angles_fr}
	 \end{subfigure} 
	 \hfill
    \begin{subfigure}{.49\textwidth}
	    \begin{tikzpicture}

			\pgfplotsset{
				width=.99\textwidth,
				height=0.9\textwidth
			}

		  \begin{axis}[
	    ymode=log,
	    grid=both,
	    grid style={line width=.01pt, draw=gray!10},
	    major grid style={line width=.2pt,draw=gray!50},
	    minor tick num=2,
	    xtick={0, 15, 30, 45, 60, 75},
	    xlabel={Déplacement moyen (mètres)},
	    ylabel={SER},
	    legend style={at={(0.03, 0.6)},anchor=south west},
		ymax = 1e-1,	    
	    ymin = 3e-4,
	    legend cell align={left},
	  ]
	    \addplot[blue, mark=diamond*] table [x=meters, y=mmse, col sep=comma] {Chapter1/figs_fr/meters_75.csv};
	    \addplot[violet, mark=triangle*] table [x=meters, y=oamp, col sep=comma] {Chapter1/figs_fr/meters_75.csv};
	    \addplot[red, mark=*] table [x=meters, y=hg, col sep=comma] {Chapter1/figs_fr/meters_75.csv};
	    \addplot[black, mark=square*] table [x=meters, y=ml, col sep=comma] {Chapter1/figs_fr/meters_75.csv};

	  \addlegendentry{LMMSE}
	  \addlegendentry{OAMPNet}
	  \addlegendentry{HyperMIMO}
	  \addlegendentry{Max. Vraisemblance}
		\end{axis}
		\end{tikzpicture}
	    \caption{Mobilité aléatoire 2D.}
	    \label{fig:chp1_distance_fr}
	\end{subfigure}

	\caption{SER obtenu par les approches comparées en cas de mobilité.}
	\label{fig:chp1_mobility_fr}

\end{figure}
	
\bigskip
\tocless\subsection{Résultats des Simulations}

On considère le modèle de diffusion locale avec corrélation spatiale présenté dans~\cite[Ch.~2.6]{massivemimobook}.
On suppose une répartition parfaite de la puissance, ce qui fait que tous les utilisateurs sont à la même distance $r$ de la BS et que le gain moyen est de un. 
Le rapport signal/bruit (signal-to-noise ration, SNR) de la transmission est défini par la $\mathrm{SNR}=\frac{\mathbb{E}\left[\frac{1}{N_{r}}\|\mathbf{y}\|_{2}^{2}\right]}{\sigma^{2}}=\frac{1}{\sigma^{2}}$.
Le nombre d'antennes qui équipent la BS a été fixé à $L = 12$, et le nombre d'utilisateurs à $K = 6$.
Pour une certaine répartition des utilisateurs, HyperMIMO a été entraîné en échantillonnant de manière aléatoire les matrices de canal $\Hm$, les SNRs dans la plage [0,10]~\si{dB}, et les symboles d'une constellation de modulation QPSK pour chaque utilisateur.

La Figure~\ref{fig:chp1_snr_fr} montre le taux d'erreur de symboles (symbol error rate, SER) obtenu par HyperMIMO, LMMSE, OAMPNet avec 10 itérations, MMNet avec 10 itérations et entraîné pour chaque réalisation de canal, et le détecteur de maximum de vraisemblance.
Comme prévu, MMNet, lorsqu'il est entraîné pour chaque réalisation de canal, atteint une performance proche de celle du maximum de vraisemblance.
HyperMIMO atteint un SER légèrement supérieur à MMNet, mais inférieur à OAMPNet et LMMSE.
La robustesse d'HyperMIMO à la mobilité des utilisateurs a été testée en évaluant le SER obtenu lorsque les utilisateurs subissent une mobilité angulaire (Figure~\ref{fig:chp1_angles_fr}) ou se déplacent dans des directions 2D aléatoires (Figure~\ref{fig:chp1_distance_fr}) à partir des positions pour lesquelles le système a été entraîné.
On peut voir que le SER obtenu par HyperMIMO se dégrade gracieusement lorsque le déplacement angulaire augmente, et ne devient jamais pire que LMMSE ou OAMPNet.
De même, le SER obtenu par HyperMIMO se dégrade gracieusement lorsque la distance de déplacement augmente.
Ces résultats montrent que, bien qu'ayant été entraîné pour un ensemble particulier de positions d'utilisateurs, HyperMIMO reste relativement robuste à la mobilité.

\bigskip
\tocless\subsection{Conclusion}

Dans cette section, nous avons proposé de tirer parti de l'idée des hyper-réseaux pour éviter de devoir réentraîner un détecteur MMNet pour chaque réalisation de canal tout en obtenant des performances compétitives.
Pour réduire la complexité de l'hyper-réseau, MMNet a été modifié pour diminuer le nombre de paramètres entraînables, et une forme de partage des poids a été utilisée.
Les simulations ont révélé que l'architecture HyperMIMO résultante atteint des performances proches de l'état de l'art dans des canaux hautement corrélés lorsqu'elle est entraînée et évaluée avec le même nombre d'utilisateurs et avec des statistiques de canal fixes.
Cependant, bien que ces détecteurs basés sur des NNs représentent des améliorations prometteuses par rapport aux algorithmes traditionnels, leur optimisation basée sur les blocs ne garantit toujours pas l'optimalité globale du récepteur.
Un autre inconvénient est que la plupart des détecteurs basés sur des NNs nécessitent des estimations parfaites du canal lors de l'entraînement, mais elles ne sont généralement pas disponibles dans la pratique.
Pour ces raisons, nous nous concentrons dans le chapitre suivant sur l'optimisation de bout en bout du couple émetteur-récepteur, mais uniquement pour les systèmes à entrée unique et sortie unique (single-input single-output, SISO).

\newpage
\tocless\section{Apprentissage de Formes d'Onde OFDM avec des Contraintes de PAPR et d'ACLR}

\tocless\subsection{Motivation} 

Le multiplexage spatial ne peut pas être exploité dans les systèmes SISO, et d'autres approches doivent donc être envisagées. 
L'une d'entre elles consiste à concevoir de nouvelles formes d'onde qui répondent à des exigences plus strictes concernant les caractéristiques du signal. 
Parmi les solutions possibles, l'OFDM est déjà utilisé dans la plupart des systèmes de communication modernes, mais il souffre d'un PAPR et d'un ACLR élevés, ce qui pourrait entraver son utilisation dans les systèmes post-5G.
Comme les futures BS et les équipements utilisateurs devraient être équipés d'accélérateurs DL dédiés, de nombreux travaux ont proposé de concevoir des émetteurs-récepteurs basés sur des NNs et destinés à différents canaux. 
Par exemple, l'apprentissage des géométries de constellation pour réaliser des communications sans pilote et sans préfixe cyclique (cyclic prefix, CP) sur des canaux OFDM a été effectué dans~\cite{pilotless20}, et la conception de systèmes basés sur des NNs pour les canaux multiporteuse avec \emph{fading} et les communications par fibre optique ont été étudiés respectivement dans~\cite{9271932} et~\cite{8433895}.

Dans ce qui suit, nous utilisons la stratégie d'optimisation de bout en bout pour concevoir des formes d'onde OFDM, approche que nous avons publié dans~\cite{goutay2021learning}.
Plus précisément, cette stratégie est basée, d'une part, sur des émetteurs-récepteurs basés sur des réseaux de neurones convolutionels (convolutional neural networks, CNNs) et, d'autre part, sur une procédure d'apprentissage qui permet à la fois de maximiser un taux d'information réalisable tout en satisfaisant des contraintes de PAPR et d'ACLR. 
Le système de bout en bout est comparé à une implémentation presque idéale d'un système à réservation de tonalité (tone reservation, TR), dans lequel un certain nombre de sous-porteuses sont utilisées pour générer des signaux de réduction de puissance de crête. 
Les deux systèmes sont évalués sur un modèle de canal conforme aux modèles 3GPP. 
Les résultats de l'évaluation montrent que le système basé sur l'apprentissage permet d'atteindre les objectifs de PAPR et d'ACLR et tout en permettant des gains de débit allant de 3\% à 30\% par rapport au système de référence avec des caractéristiques similaires.
À notre connaissance, cette méthode est la première approche basée sur du DL qui permet à la fois de maximiser un taux d'information des transmissions OFDM et de fixer des objectifs de PAPR et d'ACLR.

\bigskip
\tocless\subsection{Description du Problème}

\begin{figure}
    \centering
    \includegraphics[width=0.27\textwidth]{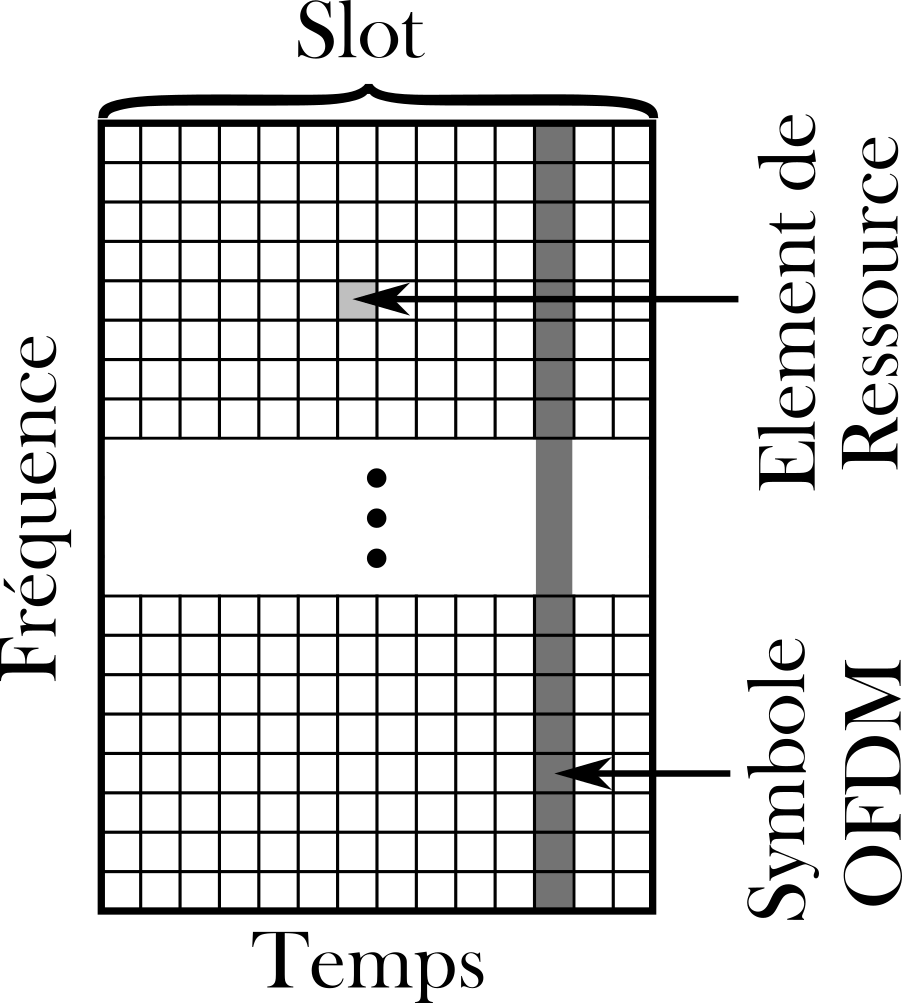}
    \caption{Une grille de ressources OFDM.}
    \label{fig:bkg_rg_fr_fr}
\end{figure}
 
Un canal OFDM est considéré, avec $N$ sous-porteuses et un \emph{slot} temporel, qui se compose de $M = 14 $ symboles OFDM adjacents (Figure~\ref{fig:bkg_rg_fr_fr}). Dans cette section, les sous-porteuses sont indexées par l'ensemble $\mathcal{N}= \LP -\frac{N -1}{2}, \cdots, \frac{N -1}{2} \RP $, $N$ étant supposé impair par commodité. 
Si l'on considère la grille de ressources (resource grid, RG) en entier, le canal OFDM peut être exprimé comme suit
\begin{align}
    \Ym = \Hm \odot \Xm + \Nm
    \label{eq:chp2_OFDM_channel_fr}
\end{align}
où $\Xm \in \mathbb{C}^{M \times N}$ et $\Ym \in \mathbb{C}^{M \times N}$ représentent respectivement la matrice des symboles en bande de fréquence de base (frequency baseband symbol, FBS) envoyés et reçus,  $\Hm \in \mathbb{C}^{M \times N}$ est la matrice des coefficients de canal, et {$\Nm~\in~\mathbb{C}^{M \times N}$} est la matrice de bruit gaussien additif telle que chaque élément a une variance $\sigma^2$. 
Nous considérons un environnement qui varie lentement de sorte que le canal peut être supposé constant sur la durée d'un slot. 
La matrice des bits à transmettre sur le symbole OFDM m est notée $\Bm_{m} = \left[  \bv_{m,1}, \cdots, \bv_{m, N} \right]\tp$, où $\bv_{m, n}\in \{0,1\}^{Q}, 1 \leq m \leq M, 1 \leq n \leq N$, est un vecteur de bits à envoyer et $Q$ est le nombre de bits par utilisation du canal. 
L'émetteur module chaque $\Bm_{m}$ sur les FBS $\xv_{m} \in \CC^{N}$, qui sont mappés sur les sous-porteuses orthogonales. Lorsque les CPs d'une durée $T^{\text{CP}}$  sont ajoutés aux symboles OFDM de durée $T$, le spectre du signal devient
\begin{align}
    \label{eq:chp2_s_cp_fr}
    S_{m}^{\text{CP}}(f) = \sum_{n \in \mathcal{N}} x_{m, n} \frac{1}{\sqrt{\Delta_f^{\text{CP}}}}\text{sinc} \left( \frac{f-n\Delta_f}{\Delta_f^{\text{CP}}} \right)
\end{align}
où $\Delta_f$ est l'espacement entre les sous-porteuses, et $\Delta_f^{\text{CP}} = \frac{1}{T+T^{\text{CP}}}$.
Le signal correspondant est 
\begin{align}
    s(t)=\sum_{m=0}^{M-1} s_{m}(t)=\sum_{m=0}^{M-1} \sum_{n _in \mathcal{N}} x_{m,n} \phi_{n}(t-mT^{\text{tot}})
\end{align}
où $T^{\text{tot}} = T+T^{\text{CP}}$ et les filtres de transmission $\phi_{n}(t), n\in \Nc $ sont définis comme :
\begin{align}
    \phi_{n}(t) = \frac{1}{\sqrt{T^{\text{tot}}}} \text{rect}\LB \frac{t}{T^{\text{tot}}} -\frac{1}{2} \RB e^{j 2 \pi n \frac{ t-T^{\text{CP}}}{T}}.
\end{align}

Les formes d'onde OFDM présentent, entre autres, deux inconvénients majeurs.
Le premier est leurs pics de forte amplitude, qui créent des distorsions dans le signal de sortie en raison de la saturation des amplificateurs de puissance. 
Désignons par $\nu(t) = \frac{|s(t)|^2}{\EE \left[|s(t)|^2 \right]} $ le rapport entre la puissance instantanée et la puissance moyenne d'un signal. 
Nous définissons le $\text{PAPR}_{\epsilon}$ comme le plus petit $e\geq 0$, tel que la probabilité que $\nu(t)$ soit plus grand que $e$ soit inférieure à un seuil $\epsilon \in \left( 0, 1 \right)$ :
\begin{align}
    \label{eq:chp2_papr_fr}
    \text{PAPR}_{\epsilon}  \coloneqq \mathrm{min} \; e, \;\; \text{s. t.} \;\;  P\left(\nu(t) > e  \right) \leq \epsilon.
\end{align}
Le réglage $\epsilon=0$ conduit à la définition plus conventionnelle du PAPR $\frac{\max{|s(t)|^2}}{\EE \left[|s(t)|^2 \right]}$.

Le deuxième inconvénient de l'OFDM est son faible confinement spectral.
Cette caractéristique est généralement mesurée par l'ACLR, qui est le rapport entre l'espérance de l'énergie hors bande $\EE_{\xv_m}\left[ E_{O_m} \right]$ et l'eespérance de l'énergie dans la bande $\EE_{\xv_m}\left[ E_{I_m}\right]$ :
\begin{align}
    \label{eq:chp2_aclr_fr}
    \text{ACLR} \coloneqq  \frac{\EE_{\xv_m} \left[ E_{O_m} \right]}{\EE_{\xv_m} \left[ E_{I_ m}\right]} 
     =   \frac{\EE_{\xv_m} \left[ E_{A_m} \right]}{\EE_{\xv_m} \left[ E_{I_m} \right]}-1 
\end{align}
où $E_{O_m}$, $E_{I_m}$, et $E_{A_m} = E_{O_m} + E_{I_m}$  sont respectivement l'énergie hors bande, dans la bande et l'énergie totale du symbole OFDM $m$.
L'énergie dans la bande $E_{I_m}$ est donnée par
\begin{equation}
\begin{split}
E_{I_m} & \coloneqq \int_{-\frac{N \Delta_f}{2}}^{\frac{N \Delta_f}{2}} \left| S_m (f)\right|^2 df = \xv_m ^H \Jm \xv_m\\
\end{split} 
\end{equation}
où chaque élément $j_{a, b}$ de la matrice $\Jm \in \RR^{N \times N}$ est
\begin{equation}
j_{a, b} = \frac{1}{\Delta_f^{\text{CP}}} \int_{-\frac{N \Delta_f}{2}}^{\frac{N \Delta_f}{2}}  \text{sinc} \left(\frac{f - a \Delta_f}{\Delta_f^{\text{CP}}} \right) \text{sinc} \left(\frac{f - b \Delta_f}{\Delta_f^{\text{CP}}} \right) df, \quad a, b \in \Nc.
\end{equation}
Enfin, l'énergie totale peut être calculée plus facilement dans le domaine temporel :
\begin{align}
E_{A_m} \coloneqq \int_{ -\frac{T}{2}}^{\frac{T}{2}} \left| s (t)\right|^2 dt = \xv_m ^H \Km \xv_m
\end{align}
où $\Km \in \RR^{N \times N}$ a pour éléments
\begin{equation}
k_{a, b} = \frac{1}{T} \int_{-\frac{T}{2}}^{\frac{T}{2}}  e^{i 2 \pi (a-b) t /T} dt, \quad a, b \in \Nc .
\end{equation}

\subsubsection{\textit{\textmd{\selectfont{Système de référence}}}}

L'une des techniques permettant de réduire le PAPR d'un signal OFDM est la TR, dans laquelle un sous-ensemble de tonalités (sous-porteuses) $R$ est utilisé pour créer des signaux de réduction de crête.
Ces sous-porteuses sont appelées tonalités de réduction de crête (peak reduction tones, PRTs), et les sous-porteuses restantes sont utilisées pour la transmission des données et des pilotes.
Dans cette thèse, nous utilisons un solveur complexe pour trouver les signaux de réduction de crête optimaux pour chaque symbole OFDM. 
Plus de détails le systeme de référence utilisant la TR peuvent être trouvés dans la Section~4.2.2.

\tocless\subsection{Apprentissage d'une Modulation de Grande Dimension} 

\begin{figure}[t]
    \centering
    \includegraphics[width=0.55\textwidth]{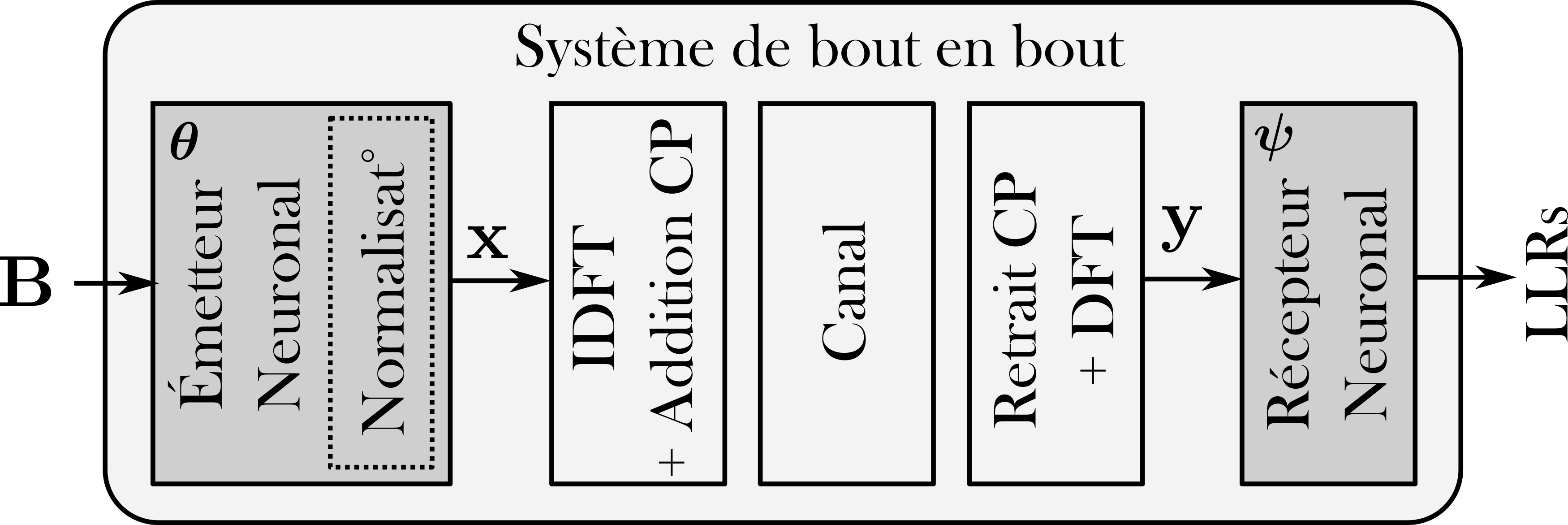}
    \caption{Système entraînable, où les blocs grisés représentent les composants entraînables.}
    \label{fig:chp2_E2E_system_fr}
\end{figure}


Dans ce qui suit, nous formons un émetteur et un récepteur basés sur des NNs de bout en bout.
Ce système est appelé système \emph{end-to-end}, ou "E2E" par souci de concision, et est représenté schématiquement sur la Figure~B.9.
Nous cherchons à trouver une modulation de grande dimension et un détecteur associé qui maximisent le taux d'information pour la transmission et satisfont des contraintes sur le PAPR et l'ACLR du signal.
L'émetteur et le récepteur du système fonctionnent au-dessus de l'OFDM, c'est-à-dire qu'une IDFT (DFT) est effectuée et qu'un préfixe cyclique est ajouté (supprimé) avant (après) la transmission (voir la Figure~B.9). 
Nous considérons un taux d'information réalisable \cite{pilotless20} qui dépend des paramètres entraînables de l'émetteur et du récepteur respectivement désignés par $\thetav$ et $\psiv$ :
\begin{align}
    \label{eq:chp2_rate0_fr}
    C(\thetav, \psiv) & = \frac{1}{MN}\sum_{m \in \mathcal{M}}\sum_{n \in \mathcal{N}}\sum_{q=0}^{Q-1}I\left( b_{m,n,q}; \yv_m \right | \thetav)\\
    & - \frac{1}{MN}\sum_{m\in\mathcal{M}}\sum_{n \in \mathcal{N}}\sum_{q=0}^{Q-1}\EE_{\yv_m}\left[\text{D}_\text{KL} \left( P (b_{m,n,q}|\yv_m)|| \widehat{P}_{\psiv}(b_{m,n,q}| \yv_m ) \right)\right] \nonumber
\end{align}
où le premier terme est l'information mutuelle entre $b_{m,n,q}$ et $\yv_m$, et le deuxième terme est la divergence K-L entre entre les vraies probabilités sur les bits et celles estimées par le démappeur basé sur un NN.
Pour garantir une énergie moyenne unitaire par élément de ressource (resource element, RE), une couche de normalisation est ajoutée à l'émetteur (voir la Figure~\ref{fig:chp2_E2E_system_fr}). 
Nous pouvons désormais formuler le problème d'optimisation sous contrainte que nous voulons résoudre :
\begin{subequations}
    \label{eq:chp2_rate_fr}
     \begin{align}
    \underset{\thetav, \psiv}{\text{maximize}} & \quad\quad C(\thetav, \psiv) \label{eq:chp2_rate1_fr} \\
   \text{subject to}  & \quad\quad \text{PAPR}_{\epsilon}(\thetav) = \gamma_{\text{peak}} \label{eq:chp2_rate3_fr} \\
    &  \quad\quad \text{ACLR}(\thetav) \leq \beta_{\text{leak}} \label{eq:chp2_rate4_fr} 
     \end{align}
\end{subequations}
où $\gpeak$ et $\bleak$ désignent respectivement le PAPR et l'ACLR cibles. 
Notez que le PAPR et l'ACLR dépendent des paramètres de l'émetteur $\thetav$.

L'un des principaux avantages d'une implémentation de la paire émetteur-récepteur en tant que système E2E est qu'elle permet l'optimisation des paramètres entraînables par descente de gradient stochastique (stochastic gradient descent, SGD).
Cela nécessite une fonction de perte différentiable afin que les gradients puissent être calculés et rétro-propagés à travers le système E2E.
Dans ce qui suit, nous exprimons donc l'objectif (B.17a) et les contraintes~(B.17b) et~(B.17c) comme des fonctions différentiables qui peuvent être évaluées pendant l'apprentissage et minimisées avec la SGD.
Tout d'abord, le taux réalisable (B.17a) peut être exprimé de manière équivalente en utilisant le système d'entropie croisée binaire (binary cross-entropy, BCE)~\cite{9118963}, qui est souvent utilisé dans ce genre de problèmes de classification binaire :
\begin{align}
    \label{eq:chp2_CE_fr}
    L_C(\thetav, \psiv) &:= - \frac{1}{MN}\sum_{m\in\mathcal{M}}\sum_{n\in\mathcal{N}}\sum_{q=0}^{Q-1} \EE_{\yv_m} \left[ \text{log}_2 \left(\widehat{P}_{\psiv} (b_{m,n,q}| \yv_m ) \right) \right]  \\
    & = Q -  C(\thetav, \psiv). 
\end{align}
Pour surmonter la complexité associée au calcul de la valeur attendue, une approximation est généralement obtenue par un échantillonnage de Monte Carlo :
\begin{align}
    \label{eq:chp2_CE_batch_fr}
    L_C(\thetav, \psiv) \approx & - \frac{1}{MNB_S}\sum_{m\in\mathcal{M}}\sum_{n\in\mathcal{N}}\sum_{q=0}^{Q-1}\sum_{i=0}^{B_S-1}  \text{log}_2 \left( \widehat{P}_{\psiv} \left( b_{m,n,q}^{[i]}| \yv_m ^{[i]} \right) \right).
\end{align}

Deuxièmement, l'évaluation de la contrainte (B.17b) nécessite le calcul de la probabilité $P(\frac{|s(t)|^2}{\EE \left[|s(t)|^2 \right]}  > e)$, où $e$ est le seuil d'énergie défini dans (B.10). 
Cependant, le calcul de cette probabilité est d'une complexité prohibitive en raison du grand nombre de symboles OFDM possibles. 
Pendant l'apprentissage, nous appliquons donc cette contrainte  avec $\epsilon=0$ et en pénalisant tous les signaux dont l'amplitude au carré dépasse $\gpeak$.
Avec $\epsilon=0$, la contrainte (B.17b) est équivalente à imposer $L_{\gamma_{\text{peak}}}(\thetav) =0$, avec
\begin{align}
    \label{eq:chp2_loss_papr_fr}
    L_{\gamma_{\text{peak}}}(\thetav)  = \EE_m \left[ \int_{-\frac{T}{2}}^{\frac{T}{2}} \left(|s_m(t)|^2-\gamma_{\text{peak}} \right)^+ dt \right]
\end{align}
où $(x)^+$  désigne la partie positive de $x$, c'est-à-dire $ (x)^+= \text{max}(0, x)$.
Pour évaluer $L_{\gamma_{\text{peak}}}(\thetav)$ pendant l'entraînement, la valeur de l'espérance peut être obtenue par échantillonnage de Monte Carlo, et l'intégrale peut être approximée en utilisant une somme de Riemann :
\begin{align}
    L_{\gamma_{\text{peak}}}(\thetav)  \approx \frac{T}{B_S N O_S} \sum_{i=0}^{B_S-1} \sum_{t=-\frac{NO_S-1}{2}}^{\frac{NO_S-1}{2}}  \left(\left| \underline{z}_{m,t}^{[i]} \right| ^2 - \gamma_{\text{peak}} \right)^+
    \label{eq:chp2_loss_papr2_fr}
\end{align}
où $\underline{\zv}_m = \Fm\htp\xv_m \in \CC^{NO_S}$ est le vecteur du signal temporel suréchantillonné correspondant à la sortie $\xv_m$ de l'émetteur neuronal, avec $\Fm\htp \in \CC^{N O_S \times N}$ la matrice d'IDFT.

Troisièmement, la contrainte d'inégalité (B.17c) peut être convertie en une contrainte d'égalité $\text{ACLR}(\thetav) - \beta_{\text{leak}}  = -v$, où $v\in\RR_+$ est une variable libre (\emph{slack variable}).
Cette contrainte d'égalité est alors appliquée en minimisant $L_{\beta_{\text{leak}}}(\thetav) + v$, avec
\begin{align}
L_{\beta_{\text{leak}}}(\thetav)  & =  \frac{ \EE \left[ E_A \right]}{ \EE \left[ E_I \right]}-1   - \beta_{\text{leak}} \\
& \approx \frac{  \frac{1}{B_S} \sum_{i=0}^{B_S-1}  \xv^{[i]^{\mathsf{H}}} \Wm \xv^{[i]}}{ \frac{1}{B_S} \sum_{i=0}^{B_S-1}  \xv^{[i]^{\mathsf{H}}} \Vm \xv^{[i]}} -1   - \beta_{\text{leak}} .
\end{align}

Enfin, pour $\epsilon=0$, le problème (B.17) peut être reformulé comme suit
\begin{subequations}
    \label{eq:chp2_pb_fr}
     \begin{align}
    \underset{\thetav, \psiv}{\text{minimize}} & \quad\quad L_C(\thetav, \psiv) \label{eq:chp2_pb1_fr} \\
    \text{subject to} & \quad\quad L_{\gamma_{\text{peak}}}(\thetav)  = 0 \label{eq:chp2_pb2_fr} \\
    &  \quad\quad L_{\beta_{\text{leak}}}(\thetav) + v = 0 \label{eq:chp2_pb3_fr} 
     \end{align}
\end{subequations}
où l'objectif et les contraintes sont différentiables et peuvent être estimés lors de l'entraînement. 
Pour entraîner le système de bout en bout, nous utilisons ensuite la méthode du Lagrangien augmentée, qui utilise deux types d'hyperparamètres qui sont mis à jour de manière itérative pendant l'entraînement. 
Désignons par  $\mu_p > 0$ et $\mu_l>0$ les paramètres de pénalité et par $\lambda_p$ et $ \lambda_l$ les multiplicateurs de Lagrange pour les fonctions de contrainte $L_{\gamma_{\text{peak}}}(\thetav)$ et $L_{\beta_{\text{leak}}}(\thetav)$, respectivement.
Le Lagrangien augmenté correspondant est défini comme suit~\cite{bertsekas2014constrained}
\begin{align}
    \label{eq:chp2_lagrange_fr}
    \overline{L} (\thetav, \psiv, \lambda_p, \lambda_l, & \mu_p, \mu_l)  = L_C(\thetav, \psiv) \nonumber \\
    & + \lambda_p L_{\gamma_{\text{peak}}}(\thetav) + \frac{1}{2} \mu_p |L_{\gamma_{\text{peak}}}(\thetav)|^2 \\
    & + \frac{1}{2\mu_l} \left( \text{max}(0, \lambda_l + \mu_l L_{\beta_{\text{leak}}}(\thetav) )^2 - \lambda_l^2 \right) \nonumber
\end{align}
où la minimisation par rapport à $v$ a été effectuée pour chaque paire fixe de $\{\thetav, \psiv\}$~\cite{bertsekas2014constrained}.
Chaque itération d'apprentissage comprend de multiples étapes de SGD sur le Lagrangien augmenté (B.26) suivies d'une mise à jour des hyperparamètres. 
La procédure d'optimisation est détaillée dans l'Algorithme 2 présent dans la Section~4.3.1.

\begin{figure}
    \centering
    \begin{subfigure}{.45\textwidth}
        \centering
        \includegraphics[height=170pt]{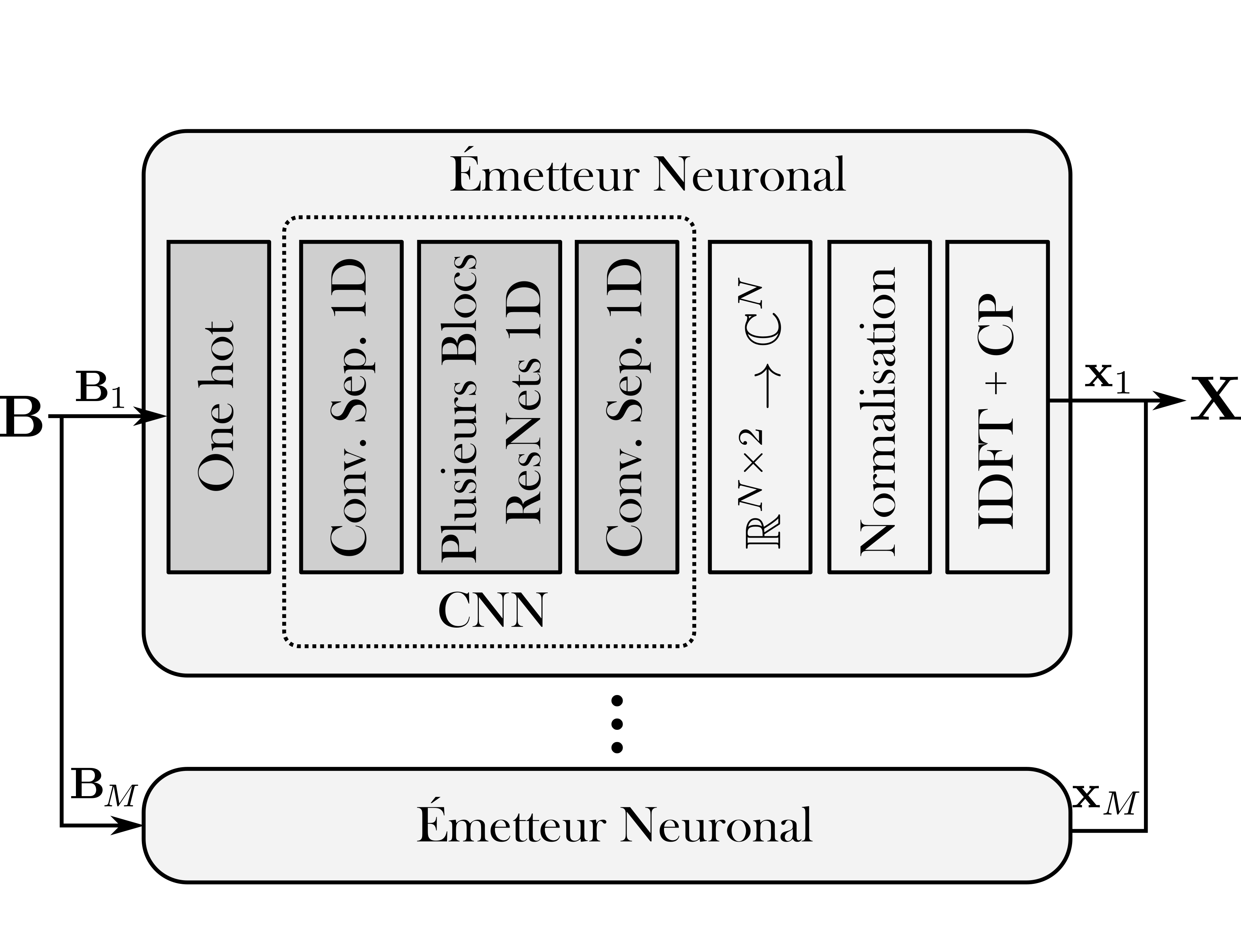}
        \caption{Transmetteur neuronal.}
        \label{fig:chp2_nn_tx_fr}
      \end{subfigure}
      \hfill
      \begin{subfigure}{.2\textwidth}
        \centering
        \includegraphics[height=170pt]{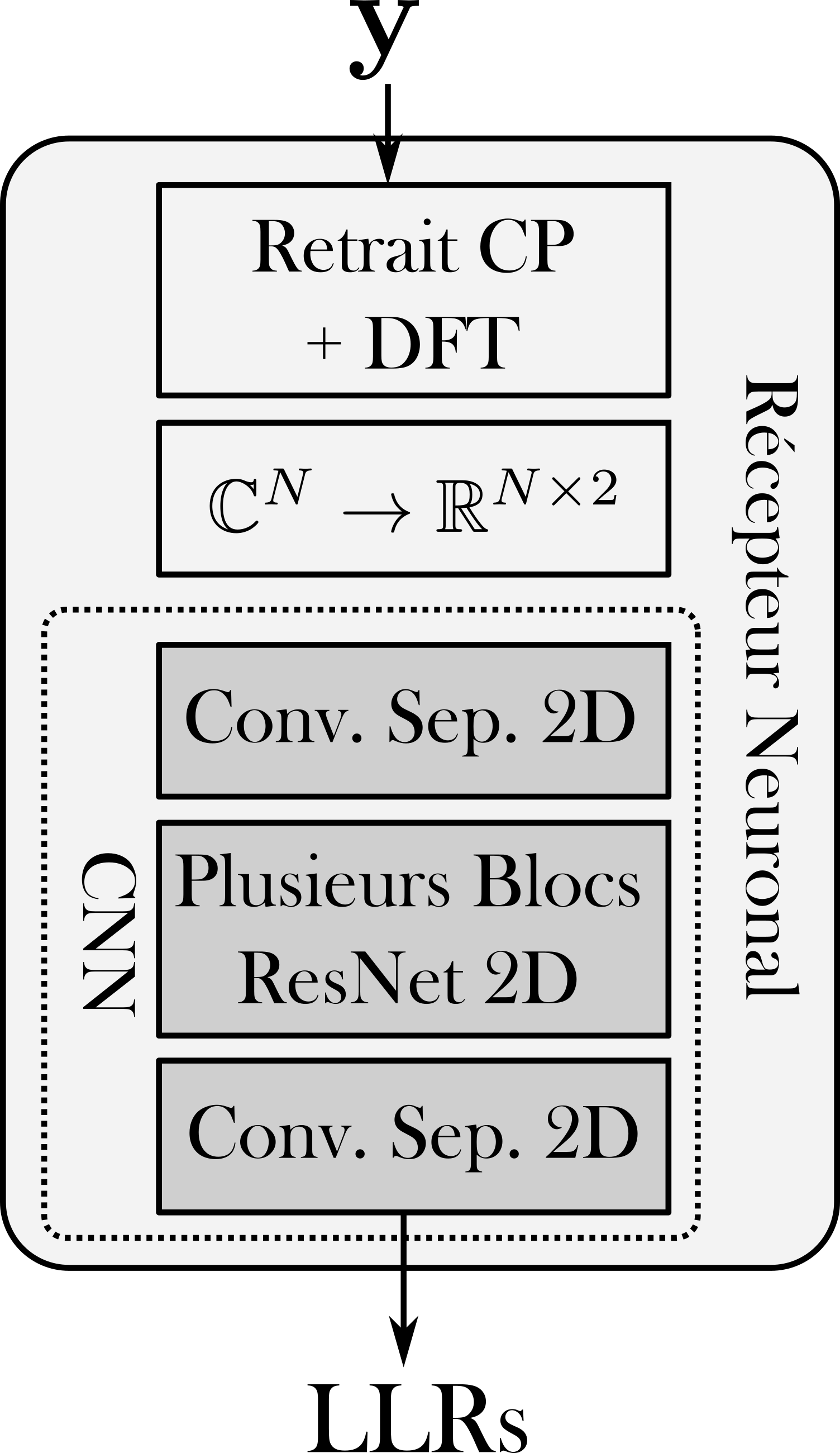}
        \caption{Récepteur neuronal.}
        \label{fig:chp2_nn_rx_fr}
      \end{subfigure}%
      \hfill
    \begin{subfigure}{.2\textwidth}
      \centering
      \includegraphics[height=170pt]{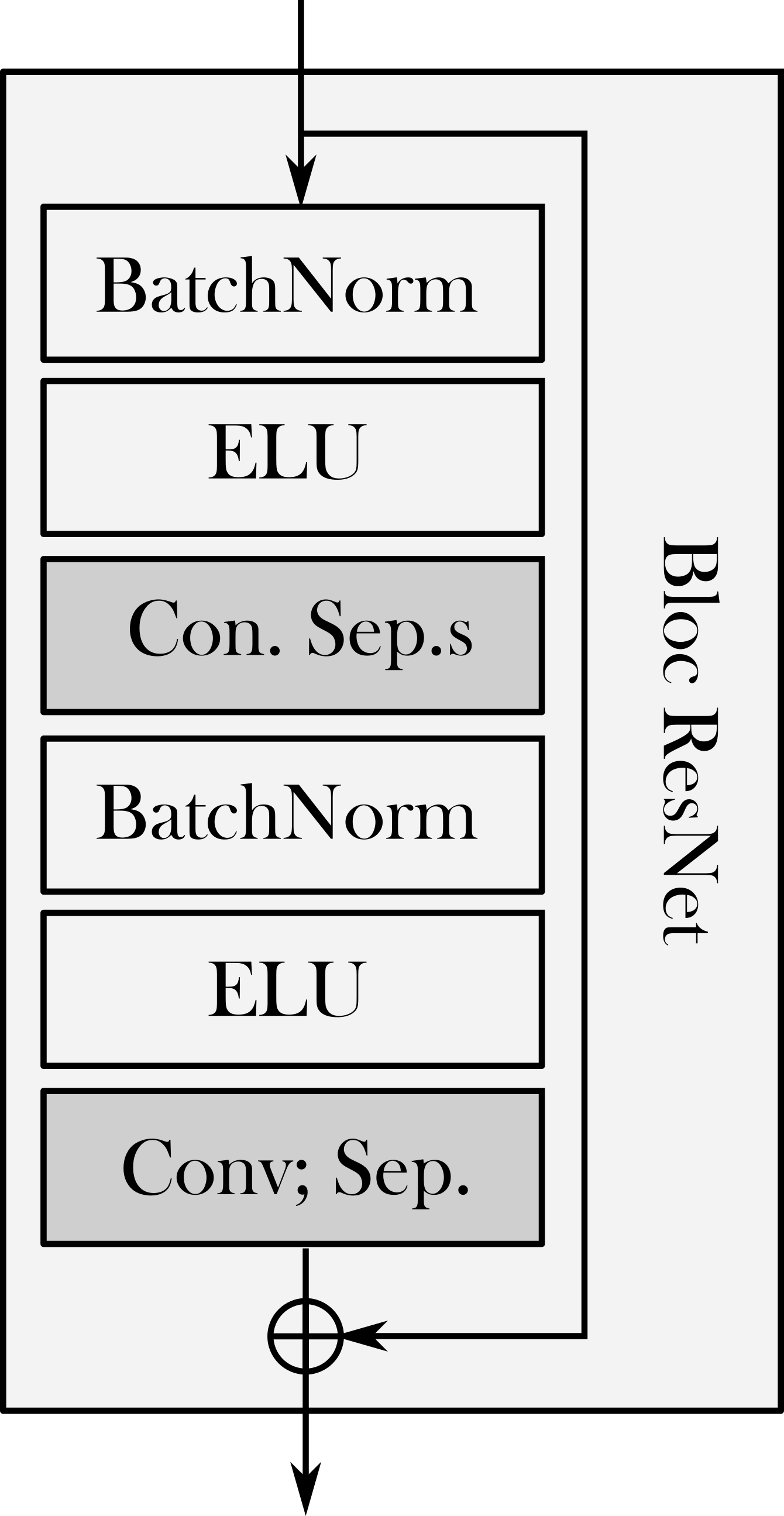}
      \caption{Bloc ResNet.}
      \label{fig:chp2_resnet_fr}
    \end{subfigure}%
    
    \caption{Différentes parties du système de bout en bout, où les blocs grisés sont des composants entraînables.}
    \label{fig:chp2_components_fr}
\end{figure}

\subsubsection{\textit{\textmd{\selectfont{Architecture du système}}}}

L'émetteur et le récepteur neuronaux sont basés sur des architectures similaires, représentées schématiquement dans la Figure~B.10. 
L'élément central est un bloc \emph{ResNet}, qui est constitué de deux séquences identiques de couches suivies de l'ajout de l'entrée, comme illustré dans la Figure~B.10c. 
Chaque séquence est composée d'une couche de normalisation par batch, d'une fonction d'activation ELU et d'une convolution séparable.
Les architectures précises de l'émetteur et du récepteur basés sur des CNNs sont détaillées dans la Section~4.3.2. 
Notez qu'aucun pilote n'est utilisé car il a été démontré dans~\cite{pilotless20} que la communication sans pilote est possible sur les canaux OFDM lorsque des récepteurs neuronaux sont utilisés.

\tocless\subsection{Résultats des Simulations}

Des ensembles de données distincts ont été utilisés pour l'entraînement et le test, tous deux générés à l'aide d'un mélange de modèles à visibilité directe (line-of-sight, LOS) et à visibilité indirecte (non-LOS) conformes aux normes 3GPP-UMi (urban microcell).
Nous avons supposé un contrôle parfait de la puissance, de sorte que l'énergie moyenne du canal par RE était de un, c'est-à-dire $\EE \left[ |h_{m, n}|^2 \right]= 1$.
Le système considéré comporte $N = 75$ sous-porteuses avec $M = 14$ symboles OFDM en utilisant des CPs de longueurs suffisantes pour que le canal puisse être représenté par (B.6) dans le domaine fréquentiel.
La fréquence centrale de la porteuse et l'espacement entre les sous-porteuses ont été fixés respectivement à 3.5~\si{GHz} et 30~\si{kHz}, et le nombre de bits par utilisation du canal a été fixé à $Q = 4$.
Les comparaisons du taux d'erreur sur les bits codés (coded BER) ont été effectuées à l'aide d'un code LDPC (low-density parity-check) standard conforme à la norme 5G, d'une longueur de 1024 et de taux de codage $\eta = \frac{1}{2}$.
Le système TR de référence a été évalué pour $R\in\{0, 2, 4, 8, 16, 32\}$ PRT et le système E2E a été entraîné pour atteindre des ACLRs de $\bleak \in \{-20, -30, -40\}$~\si{dB} et des PAPRs de $\gpeak \in \{ 4, 5, 6, 7, 8, 9 \}$~\si{dB}.
Finallement, le $\text{SNR} =  \frac{\EE_{h_{m,n}} \left[ |h_{m,n}|^2 \right]}{\sigma^2}  =  \frac{1}{\sigma^2}$ a été choisi au hasard dans l'intervalle $[10, 30]$~\si{dB} pour chaque RG du batch pendant l'entraînement.

Les taux de transmission moyens par RE obtenus par le système de référence et les systèmes E2E sont présentés dans la Figure~B.11 où les chiffres à côté des points de données sont les ACLR correspondants. 
Tout d'abord, on peut voir qu'au PAPR maximal d'environ $8.5$~\si{dB}, le système E2E entraîné avec $\bleak = -20$~\si{dB} atteint un débit supérieur de 3\% à celui du système de référence sans PRTs. 
Cela peut s'expliquer par la perte de débit due à la présence de pilotes dans le système de référence, qui ne transportent pas de données et représentent environ 4\% du nombre total de REs. 
Ensuite, à des PAPR plus faibles, les taux de transmisison obtenus par le système E2E entraîné avec $\bleak\in\{-20, -30\}$ sont significativement plus élevés que ceux obtenus par le système de référence. 
Enfin, les systèmes E2E sont capables d'atteindre leurs objectifs respectifs de PAPR et d'ACLR.

\begin{figure}[t]
	\centering
	\input{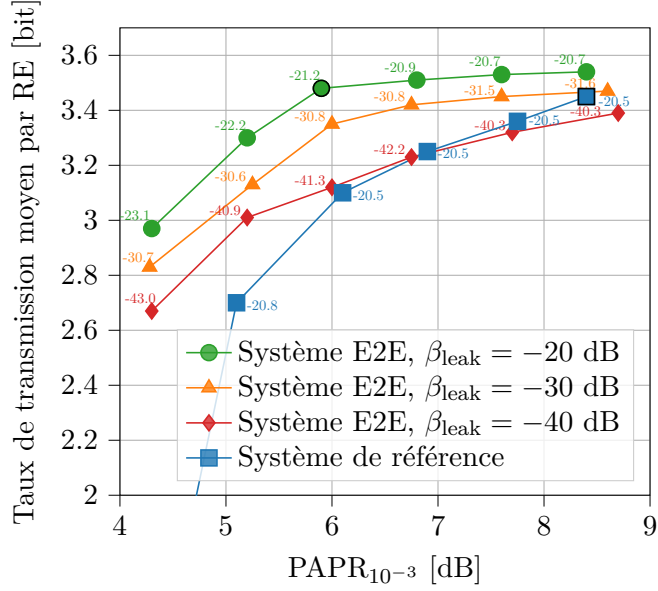}
	\caption{Taux de transmission obtenus pour les systèmes comparés. Les nombres près des points de données indiquent les ACLRs correspondants.}
	\label{fig:chp2_rates_fr}
\end{figure}

\begin{figure}
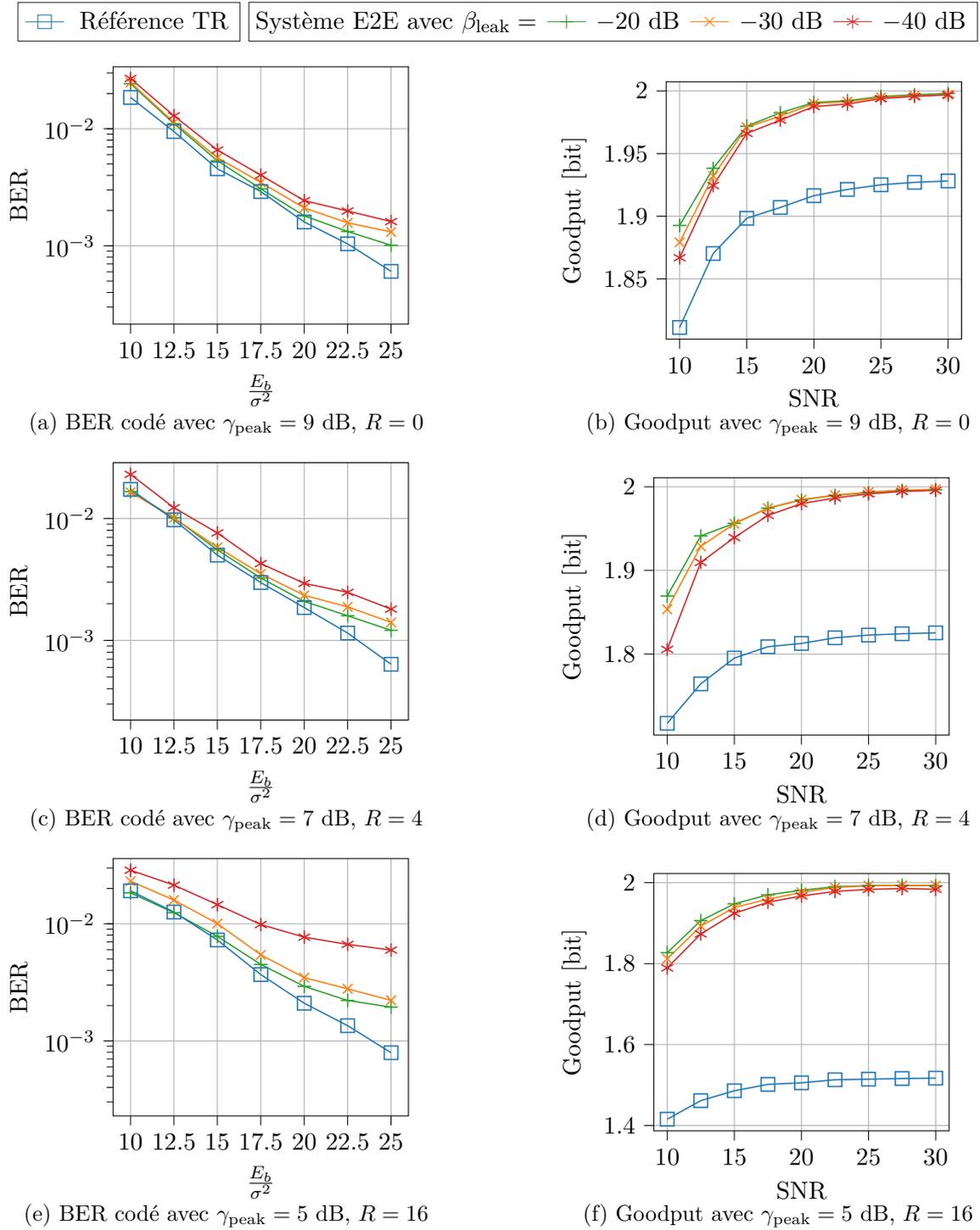


	\centering

	\begin{subfigure}[b]{0.2\textwidth}
		\begin{tikzpicture} 
		
			\definecolor{color0}{rgb}{0.12156862745098,0.466666666666667,0.705882352941177}
			\definecolor{color1}{rgb}{1,0.498039215686275,0.0549019607843137}
			\definecolor{color2}{rgb}{0.172549019607843,0.627450980392157,0.172549019607843}
			\definecolor{color3}{rgb}{0.83921568627451,0.152941176470588,0.156862745098039}
		
				\begin{axis}[%
				hide axis,
				xmin=10,
				xmax=50,
				ymin=0,
				ymax=0.4,
				legend columns=4, 
				legend style={draw=white!15!black,legend cell align=left,column sep=1.0ex}
				]
				\addlegendimage{color0, mark=square, mark size=3}
				\addlegendentry{Référence TR};
				\end{axis}
		\end{tikzpicture}
	\end{subfigure}
	\hspace{1pt}
	\begin{subfigure}[b]{0.75\textwidth}
	\centering
	\begin{tikzpicture} 
	
	\definecolor{color0}{rgb}{0.12156862745098,0.466666666666667,0.705882352941177}
	\definecolor{color1}{rgb}{1,0.498039215686275,0.0549019607843137}
	\definecolor{color2}{rgb}{0.172549019607843,0.627450980392157,0.172549019607843}
	\definecolor{color3}{rgb}{0.83921568627451,0.152941176470588,0.156862745098039}

    	\begin{axis}[%
    	hide axis,
    	xmin=10,
   	 xmax=50,
    	ymin=0,
    	ymax=0.4,
    	legend columns=4, 
    	legend style={draw=white!15!black,legend cell align=left,column sep=.5ex}
    	]
    	\addlegendimage{white ,mark=*, mark size=2}
    	\addlegendentry{\hspace{-0.9cm} Système E2E avec $\bleak=$};
    	\addlegendimage{color2, mark=+, mark size=3}
    	\addlegendentry{$-20$~\si{dB}};
    	\addlegendimage{color1, mark=x, mark size=3}
    	\addlegendentry{$-30$~\si{dB}};
    	\addlegendimage{color3, mark=asterisk, mark size=3}
    	\addlegendentry{$-40$~\si{dB}};
    	\end{axis}
	\end{tikzpicture}
\end{subfigure}

	\vspace{10pt}

	  	\begin{subfigure}[b]{0.45\textwidth}
	  		\input{Chapter2/eval_fr/ber_20.tex}
			\vspace{-8pt}
	  		\caption{BER codé avec $\gpeak=9$~\si{dB}, $R=0$}
	  		\label{fig:chp2_ber_1_fr}
		\end{subfigure}%
		\hfill
		\begin{subfigure}[b]{0.45\textwidth}
			\input{Chapter2/eval_fr/gp_20.tex}
			\vspace{-8pt}
			\caption{Goodput avec $\gpeak=9$~\si{dB}, $R=0$}
			\label{fig:chp2_goodput_1_fr}
		\end{subfigure}

		\vspace{10pt}

		\begin{subfigure}[b]{0.45\textwidth}
			\input{Chapter2/eval_fr/ber_30.tex}
			\vspace{-8pt}
			\caption{BER codé avec $\gpeak=7$~\si{dB}, $R=4$}
			\label{fig:chp2_ber_2_fr}
		\end{subfigure}%
		\hfill
	  	\begin{subfigure}[b]{0.45\textwidth}
	  		\input{Chapter2/eval_fr/gp_30.tex}
			\vspace{-8pt}
			\caption{Goodput avec $\gpeak=7$~\si{dB}, $R=4$}
	  		\label{fig:chp2_goodput_2_fr}
		\end{subfigure}%

		\vspace{10pt}

		\begin{subfigure}[b]{0.45\textwidth}
			\input{Chapter2/eval_fr/ber_40.tex}
			\vspace{-8pt}
			\caption{BER codé avec $\gpeak=5$~\si{dB}, $R=16$}
			\label{fig:chp2_ber_3_fr}
		\end{subfigure}
		\hfill
		\begin{subfigure}[b]{0.45\textwidth}
			\input{Chapter2/eval_fr/gp_40.tex}
			\vspace{-8pt}
			\caption{Goodput avec $\gpeak=5$~\si{dB}, $R=16$}
			\label{fig:chp2_goodput_3_fr}
		\end{subfigure}%
	
	\caption{BERs et goodputs obtenus par les différents systèmes.}
	\label{fig:chp2_ber_goodput_fr}
\end{figure}

Comme le système de référence transmet des pilotes et des signaux de réduction en plus des signaux de données, l'énergie par bit transmis est plus élevée que celle des systèmes E2E.
Pour refléter cette caractéristique, nous définissons le rapport énergie par bit sur densité spectrale du bruit comme suit
\begin{align}
	\frac{E_b}{\sigma^2} = \frac{\EE_{x_{m,n}} \left[ |x_{m,n}|^2 \right]}{\rho K \sigma^2} = \frac{1}{\rho Q \sigma^2}
\end{align}
où $\rho$ est le rapport entre le nombre de RE transportant des signaux de données et le nombre total de REs dans la RG.
Les BER codés du système de référence et des trois systèmes E2E sont présentés dans la colonne de gauche de la Figure~B.12. 
On peut voir que le système de base obtient systématiquement un BER légèrement inférieur à celui des systèmes E2E.
Ceci est attendu car le système de base transmet également moins de bits par RG, du fait que certains REs sont utilisés pour transmettre des pilotes ou des signaux de réduction.

Pour comprendre les avantages offerts par l'approche E2E, la deuxième colonne de la Figure~B.12 présente les \emph{goodputs} obtenus par chaque système comparé. 
Le goodput est défini comme le nombre moyen de bits d'information qui ont été reçus avec succès dans un RE :
\begin{align}
	\text{Goodput} = \eta \rho Q (1-\text{BER}),
\end{align}
Les résultats de l'évaluation montrent que les goodputs obtenus par tous les systèmes entraînés, y compris ceux entraînés pour des ACLR plus faibles, sont nettement supérieurs à ceux du système de référence. 
Ces gains sont rendus possibles par le fait que le système E2E n'utilise pas de pilotes, par le schéma efficace de réduction de l'ACLR appris par la procédure d'optimisation proposée, et par le fait que tous les REs peuvent être utilisés pour transmettre des données.

Des observations d'évaluation disponibles dans la Section~4.4.2 révèlent que l'émetteur neuronal parvient à une réduction du PAPR et de l'ACLR grâce à un filtrage dépendant de la sous-porteuse, une distribution inégale de l'énergie sur la sous-porteuse et un réajustement de la position de chaque point dans la constellation.
D'autre part, le récepteur neuronal est capable d'égaliser le canal sans aucun pilote transmis grâce aux constellations asymétriques apprises.
Cela se traduit directement par des gains de débit puisque les transmissions additionnelles associées à l'envoi des pilotes sont supprimée.

\tocless\subsection{Conclusion} 

Dans ce chapitre, nous avons proposé une approche d'apprentissage pour concevoir des formes d'onde OFDM qui répondent à des contraintes spécifiques sur l'enveloppe et les caractéristiques spectrales.
Nous avons exploité la stratégie de bout en bout pour modéliser l'émetteur et le récepteur comme deux CNNs qui effectuent respectivement une modulation et une démodulation de grande dimension.
La procédure d'apprentissage associée exige d'abord que toutes les contraintes d'optimisation soient exprimées sous forme de fonctions différentiables qui peuvent être minimisées par SGD.
Ensuite, un problème d'optimisation sous contrainte est formulé et résolu à l'aide de la méthode du Lagrangien augmenté.
Les résultats des simulations montrent que la procédure d'optimisation permet de concevoir des formes d'onde qui satisfont des contraintes de PAPR et d'ACLR.
De plus, le système de bout en bout permet un débit jusqu'à 30\% plus élevé qu'une implémentation quasi optimale d'un système TR avec un ACLR et un PAPR similaires.
Il est intéressant de noter que ces améliorations sont possibles parce que le système de communication est entièrement conçu grâce au DL. 
Les normes actuelles exigent toutefois l'utilisation de modulations et de placement de pilotes bien connus afin de garantir la compatibilité avec une large gamme de matériel. 
Les implémentations à court terme de systèmes de bout en bout sont donc peu probables, et des solutions plus pratiques doivent être trouvées pour libérer le potentiel du DL dans un avenir proche.

\newpage

\tocless\section{Récepteur Amélioré par DL pour les Systèmes MU-MIMO}

\tocless\subsection{Motivations} 

Les stratégies d'optimisation basées sur les blocs et de bout en bout présentent toutes deux des avantages et des inconvénients qui ont été examinés respectivement dans les Sections~B.2 et~B.3.
D'une part, la stratégie basée sur les blocs pour la détection MU-MIMO a l'avantage de rester interprétable et adaptable à un nombre variable d'utilisateurs, mais elle n'est pas entraînée à optimiser les performances de bout en bout. 
D'autre part, la stratégie de bout en bout permet l'émergence de formes d'onde profondément optimisées, mais les émetteurs-récepteurs basés sur des NNs sont coûteux en calcul et ne sont pas adaptés aux transmissions MU-MIMO. 
Comme nous l'avons vu dans la Section~B.2, la plupart des travaux traitant de la détection MU-MIMO améliorée par DL ne prennent en compte que l'étape d'égalisation. 
Cependant, il a été proposé dans~\cite{korpi2020deeprx} de modéliser tout un récepteur à utilisateur unique  MIMO (SU-MIMO) comme un seul NN grâce à une couche dite `de transformation'. 
Cette solution, appelée DeepRX MIMO, présente des gains importants mais reste très coûteuse en termes de calcul. 
Les principaux inconvénients de ces récepteurs MIMO basés sur des NNs restent leur manque d'interprétabilité et d'adaptabilité à un nombre variable d'utilisateurs.

Dans ce contexte, nous avons introduit une nouvelle approche hybride dans laquelle plusieurs composants DL sont entraînés de bout en bout pour améliorer un récepteur MU-MIMO classique~\cite{goutay2020machine}. 
L'objectif est de combiner l'interprétabilité de la première stratégie, l'efficacité de la seconde et la flexibilité des récepteurs traditionnels. 
Notre solution améliore la prédiction des statistiques d'erreur d'estimation du canal et utilise un démappeur basé sur un CNN pour affiner les estimations de la probabilité des bits. 
L'architecture proposée est évaluée sur des modèles de canaux conformes à la norme 3GPP pour les transmissions en liaison montante et descendante. 
Deux systèmes de références ont été implémentés, le premier étant un récepteur conventionnel, et le second étant le même récepteur mais avec une connaissance parfaite du canal aux positions pilotes et des statistiques d'erreur d'estimation du canal. 
Nos résultats montrent que les gains apportés par le récepteur amélioré par DL augmentent avec la vitesse de l'utilisateur, avec de petites améliorations du BER à faible vitesse mais des améliorations significatives à grande vitesse.

\bigskip
\tocless\subsection{Modélisation du Système}

 \begin{figure*}[t!]
  	\begin{subfigure}{0.20\textwidth}
    	\includegraphics[height=140pt]{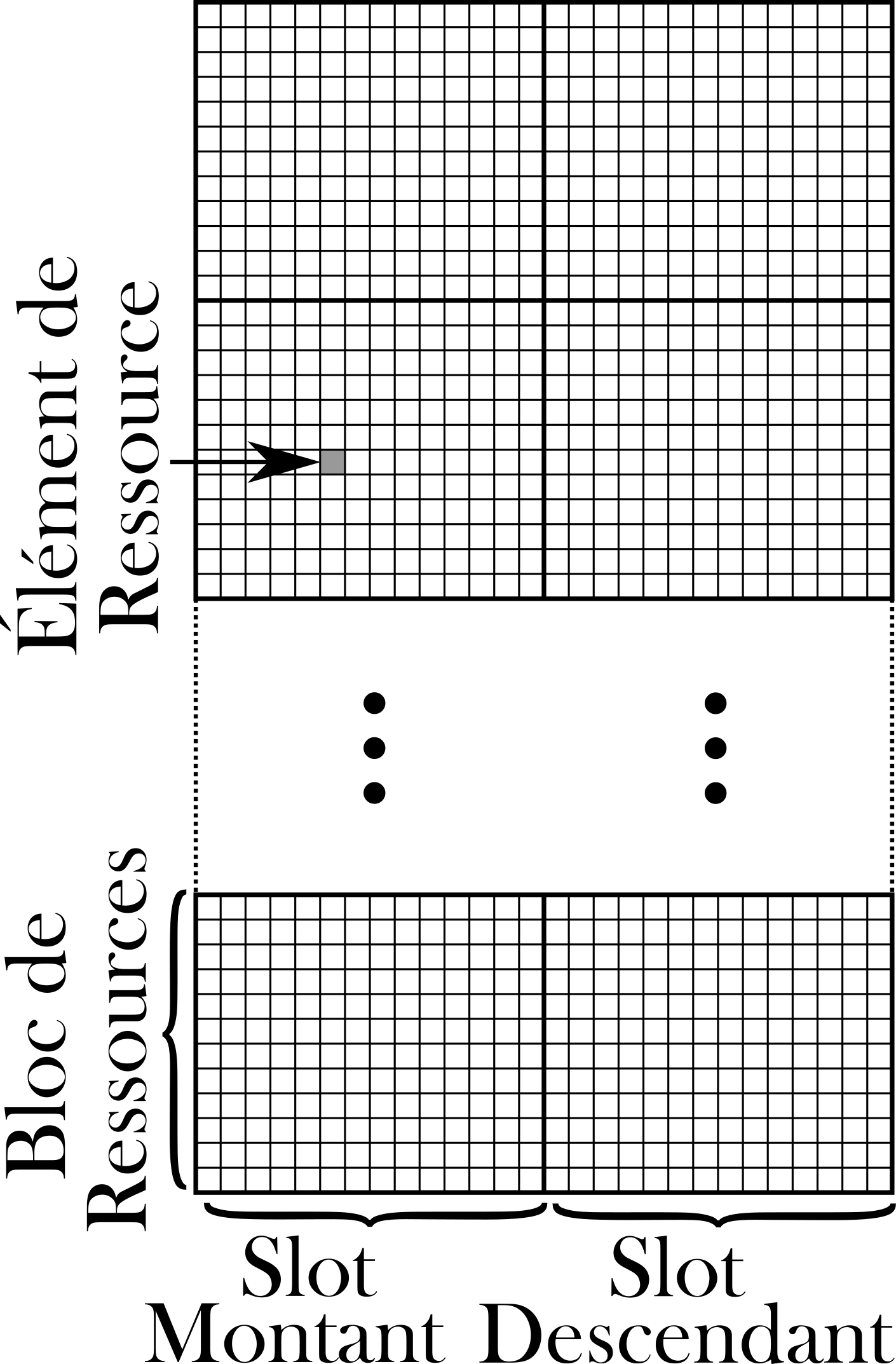} 
  	\caption{Grille de resources.}
  	\label{fig:chp3_resource_grid_fr}
	\end{subfigure}%
\hspace{13pt}
  	\begin{subfigure}{0.33\textwidth}
  	\includegraphics[height=140pt]{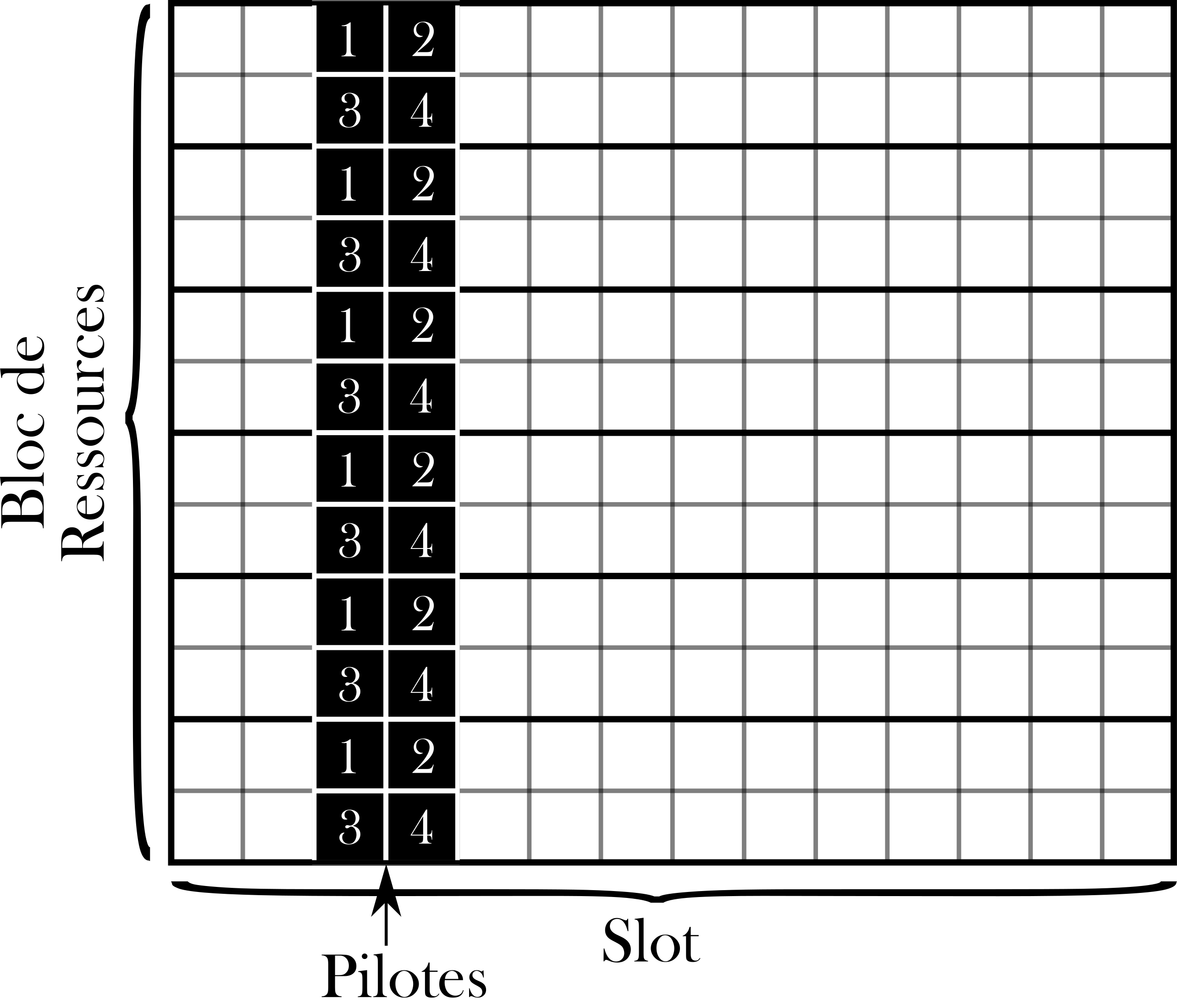}
  	\caption{Configuration de pilotes 1P pour $K=4$ utilisateurs.}
  	\label{fig:chp3_1P_pattern_fr}
	\end{subfigure}%
\hspace{23pt}
  	\begin{subfigure}{0.33\textwidth}
  	\includegraphics[height=140pt]{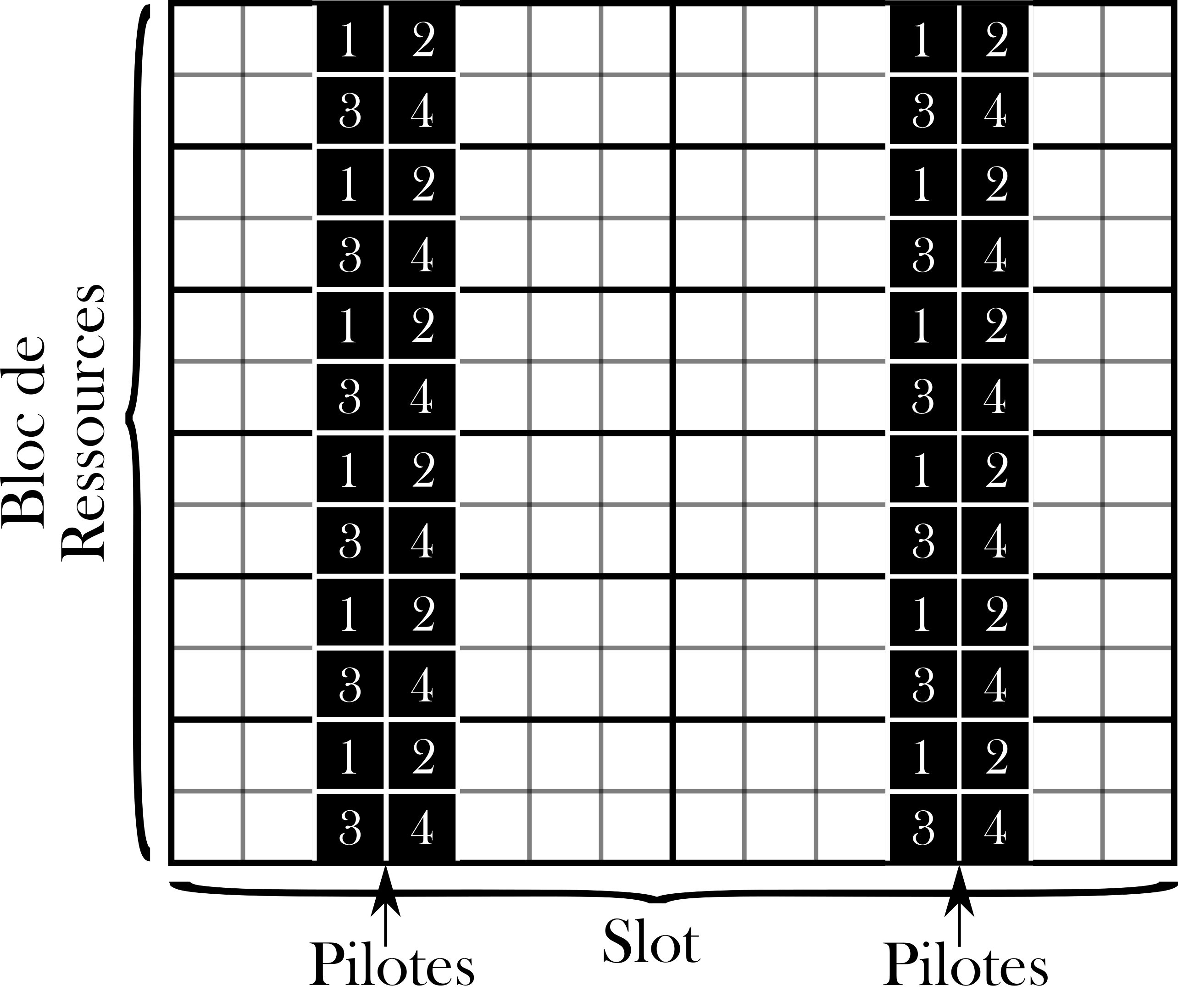}
  	\caption{Configuration de pilotes 2P pour $K=4$ utilisateurs.}
  	\label{fig:chp3_2P_pattern_fr}
	\end{subfigure}%
\caption{Les pilotes sont disposés sur la RG selon deux configurations différentes, où chaque numéro correspond à un émetteur différent.}
\label{fig:chp3_channel_model_fr}
\end{figure*}

Nous considérons un système MU-MIMO, dans lequel $K$ utilisateurs à antenne unique communiquent avec une BS équipée de $L$ antennes sur les liaisons montante et descendante. 
On considère des transmissions OFDM, et la RG est divisé en blocs de ressources constitués de 12 sous-porteuses adjacentes (Figure~B.13a). 
Des modulations d'amplitude en quadrature (quadrature amplitude modulation, QAM) d'ordre $2^Q$ sont utilisées pour transmettre les données. 
Les coefficients de canal forment un tenseur à 4 dimensions noté $\Hm \in \CC^{2 M \times N \times L \times K}$, tel que $\Hm_{m, n} \in \CC^{L \times K}$ est la matrice de canal pour le RE $(m,n)$, et $\hv_{m, n, k}\in\CC^{L}$ est le vecteur de canal pour le RE $(m,n)$ et pour l'utilisateur $k$. 
Le duplexage est réalisé par division dans le temps (time division duplex, TDD), de sorte qu'un slot est attribué alternativement à la liaison montante ou à la liaison descendante, comme l'illustre la Figure~B.13B. 
Deux configurations de pilotes différentes sont considérées, appelées configurations de pilotes 1P et 2P, et contiennent respectivement des pilotes sur deux ou quatre symboles OFDM dans un slot. 
Les Figures B.13b et B.13c montrent respectivement les configurations de pilotes 1P et 2P pour 4 utilisateurs.

\sloppy L'ensemble des REs portant les pilotes correspondant à un utilisateur $k \in \LP 1, \dots, K \RP$ est désigné par $\Pc^{(k)}$ et les nombres de symboles et de sous-porteuses portant les pilotes sont respectivement désignés par $|\Pc_M|$ et $|\Pc_N|$. 
A titre d'exemple, si la configuration de pilotes 1P représentée dans la Figure~\ref{fig:chp3_1P_pattern_fr} est utilisée avec $N = 12$, les positions $(\text{symboles}, \text{sous-porteuses})$ de tous les REs portant des pilotes pour l'utilisateur 1 sont désignées par $\mathcal{P}^{(1)} = \{(3, 1), (3, 3), (3, 5), (3, 73), (3, 93), (3, 11)\}$, ce qui donne $|\Pc_M| = 1$ et $|\Pc_N| = 6$.
On notera que les pilotes ne subissent aucune interférence. 
La puissance du bruit est désignée par $\sigma^2$ et est supposée égale pour tous les utilisateurs et tous les REs. 
Dans ce qui suit, nous supposons un contrôle parfait de la puissance sur la RG de sorte que l'énergie moyenne correspondant à une seule antenne de BS et à un seul utilisateur est égale à un, c'est-à-dire, $\EE\LSB |h_{m,n,l,k}|^2 \RSB = 1$.
Le rapport signal/bruit de la transmission est défini comme suit
\begin{equation}
\label{eq:chp3_snr_fr}
\text{SNR} = 10 \log{\frac{\EE\LSB |h_{m,n,l,k}|^2 \RSB}{\sigma^2}} = 10 \log{\frac{1}{\sigma^2}} \, [\si{dB}].
\end{equation}

En liaison montante, la BS vise à récupérer les bits transmis simultanément par les $K$ utilisateurs sur les REs transportant les données. 
Les tenseurs des signaux émis et reçus de tous les utilisateurs sont respectivement notés $\Xm \in \CC^{2M \times N \times K}$ et $\Ym \in \CC^{2M \times N \times  L}$, et la fonction de transfert sur la liaison montante est $\yv_{m,n} = \Hm_{m,n} \xv_{m,n} + \nv_{m,n}$, où $\nv_{m,n} \sim \Cc\Nc \LB \zerov, \sigma^2\Id_{L} \RB$ est le vecteur de bruit.
Dans ce scénario , seul le créneau de liaison montante est utilisé et, par conséquent, tous les signaux ayant des indices $m > M$ sont ignorés, c'est-à-dire que les valeurs correspondantes sont fixées à 0.
L'architecture du système en liaison montante est illustrée dans la Figure~B.14, où une IDFT (DFT) et l'ajout (suppression) du préfixe cyclique avant (après) le canal ne sont pas représentés pour plus de clarté. 
Les étapes d'estimation du canal, d'égalisation et de démappage du système de référence sont expliquées dans ce qui suit.

\begin{figure*}[!t]
    \centering
    \includegraphics[width=1\textwidth]{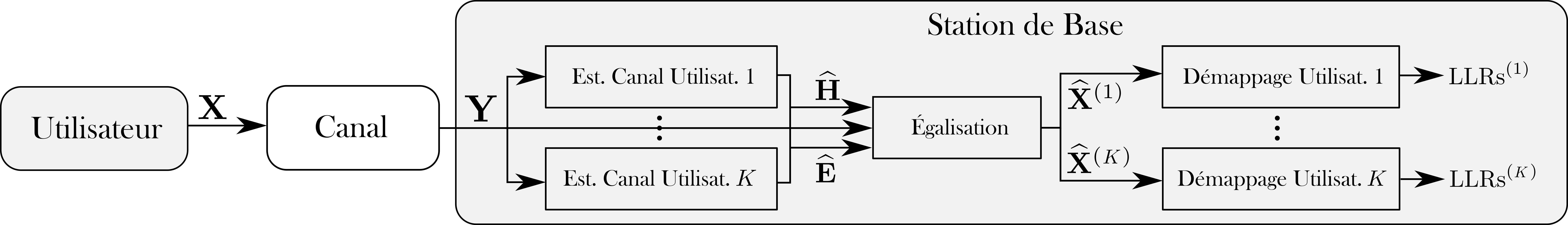}
    \caption{Architecture du système de communication pour la liaison montante.}
    \label{fig:chp3_uplink_fr}
\end{figure*}

\bigskip
\subsubsection{\textit{\textmd{\selectfont{Estimation du canal pour la liaison montante}}}}

Comme les pilotes sont supposés être orthogonaux, l'estimation du canal LMMSE peut être effectuée pour chaque utilisateur indépendamment. 
La matrice de covariance du canal fournissant les corrélations spatiales, temporelles et spectrales entre tous les REs portant des pilotes est notée $\Sigmam \in \CC^{|\Pc_M| \cdot |\Pc_N|\cdot  L \times |\Pc_m| \cdot |\Pc_n| \cdot L}$.
Pour un utilisateur $k \in \{1,\dots,K\}$, l'estimation du canal LMMSE au niveau des REs portant des pilotes est notée par $\widehat{\Hm}^{(k)}_{\mathcal{P}^{(k)}} \in \CC^{|\Pc_M| \times |\Pc_N|  \times L}$.
Inspiré par les directives 3GPP~\cite{std3gpp}, les estimations de canal pour tous les REs sont calculées en interpolant d'abord linéairement les estimations des REs portant des pilotes dans la dimension fréquentielle, puis en utilisant l'estimation au RE interpolé le plus proche (nearest interpolated resource element, NIRE) sur les REs voisins.

Il est également possible d'exploiter l'interpolation linéaire temporelle entre les symboles OFDM portant les pilotes lorsque la configuration de pilotes 2P est utilisée. 
Le tenseur d'estimations de canal ainsi obtenu est noté $\widehat{\Hm}^{(k)} \in \CC^{2M \times  N \times L}$. 
L'estimation globale du canal pour tous les utilisateurs $\widehat{\Hm} \in \CC^{2M \times N \times  L \times K} $ est obtenue en concaténant les estimations de canal de tous les utilisateurs.
Puisque seul le slot de liaison montante est considéré ici, les estimations de canal pour les $M$ derniers symboles (slot de liaison descendante) sont nulles. 
L'erreur d'estimation du canal est notée $\widetilde{\Hm}$ et est telle que $\Hm = \widehat{\Hm} + \widetilde{\Hm}$.
Pour un RE $(m,n)$, nous définissons
\begin{equation}
\label{eq:chp3_E_fr}
\Em_{m,n} \coloneqq \EE \LSB \widetilde{\Hm}_{m, n} \widetilde{\Hm}_{m, n}\htp \RSB
\end{equation}
comme la somme des matrices de covariance des erreurs d'estimation \emph{spatiales} du canal de tous les utilisateurs. L'estimation de ces matrices de covariance, utilisant une approximation NIRE similaire, est détaillée dans la Section~5.4.2.

\begin{figure}
	\center
\includegraphics[width=0.85\textwidth]{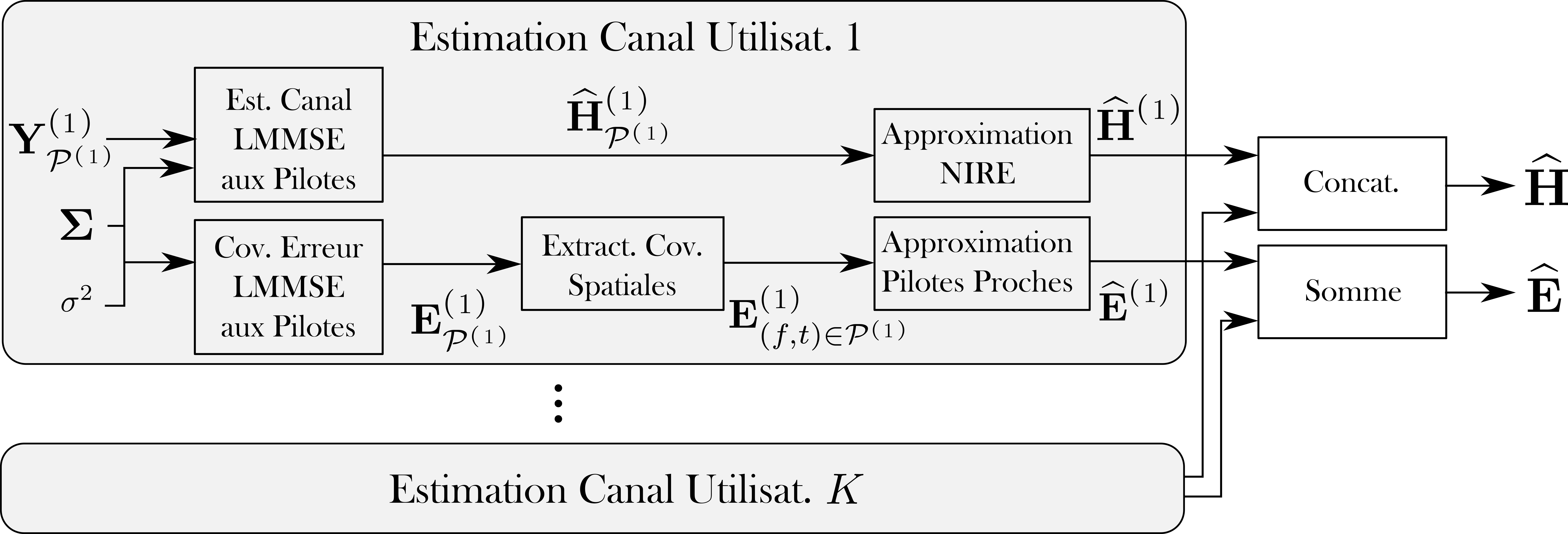}
  \caption{Estimation du canal pour la liaison montante.}
    \label{fig:chp3_ch_est_tradi_fr_fr}
\end{figure}

\subsubsection{\textit{\textmd{\selectfont{Égalisation de la liaison montante}}}}

Nous utilisons l'égaliseur LMMSE, très répandu car il maximise le rapport signal/bruit après égalisation.
Cependant, étant donné que le calcul d'un opérateur LMMSE dédié pour chaque RE est irréalisable en pratique en raison de sa complexité prohibitive, nous avons recours à un égaliseur LMMSE groupé, c'est-à-dire qu'un seul opérateur LMMSE est appliqué à un groupe de REs adjacents couvrant plusieurs symboles et sous-porteuses.
Les symboles égalisés de l'utilisateur $k$ sont notés $\widehat{\Xm}^{(k)} \in \mathbb{C}^{M \times N}$, comme illusté dans la Figure~\ref{fig:chp3_uplink_fr}.
Une dérivation de l'estimateur LMMSE groupé est donnée en Appendix A.

\subsubsection{\textit{\textmd{\selectfont{Démappage de la liaison montante}}}}

Après égalisation, le canal de liaison montante peut être considéré comme $MNK$ canaux de bruit additif parallèles qui peuvent être démodulés indépendamment pour chaque RE et chaque utilisateur.
Pour un RE $(m,n)$ et un utilisateur $k$, le canal post-égalisation s'exprime par  $\hat{x}_{m, n, k}= x_{m, n, k} + \zeta_{m, n, k}$, où le bruit $\zeta_{m, n, k}$ comprend à la fois les interférences et le bruit subi par l'utilisateur $k$. 
Sa variance est donnée par $\rho_{m, n, k}^2  =  \EE\LSB \zeta_{m, n, k}^* \zeta_{m, n, k} \RSB$, et son calcul est donné dans la Section~5.2.2. 
Un démappage standard pour un bruit blanc additif gaussien (additive white Gaussian noise, AWGN) est finalement effectué sur chaque symbole  $\hat{x}_{m, n, k}$ en utilisant sa variance de bruit correspondante $\rho^2_{m, n, k}$.

\subsubsection{\textit{\textmd{\selectfont{Système de référence pour la liaison descendante}}}}

Dans la liaison descendante, la BS doit transmettre simultanément à $K$ utilisateurs sur tous les REs du slot descendant.
Désignons par $\Sm \in \mathcal{C}^{2M \times N \times  K}$ et par  $\Tm \in \CC^{2M \times N \times  L}$  les tenseurs des symboles non précodés et précodés, respectivement.
Nous désignons par $\Um \in \CC^{ 2 M \times N \times K}$  le tenseur des symboles reçus par les $K$ utilisateurs.
Ces quantités ne sont pertinentes que sur le slot descendant et sont donc considérées comme nulles sur les $M$ premiers symboles. 
La fonction de transfert du canal en liaison descendante pour un RE $(m,n)$ est
\begin{equation}
\label{eq:chp3_down_chan_fr}
\uv_{m,n} = \Hm_{m,n}\htp \tv_{m,n} + \qv_{m,n}
\end{equation}
où $\qv_{m,n} \sim \Cc\Nc(\zerov,\sigma^2 \Id_{K})$ est le vecteur de bruit, considéré nul sur les $M$ premiers symboles.  
Les étapes de précodage, d'estimation et d'égalisation des canaux et de démappage de la liaison descendante sont détaillées dans la Section~5.2.3.

\tocless\subsection{Architecture du Récepteur Amélioré par DL}

Les systèmes de référence présentés dans la section précédente présentent plusieurs limites. 
En particulier, l'approximation NIRE conduit à des erreurs d'estimation de canal élevées pour les REs qui sont éloignés des pilotes. 
De même, l'égalisation groupée peut être inexacte pour ces REs. 
Cette section détaille l'architecture d'un récepteur qui s'appuie sur le système de référence mais utilise plusieurs CNNs pour améliorer ses performances.

\subsubsection{\textit{\textmd{\selectfont{Entraînement du récepteur}}}}

\begin{figure}
    \centering
    \includegraphics[width=1\textwidth]{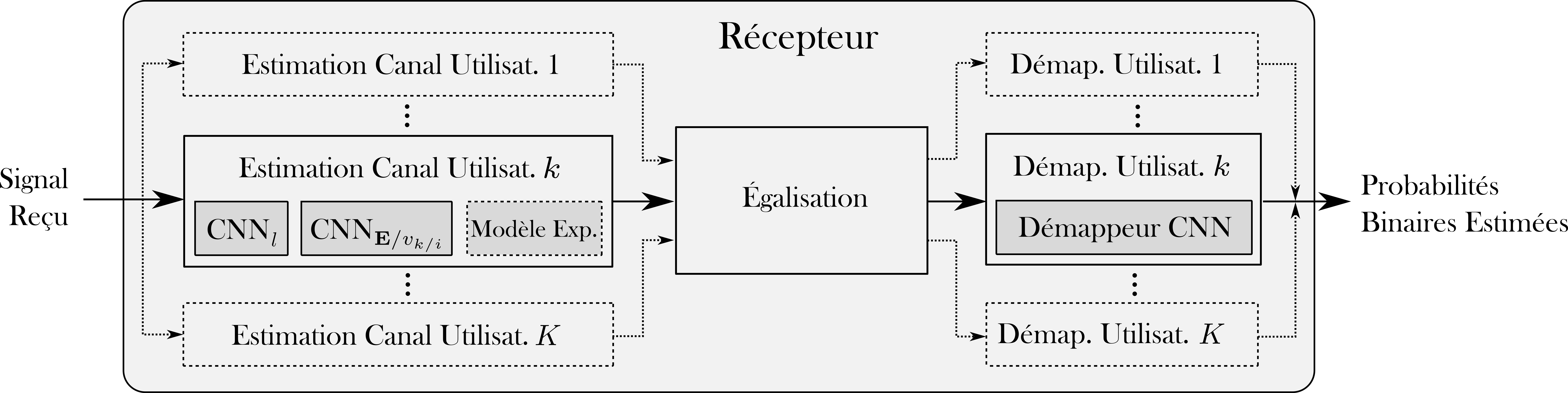}
    \caption{Architecture de récepteur améliorée par DL. Les éléments en pointillés ne sont présents que sur la liaison montante, où la station de base traite tous les utilisateurs. }
    \label{fig:chp3_ml_receiver_fr}
	\vspace{-10pt}
\end{figure}

L'architecture du récepteur amélioré par DL est illustrée dans la Figure~B.16, où les composants entraînables sont représentés en gris foncé. 
Dans la liaison descendante, chaque utilisateur $k$ effectue uniquement l'estimation de canal, l'égalisation et le démappage de son propre signal, et les composants correspondantes sont illustrées par des contours continus. 
En revanche, sur la liaison montante, la station de base traite tous les utilisateurs en parallèle, et les composants supplémentaires sont délimitées par des lignes pointillées. 
L'adaptabilité au nombre d'utilisateurs est obtenue en utilisant différentes copies des mêmes composants DL pour chaque utilisateur, toutes les copies partageant le même ensemble de paramètres entraînables.
Nous proposons d'optimiser conjointement tous ces composants en se basant uniquement sur les probabilités binaires estimées, et non en entraînant chacun d'entre eux individuellement.
Cette approche est pratique car elle ne suppose pas la connaissance des coefficients du canal lors de l'entraînement. Désignons par $\thetav$  l'ensemble des paramètres entraînables du récepteur amélioré par DL, et par  $b_{m,n,k,q}$ le bit $(m,n,k,q)$ envoyé. 
Ces paramètres sont optimisés pour minimiser la BCE totale :
\begin{align}
  \label{eq:chp3_bce_fr}
  \mathcal{L} \triangleq & - \sum_{k=1}^{K} \sum_{(m,n)\in \mathcal{D}} \sum_{q=0}^{Q-1} \EE_{b, \Ym} \left[  \text{log}_2 \LB 
  \widehat{P}_{\thetav} \LB b_{m,n,k,q} | \Ym \RB \RB   \right]
  \end{align}
où $\mathcal{D}$ désigne l'ensemble des REs transportant des données et $\widehat{P}_{\thetav} \LB \cdot | \Ym \RB $ est l'estimation du récepteur de distribution postérieure sur les bits sachant $\Ym$.
L'espérance dans (B.32) est estimée par échantillonnage de Monte Carlo en utilisant des batchs contenant $B_S$ examples :
\begin{align}
\label{eq:chp3_loss_mc_fr}
\mathcal{L} \approx & - \frac{1}{B_S} \sum_{i=1}^{B_S} \sum_{k=1}^{K} \sum_{(m,n)\in \mathcal{D}} \sum_{q=0}^{Q-1} \LB  \text{log}_2 \LB 
\widehat{P}_{\thetav} \LB b_{m,n,k,q}^{[i]}| \Ym^{[i]} \RB \RB \RB
\end{align}
où l'exposant $[i]$ désigne le $i^{\text{ième}}$ example dans le batch.
En suivant une dérivation similaire à celle proposée dans la Section~2.3.3, la perte (B.32) peut être redéfinie comme
\begin{equation}
\mathcal{L} = \sum_{k=1}^{K} \LB \text{Card}(\mathcal{D}) Q - C_k \RB
\end{equation}
où $\text{Card}(\mathcal{D})$ est le nombre de REs transporant des données, $\text{Card}(\mathcal{D}) Q $ est le nombre de bits transmis par un utilisateur, et $C_k$ est un taux de transmission atteignable pour l'utilisateur $k$.

\subsubsection{\textit{\textmd{\selectfont{Estimateur de canal amélioré par DL}}}}

\begin{figure}
    \centering
    \includegraphics[width=0.9\textwidth]{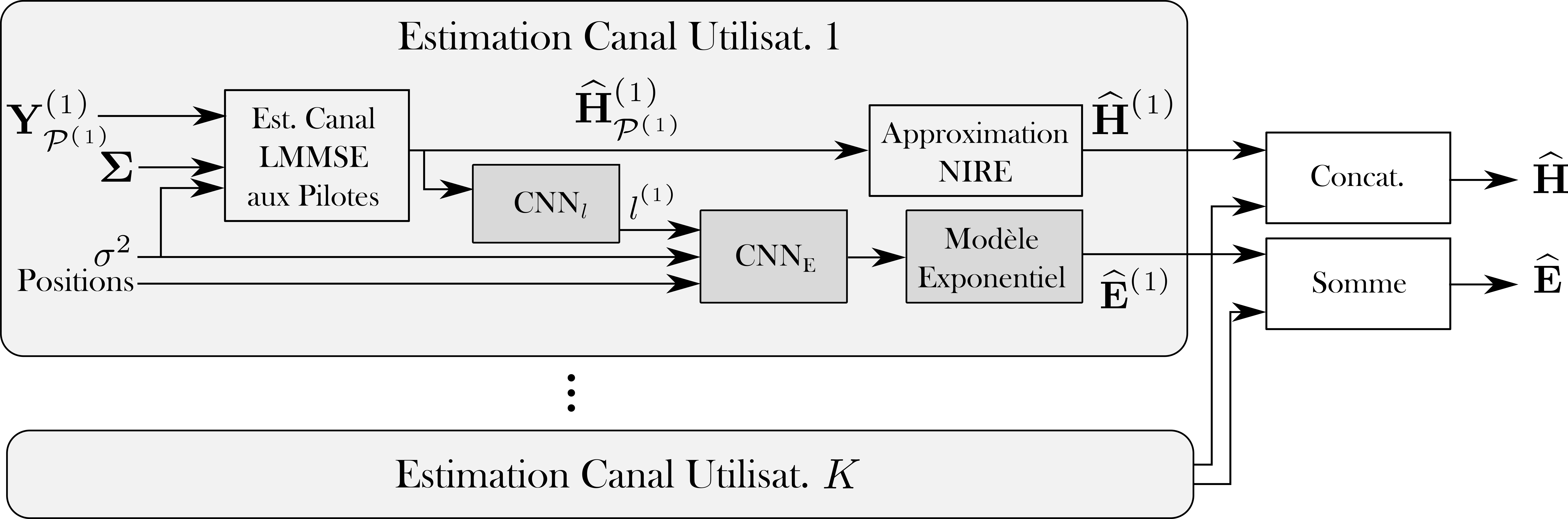}
    \caption{Estimation du canal de la liaison montante améliorée par DL.}
    \label{fig:chp3_ch_est_ml_fr}
\end{figure}

Dans le récepteur conventionnel, les statistiques d'erreur d'estimation du canal ne peuvent être obtenues que pour les REs qui transportent des pilotes, ce qui signifie que la précision de l'estimation diminue à mesure que l'on s'en éloigne. 
Dans ce qui suit, nous présentons des CNNs qui estiment les matrices de covariance des erreurs d'estimation du canal sur la liaison montante et les variances des erreurs d'estimation sur la liaison descendante. 
Sur la liaison montante (Figure~B.17), les matrices de covariance des erreurs d'estimation spatiales du canal $\Em_{m,n}$ sont nécessaires pour calculer les symboles égalisés et la variance du bruit post-égalisation. 
Pour prédire chaque élément de $\widehat{\Em}^{(k)}$ pour l'utilisateur $k$, un CNN conçu naïvement produirait $M  N  L ^2$ paramètres complexes. 
Cela serait d'une complexité prohibitive pour tout grand nombre de sous-porteuses, de symboles ou d'antennes de réception. 
Pour cette raison, nous approximons chaque élément $(x,y)$ de $\widehat{\Em}_{m,n}^{(k)}$ avec le modèle exponentiel suivant :
\begin{equation}
\label{eq:chp3_exp_decay_fr}
\hat{e}^{(k)}_{m,n,a,b}= \alpha_{m,n} \beta_{m,n}^{|b-a|} \exp{j \gamma (b-a)}
\end{equation}
où $b - a$ est la différence de position horizontale entre cet élément et la diagonale, et $\alpha_{m,n}$, $\beta_{m,n}$, et $\gamma$ sont des paramètres de ce modèle. 
Les paramètres $\alpha_{m,n}$ et $\beta_{m,n}$ contrôlent respectivement l'amplitude et la décroissance du modèle, et dépendent du RE $(m,n)$.
Pour estimer ces deux paramètres pour chaque RE, nous utilisons un CNN, désigné par $\text{CNN}_{\mathbf{E}}$, qui admet quatre entrées, chacune de taille $M \times N$, pour une dimension d'entrée totale de $M \times N \times 4$. 
Les deux premières entrées fournissent l'emplacement de chaque RE dans la RG, et la troisième entrée fournit le SNR de la transmission.
Enfin, la quatrième entrée est une caractéristique $f^{(k)} \in \RR$ fournie par un autre CNN, désigné par $\text{CNN}_{f}$, qui a été conçu avec l'intuition de prédire la variabilité temporelle du canal expérimentée par l'utilisateur $k$. 
Pour ce faire, $\text{CNN}_{f}$ utilise les estimations du canal au niveau des REs portant des pilotes pour estimer l'étalement Doppler. 
Il produit le scalaire $f^{(k)}$, qui est transmis à $\text{CNN}_{\mathbf{E}}$  sous forme de matrice $f^{(k)}\cdot \mathds{1}_{M \times N}$.
De plus amples détails sont donnés dans la Section~5.3.2.

Dans la liaison descendante, les variances des erreurs d'estimation des canaux sont estimées de manière similaire.
Cependant, chaque utilisateur estime à la fois son propre canal et les canaux interférents des autres utilisateurs.
Cela conduit à l'utilisation de deux CNNs distincts au lieu du $\text{CNN}_{\Em}$ unique utilisé sur la liaison montante. 
La Section~5.3.2 donne plus d'information sur l'utilisation de ces deux CNNs.

\begin{figure*}[b!]
\centering
\begin{minipage}[b]{0.6\textwidth}
\centering
\includegraphics[height=115pt]{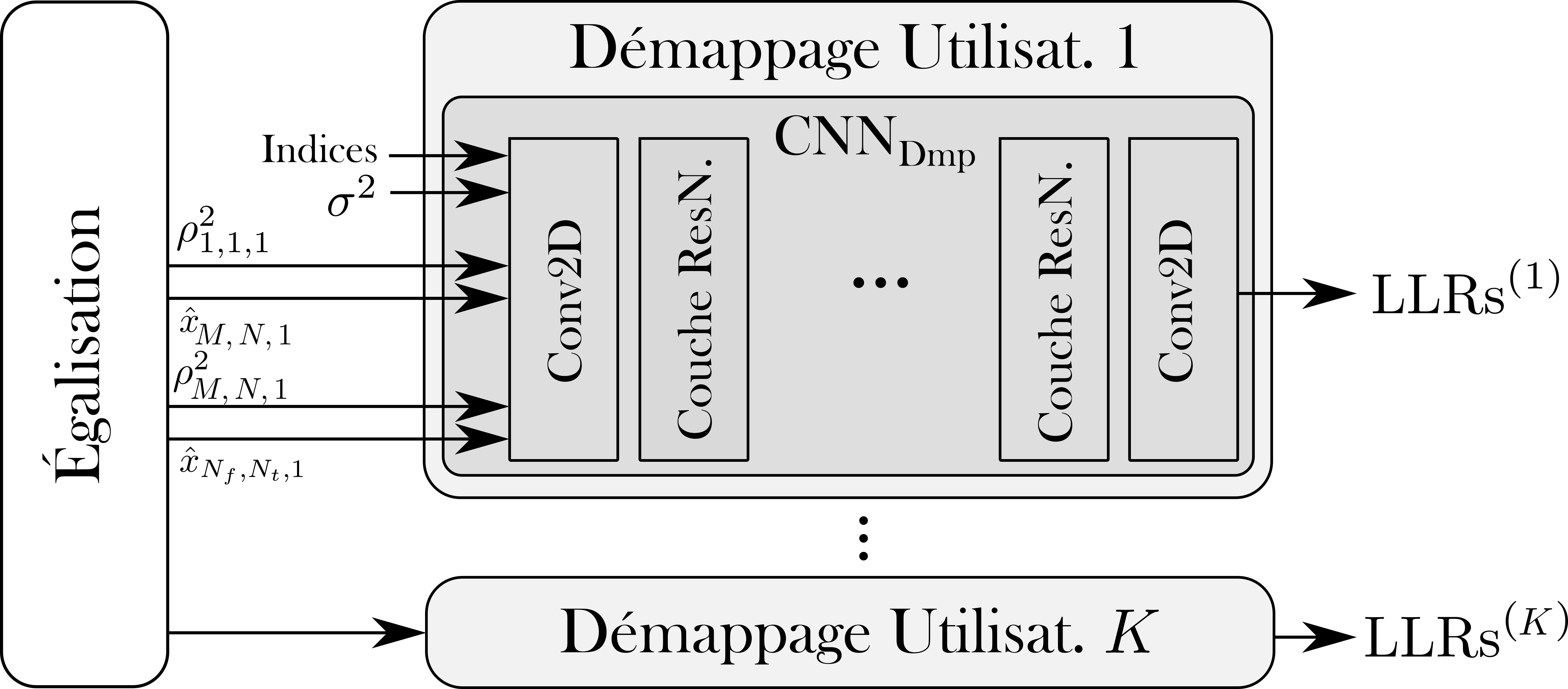}
  \captionof{figure}{Démappage neuronal en liaison montante.}
  \label{fig:chp3_demapper_ml_fr}
\end{minipage}
\hfill
\begin{minipage}[b]{0.35\textwidth}
\centering
\includegraphics[height=100pt]{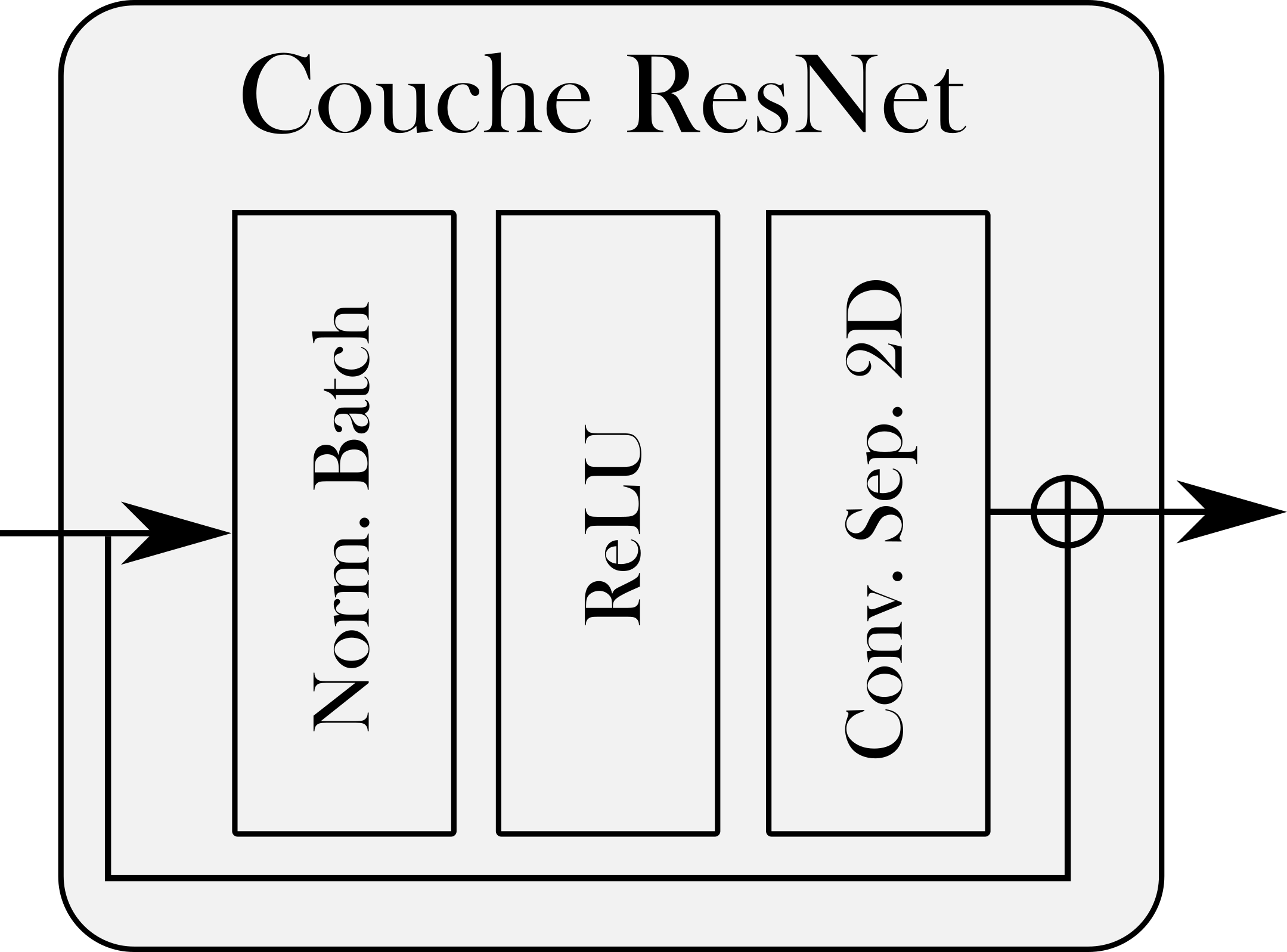}
\captionof{figure}{Illustration d'une couche ResNet.}
  \label{fig:chp3_resnet_fr}
\end{minipage}
\end{figure*}

\subsubsection{\textit{\textmd{\selectfont{Démappage amélioré par DL}}}}

Une conséquence de l'estimation et de l'égalisation imparfaites est le vieillissement du canal, qui entraîne des  distorsions résiduelles sur les signaux égalisés. 
Un démappeur traditionnel opère indépendamment sur chaque RE et ne voit donc qu'un seul symbole égalisé à la fois. 
En revanche, nous proposons d'utiliser un CNN, appelé $\text{CNN}_{\text{Dmp}}$, pour effectuer un démappage conjoint de l'ensemble de la RG. 
En traitant conjointement tous les symboles égalisés, ce CNN peut estimer et corriger les effets du vieillissement du canal pour calculer de meilleurs rapports de log-vraisemblance (log-likelihood ratios, LLRs). 
L'entrée de $\text{CNN}_{\text{Dmp}}$ est de dimension $M \times N \times 6$ et contient les indices des symboles et des sous-porteuses pour chaque RE, le SNR, les parties réelles et imaginaires des symboles égalisés, et les variances du bruit du canal après égalisation. 
La sortie de $\text{CNN}_{\text{Dmp}}$ est de dimension $M \times N \times Q$ et correspond aux $\text{LLRs}^{(k)}$ prédits sur la RG pour un utilisateur $k$. 
Comme avec un récepteur conventionnel, le démappage est effectué indépendamment pour chaque utilisateur afin de rendre l'architecture facilement adaptable à un nombre variable d'utilisateurs. 
Le démappeur $\text{CNN}_{\text{Dmp}}$ est présenté dans la Figure~B.18, qui décrit le processus de démappage de la liaison montante.

\subsubsection{\textit{\textmd{\selectfont{Architecture des CNNs}}}}

Tous les CNN présentés ci-dessus partagent les mêmes blocs : couches convolutionelles 2D, couches denses et couches ResNet personnalisées.
Ces couches ResNet sont constituées d'une couche de normalisation par batch, d'une fonction d'activation ReLU, d'une couche convolutive séparable 2D et enfin de l'ajout de l'entrée, comme le montre la Figure~B.19.
Les convolutions séparables sont moins coûteuses en termes de calcul tout en maintenant des performances similaires à celles des couches convolutionnelles classiques \cite{howard2017mobilenets}.
Les détails des architectures de tous les CNNs sont donnés dans le Tableau~5.1.

\tocless\subsection{Évaluations} 

\subsubsection{\textit{\textmd{\selectfont{Dispositif d'entraînement et d'évaluation}}}}

Pour un entraînement et une évaluation réalistes, les réalisations de canal ont été générées avec QuaDRiGa version 2.0.0 \cite{quadriga}. 
Le nombre d'utilisateurs a été fixé à $K = 4$, sauf pour les liaisons descendantes à haute vitesse où il a été réduit à $K = 2$, et le nombre d'antennes de la BS a été fixé à $L = 16$. 
Les RGs étaient composées de $N = 72$ sous-porteuses, avec une fréquence centrale de \SI{3.5}{\GHz} et un espacement des sous-porteuses de \SI{15}{kHz}.
Une QAM utilisant un code de Gray a été utilisée avec $Q = 4$ bits par utilisation du canal sur la liaison montante et $Q = 2$ bits par utilisation du canal sur la liaison descendante. 
Les récepteurs ont été entraînés à des vitesses allant de \SIrange[range-units=single, range-phrase= \text{ à }]{0}{45}{\km\per\hour} et de \SIrange[range-units=single, range-phrase= \text{ à }]{50}{130}{\km\per\hour}.
Trois gammes de vitesses faibles ont été considérées pour les essais avec la configuration de pilotes 1P : \SIrange[range-units=single, range-phrase= \text{ à }]{0}{15}{\km\per\hour}, \SIrange[range-units=single, range-phrase= \text{ à }]{15}{30}{\km\per\hour}, et \SIrange[range-units=single, range-phrase= \text{ à }]{30}{45}{\km\per\hour}.
De même, trois gammes de vitesses élevées ont été considérées avec la configuration de pilotes 2P : \SIrange[range-units=single, range-phrase= \text{ à }]{50}{70}{\km\per\hour}, \SIrange[range-units=single, range-phrase= \text{ à }]{80}{100}{\km\per\hour}, et  \SIrange[range-units=single, range-phrase= \text{ à }]{110}{130}{\km\per\hour}.
Les évaluations ont été effectuées séparément pour chaque plage de vitesse en utilisant un code LDPC standard IEEE 802.11n de longueur 1296 bits. 
Des taux de codage de  $\frac{1}{2}$ et $\frac{1}{3}$ ont été utilisés respectivement pour les transmissions en liaison montante et en liaison descendante. 
Chaque entraînement a été effectué à l'aide de batchs de taille $B_S=27$  de sorte que le nombre total de bits envoyés pour chaque utilisateur soit un multiple de la longueur du code.

\subsubsection{\textit{\textmd{\selectfont{Résultats des simulations sur la liaison montante}}}}

\begin{figure}
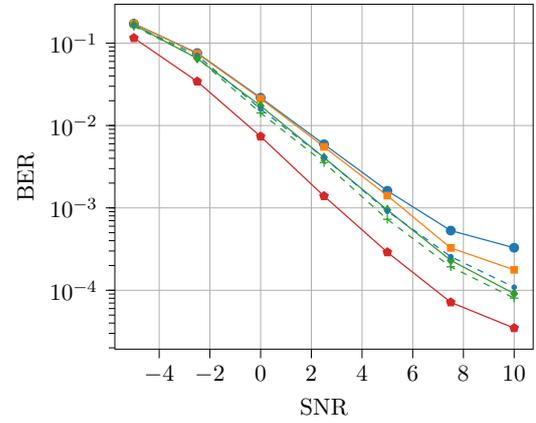
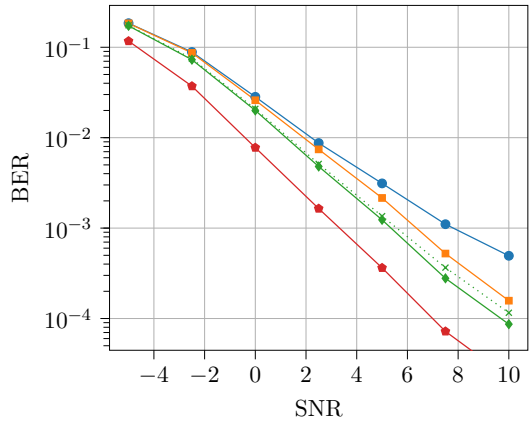
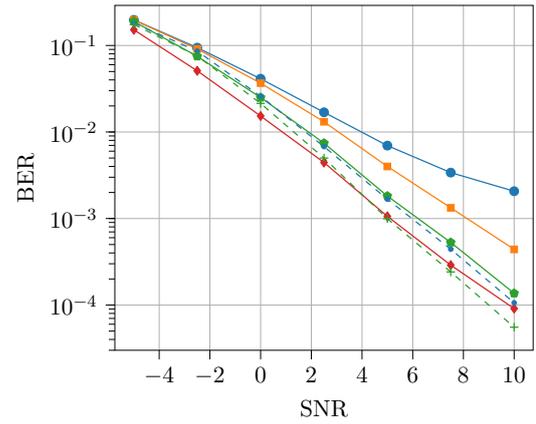
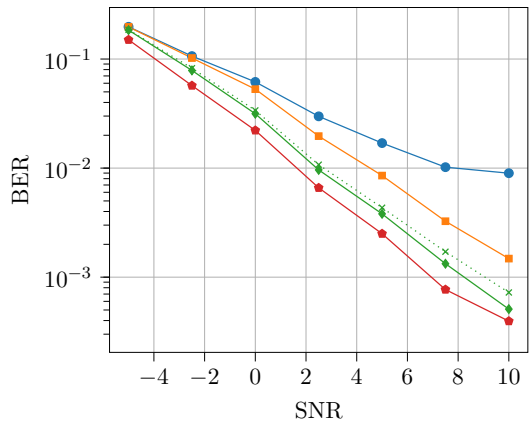
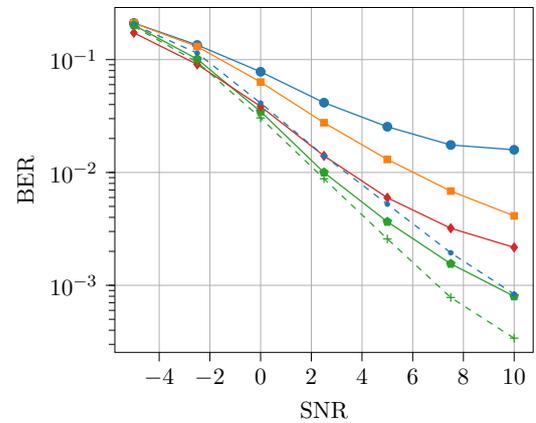

	\begin{subfigure}[l]{1\textwidth}
	\begin{tikzpicture} 
	
	\definecolor{color0}{rgb}{0.12156862745098,0.466666666666667,0.705882352941177}
	\definecolor{color1}{rgb}{1,0.498039215686275,0.0549019607843137}
	\definecolor{color2}{rgb}{0.172549019607843,0.627450980392157,0.172549019607843}
	\definecolor{color3}{rgb}{0.83921568627451,0.152941176470588,0.156862745098039}

	\begin{axis}[%
		hide axis,
		xmin=10,
		xmax=50,
		ymin=0,
		ymax=0.4,
		legend columns=5, 
		legend style={draw=white!15!black,legend cell align=left,column sep=1ex}
		]
		\addlegendimage{white,mark=square*}
		\addlegendentry{\hspace{-1cm} Interp. spectrale :};
		\addlegendimage{color0,mark=square*}
		\addlegendentry{\hspace{-0.25cm} Référence};
		\addlegendimage{color1, mark=*}
		\addlegendentry{\hspace{-0.25cm} Est. canal DL };
		\addlegendimage{color2, mark=pentagon*}
		\addlegendentry{\hspace{-0.25cm} Récepteur DL };
		\addlegendimage{color3, mark=diamond*}
		\addlegendentry{\hspace{-0.25cm} CSI Parfait};
		
		\end{axis}
	\end{tikzpicture}
	\end{subfigure}%
	\vspace{1pt}
	
	  \begin{subfigure}[c]{1\textwidth}
	  
	\begin{tikzpicture} 
	
	\definecolor{color2}{rgb}{0.172549019607843,0.627450980392157,0.172549019607843}

		\begin{axis}[%
		hide axis,
		xmin=10,
	   xmax=50,
		ymin=0,
		ymax=0.4,
		legend columns=2, 
		legend style={draw=white!15!black,legend cell align=right,column sep=1.9ex}
		]
		\addlegendimage{white,mark=square*, mark size=2}
		\addlegendentry{\hspace{-1.15cm} Interp. spectrale (config. 1P uniquement) :};
		\addlegendimage{color2, dotted, mark=x, mark size=2, mark options={solid}}
		\addlegendentry{DL receiver trained with $K=2$ \hspace{0.92cm} \quad};
		\end{axis}
	\end{tikzpicture}
	
	\end{subfigure}
	
	\vspace{1pt}

	  \begin{subfigure}[c]{1\textwidth}
	  
	\begin{tikzpicture} 
	
	\definecolor{color0}{rgb}{0.12156862745098,0.466666666666667,0.705882352941177}
	\definecolor{color2}{rgb}{0.172549019607843,0.627450980392157,0.172549019607843}

		\begin{axis}[%
		hide axis,
		xmin=10,
		xmax=50,
		ymin=0,
		ymax=0.4,
		legend columns=3, 
		legend style={draw=white!15!black,legend cell align=left, column sep=1.9ex}
		]
		\addlegendimage{white,mark=square*, mark size=2}
		\addlegendentry{\hspace{-1.15cm} Interp. spectrale + temporellle  (config 2P) :};
		\addlegendimage{color0, dashed, mark=*, mark size=1, mark options={solid}}
		\addlegendentry{\hspace{-0.15cm}  Référence};
		\addlegendimage{color2, dashed, mark=+, mark size=2, mark options={solid}}
		\addlegendentry{\hspace{-0.15cm}  Récepteur DL  \hspace{0.73cm} \quad };
		\end{axis}
	\end{tikzpicture}
	\end{subfigure}%
	
	\vspace{10pt}

	\centering
	\begin{subfigure}[b]{0.45\textwidth}
	  \begin{adjustbox}{width=\linewidth} 
		  \input{Chapter3/eval_fr/UL_1P_S.tex}
	\end{adjustbox}    
	\caption{Configuration 1P de \SIrange[range-units=single, range-phrase= \text{ à }]{0}{5}{\km\per\hour}.}
	\label{fig:chp3_UL_1P_L_fr}
  \end{subfigure}%
 \hfill
	\begin{subfigure}[b]{0.45\textwidth}
	\begin{adjustbox}{width=\linewidth} 
		\input{Chapter3/eval_fr/UL_2P_S.tex}
	\end{adjustbox} 
	\caption{Configuration 2P de \SIrange[range-units=single, range-phrase= \text{ à }]{40}{60}{\km\per\hour}.}
	\label{fig:chp3_UL_2P_L_fr}
  \end{subfigure}%
  
\vspace{15pt}
\centering
	\begin{subfigure}[b]{0.45\textwidth}
	  \begin{adjustbox}{width=\linewidth} 
		  \input{Chapter3/eval_fr/UL_1P_M.tex}
	\end{adjustbox}     
	\caption{Configuration 1P de \SIrange[range-units=single, range-phrase= \text{ à }]{10}{20}{\km\per\hour}.}  %
	\label{fig:chp3_UL_1P_M_fr}
  \end{subfigure}%
 \hfill
	\begin{subfigure}[b]{0.45\textwidth}
	\begin{adjustbox}{width=\linewidth} 
		\input{Chapter3/eval_fr/UL_2P_M.tex}  		
	\end{adjustbox} 
	\caption{Configuration 2P de \SIrange[range-units=single, range-phrase= \text{ à }]{70}{90}{\km\per\hour}.}
	\label{fig:chp3_UL_2P_M_fr}
  \end{subfigure}%
  
\vspace{15pt}
\centering

	\begin{subfigure}[b]{0.45\textwidth}
	  \begin{adjustbox}{width=\linewidth} 
		  \input{Chapter3/eval_fr/UL_1P_H.tex}
	\end{adjustbox}    
	\caption{Configuration 1P de \SIrange[range-units=single, range-phrase= \text{ à }]{25}{35}{\km\per\hour}.}   %
	\label{fig:chp3_UL_1P_H_fr}
  \end{subfigure}%
 \hfill
	\begin{subfigure}[b]{0.45\textwidth}
	\begin{adjustbox}{width=\linewidth} 
		\input{Chapter3/eval_fr/UL_2P_H.tex}
	\end{adjustbox} 
	\caption{Configuration 2P de \SIrange[range-units=single, range-phrase= \text{ à }]{110}{130}{\km\per\hour}.}
	\label{fig:chp3_UL_2P_H_fr}
  \end{subfigure}%

\vspace{-5pt}
\caption{BERs en liaison montante obtenus avec les configurations de pilotes 1P et 2P.}
\label{fig:chp3_eval_UL_fr}
\end{figure}

Différents systèmes ont été comparés dans les simulations de liaison montante. 
Le premier est le système de référence pour la liaison montante présenté dans la Section~B.4.2.
Le deuxième, appelé `estimateur de canal DL', utilise l'estimateur de canal amélioré par DL mais est entraîné et testé avec un démappeur standard. 
Le troisième est le récepteur amélioré par DL, qui utilise à la fois l'estimateur de canal amélioré et le démappeur amélioré.
Nous l'appelons `récepteur DL'. 
L'évaluation séparée de l'estimateur de canal DL nous permet de mieux comprendre le rôle des deux composants dans les gains observés. 
Une système de référence idéal avec une connaissance parfaite du canal au niveau des REs transportant les pilotes et de $\Em$ est également considéré, et est désigné sous le nom de `Parfaite connaissance des informations sur l'état du canal (channel state information, CSI)', ou `CSI parfait'. 
Tous les systèmes ont utilisé une interpolation spectrale suivie d'une approximation NIRE pour l'estimation du canal. 
Des simulations supplémentaires ont été effectuées pour la configuration de pilotes 2P en utilisant l'interpolation spectrale et temporelle pour le système de référence et le récepteur DL. 
Enfin, un récepteur DL entraîné avec seulement $K = 2$ utilisateurs a également été évalué.

Les résultats de la simulation pour la configuration 1P sont présentés dans la première colonne de la Figure~B.20 pour les trois plages de vitesse différentes. 
Aux vitesses comprises entre \SIrange[range-units=single, range-phrase= \text{ et }]{0}{15}{\km\per\hour}, on constate que le récepteur DL n'apporte aucune amélioration. 
Dans la plage de vitesse la plus élevée (Figure~B.20e), le récepteur DL permet des gains de \SI{3}{\dB} par rapport au système de référence pour un BER codé de  $10^{-2}$. 
On constate que le récepteur DL entraîné avec seulement deux utilisateurs est presque aussi performant que le récepteur entraîné avec quatre utilisateurs pour toutes les vitesses considérées, ce qui démontre l'adaptabilité du système proposé vis-à-vis du nombre d'utilisateurs. 
Les résultats pour la configuration 2P sont présentés à la deuxième colonne de la Figure~B.20. 
Les gains fournis par le récepteur DL suivent la même tendance que pour la configuration 1P. 
Plus précisément, le récepteur DL est le seul système qui permet d'obtenir un BER codé de $10^{-3}$ pour les vitesses les plus élevées avec l'interpolation spectrale uniquement. 
En effet, à des vitesses élevées, le démappeur appris est encore capable d'atténuer les effets du vieillissement du canal, alors que même le système de référence avec CSI parfait souffre de signaux égalisés fortement perturbés.
L'utilisation des interpolations spectrales et temporelles réduit les gains fournis par le récepteur DL, ce qui peut s'expliquer par les meilleures estimations du canal conduisant à un moindre vieillissement du canal, mais ils représentent toujours un écart de \SI{2.2}{\dB} à un BER codé de $10^{-3}$ pour les vitesses les plus élevées. 
Dans l'ensemble, on peut voir que seule la combinaison d'une estimation des statistiques d'erreur d'estimation du canal basée sur des CNNs et d'un démappage basé sur un CNN permet d'obtenir des gains pour les deux configurations de pilotes et pour toutes les intervalles de vitesse.

\subsubsection{\textit{\textmd{\selectfont{Résultats des simulations sur la liaison descendante}}}}

\begin{figure}
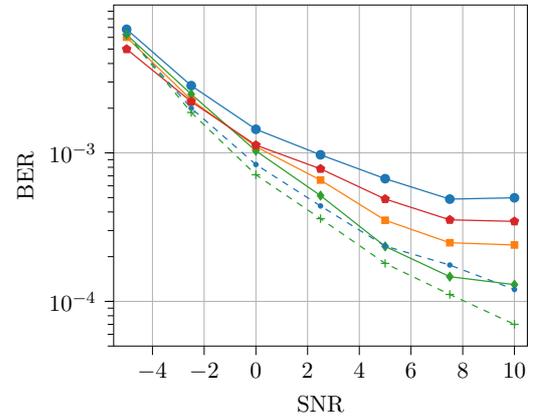
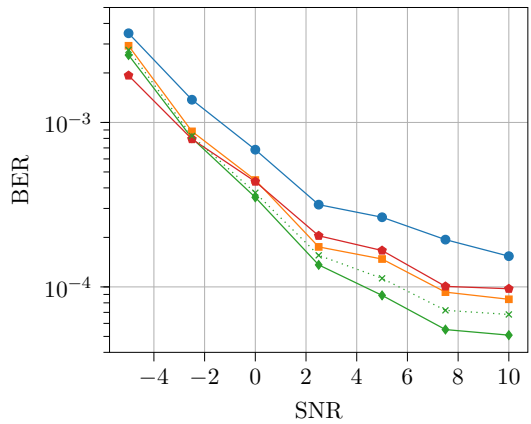
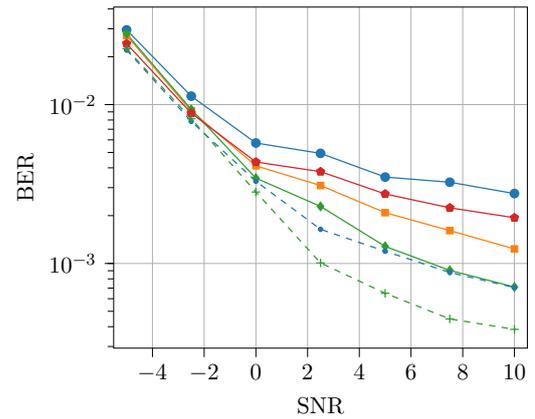
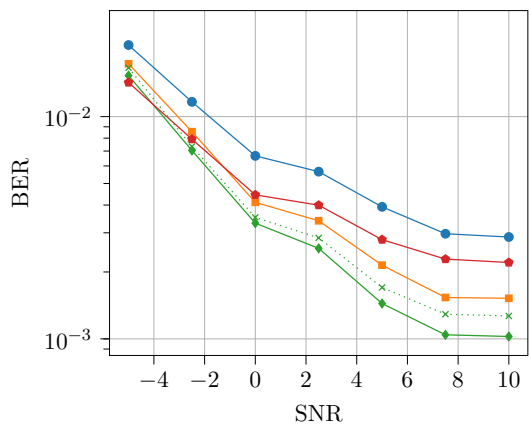
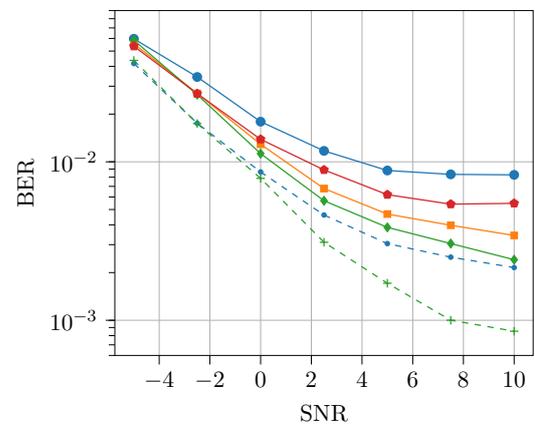

		\begin{subfigure}[l]{1\textwidth}
		\begin{tikzpicture} 
		
		\definecolor{color0}{rgb}{0.12156862745098,0.466666666666667,0.705882352941177}
		\definecolor{color1}{rgb}{1,0.498039215686275,0.0549019607843137}
		\definecolor{color2}{rgb}{0.172549019607843,0.627450980392157,0.172549019607843}
		\definecolor{color3}{rgb}{0.83921568627451,0.152941176470588,0.156862745098039}
	
		\begin{axis}[%
			hide axis,
			xmin=10,
			xmax=50,
			ymin=0,
			ymax=0.4,
			legend columns=5, 
			legend style={draw=white!15!black,legend cell align=left,column sep=1ex}
			]
			\addlegendimage{white,mark=square*}
			\addlegendentry{\hspace{-1cm} Interp. spectrale :};
			\addlegendimage{color0,mark=square*}
			\addlegendentry{\hspace{-0.25cm} Référence};
			\addlegendimage{color1, mark=*}
			\addlegendentry{\hspace{-0.25cm} Est. canal DL };
			\addlegendimage{color2, mark=pentagon*}
			\addlegendentry{\hspace{-0.25cm} Récepteur DL };
			\addlegendimage{color3, mark=diamond*}
			\addlegendentry{\hspace{-0.25cm} CSI Parfait};
			
			\end{axis}
		\end{tikzpicture}
		\end{subfigure}%
		\vspace{1pt}
		
		  \begin{subfigure}[c]{1\textwidth}
		  
		\begin{tikzpicture} 
		
		\definecolor{color2}{rgb}{0.172549019607843,0.627450980392157,0.172549019607843}
	
		\begin{axis}[%
			hide axis,
			xmin=10,
		   xmax=50,
			ymin=0,
			ymax=0.4,
			legend columns=2, 
			legend style={draw=white!15!black,legend cell align=right,column sep=1.9ex}
			]
			\addlegendimage{white,mark=square*, mark size=2}
			\addlegendentry{\hspace{-1.15cm} Interp. spectrale (config. 1P uniquement) :};
			\addlegendimage{color2, dotted, mark=x, mark size=2, mark options={solid}}
			\addlegendentry{ DL receiver trained with $K=2$ \hspace{0.92cm} \quad};
			\end{axis}
		\end{tikzpicture}
		
		\end{subfigure}
		
		\vspace{1pt}
	
		  \begin{subfigure}[c]{1\textwidth}
		  
		\begin{tikzpicture} 
		
		\definecolor{color0}{rgb}{0.12156862745098,0.466666666666667,0.705882352941177}
		\definecolor{color2}{rgb}{0.172549019607843,0.627450980392157,0.172549019607843}
	
		\begin{axis}[%
			hide axis,
			xmin=10,
			xmax=50,
			ymin=0,
			ymax=0.4,
			legend columns=3, 
			legend style={draw=white!15!black,legend cell align=left, column sep=1.9ex}
			]
			\addlegendimage{white,mark=square*, mark size=2}
			\addlegendentry{\hspace{-1.15cm} Interp. spectrale + temporellle  (config 2P) :};
			\addlegendimage{color0, dashed, mark=*, mark size=1, mark options={solid}}
			\addlegendentry{\hspace{-0.15cm}  Référence};
			\addlegendimage{color2, dashed, mark=+, mark size=2, mark options={solid}}
			\addlegendentry{\hspace{-0.15cm}  Récepteur DL  \hspace{0.73cm} \quad };
			\end{axis}
		\end{tikzpicture}
		\end{subfigure}%
		
		\vspace{10pt}
		\centering
		\begin{subfigure}[b]{0.45\textwidth}
		  \begin{adjustbox}{width=\linewidth} 
			  \input{Chapter3/eval_fr/DL_1P_S.tex}
		\end{adjustbox}    
		\caption{Configuration 1P de \SIrange[range-units=single, range-phrase= \text{ à }]{0}{5}{\km\per\hour}.}
		\label{fig:chp3_DL_1P_L_fr}
	  \end{subfigure}%
	 \hfill
		\begin{subfigure}[b]{0.45\textwidth}
		\begin{adjustbox}{width=\linewidth} 
			\input{Chapter3/eval_fr/DL_2P_S.tex}
		\end{adjustbox} 
		\caption{Configuration 2P de \SIrange[range-units=single, range-phrase= \text{ à }]{40}{60}{\km\per\hour}.}
		\label{fig:chp3_DL_2P_L_fr}
	  \end{subfigure}%
	  
  \vspace{15pt}
  \centering
		\begin{subfigure}[b]{0.45\textwidth}
		  \begin{adjustbox}{width=\linewidth} 
			  \input{Chapter3/eval_fr/DL_1P_M.tex}
		\end{adjustbox}     
		\caption{Configuration 1P de \SIrange[range-units=single, range-phrase= \text{ à }]{10}{20}{\km\per\hour}.}  %
		\label{fig:chp3_DL_1P_M_fr}
	  \end{subfigure}%
	 \hfill
		\begin{subfigure}[b]{0.45\textwidth}
		\begin{adjustbox}{width=\linewidth} 
			\input{Chapter3/eval_fr/DL_2P_M.tex}  		
		\end{adjustbox} 
		\caption{Configuration 2P de \SIrange[range-units=single, range-phrase= \text{ à }]{70}{90}{\km\per\hour}.}
		\label{fig:chp3_DL_2P_M_fr}
	  \end{subfigure}%
	  
  \vspace{15pt}
  \centering
  
		\begin{subfigure}[b]{0.45\textwidth}
		  \begin{adjustbox}{width=\linewidth} 
			  \input{Chapter3/eval_fr/DL_1P_H.tex}
		\end{adjustbox}    
		\caption{Configuration 1P de \SIrange[range-units=single, range-phrase= \text{ à }]{25}{35}{\km\per\hour}.}   %
		\label{fig:chp3_DL_1P_H_fr}
	  \end{subfigure}%
	 \hfill
		\begin{subfigure}[b]{0.45\textwidth}
		\begin{adjustbox}{width=\linewidth} 
			\input{Chapter3/eval_fr/DL_2P_H.tex}
		\end{adjustbox} 
		\caption{Configuration 2P de \SIrange[range-units=single, range-phrase= \text{ à }]{110}{130}{\km\per\hour}.}
		\label{fig:chp3_DL_2P_H_fr}
	  \end{subfigure}%
  
  \vspace{-5pt}
  \caption{BERs en liaison descendante obtenus avec les config. de pilotes 1P et 2P.}
  \label{fig:chp3_eval_DL_fr}
  \end{figure}

Quatre systèmes ont été évalués sur la liaison descendante pour les intervalles de vitesse considérées.
Le premier est le système de référence présenté dans la Section~B.4.2. 
Dans le second, appelé `estimateur de canal DL', chaque utilisateur utilise l'estimateur de canal amélioré par DL et est entraîné et testé avec un démappeur standard.
Le troisième est le récepteur amélioré par DL qui utilise à la fois l'estimateur de canal amélioré et le démappeur DL pour chaque utilisateur, et est appelé `récepteur DL'. 
Le dernier schéma a un CSI parfait, c'est-à-dire que tous les utilisateurs connaissent parfaitement à la fois le canal aux REs portant des pilotes et les variances des erreurs d'estimation partout. 
Comme pour la liaison montante, tous les systèmes ont utilisé l'interpolation spectrale avec approximation NIRE pour l'estimation du canal, mais des simulations supplémentaires ont été effectuées pour la configuration de pilotes 2P en utilisant l'interpolation spectrale et temporelle.

Les évaluations de la configuration de pilotes 1P en liaison descendante sont présentées dans la première colonne de la Figure~B.21. 
Comme prévu, les gains permis par l'estimateur de canal et le démappeur appris augmentent avec la vitesse. 
On peut également observer que le récepteur DL entraîné avec seulement $K = 2$ utilisateurs est capable de s'approcher des performances du récepteur DL entraîné avec quatre utilisateurs.
Les évaluations de la configuration de pilotes 2P sont présentées sur la deuxième colonne de la Figure~B.21 pour des vitesses plus élevées. 
En utilisant les interpolations spectrales et temporelles, le récepteur DL offre toujours des gains significatifs à des vitesses élevées, étant le seul à atteindre un BER de $10^{-3}$ dans l'intervalle de vitesse de \SI{110}{} à \SI{130}{\km\per\hour}. 
Dans tous les scénarios de liaison descendante, à l'exception de l'intervalle de vitesse le plus lent, l'estimateur de canal DL est plus performant que le système de référence avec un CSI parfait. 
Nous supposons que cela est dû au fait que le système de référence suppose que le bruit de la liaison descendante après égalisation est distribué de manière gaussienne et est non corrélé au signal transmis, ce qui est généralement faux.

\tocless\subsection{Conclusion}

Le présent chapitre propose une nouvelle stratégie hybride pour la conception de récepteurs MU-MIMO améliorés par des composants DL.
Notre architecture s'appuie sur un récepteur MU-MIMO traditionnel et l'améliore avec des composants DL qui sont entraînés pour maximiser les performances de bout en bout du système. 
Cette approche ne nécessite aucune connaissance du canal pour l'entraînement, elle est interprétable et facilement adaptable à un nombre quelconque d'utilisateurs. 
Des évaluations en liaison montante et descendante ont été réalisées avec plusieurs intervalles de vitesses et deux configurations de pilotes différentes sur des modèles de canal conformes à la norme 3GPP. 
Les résultats révèlent que l'architecture proposée exploite efficacement la structure OFDM pour obtenir des gains tangibles à faible vitesse et des gains importants à vitesse élevée par rapport à un récepteur traditionnel.
Nous pensons que de telles architectures, permettant d'améliorer les performances tout en restant conformes aux standards et raisonnablement complexes, pourraient être déployées dans les BS de la prochaine génération de systèmes de communication sans fil. 
Cependant, il a été démontré que des gains supplémentaires peuvent encore être obtenus avec un récepteur entièrement basé sur un NN qui traitent conjointement tous les utilisateurs~\cite{korpi2020deeprx}, contrairement à notre approche dans laquelle ils sont traités indépendamment. 
Débloquer ces gains tout en préservant l'adaptabilité et l'interprétabilité des architectures conventionnelles reste un défi important qui n'a pas encore été résolu.

\newpage
\tocless\section{Réflexions Finales}

Au cours de ce manuscrit, nous avons étudié trois stratégies qui visent à apporter les avantages du DL à la couche physique des systèmes de communication.
La première stratégie, que nous avons appelée l'optimisation par blocs basée sur des NN, a été utilisée pour concevoir un NN qui pourrait remplacer les détecteurs MU-MIMO actuels. 
Mais ces détecteurs basés sur un NN supposent souvent un CSI parfait et sont toujours entraînés à optimiser leurs propres performances uniquement, ce qui ne garantit pas l'optimalité au niveau du système en entier. 
La deuxième stratégie, dans laquelle le système est optimisé de bout en bout en implémentant les émetteurs-récepteurs en tant que NNs, a été discutée dans la Section~B.3. 
Elle permet une optimisation plus poussée du traitement du signal en émission et en réception puisqu'elle ne doit pas se conformer aux protocoles existants. 
Bien que prometteurs, l'émetteur et le récepteur correspondants sont trop complexes pour une implémentation à court terme, ne répondent pas aux standards actuels et sont limités aux transmissions SISO. 
Enfin, une troisième stratégie hybride a été proposée dans la Section~B.4. 
Elle consiste à insérer des composants basés sur du DL dans une architecture traditionnelle qui est entraînée de bout en bout. 
Nous avons utilisé cette stratégie pour améliorer un récepteur MO-MIMO classique avec plusieurs CNNs. Les résultats indiquent que notre récepteur réalise des gains dans tous les scénarios, tant en liaison montante qu'en liaison descendante, et surtout à des vitesses élevées. 
Plus important encore, il préserve l'adaptabilité des architectures conventionnelles à un nombre variable d'utilisateurs et pourrait devenir conforme aux standards.
Globalement, cette stratégie semble être la plus adaptée à une utilisation dans un avenir proche.

Le déploiement du DL dans la couche physique se fera probablement en plusieurs phases. 
La première correspond à l'intégration de blocs basés sur du DL dans les récepteurs traditionnels de la BS, entraînés selon la stratégie basée sur les blocs ou la stratégie hybride. 
Comme le DL est avant tout un problème de données, les gains apportés par l'intégration du DL sont liés à notre capacité à constituer des ensembles de données suffisants. 
La collecte de données représentatives d'environnements divers et variés est donc essentielle pour garantir la fiabilité de ces composants. 
Ces défis sont exacerbés sur les appareils mobiles, dans lesquels l'intégration du DL pourrait correspondre à une deuxième phase de déploiement.
Les gains devraient s'en trouver améliorés puisque les modèles DL pourront prendre en compte les distorsions et les spécificités des canaux qui ne sont pas présentes dans les modèles traditionnels. 
Cette flexibilité est particulièrement intéressante pour les futurs réseaux 6G, qui devraient prendre en charge un large éventail de cas d'utilisation, comme les communications entre véhicules ou les sous-réseaux de petite échelle. 
La disponibilité croissante d'accélérateurs DL aux deux extrémités de la communication devrait permettre à terme l'émergence d'émetteurs-récepteurs entièrement basés sur des NNs, entraînés en utilisant la stratégie de bout en bout. 
Cependant, l'apprentissage conjoint d'émetteurs et de récepteurs conçus par différents fabricants représente un défi majeur pour l'industrie des télécommunications. 
Les futurs standards devront donc passer d'une réglementation du comportement des algorithmes de communication à une optimisation de bout en bout des systèmes basés sur des NNs.
Les prochaines phases du déploiement du DL dans les systèmes de communication pourraient finalement se concentrer sur l'apprentissage des couches physique et de contrôle d'accès au support (medium access control, MAC), donnant finalement naissance aux premières radios entièrement concues et controllées par intelligence artificielle.

\end{otherlanguage}

\backmatter

\begin{nopageskip}
\chapter*{\centering \fontsize{30}{30}\selectfont{Acronyms, Symbols, Figures and Tables}}
\addcontentsline{toc}{chapter}{Acronyms, Symbols, Figures and Tables}
\printglossary[type=\acronymtype,title=List of acronyms, nonumberlist]
\vfill
\pagebreak
\end{nopageskip}

\begin{nopageskip}
  \chapter*{List of Symbols}

\begin{longtable}[l]{  m{6em}  m{15cm} }
  
  $\Am^{(i)}$ & Matrix used in $i^{\text{th}}$ iteration of an iterative decoder(Chapter 3)\\ 
          


  $\Bm$ & Tensor of bits \\

  $B_S$ & Batch size \\
  
  $\Cm$ & Downlink normalization precoding downlink matrix (Chapter 5) \\
    
  $\mathcal{C}$ & Constellation \\ 
  
  $C$ & Achievable rate \\ 
    
  $\dv_m, d_{m,n\in \mathcal{D}}$ & (Vector of) FBSs carrying data signals (Chapter 4)\\
    
  $\Dm$ & Uplink post-equalization rescaling matrix (Chapter 5)\\ 
  
  $\mathcal{D}, D$ & Set and number of subcarriers carrying data (Chapter 4)\\ 
  
  $\Em$ & Channel estimation error spatial covariance (Chapter 5)\\ 
  
  $E_{A_m}, E_{I_m}$ & In-band and total energies of an OFDM symbol\\ 
  
  $\Fm\htp$ & Inverse Fourier Transform \\ 
  
  $\Gm$ & Equivalent downlink channel (Chapter 5) \\ 
  
  $\Hm$ & Tensor/matrix of channel coefficients \\
  
  $\Jm$ & Matrix for the in-band energy of an OFDM symbol (Chapter 4)\\
  
  $\Km$ & Matrix for the total energy of an OFDM symbol (Chapter 4) \\
  
  $K$ & Number of users in MU-MIMO systems \\

  $k$ & User index \\
  
  $\Lm$ & Number of antennas at the BS \\
  
  $l$ & BS antenna index \\
  
  $\mathcal{M}, M$ & Set and number of OFDM symbols\\
  
  $m$ & OFDM symbol index \\

  $\Nm$ & Noise tensor/matrix\\

  $\mathcal{N}, N$ & Subset and number of subcarriers \\
  
  $n$ & Subcarrier index \\
              
  $|\Pc_M|$, $|\Pc_N|$ & Number of pilots in the time and frequency domain\\
  
  $\widehat{P}(\cdot), P(\cdot)$ & (Predicted) probability\\
  
  $\mathcal{P}^{(k)}$ & Pilot pattern for the $k^{\text{th}}$ user (Chapter 5)\\
  
  $\pv$ & Pilots \\ 
  
  $Q$ & Number of bits per channel uses\\

  $q$ & Bit index \\ 
   
  $\qv_{m,n}$ & Downlink noise vector (Chapter 5)\\ 

  $\Qm$, $\Qm_{\Am}$, $\Qm_{\Bm}$ & QR-decomposition \\ 
  
  $\rv_m, r_{m,n\in \mathcal{R}}$ & (Vector of) FBS carrying peak-reduction signal \\
  
  $\Rm$, $\Rm_{\Am}$ & QR-decomposition \\ 
  
  $\mathcal{R}, R$ & Set of number of subcarriers carrying peak-reduction signals (Chapter 4)\\ 
  
  $\Sm$ & Tensor of unprecoded downlink FBSs (Chapter 5)\\ 
     
  $S_m(f)$ & Baseband spectrum\\
   
  $s_m(t)$ & Time-domain signal\\
   
  $\Tm$ & Tensor of precoded downlink symbols (Chapter 5)\\ 
  
  $T, T^{\text{CP}}, T^{\text{tot}}$ & Duration of an OFDM symbol, of its CP, and total duration\\ 
      
  $\Um$ & Downlink received signal (Chapter 5)\\
  
  $\Vm^{(k)}$  & Equivalent downlink channel estimation error variances (Chapter 5) \\ 
    
  $\Wm$  & LMMSE matrix  \\ 
    
  $\hat{x}$ & Estimated symbols after equalization \\ 
  
  $\doublehat{x}$  & Hard decision on the estimated symbols ($\doublehat{x}\in\Cc$)  \\
  
  $\zv$  & Discrete time-domain signal (Chapter 4) \\ 
  
  $\thetav$, $\psiv$ & Set of trainable parameters\\ 
    
  $\nu(t)$ & Ration between the instantaneous and average power of a signal (Chapter 4) \\ 
  
  $\alpha, \beta, \gamma$ & Parameters of the exponential decay model (Chapter 5)\\  
  
  $\sigma^2$ & General notation for a noise variance \\ 

  $\tau^2$ & Variance of the post-equalization downlink noise (Chapter 5)\\ 
      
  $\epsilon$ &  PAPR threshold \\

  $\eta$ &  Learning rate (Chapter 2) or code rate (Chapter 3, 4, and 5)\\

  $\rho^2$ & Variance of the post-equalization uplink noise (Chapter 5)\\ 

  $\xi_{m,n,k}$ & Post-equalization downlink noise (Chapter 5)\\

  $\zeta_{m,n,k}$ & Post-equalization downlink noise (Chapter 5)\\
  
  $\mathbf{\Theta}$ & Matrix to be optimized in the iterative detection algorithm (Chapter 3)\\ 
  
  $\Sigmam$ & General notation for a channel covariance matrix \\ 
  
  $\Omegam$ & Covariance matrix for the main downlink channel (Chapter 5)\\ 
  
  $\Psim$ & Covariance matrix for the interfering downlink channels (Chapter 5) \\

\end{longtable}

  \vfill
  \pagebreak
\end{nopageskip}

\begin{nopageskip}
  \listoffigures
  \vfill
  \pagebreak
\end{nopageskip}
\begin{nopageskip}
  \listoftables
  \vfill
  \pagebreak
\end{nopageskip}

\addcontentsline{toc}{chapter}{Bibliography}
\chapter*{\centering \fontsize{30}{30}\selectfont{Bibliography}}
\printbibliography[heading=none]

\end{refsection}


\end{document}